\documentclass[table]{ai2style/ai2}

\usepackage[utf8]{inputenc}
\usepackage[T1]{fontenc}
\usepackage{amsmath,amssymb,amsfonts}
\usepackage{needspace}
\usepackage{booktabs,tabularx,longtable,array,makecell}
\usepackage{graphicx}
\usepackage{enumitem}
\usepackage{listings}
\usepackage{xcolor}
\usepackage{caption}
\usepackage{subcaption}
\usepackage[figuresleft]{rotating}
\usepackage{placeins}
\usepackage{float}
\usepackage{todonotes}
\usepackage{url}
\usepackage{fontawesome5}%
\usepackage{hyperref}

\microtypesetup{expansion=false}
\makeatletter
\renewcommand{\BreakableUnderscore}{\leavevmode%
  \prw@zbreak\textunderscore\discretionary{}{}{}\prw@zbreak}
\makeatother
\let\originalleft\left
\let\originalright\right
\renewcommand{\left}{\mathopen{}\mathclose\bgroup\originalleft}
\renewcommand{\right}{\aftergroup\egroup\originalright}

\newcommand{\defeq}{\mathrel{:\mkern-0.25mu=}}

\def\eqref#1{eq.~\ref{#1}}

\def\Partref#1{Part~\ref{#1}}

\DeclareMathAlphabet{\mathsfit}{\encodingdefault}{\sfdefault}{m}{sl}
\SetMathAlphabet{\mathsfit}{bold}{\encodingdefault}{\sfdefault}{bx}{n}

\tcbset{olmo-callout/.style={
  enhanced,
  breakable,
  colback=ai2background!72!white,
  colframe=ai2accent!85!black,
  coltext=ai2foregroundDark,
  boxrule=0.45pt,
  arc=3pt,
  left=11pt,
  right=11pt,
  top=9pt,
  bottom=9pt,
  before skip=11pt,
  after skip=11pt,
  borderline west={2.4pt}{0pt}{ai2pink},
}}

\newtcolorbox{keyidea}[1][Key idea]{
  olmo-callout,
  breakable=false,
  title={#1},
  detach title,
  fonttitle=\sffamily\bfseries,
  coltitle=ai2accent!70!black,
  before upper={\noindent\tcbtitle\par\smallskip},
}

\newtcolorbox{definitionbox}[1][Definitions]{
  olmo-callout,
  breakable=false,
  title={#1},
  detach title,
  fonttitle=\sffamily\bfseries,
  coltitle=ai2accent!70!black,
  before upper={\noindent\tcbtitle\par\smallskip},
}

\newtcolorbox{sectiontakeaway}[1][Section takeaway]{
  enhanced,
  breakable=false,
  colback=ai2accent!6!white,
  colframe=ai2accent!80!black,
  coltext=ai2foregroundDark,
  boxrule=0.45pt,
  arc=2pt,
  left=10pt,
  right=10pt,
  top=7pt,
  bottom=7pt,
  before skip=14pt,
  after skip=4pt,
  borderline west={2.4pt}{0pt}{ai2accent},
  fontupper=\small\sffamily,
  before upper={\noindent{\bfseries\color{ai2accent!70!black}#1.}\enspace\ignorespaces},
}

\newcommand{\conceptlead}[1]{%
  \noindent{\sffamily\bfseries\color{ai2accent!80!black}%
    \spaceskip=\fontdimen2\font plus\fontdimen3\font\relax #1}\enspace
}

\newcommand{\paperTitle}{Supercharging Olmo-core for Efficient and Scalable MoE Training}
\newcommand{\githubicon}{%
  \raisebox{-1.5pt}{\includegraphics[height=1.05em]{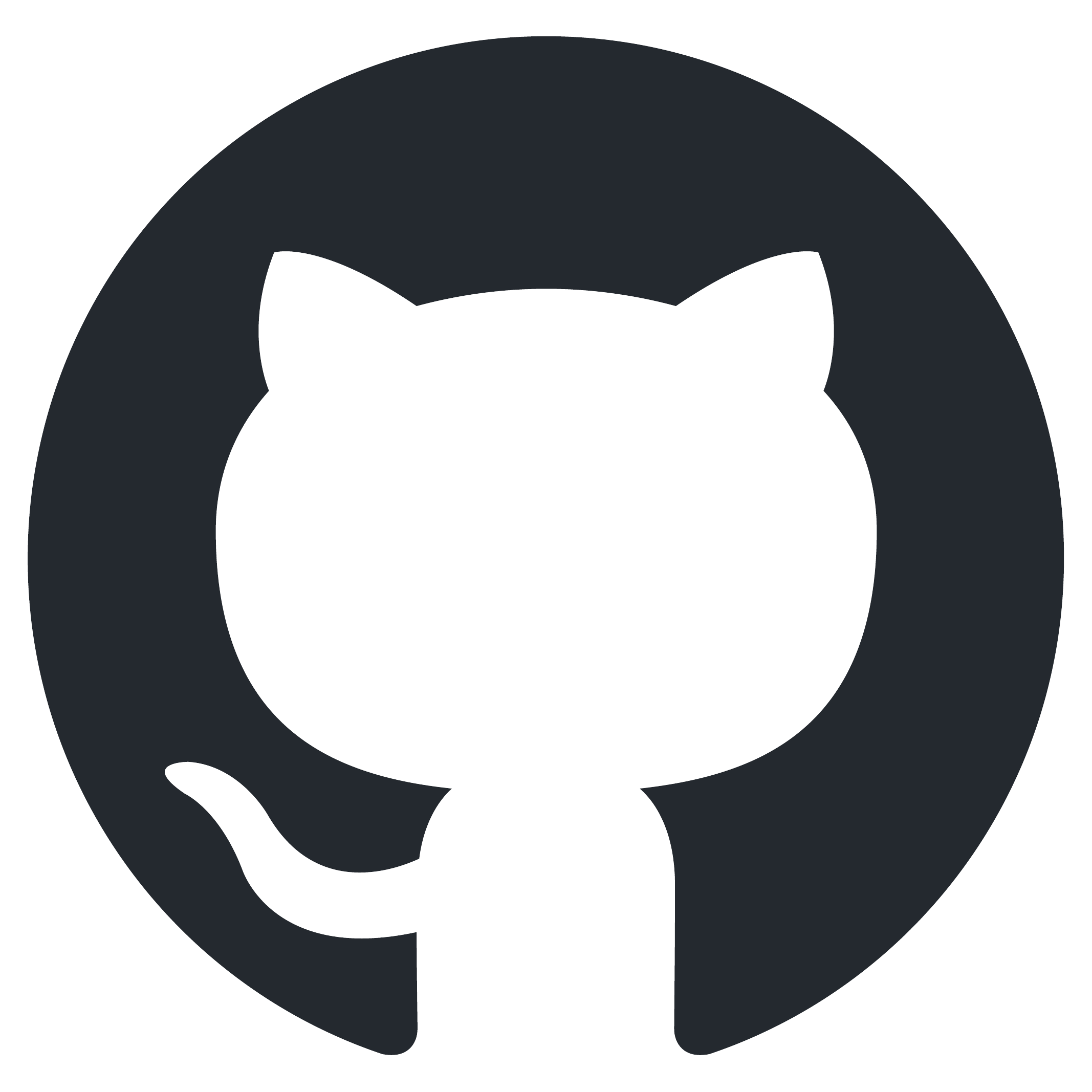}}}
\hypersetup{pdftitle={Supercharging Olmo-core for Efficient and Scalable MoE Training},
  pdfauthor={Tianhua Tao, Akshita Bhagia, Dirk Groeneveld, Pete Walsh, Tyler Romero,
    Yashas Samaga, Jacob Morrison, Iz Beltagy, Taira Anderson, Noah A. Smith,
    Hannaneh Hajishirzi}}

\title{\paperTitle}
\authorOne[$^{1,2,\dagger,*}$]{Tianhua Tao}
\authorOne[$^{1}$]{Akshita Bhagia}
\authorOne[$^{1,\dagger}$]{Dirk Groeneveld}
\authorOne[$^{1,\dagger}$]{Pete Walsh}
\authorOne[$^{1,\dagger}$]{Tyler Romero}
\authorTwo[$^{1,2}$]{Yashas Samaga}
\authorTwo[$^{1,2}$]{Jacob Morrison}
\authorTwo[$^{1}$]{Iz Beltagy}
\authorTwo[$^{1}$]{Taira Anderson}
\authorThree[$^{1,2}$]{Noah A. Smith}
\authorThree[$^{1,2,\dagger}$]{Hannaneh Hajishirzi}
\affiliation[$^1$]{Allen Institute for AI}
\affiliation[$^2$]{University of Washington}
\contribution[*]{Lead author and primary developer}
\contribution[\dagger]{Work done while at the Allen Institute for AI}
\contribution[]{see Appendix~\ref{app:author-contributions} for author contributions.}
\correspondence{\email{tianhuat@cs.washington.edu}; \email{olmo@allenai.org}}
\metadata[\githubicon\ Code:]{%
  \href{https://github.com/allenai/olmo-core}{\texttt{Olmo-core}}\quad
  {\sffamily\bfseries\faGlobe\ Web version:}
  \href{https://www.taotianhua.com/olmo-core-moe/}{\texttt{taotianhua.com/olmo-core-moe}}\quad
  {\sffamily\bfseries\faPlayCircle\ Video:}
  \href{https://www.taotianhua.com/olmo-core-moe/about/speed-read/}{\texttt{4-minute overview}}}
\date{October 2026}
\abstract{
\setlength{\parskip}{4pt}%
Mixture-of-Experts (MoE) models offer more total parameters without
proportionally increasing per-token compute: each token activates only a few
experts. However, training cost depends not only on active compute but also
on many other factors, including bytes moved, memory held, and kernels
launched; many of those costs follow total parameter size rather than
active compute. On infrastructure built for dense models, this mismatch is
expensive: in our early baselines, throughput fell as stored expert parameter size
grew, even though the useful work per token stayed nearly fixed.

To close this gap, we present a redesigned training stack in Olmo-core, which
is optimized for efficient and scalable MoE training. We switch from the Fully
Sharded Data Parallel (FSDP) used by previous Olmo-core versions to a
parallelization architecture built on Distributed Data Parallel (DDP). On top
of it, we add Expert Parallelism (EP), Pipeline Parallelism (PP), a
distributed optimizer, and activation recompute to support large-scale models.

We bring together a suite of optimizations to make MoE training more
efficient. For MoE layers, we propose NVSHMEM-based rowwise EP, which
writes each token directly into its expert's buffer, avoiding permutation
overhead in dispatch and combine operations. We make EP fully
synchronization-free by keeping routing metadata on the GPU and using
device-scheduled grouped GEMMs. Closely integrated MXFP8 raises throughput by
cutting compute, activation-memory, and transport costs, and topology-agnostic
checkpoints let the parallel layout change between runs.

On NVL8 B300 nodes, the stack supports training configurations from
\textbf{12.9 billion} to \textbf{1.2 trillion} total parameters on up to 512
GPUs, reaching 858 useful-model TFLOP/s/GPU at 1.2 trillion with MXFP8 and
per-layer recompute; an experimental capacity test with the optional DeepEP
v2 backend reaches \textbf{2.38 trillion}.

We also explain the reasoning behind each design choice and share findings
that can inform other training systems. For example, we identify a failure
pattern we call Token Gerrymandering, in which the MoE router can learn to hack
the load-balancing loss, and find that overlapping communication with
computation can slow the overlapped kernels by more than it hides. These are
two of many findings in the report. Together, the stack and these findings offer practical insights into building efficient, scalable MoE training systems.
}

\begin{document}
\maketitle

\clearpage
\section*{Statement of AI Assistance}
\label{sec-ai-assistance-statement}

We used AI tools to assist with developing the codebase and kernels,
drafting and revising this report, managing experiments, and creating
visualizations. The authors retain full responsibility for the report's
content. See Appendix~\ref{app:ai-assistance} for the role of these tools
and our verification approach.

\clearpage
\section*{Speed-Read}
\addcontentsline{toc}{section}{Speed-Read}
\label{sec-speed-read}

This report describes the MoE training stack in Olmo-core, the measurements
that guided its design, and the approaches we tried but did not adopt.  This
guide summarizes the main results; the table at the end points to sections
for readers interested in a particular problem.

\conceptlead{Sparse compute leaves substantial costs that scale with total parameter size.}
An MoE model stores many expert networks but sends each token through only a
few of them, so the compute per token stays small while the stored model
grows.  Weight storage, gradient synchronization, and optimizer updates still
depend on total parameter count.  The fully sharded data parallel
(FSDP) with ``full-reshard'' configuration used for dense Olmo also gathers and frees weights around every
microbatch.  In our early baselines, adding experts at fixed active size
therefore reduced throughput even though the useful work per token barely changed
(Figure~\ref{fig-intro-fsdp-tps-capacity}).  Section~\ref{sec-intro-dense-assumptions}
names this \textbf{capacity tax} and separates it from the \textbf{runtime
tax} of sparse routing and execution; Section~\ref{sec-olmo-ddp-training-stack}
introduces the changes we made to reduce both.

\conceptlead{DDP keeps model partitions resident; EP moves selected token rows.}
The stack replaces the full-reshard FSDP configuration used for dense Olmo
with a custom distributed data
parallel (DDP) path: each rank keeps its partition of the model in place for
the whole accumulation window and synchronizes gradients once per optimizer
step, with a distributed optimizer sharding the FP32 main weights and optimizer states
(Section~\ref{sec-ddp-vs-fsdp}).  Expert parallelism (EP) shards the routed
experts across ranks and dispatches only the selected token rows to the rank
that stores their expert.  Our rowwise intra-node transport writes each token
row directly into its destination buffer using GPU-initiated one-sided
communication through NVSHMEM, avoiding the host-visible split lists that the
block all-to-all path needs.  A block-oriented DeepEP v2 backend supports
expert groups that span nodes (Section~\ref{sec-expert-parallelism}).  Pipeline
parallelism (PP) divides layers when one rank cannot hold them all
(Section~\ref{sec-pipeline-parallelism}).  Section~\ref{sec-parallelism-overview}
shows how DDP, EP, and PP combine, and Section~\ref{sec-hardware-topology}
maps their rank groups onto NVLink and InfiniBand.

\conceptlead{GPU-resident metadata removes recurring host waits.}
The original MoE path copied routing counts to the CPU for both all-to-all
split sizes and expert matrix-multiplication group sizes.  Keeping this
metadata on the device removes those synchronization points from the
steady-state loop
(Section~\ref{sec-sync-free-training}).  Device-scheduled
\textbf{grouped general matrix multiplication (grouped GEMM)} executes the
uneven per-expert shapes produced by routing without padding every expert
batch to a common capacity.  Section~\ref{sec-grouped-gemm} measures where it
beats the padded alternative and where it does not.

\conceptlead{Routers can hack the balancing loss: lower loss, higher imbalance.}
Section~\ref{sec-routing-load-balancing} treats routing as a systems
problem: what capacity admits, what the load-balancing loss actually
optimizes, and how it can reward a failure we call \textbf{Token
Gerrymandering}, in which the loss improves while imbalance grows
(an analytic example and one checkpoint-branched run,
Figure~\ref{fig-token-gerrymandering-intro}).

\conceptlead{MXFP8 reduces compute and data-movement costs.}
\textbf{MXFP8}, an
eight-bit format with shared block scales, can cut the time of large GEMMs,
the size of selected saved activations, and dispatch payloads, while one FP32
main weight remains the source of truth
(Section~\ref{sec-mxfp8-training}).

\conceptlead{Checkpoints preserve the model independently of its parallel layout.}
Checkpoints describe every saved chunk by its
place in the global FP32 tensor, so a run can resume on a
different topology (Section~\ref{sec-checkpointing}).

\conceptlead{Recompute lets larger recipes fit in memory.}
Activation recompute reduces the activations retained for the backward pass,
lowering peak high-bandwidth memory (HBM) usage.  In our controlled comparison,
recomputing every block removes nearly two thirds of activation memory and reduces peak memory by
about a quarter, at the cost of about a fifth of throughput.  We use it in the larger recipes
that would otherwise exceed device memory (Section~\ref{sec-recompute}).

\conceptlead{The training configurations span 16 to 512 GPUs.}
Section~\ref{sec-production-throughput} lists measured operating points,
from a 12.9-billion-parameter model on 16 GPUs to a
1.2-trillion-parameter model on 512 GPUs, with a short
2.38-trillion-parameter extension.  These configurations differ in batch,
parallelism, precision, and recompute, so their rates are not a controlled
scaling curve.  Table~\ref{tab-production-throughput-headline} states the
batch, topology, precision, and recompute policy needed
to interpret each number.

\conceptlead{Some optimizations traded one cost for another.}
Section~\ref{sec-optimization-attempts} describes approaches we did not adopt:
CPU activation offload, two overlap schemes, a persistent MoE megakernel, a
natural-layout MXFP8 weight-gradient prototype, and CUDA Graph replay.
The section quantifies their trade-offs between concurrency, kernel
efficiency, and memory use; notably, overlapping communication with compute
can slow concurrent kernels by more than it hides.

\conceptlead{The experiments also lead to training-recipe findings.}  Critical
batch size says when a larger batch still helps but not which learning rate to
pair with it, and in our internal tuning runs, square-root scaling needed a
small empirical correction
(Section~\ref{sec-lr-batch-scaling}); in our sweep, scaling expert learning
rates by sparsity did not help (Section~\ref{sec-lr-sparsity-scaling}).  We train from
scratch rather than upcycle because, in the published OLMoE study, a dense
model's head start did not persist over a longer run
(Section~\ref{sec-upcycling}).  Several ungated shared experts are
mathematically one wider shared expert
(Section~\ref{sec-ungated-shared-experts}),
and the rule that merges routed and shared outputs changes the
initialization scale
(Section~\ref{sec-merge-expert-activations}).

\conceptlead{Beyond pretraining, and comparisons with other systems.}
Section~\ref{sec-beyond-pretraining} discusses how the same mechanisms could
support supervised fine-tuning, preference training, and reinforcement
learning; it is a design discussion rather than a post-training benchmark.
Section~\ref{sec-related-work} compares the design against Megatron-Core,
DeepSpeed-MoE, and DeepEP in terms of parameter movement, communication, and
execution costs.  Section~\ref{sec-conclusion} summarizes the implications
for memory use, throughput, and measurement.

\begin{table}[!htb]
\centering
\small
\renewcommand{\arraystretch}{1.22}
\begin{tabularx}{\linewidth}{>{\raggedright\arraybackslash}p{0.34\linewidth} X}
\toprule
\textbf{If you want to} & \textbf{Read} \\
\midrule
See the headline throughput and exactly what produced it
  & The panel at the end of Section~\ref{sec-olmo-ddp-training-stack}
    (p.~\pageref{panel-selected-throughput}), then
    Section~\ref{sec-production-throughput} (p.~\pageref{sec-production-throughput})
    and Appendix~\ref{app-experiment-configurations} \\
\arrayrulecolor{black!22}\hline\arrayrulecolor{black}
Understand why FSDP was replaced for MoE
  & Sections~\ref{sec-intro-dense-assumptions},
    \ref{sec-parallelism-overview}, and \ref{sec-ddp-vs-fsdp}
    (pp.~\pageref{sec-intro-dense-assumptions}--\pageref{sec-ddp-vs-fsdp-end}) \\
\arrayrulecolor{black!22}\hline\arrayrulecolor{black}
Build or debug an MoE trainer
  & Sections~\ref{sec-expert-parallelism}, \ref{sec-sync-free-training},
    \ref{sec-grouped-gemm}, and \ref{sec-checkpointing}
    (from p.~\pageref{sec-expert-parallelism}), with
    Appendix~\ref{app-kernel-interface-reference} as the interface reference \\
\arrayrulecolor{black!22}\hline\arrayrulecolor{black}
Fit a larger model into memory
  & Sections~\ref{sec-pipeline-parallelism}, \ref{sec-recompute}, and
    \ref{sec-production-throughput}, plus the offload budget in
    Section~\ref{sec-optimization-attempts} (p.~\pageref{sec-optimization-attempts}) \\
\arrayrulecolor{black!22}\hline\arrayrulecolor{black}
Adopt low-precision training
  & Section~\ref{sec-mxfp8-training} (p.~\pageref{sec-mxfp8-training}), then
    Sections~\ref{sec-gemm-operand-values} and \ref{sec-optimization-attempts} \\
\arrayrulecolor{black!22}\hline\arrayrulecolor{black}
Diagnose routing or load-balancing behavior
  & Section~\ref{sec-routing-load-balancing} (p.~\pageref{sec-routing-load-balancing}),
    with the rank-capacity policy in Section~\ref{sec-routing-rank-capacity} \\
\arrayrulecolor{black!22}\hline\arrayrulecolor{black}
Choose the model recipe: learning rate, upcycling, shared experts, merge rule
  & Sections~\ref{sec-lr-batch-scaling}, \ref{sec-lr-sparsity-scaling},
    \ref{sec-upcycling}, \ref{sec-ungated-shared-experts}, and
    \ref{sec-merge-expert-activations} (from p.~\pageref{sec-lr-batch-scaling}) \\
\arrayrulecolor{black!22}\hline\arrayrulecolor{black}
Reuse the stack for post-training (SFT, preference training, RL)
  & Section~\ref{sec-beyond-pretraining} (p.~\pageref{sec-beyond-pretraining}) \\
\arrayrulecolor{black!22}\hline\arrayrulecolor{black}
Compare with Megatron-Core, DeepSpeed-MoE, or DeepEP
  & Section~\ref{sec-related-work} (p.~\pageref{sec-related-work}) \\
\arrayrulecolor{black!22}\hline\arrayrulecolor{black}
Reproduce a measurement
  & Appendix~\ref{app-experiment-configurations}
    (p.~\pageref{app-experiment-configurations}) for configurations and metrics;
    Appendix~\ref{app-kernel-interface-reference} for profiling method \\
\bottomrule
\end{tabularx}
\caption*{\textbf{Reading paths through the report.}
Part~\ref{part-foundations} introduces the problem and the training stack,
Part~\ref{part-parallelization} covers placement and communication,
Part~\ref{part-efficient-training} the execution loop and its measurements,
and Part~\ref{part-outlook} the recipe findings and the outlook.}
\end{table}

\clearpage

\setcounter{tocdepth}{2}
\tableofcontents
\clearpage

\part{Introduction}
\label{part-foundations}
\section{Why the Dense Training Assumptions Break}

\subsection{Mixture of Experts (MoE): Sparse Activation}

\conceptlead{MoE separates total and active parameters.} In a
dense Transformer block %
\citep{vaswani2017attention}, every token\textemdash{}one position in the input sequence\textemdash{}is
processed by the same feed-forward network (FFN). Making that FFN larger gives
the model more capacity and increases computation for every token. MoE adds
parameters while using only a subset of them for each token.

\begin{figure}[!htb]
\centering
\includegraphics[width=\linewidth]{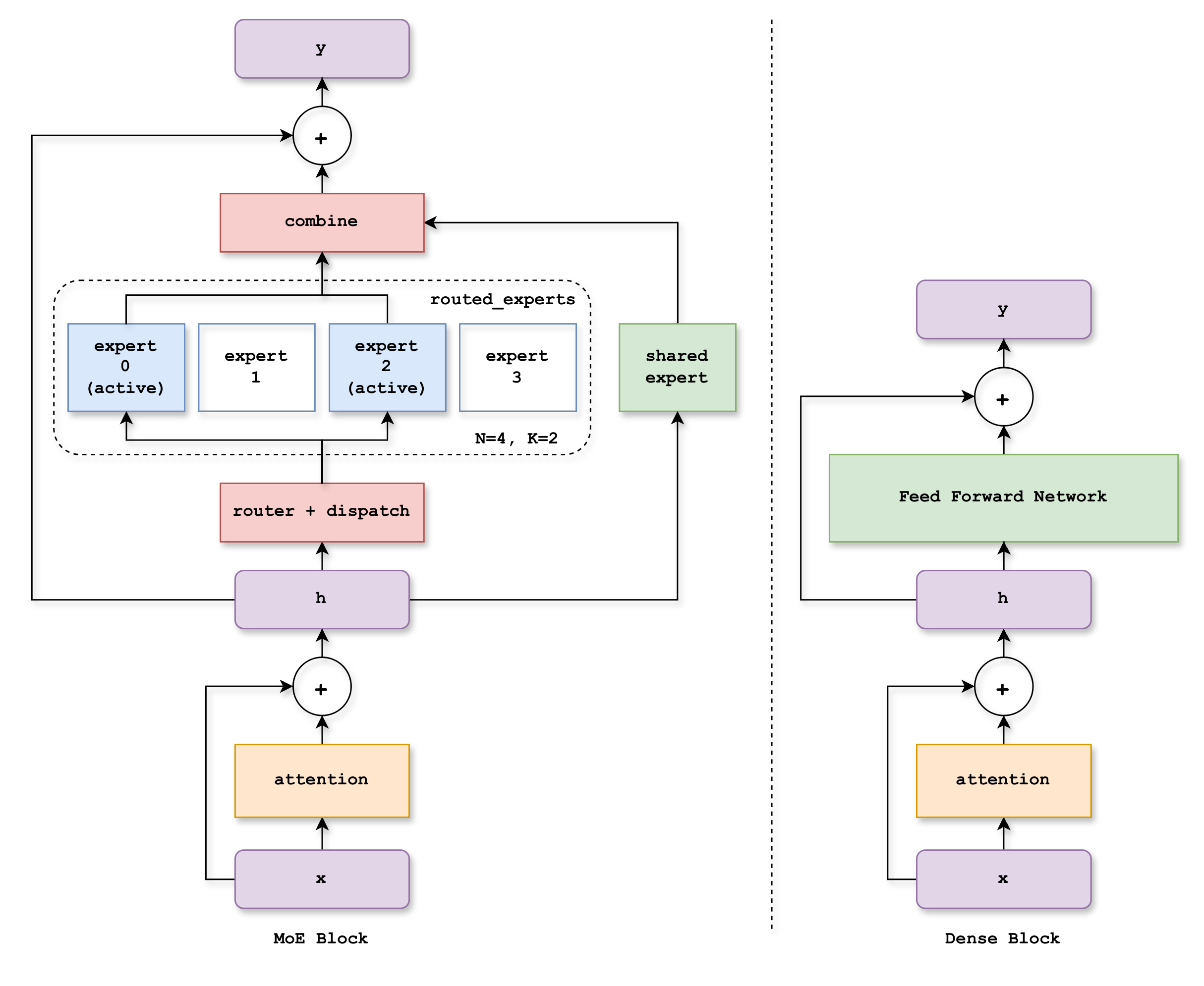}
\caption{\textbf{MoE and Dense Transformer blocks.} Both blocks retain attention and
residual connections. The dense block (right) applies the same feed-forward
network to every token. The MoE block (left) illustrates one token routed to
$K=2$ of $N=4$ routed experts\textemdash{}experts 0 and 2 in this example\textemdash{}and combines
their outputs with an optional always-active shared expert before the residual
addition. Here, $K=2$ is a per-token choice. Other tokens can select different
experts, so a training batch can collectively use all $N$ experts.}
\label{fig-intro-moe-architecture}
\end{figure}

\begin{samepage}
MoE changes this part of the block. Instead of one FFN, an MoE layer contains a
pool of separately parameterized FFNs called \textbf{experts}.
A small learned \textbf{router}, or gate, decides which experts should process
each token.
\end{samepage}

\conceptlead{Each token selects a few experts.} The router scores the $N$ available experts
and chooses the top $K$. The runtime sends\textemdash{}or \textbf{dispatches}\textemdash{}the token's
vector to those selected experts. Each selected expert applies its FFN to the
vector. The router supplies a combination weight for each selected expert, and
the \textbf{combine} operation multiplies the expert outputs by those weights,
sums them, and restores the result to the token's original sequence position.
The router and experts learn
together with the rest of the model. This sparse-gated pattern was established
by early large-scale MoE work and later simplified by the Switch Transformer
\citep{shazeer2017outrageously,fedus2022switch}.

For token representation $x_t$, let
$S_t$ be the selected expert set, let $p_{t,e}$ be expert $e$'s combination
weight, and let $f_e$ be that expert's FFN. Then

\begin{equation}
y_t^{\mathrm{routed}}
= \sum_{e \in S_t} p_{t,e} f_e(x_t),
\qquad |S_t|=K \ll N.
\label{eq-routed-moe}
\end{equation}

Some MoE blocks also include
a \textbf{shared expert}, an FFN that every token uses alongside its selected
routed experts. Attention, embeddings, normalization, the router, and any
shared expert also remain active when most routed experts do not. And because
the router's choices depend on the input, experts do not necessarily receive
equal numbers of tokens. Training systems use load-balancing objectives and
capacity limits to keep one expert from becoming an unbounded hotspot
\citep{shazeer2017outrageously,fedus2022switch,dai2024deepseekmoe}.

Consider a layer with $N=128$ routed
experts and $K=4$. The model has 128 routed expert FFNs, yet a
single token uses only four of them. We therefore distinguish the model's
total parameter count from the number of parameters active for one token.

\begin{keyidea}[Sparse Capacity Without Dense Computation]
Here, $N$ is the total number of routed experts and $K$ is the number selected
for one token. Let $P_{\mathrm{always}}$ denote the parameters used by every
token\textemdash{}attention, embeddings, routers, norms, shared experts, and the
language-model output head\textemdash{}and let
$P_{\mathrm{expert}}$ denote the parameters in one routed expert. In a
simplified per-token accounting,

\begin{equation}
\begin{aligned}
P_{\mathrm{total}} &= P_{\mathrm{always}} + N P_{\mathrm{expert}}, \\
P_{\mathrm{active}} &= P_{\mathrm{always}} + K P_{\mathrm{expert}}.
\end{aligned}
\label{eq-intro-active-total-parameters}
\end{equation}

We use two activation ratios:

\begin{equation}
\begin{aligned}
\alpha_E &= \frac{K}{N}
&& \text{(expert activation ratio)}, \\
\alpha_P &= \frac{P_{\mathrm{active}}}{P_{\mathrm{total}}}
&& \text{(parameter activation ratio)}.
\end{aligned}
\label{eq-intro-activation-ratios}
\end{equation}

In the $N=128$, $K=4$ example, a token uses $3.125\%$ of the routed experts\textemdash{}not
$3.125\%$ of the whole model, because the always-active path still participates.
Holding expert shape fixed, routed parameter size grows with $N$, while routed expert
computation per token grows with $K$.
\end{keyidea}

This per-token accounting only approximates total model FLOPs, because
attention, routing, shared branches, and auxiliary work also contribute. Sparse activation still allows
substantially more parameters without a proportional increase in
per-token computation. Different token paths may also encourage experts
to learn different functions, although useful specialization is neither
guaranteed nor assigned by hand.

\conceptlead{Models are getting larger\textemdash{}and sparser.} A visible design trend
among recent high-capacity MoEs is to grow total expert parameter size faster than
the active parameter count. The cleanest evidence comes from comparisons
within one model family: from DeepSeek-V3 to DeepSeek-V4-Pro, and from
Qwen3-235B-A22B to Qwen3.5-397B-A17B, total parameters grow substantially while
the active count grows far less or even falls, so the advertised parameter
activation ratio drops
\citep{deepseekai2024deepseekv3,deepseekai2026deepseekv4,yang2025qwen3,qwenteam2026qwen35modelcard}.

Table~\ref{tab-intro-modern-moe-sparsity} places those examples among their
contemporaries. It is deliberately a design survey rather than a leaderboard: model
objectives, layer types, and parameter-counting conventions differ, and neither
activation ratio alone measures end-to-end efficiency.

\begin{table}[!htb]
\centering
\scriptsize
\setlength{\tabcolsep}{3.2pt}
\renewcommand{\arraystretch}{1.12}
\begin{tabularx}{\linewidth}{>{\raggedright\arraybackslash}X c c c c c}
\toprule
\textbf{Model and primary source} & \textbf{Year} &
\makecell{\textbf{Routed experts}\\$N/K$} & $\boldsymbol{\alpha_E}$ &
\makecell{\textbf{Parameters}\\total/active} & $\boldsymbol{\alpha_P}$ \\
\midrule
Mixtral 8x7B %
\citep{jiang2024mixtral,mistralai2023mixtralrelease} & 2023 & 8/2 & $25.0\%$ & 47B/13B & $27.7\%$ \\
DBRX %
\citep{databricks2024dbrx} & 2024 & 16/4 & $25.0\%$ & 132B/36B & $27.3\%$ \\
Snowflake Arctic %
\citep{snowflake2024arctic} & 2024 & 128/2 & $1.56\%$ & 480B/17B & $3.54\%$ \\
OLMoE-1B-7B %
\citep{muennighoff2024olmoe} & 2024 & 64/8 & $12.5\%$ & 6.9B/1.3B & $18.8\%$ \\
DeepSeek-V3 %
\citep{deepseekai2024deepseekv3} & 2024 & 256/8 $+1\mathrm{S}$ & $3.13\%$ & 671B/37B & $5.51\%$ \\
Qwen3-235B-A22B %
\citep{yang2025qwen3} & 2025 & 128/8 & $6.25\%$ & 235B/22B & $9.36\%$ \\
Llama 4 Maverick %
\citep{meta2025llama4} & 2025 & 128/1 $+1\mathrm{S}$ & $0.78\%$ & 400B/17B & $4.25\%$ \\
gpt-oss-20b %
\citep{openai2025gptoss} & 2025 & 32/4 & $12.5\%$ & 20.91B/3.61B & $17.3\%$ \\
gpt-oss-120b %
\citep{openai2025gptoss} & 2025 & 128/4 & $3.13\%$ & 116.8B/5.13B & $4.39\%$ \\
DeepSeek-V4-Pro %
\citep{deepseekai2026deepseekv4} & 2026 & 384/6 $+1\mathrm{S}$ & $1.56\%$ & 1.60T/49B & $3.06\%$ \\
Gemma 4 26B A4B %
\citep{gemmateam2026gemma4} & 2026 & 128/8 $+1\mathrm{S}$ & $6.25\%$ & 25.2B/3.8B & $15.1\%$ \\
GLM-5 %
\citep{glm5team2026glm5} & 2026 & 256/8 $+1\mathrm{S}$ & $3.13\%$ & 744B/40B & $5.38\%$ \\
Kimi K2.5 %
\citep{moonshotai2026kimik25modelcard} & 2026 & 384/8 $+1\mathrm{S}$ & $2.08\%$ & 1.00T/32B & $3.20\%$ \\
MiMo-V2.5-Pro %
\citep{mimo2026v25pro} & 2026 & 384/8 & $2.08\%$ & 1.02T/42B & $4.12\%$ \\
MiniMax-M3 %
\citep{minimax2026m3} & 2026 & 128/4 $+1\mathrm{S}$ & $3.13\%$ & $\sim$428B/$\sim$23B & ${\sim}5.37\%$ \\
Nemotron 3 Ultra %
\citep{nvidia2026nemotron3ultra} & 2026 & 512/22 $+1\mathrm{S}$ & $4.30\%$ & 550B/55B & $10.0\%$ \\
Qwen3.5-397B-A17B %
\citep{qwenteam2026qwen35modelcard} & 2026 & 512/10 $+1\mathrm{S}$ & $1.95\%$ & 397B/17B & $4.28\%$ \\
Qwen3.5-35B-A3B %
\citep{qwenteam2026qwen35a3bmodelcard} & 2026 & 256/8 $+1\mathrm{S}$ & $3.13\%$ & 35B/3B & $8.57\%$ \\
GLM-5.2 %
\citep{zai2026glm52modelcard} & 2026 & 256/8 $+1\mathrm{S}$ & $3.13\%$ & 744B/40B & $5.38\%$ \\
Kimi K3 %
\citep{kimiteam2026kimik3} & 2026 & 896/16 $+2\mathrm{S}$ & $1.79\%$ & 2.80T/104B & $3.71\%$ \\
\bottomrule
\end{tabularx}
\caption{\textbf{Selected publicly documented MoE models, ordered by initial
release year.} The $N/K$ column gives total and active \emph{routed} experts;
$+n\mathrm{S}$ denotes $n$ additional always-active shared experts.
$\alpha_E=K/N$, while $\alpha_P$ uses the creators' published active and total
parameter counts. B and T denote $10^9$ and $10^{12}$ parameters. Parameter
counts are often rounded, and conventions for embeddings, output heads,
multimodal towers, MTP heads, shared experts, and dense layers are not fully
standardized; $\alpha_P$ is therefore approximate even when no approximation
symbol is shown.}
\label{tab-intro-modern-moe-sparsity}
\end{table}

For the
routed language models they study, %
\citet{clark2022unified} show that total
parameter count
and per-token computation form distinct scaling axes. Fitted scaling laws for fine-grained MoEs further
predict that compute-optimal MoE configurations reach lower validation loss
than dense Transformers at the same training-FLOP budget, with the predicted
advantage widening at larger budgets %
\citep{ludziejewski2024scaling}. Yet more
sparsity is not automatically better: the best sparsity level depends on
constraints such as total parameter count and training compute
\citep{abnar2025parameters}. These results indicate that well-configured MoEs can improve the
compute\textendash{}quality frontier, although a given MoE need not beat a
dense model at matched FLOPs.

\subsection{Lower FLOPs Do Not Guarantee Lower Wall Time}
\label{subsec-intro-walltime}

Useful model FLOPs account for arithmetic but leave out much of the work
required to train an MoE. The training system must also move
weights and token activations, synchronize ranks, and absorb load imbalance.
Those costs determine how many GPU-hours it actually takes to reach the desired
quality.

Let $F$ be the useful model FLOPs needed to reach a declared training or quality
target, $R$ the realized throughput in useful model FLOPs per GPU-hour under the
same accounting convention, and $H$ the GPU-hours consumed. We compare the
GPU-hours required by MoE with those required by dense training to reach the
same target:

\begin{equation}
\rho_H
\equiv \frac{H_{\mathrm{MoE}}}{H_{\mathrm{dense}}}
=
\frac{F_{\mathrm{MoE}}}{F_{\mathrm{dense}}}
\frac{R_{\mathrm{dense}}}{R_{\mathrm{MoE}}}.
\label{eq-flops-to-wall-time}
\end{equation}

Here $\rho_H<1$ means MoE uses fewer GPU-hours than dense training;
for example, $\rho_H=0.5$ means half as many. For runs using the same number
of GPUs, $\rho_H$ is also the elapsed wall-time ratio.

\begin{figure}[!htb]
\centering
\includegraphics[width=\linewidth]{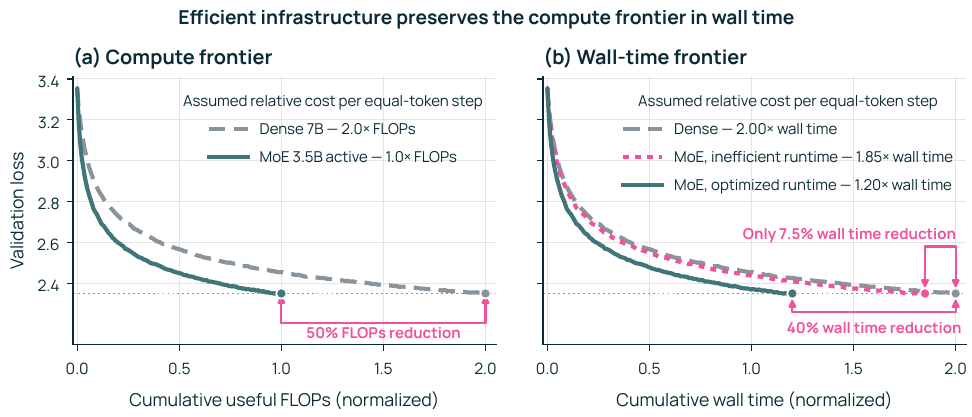}
\caption{\textbf{A FLOP advantage can shrink or survive in wall time.} This
illustrative, simplified comparison uses two power-law-shaped loss traces that
run for 10,000 equal-token steps and end at the same loss.
Panel~(a) assigns the 3.5B-active/13B-total MoE a normalized
useful-FLOP cost of $1$ per step, versus $2$ for the 7B dense model. Panel~(b) maps the
same MoE loss samples onto two hypothetical runtimes: normalized wall-time
costs of $1.85$ for inefficient infrastructure and $1.2$ for optimized
infrastructure, versus $2$ for dense. The corresponding time-to-target
advantages are only $2/1.85\approx1.08\times$ and $2/1.2\approx1.67\times$.
All costs and loss curves are illustrative assumptions, not measurements or
predictions for an Olmo run.}
\label{fig-intro-flops-wall-time-frontiers}
\end{figure}

The first factor captures \textbf{algorithmic efficiency}: a smaller
$F_{\mathrm{MoE}}/F_{\mathrm{dense}}$ means fewer useful FLOPs are needed to
reach the target. The scaling results above suggest that a well-configured
MoE can reduce this factor in the regimes they study. The second factor
captures the \textbf{throughput penalty}: when MoE sustains lower useful-FLOP
throughput than dense training, $R_{\mathrm{dense}}/R_{\mathrm{MoE}}>1$.
This penalty can erase the algorithmic gain
(Figure~\ref{fig-intro-flops-wall-time-frontiers}).

\begin{figure}[!htb]
\centering
\includegraphics[width=\linewidth]{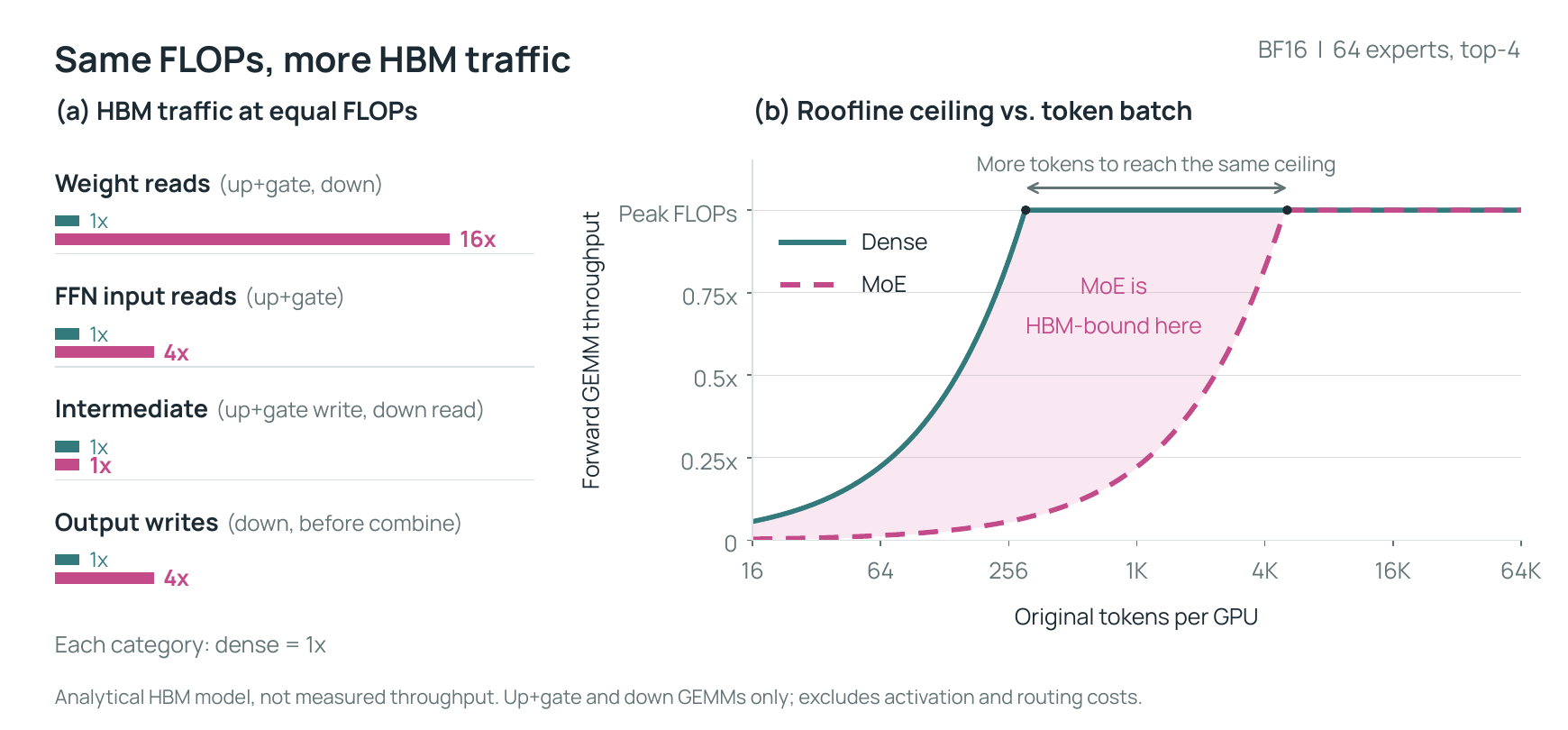}
\caption{\textbf{Equal FLOPs do not imply equal HBM traffic.}
The forward fused up/gate and down GEMMs of a dense FFN with intermediate width $Kh$
are compared with $N$ experts of intermediate width $h$, selecting $K$ per token.
Here $N=64$, $K=4$, and the model width and expert intermediate width are $d=h=4096$.
\textbf{(a)} GEMM-side HBM byte categories, each normalized separately to dense.
These ratios are not additive: each category contributes according to its
absolute byte volume, not equally.
The intermediate row counts up/gate output writes and down input reads,
excluding the pointwise activation's own accesses.
\textbf{(b)} Analytical ceilings in bfloat16 (BF16), normalized to peak FLOPs,
with balanced routes and one HBM read/write per operand/result. Each curve
adds the two GEMM times; pointwise activation, packing, combine, network,
and backward costs are excluded. Conceptual illustration with no measured
quantities; assumptions are detailed in Appendix~\ref{app-expert-roofline}.}
\label{fig-dense-moe-roofline}
\end{figure}

\conceptlead{Equal FLOPs can require more HBM traffic.}
Compared with a dense FFN of equal active compute, an MoE reads more weights
from high-bandwidth memory (HBM), each reused by fewer tokens.
Materialized top-$K$ routes also give grouped GEMM $K$ input rows and
$K$ final output rows per original token; combine happens after those
outputs are written. A \textbf{roofline} estimates the throughput ceiling
imposed by arithmetic and HBM traffic %
\citep{austin2025rooflines}.
Figure~\ref{fig-dense-moe-roofline} shows how the extra bytes lower MoE's
modeled ceiling at small batches, even before routing or communication time
is added. More tokens per expert amortize the weight reads: both curves
eventually reach the same compute ceiling. Appendix~\ref{app-expert-roofline}
derives the comparison; Section~\ref{subsec-grouped-gemm-crossover} measures
how actual kernel efficiency changes with rows per expert.

\textbf{Model FLOPs utilization (MFU)} reports useful model throughput as a
fraction of a stated hardware peak %
\citep{chowdhery2023palm}. Our complete-step
results in Section~\ref{sec-production-throughput} use the B300's dense BF16
Tensor Core peak as a common denominator %
\citep{nvidiaHGXB200spec}, including
for MXFP8 runs; those values are not utilization of the FP8 peak.
Appendix~\ref{app-metric-definitions} gives the accounting.

An expert unused by one token contributes no expert compute for that token,
but its parameters and optimizer states still consume memory. Depending on the parallelization
method, gradient reduction and weight synchronization can also scale with
total expert parameter size. Expert placement determines how far a routed token must
travel. Routing itself adds work: count assignments, enforce
capacity, move or permute token rows, run experts on the devices that store
them, restore token order,
and cope with imbalance. Useful model-FLOP accounting excludes most of this work.

We observed low realized throughput in two early development baselines. One used a
straightforward MoE configuration in Megatron-LM, NVIDIA's distributed
Transformer training framework %
\citep{shoeybi2019megatronlm}.\footnote{This observation concerns a specific
earlier Megatron-LM configuration, not current Megatron-Core. The recent
Megatron-Core MoE report documents DeepEP and HybridEP token-dispatch
backends, overlap of expert-parallel communication with computation from
adjacent microbatches, and reduced-precision training in FP8 and FP4
\citep{nvidia2026megatroncoremoe}.} The other used the legacy MoE path in
Olmo-core, Ai2's training library for the Olmo model family; that path inherited
fully sharded data parallel (FSDP), the parameter-sharding mechanism used for
dense Olmo~3 training %
\citep{teamolmo2025olmo3} and described in
Section~\ref{sec-ddp-vs-fsdp}. Both
suggested that low realized throughput could consume much of the advantage
predicted by active-FLOP accounting.

\begin{figure}[!htb]
\centering
\includegraphics[width=\linewidth]{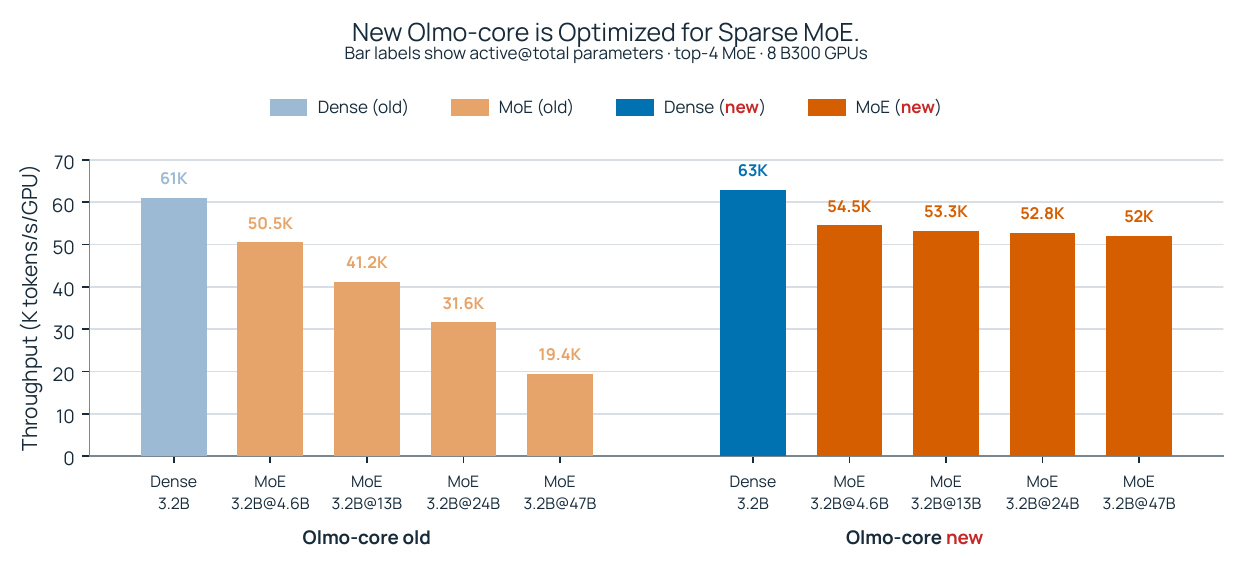}
\caption[Throughput as total expert parameter size grows at nearly fixed active FLOPs.]{\textbf{Throughput as total expert parameter size grows at nearly fixed active FLOPs.} These preliminary measurements
use an eight-layer bfloat16 (BF16) model on one 8$\times$B300 node without
pipeline parallelism (PP; Section~\ref{sec-pipeline-parallelism}); MoE uses
top-4 routing, and the new MoE path uses
EP8. The old,
dense-oriented FSDP path loses throughput as total expert parameter size grows,
whereas the new DDP+EP path keeps MoE throughput within a modest margin of its
dense baseline. This is a composed-system result, not an isolated
FSDP-versus-DDP or EP ablation.}
\label{fig-intro-fsdp-tps-capacity}
\end{figure}

A controlled eight-layer experiment reproduces the Olmo-side slowdown.
Figure~\ref{fig-intro-fsdp-tps-capacity} holds top-$K$ and active size nearly
fixed while increasing total routed-expert parameter size.\footnote{For this
illustrative systems comparison, all MoE runs use random routing to approximate
balanced expert loads. Production routing is data-dependent and generally less
balanced, so its throughput can be lower than this idealized case.} The legacy FSDP-based path loses
throughput as total parameter size grows, even though useful expert compute per
token stays nearly constant. The new distributed data parallel
(DDP; Section~\ref{sec-ddp-vs-fsdp}) path with expert parallelism
(EP; Section~\ref{sec-expert-parallelism}) degree 8 retains most of its dense
throughput across the same parameter-size range. Active-FLOP accounting alone does
not predict this difference.

Figure~\ref{fig-intro-fsdp-tps-capacity} compares the old FSDP-based
Olmo training stack with the new DDP+EP stack, changing several components
together. It therefore does not isolate the benefit of DDP or EP alone, or
rank Olmo against other training frameworks. It
motivates the systems question pursued by this report: which costs continue to
grow with total expert parameter size, and how the training stack can keep them
from dominating wall time.

\begin{keyidea}[From FLOPs to GPU-Hours]
An MoE's lower useful-FLOP requirement reduces GPU-hours only if its realized
throughput is high enough to preserve that advantage.
\end{keyidea}

\subsection{Sparse Routing Breaks the Dense Training Assumptions}
\label{sec-intro-dense-assumptions}

\conceptlead{Training cost includes useful model work and systems overhead.}
To organize the costs of an optimizer step, we use the following reference
accounting:

\begin{equation}
T_{\mathrm{step}}
= T_{\mathrm{model}} + T_{\mathrm{tax}}.
\label{eq-intro-step-tax}
\end{equation}

\begin{table}[!htb]
\centering
\scriptsize
\begingroup
\renewcommand{\arraystretch}{1.0}
\setlength{\tabcolsep}{3pt}
\newcommand{\taxworkitems}[1]{%
  \begin{itemize}[
    leftmargin=1.15em,
    label=\textbullet,
    itemsep=0pt,
    topsep=0pt,
    parsep=0pt,
    partopsep=0pt]
  #1
  \end{itemize}}
\begin{NiceTabularX}{\linewidth}[cell-space-limits=3pt]{
  >{\centering\arraybackslash}m{0.045\linewidth}
  >{\centering\arraybackslash}m{0.135\linewidth}
  >{\raggedright\arraybackslash}m{0.29\linewidth}
  >{\raggedright\arraybackslash}m{0.20\linewidth}
  >{\raggedright\arraybackslash}X[m]}
\arrayrulecolor{black}
\toprule
\multicolumn{2}{c}{\cellcolor{ai2accent!7}\textbf{Component}} &
\cellcolor{ai2accent!7}\makecell[c]{\textbf{What it includes}} &
\cellcolor{ai2accent!7}\makecell[c]{\textbf{Dense}\\
  ($P_{\mathrm{active}}=P_{\mathrm{total}}$)} &
\cellcolor{ai2accent!7}\makecell[c]{\textbf{MoE}\\
  ($P_{\mathrm{active}}\ll P_{\mathrm{total}}$)} \\
\midrule
\arrayrulecolor{black!22}
\Block[C,m]{1-2}{\textbf{Useful Model FLOPs}} & &
\taxworkitems{
  \item Model forward pass, excluding recomputation
  \item Model backward pass
  \item Covers embeddings, attention, dense/expert MLPs, normalization,
        residual additions, and the LM head} &
$\propto P_{\mathrm{total}}$ &
$\propto P_{\mathrm{active}}$, not $P_{\mathrm{total}}$ \\
\hline
\Block[C,m]{6-1}{\textbf{Tax}} &
\strut\par
\textbf{Gradients}\par
(\textsection\textsection\ref{sec-ddp-vs-fsdp},~\ref{subsec-ep-shards-models}) &
\taxworkitems{
  \item Gradient accumulation
  \item Gradient reduction} &
$\propto P_{\mathrm{total}}$; $\propto$ model FLOPs &
$\propto P_{\mathrm{total}}$; $\not\propto$ model FLOPs \\
\cline{2-5}
& \strut\par\textbf{Weights}\par
(\textsection\ref{sec-ddp-vs-fsdp}) &
\taxworkitems{
  \item Optimizer-state updates
  \item Parameter updates
  \item Weight synchronization or redistribution} &
$\propto P_{\mathrm{total}}$; $\propto$ model FLOPs &
$\propto P_{\mathrm{total}}$; $\not\propto$ model FLOPs \\
\cline{2-5}
& \strut\par\textbf{Routing}\par
(\textsection\textsection\ref{sec-expert-parallelism},~\ref{sec-routing-load-balancing}) &
\taxworkitems{
  \item Router decisions and counting
  \item Router auxiliary losses
  \item Dispatch/combine communication
  \item Dispatch/combine permutations} &
Absent &
High tax: follows EP topology \\
\cline{2-5}
& \strut\par\textbf{Compute efficiency}\par
(\textsection\textsection\ref{sec-sync-free-training},~\ref{sec-grouped-gemm}) &
\taxworkitems{
  \item Small expert GEMMs
  \item Uneven or fragmented grouped GEMMs
  \item Smaller microbatches due to high memory pressure
  } &
Low tax: large, regular GEMMs &
Higher tax: smaller, uneven expert batches reduce realized FLOP/s \\
\cline{2-5}
& \strut\par\textbf{Recompute}\par
(\textsection\ref{sec-recompute}) &
\taxworkitems{
  \item Repeated forward operations to regenerate saved activations
  \item Repeated EP communication when included in the recomputed region} &
Extra work set by the recompute policy &
Can repeat dispatch/combine as well as expert compute \\
\cline{2-5}
& \strut\par\textbf{Indirect runtime}\par
(\textsection\textsection\ref{sec-expert-parallelism},~\ref{sec-sync-free-training}--\ref{sec-routing-load-balancing}) &
\taxworkitems{
  \item Device--host or stream synchronization
  \item Dynamic route and general matrix multiplication (GEMM) shapes
  \item Load imbalance
  \item Extra launches and CPU overhead
  \item Temporary buffers and memory traffic
  \item Communication--compute contention} &
Absent or minimal &
High tax: runtime dependent \\
\arrayrulecolor{black}
\bottomrule
\end{NiceTabularX}
\endgroup
\caption{\textbf{Useful model work and the systems tax.} In dense training,
many parameter-sized costs grow together with the useful model work. MoE
decouples active computation from total parameter size and adds routing-specific
costs. ``Tax'' describes the role of the work, not its instruction type:
optimizer and routing work may themselves include FLOPs.}
\label{tab-intro-useful-work-tax}
\end{table}

Table~\ref{tab-intro-useful-work-tax} lists the work included in each term.
$T_{\mathrm{model}}$ is an ideal reference time for the prescribed forward and
backward work at the hardware peaks for the precisions actually used, not
the common BF16 normalization used for MFU. $T_{\mathrm{tax}}$ is the remaining
elapsed time, including execution below those peaks and exposed overhead.
The table identifies sources of that overhead, not independent profiler
durations to add together: communication and computation can overlap.
\textbf{Activation rematerialization}, or recompute, discards selected
forward activations and recomputes them for the backward pass instead of
keeping them in memory. It adds executed FLOPs
without increasing useful model FLOPs, and a recomputed routed block can also
repeat dispatch and combine (Section~\ref{sec-recompute}). In a
well-tuned dense stack, large, regular model computations usually keep these
costs in a workable balance. In MoE, useful computation follows the active
parameters, while much of the tax follows total parameter size or routed token
assignments. Infrastructure built around dense assumptions can therefore let
the tax dominate. The systems goal is to lower it until fewer useful FLOPs also
mean less wall time.

\subsubsection{Capacity Tax}

\conceptlead{Dense compute can amortize parameter-sized work.} In a dense model,
$P_{\mathrm{active}}=P_{\mathrm{total}}$: adding parameters increases the
forward and backward compute performed for every token as well as the
parameters, gradients, and optimizer buffers that must be stored, reduced, or
updated. The tax does not disappear, but it grows in step with useful model
computation.

In MoE, the two no longer grow together. Sparse routing lets total expert
parameter size grow while each token continues to execute only its selected experts.
We call the following dimensionless ratio, compared at fixed token count,
execution schedule, and hardware, the \textbf{capacity-to-active gap}:

\begin{equation}
\Gamma_{\mathrm{capacity}}
\ \equiv\ 
\frac{P_{\mathrm{total}}}{P_{\mathrm{active}}}
= \frac{1}{\alpha_P}.
\label{eq-intro-capacity-compute-gap}
\end{equation}

For dense models, $\Gamma_{\mathrm{capacity}}=1$. For MoE, it can grow as new
routed experts increase $P_{\mathrm{total}}$ without proportionally increasing
$P_{\mathrm{active}}$. Related systems work observes the same decoupling
between total model size and per-token work
\citep{nvidia2026megatroncoremoe}.

We call the concrete costs that continue to follow total parameter size the
\textbf{capacity tax}. Weight and gradient storage, optimizer state, and some
synchronization can grow without a matching increase in useful per-token
compute. The ratio $\Gamma_{\mathrm{capacity}}$ only measures how far stored
parameters have outgrown the parameters each token uses; how much time that
separation costs depends on the implementation. As the ratio increases, parameter-sized work grows
relative to the useful computation available to amortize it.

\conceptlead{Full-reshard FSDP matches dense models better than sparse MoE.} Before model parallelism, ordinary DDP
(Section~\ref{sec-ddp-vs-fsdp}) divides the training data while every rank
stores the complete model. Adding DDP ranks therefore does not reduce per-GPU
model memory. When a full dense-model replica no longer fits, dense Olmo uses
FSDP to shard parameters, gradients, and optimizer
state. In the \textbf{full-reshard} regime used by our comparison, FSDP
all-gathers the full weights of each wrapped unit before the unit computes and
frees them after each use
\citep{zhao2023pytorchfsdp}.\footnote{FSDP2 can instead retain
unsharded parameters through backward and, with no reshard after intermediate
backwards, across an accumulation window %
\citep{pytorch2026fsdp2docs}. That policy gives up most of
full-reshard FSDP's peak unsharded-parameter memory advantage; its parameter
footprint and communication cadence then approach DDP with sharded optimizer
state. We treat it as a mechanism ablation, not a primary competing design.
The primary comparison therefore uses full reshard;
Appendix~\ref{app-fsdp-reshard-policy} discusses the other reshard policies.}
For a dense model, the parameters materialized for a wrapped unit also support
useful computation for every token. The materialization is still a tax, but
a larger dense model brings a matching increase in useful model computation.

FSDP can make a large MoE fit as well, but its weight materialization no
longer tracks useful work. In
this configuration, every microbatch materializes
the wrapped MoE unit's parameter shards according to its stored expert
parameter size, while useful expert compute follows the number of
token--expert assignments. As $P_{\mathrm{total}}/P_{\mathrm{active}}$ grows, weight movement
can therefore grow without a proportional increase in useful model work. Full
resharding solves the immediate fit problem, but its per-microbatch weight
materialization becomes a substantial part of the MoE capacity tax.

\conceptlead{Olmo uses DDP plus model parallelism for MoE.} The Olmo MoE design
composes DDP with two forms of model parallelism: EP
(Section~\ref{sec-expert-parallelism}) shards the routed experts, and pipeline
parallelism (PP; Section~\ref{sec-pipeline-parallelism}) assigns different
layers to different stages. EP and PP first make each rank's model partition
small enough to fit; DDP then synchronizes replicas of that local partition. A
distributed optimizer shards the larger FP32 main weights and optimizer
states, while
the compute weights---the BF16 or MXFP8 (Section~\ref{sec-mxfp8-training}) representations read by forward and
backward---remain static across the accumulation
window and are refreshed once per optimizer step.
Section~\ref{sec-parallelism-overview} introduces the broader parallelism
toolbox before the following sections develop these choices.

\begin{keyidea}[FSDP Moves Weights; DDP + EP Moves Activations]
In the full-reshard regime tested here, FSDP gathers
weight shards to materialize the wrapped weights around each microbatch. In
Olmo's DDP+EP design, each rank keeps its assigned expert weights in the same
GPU buffers across the accumulation window, and EP dispatches only the selected
token rows to those experts. EP also exchanges activation gradients during
backward. Both approaches synchronize parameter gradients across replicas.
Part~\ref{part-parallelization} develops this design.
\end{keyidea}

\subsubsection{Runtime Tax}

\conceptlead{Sparse execution adds work around the expert computation.} For a
fixed batch shape, a dense feed-forward layer executes a predictable sequence
of large, regular operations. A routed MoE layer must first decide which
experts will process each token. It must then arrange and move token rows to
those experts, execute the selected experts, and restore and combine their
outputs. This work repeats in every routed layer and every microbatch. Sparse execution requires this work, but it is overhead around the useful
expert computation.

The extra time appears in several forms. Routing requires decisions, counting,
and bookkeeping. Dispatch and combine add memory movement, permutations, and,
when experts are distributed, communication. Expert computation can become
less efficient when the routed batches are small or uneven. Dynamic shapes and
load imbalance can also introduce temporary buffers, extra kernel launches,
CPU overhead, synchronization, and contention between communication and
computation. The exact mixture depends on the routing decisions, the runtime
implementation, and the hardware.

We refer to these routing, compute-efficiency, and supporting runtime costs
collectively as the \textbf{MoE runtime tax}. Unlike the capacity tax, it is not
explained by model size alone: two batches with similar nominal model FLOPs can
take different amounts of time because they route tokens differently and
exercise the runtime differently. Section~\ref{sec-expert-parallelism}
examines token movement, Section~\ref{sec-sync-free-training} addresses host
submission and synchronization, Section~\ref{sec-grouped-gemm} studies expert
compute efficiency, and Section~\ref{sec-routing-load-balancing} examines
routing imbalance.

\begin{keyidea}[Capacity and Runtime Costs]
The capacity tax follows total parameter size; the runtime tax follows
sparse routing and execution. MoE's FLOP advantage becomes a wall-time
advantage only when the training system controls both.
\end{keyidea}

\subsubsection{End-to-End Optimization}

EP lowers the expert
parameters stored and synchronized by each rank, but introduces token traffic.
Recomputation lowers activation memory by repeating model work
(Section~\ref{sec-recompute}). Reduced
precision can speed up matrix multiplications and reduce communication
volume, but adds conversion work and can expose launch overhead. Its numerical
recipe also matters: with round-ceiling scaling, MXFP8 closely follows BF16's
loss and gradient-norm trajectories over our measured checkpoint-branched
continuation (Section~\ref{sec-mxfp8-scale-rounding},
Figure~\ref{fig-mxfp8-rceil-training}). Overlapping
communication with computation shortens the step only when the slowdown it
causes in the concurrent kernels is smaller than the communication time it
hides (Section~\ref{subsec-ep-overlap}). These interactions are a recurring
property of MoE systems
\citep{nvidia2026megatroncoremoe}. The report-wide matrix in
Appendix~\ref{app-system-method-tradeoffs},
Figure~\ref{fig-system-method-tradeoffs}, summarizes the first-order benefits
and costs of the parallelism, communication, scheduling, precision, runtime,
and checkpoint methods developed in the rest of this report. We place the full
matrix in the appendix because many of its columns are introduced only in
later sections.

We therefore evaluate component changes by their effect on complete-step
time and throughput for a given model, batch, and topology. An isolated
kernel or collective speedup is insufficient to establish that effect.

The next section presents how these mechanisms are integrated into one
end-to-end system: the \emph{Olmo-core DDP training stack}.

\section{Olmo-core DDP Training Stack}
\label{sec-olmo-ddp-training-stack}

Dense Olmo pretraining uses the FSDP-oriented path described by
\citet{teamolmo2025olmo3}. The \emph{Olmo-core DDP training stack} is a separate
path centered on distributed data parallel (DDP) and built for MoE. It coordinates
expert placement, token movement,
gradient synchronization, optimizer updates, memory use, and low-precision execution
across the training step.\footnote{Olmo-core continues to support MoE through
its earlier FSDP-oriented Transformer stack. We call that implementation
the \emph{legacy MoE path} only to distinguish it from the Olmo-core DDP stack
presented here.}

\conceptlead{Sparse routing adds work throughout the training step.}
It adds data movement, irregular expert matrix multiplications,
dynamic shapes, and CPU launch work. DDP provides the synchronization
foundation; expert parallelism (EP)
divides the expert parameters, pipeline parallelism (PP) divides layers, and the
distributed optimizer shards the FP32 main weights and optimizer states.

\begin{figure}[!htb]
\centering
\includegraphics[width=0.80\linewidth]{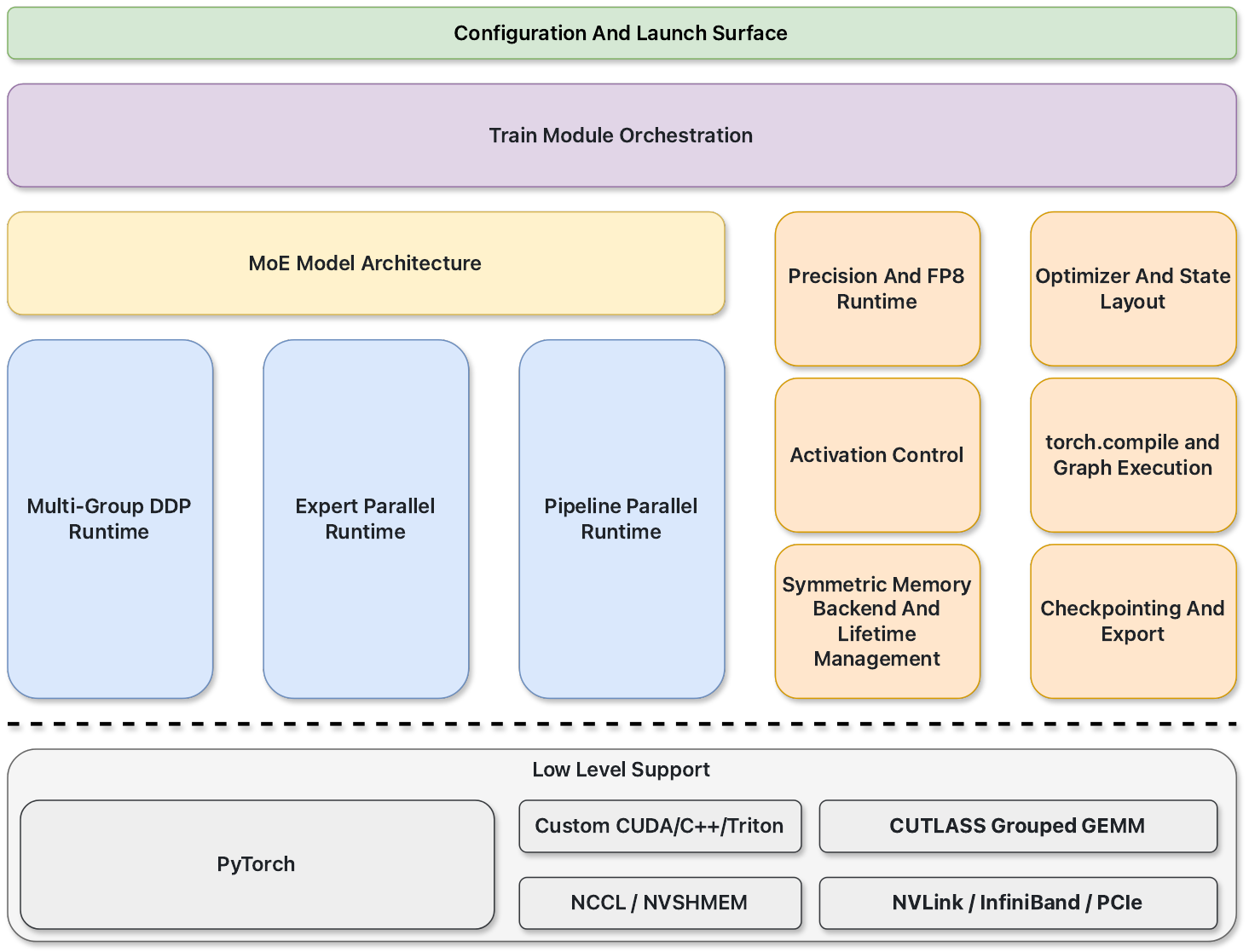}
\caption{\textbf{Subsystems of the Olmo-core DDP training stack.} Colors mark the
layers: configuration (green) and train-module orchestration (purple) sit above
the MoE model (yellow), the parallel runtimes (blue), and supporting systems
(orange). Low-level libraries, kernels, communication systems, and hardware
interconnects (gray) form the substrate below. The layout groups system
responsibilities rather than prescribing an execution order.}
\label{fig-olmo-ddp-stack-subsystems}
\end{figure}

Figure~\ref{fig-olmo-ddp-stack-subsystems} shows the scope of this design.
Configuration and train-module orchestration sit above the MoE model,
parallel runtimes, and supporting systems. A low-level software, kernel,
communication, and transport layer provides their common execution substrate.

PyTorch provides the foundation, but its standard operators cannot remove
every overhead introduced by sparse routing. Olmo-core therefore
combines PyTorch with custom CUDA, C++, and Triton kernels
\citep{tillet2019triton}; NCCL and NVSHMEM communication
\citep{nvidia2026ncclcollectives,nvidiaNVSHMEMUsing}; and CUTLASS grouped GEMM
\citep{nvidiaCUTLASSGroupedScheduler}. These components run over the hardware
links beneath them: NVLink within a node, InfiniBand
between nodes, and PCIe between each GPU and its host. InfiniBand GPUDirect
Async (IBGDA) lets GPUs initiate InfiniBand transfers directly.
We design token
layouts together with communication and expert computation, and
low-precision formats are integrated with activation storage and transport
rather than applied only to isolated GEMMs. A fast kernel is insufficient if
it requires an extra permutation, a device-to-host synchronization, or a gap
between launches elsewhere in the step.

The preliminary capacity sweep in
Figure~\ref{fig-intro-fsdp-tps-capacity} motivates this design. One
composed DDP+EP8 configuration retains most of its throughput as stored expert parameter size
grows, whereas the FSDP-oriented baseline slows. The result does not isolate the contribution of DDP, EP, or any particular
kernel, and it measures throughput rather than time to quality.

\conceptlead{The same design extends to trillion-parameter configurations.}
Because EP, PP, and the distributed optimizer each divide a different part
of the model, the amount stored by each GPU stays bounded as the global model
grows. We have run
296B-total-parameter BF16 and 1.200T-total-parameter MXFP8 configurations on
512 B300 GPUs, and an optional DeepEP v2 backend %
\citep{zhao2025deepep} extends the demonstrated
envelope to a 2.380T-total-parameter configuration. The 2.380T figure comes
from a short nine-step run and describes reachable capacity, not a sustained-throughput
or time-to-quality result at that scale;
Section~\ref{sec-production-throughput} reports production throughput
separately.

\conceptlead{Measured throughput spans scales.}
Section~\ref{sec-production-throughput} reports achieved useful-model
TFLOP/s/GPU for five model scales, which we name Tiny, Small, Medium, Large,
and Ultra in order of size, together with the active and total parameter
counts, batch, EP/PP topology, precision, and recomputation policy
needed to interpret each number. The selected headline values use random
routing as a systems reference and are not a controlled scaling curve or a
time-to-quality comparison.

\Partref{part-parallelization} covers model placement, DDP synchronization,
EP, PP, and their physical-network mapping. \Partref{part-efficient-training}
covers synchronization-free execution, grouped GEMM, routing, MXFP8,
topology-agnostic checkpointing, and recompute. The later findings examine training
recipes and interactions observed in the complete stack.

\begingroup
\newcommand{\stackresult}[7]{%
  \parbox[t]{0.184\linewidth}{%
    \centering
    {\sffamily\bfseries\fontsize{10.5}{11.5}\selectfont\mbox{#2}}\par
    \vspace{1pt}
    {\fontsize{7.2}{8}\selectfont\color{black!65}#7\% active}\par
    \vspace{5pt}
    {\sffamily\bfseries\fontsize{19}{20}\selectfont
     \color{ai2pink!85!black}#3}%
    {\sffamily\bfseries\fontsize{7.2}{8}\selectfont
     \color{ai2pink!85!black}\raisebox{1.5pt}{\,TFLOP/s}}\par
    \vspace{2pt}
    {\sffamily\fontsize{7.5}{8.5}\selectfont\textbf{#6\%} MFU}\par
    \vspace{3pt}
    {\fontsize{7.2}{8}\selectfont #4 Experts}\par
    {\fontsize{6.9}{7.7}\selectfont\color{black!65}#5}\par
    \vspace{3pt}
    {\sffamily\bfseries\fontsize{8.5}{9.5}\selectfont #1}}}

\vspace{7pt}
\noindent\begin{minipage}{\linewidth}
\phantomsection\label{panel-selected-throughput}%
{\color{ai2pink}\hrule height 1.2pt}
\vspace{4pt}
\begin{tabularx}{\linewidth}{@{}Xr@{}}
{\sffamily\bfseries\fontsize{10.5}{11.5}\selectfont Selected achieved throughput} &
{\sffamily\bfseries\fontsize{7.8}{8.5}\selectfont\color{ai2pink!85!black}
B300 NVL8} \\
{} &
{\fontsize{7.2}{8}\selectfont\color{black!62}useful-model TFLOP/s/GPU} \\
\end{tabularx}
\vspace{5pt}

\noindent
\stackresult{Tiny}{1.59B @ 12.9B}{903}{64}{\textcolor{green!55!black}{\textbf{BF16}} EP1 PP1}{40.1}{12.3}\hfill
\stackresult{Small}{4.29B @ 40.9B}{841}{64}{\textcolor{green!55!black}{\textbf{BF16}} EP8 PP1}{37.4}{10.5}\hfill
\stackresult{Medium}{7.38B @ 137B}{843}{128}{\textcolor{green!55!black}{\textbf{BF16}} EP8 PP4}{37.5}{5.38}\hfill
\stackresult{Large}{15.1B @ 296B}{768}{128}{\textcolor{green!55!black}{\textbf{BF16}} EP8 PP4}{34.1}{5.12}\hfill
\stackresult{Ultra}{58.4B @ 1.20T}{858}{128}{\textcolor{red!75!black}{\textbf{MXFP8}} EP8 PP8}{38.1}{4.86}
\par\vspace{5pt}

{\color{black!15}\hrule height 0.4pt}
\vspace{3pt}
{\fontsize{7}{8}\selectfont\color{black!62}\raggedright
The percentage under each size is the parameter activation ratio $\alpha_P$
(active over total parameters). MFU uses the B300 dense BF16
peak (2,250 TFLOP/s/GPU), also for MXFP8; it is not FP8-peak utilization
(Equation~\ref{eq-intro-mfu}). Full configurations and evidence limits:
Section~\ref{sec-production-throughput}.\par}
\end{minipage}
\endgroup

\clearpage
\part{Parallelization}
\label{part-parallelization}
\section{Parallelism Methods Divide Different Work and Storage}
\label{sec-parallelism-overview}

Section~\ref{sec-olmo-ddp-training-stack} introduced the complete Olmo-core DDP
stack. Part~\ref{part-parallelization} examines its parallel methods in terms of
what ranks divide, what one rank keeps during computation, and when ranks
communicate (Table~\ref{tab-parallelism-toolbox}).

A \textbf{rank-local model partition} is the part of the model assigned to one
rank after model-parallel placement; without model parallelism, it is the full
model. Model-parallel methods, namely expert parallelism (EP), pipeline
parallelism (PP), and tensor parallelism (TP), decide which parameters the
partition contains. Ranks that are assigned the same partition and process different
training examples are \textbf{data-parallel (DP) ranks}. Data-parallel methods
decide how those ranks store and synchronize the partition they share:
distributed data parallel (DDP) keeps a complete copy on every DP rank, whereas
fully sharded data parallel (FSDP) splits it across the DP ranks and gathers
each piece when it is needed. Context parallelism (CP) divides the sequence
rather than the model, while a distributed optimizer shards the FP32 main
weights and optimizer states without changing forward or backward model
placement.

For data-parallel and optimizer-sharding methods, we follow the standard
definitions of DDP
\citep{li2020pytorchdistributed}, ZeRO-style distributed optimization
\citep{rajbhandari2020zero}, FSDP, and hybrid sharded data parallel (HSDP)
\citep{zhao2023pytorchfsdp}. For model and activation partitioning, we follow
the standard definitions of EP
\citep{lepikhin2021gshard,rajbhandari2022deepspeedmoe}, PP
\citep{huang2019gpipe,narayanan2019pipedream}, TP
\citep{shoeybi2019megatronlm}, and CP
\citep{jacobs2023deepspeedulysses,liu2024ringattention}.

\begin{table}[!htb]
\centering
\footnotesize
\begingroup
\setlength{\tabcolsep}{4pt}
\renewcommand{\arraystretch}{1.08}
\renewcommand{\tabularxcolumn}[1]{m{#1}}
\begin{tabularx}{\linewidth}{
  >{\centering\arraybackslash\bfseries}m{0.18\linewidth}
  >{\RaggedRight\arraybackslash}m{0.21\linewidth}
  >{\RaggedRight\arraybackslash}X
  >{\RaggedRight\arraybackslash}m{0.25\linewidth}}
\arrayrulecolor{black}
\toprule
\rowcolor{ai2accent!7}
\textbf{Method} & \makecell[c]{\textbf{What it}\\\textbf{divides}} &
\makecell[c]{\textbf{What one rank keeps}\\\textbf{during computation}} &
\makecell[c]{\textbf{When ranks}\\\textbf{communicate}} \\
\midrule
\rowcolor{black!5}
\multicolumn{4}{c}{\rule[-1.2ex]{0pt}{3.8ex}\textbf{Data-parallel and optimizer-sharding methods}} \\
\arrayrulecolor{black!22}
DDP & Input examples across DP ranks & One complete copy of the
rank-local model partition & Accumulated gradients synchronize before each
optimizer update \\\hline

Distributed optimizer & FP32 main weights and optimizer states within a
replica group & Compute weights and the local optimizer shard &
Gradient intake, local shard update, and compute-weight reconstruction once
per optimizer step \\\hline

FSDP / HSDP & Persistent training state---parameters, gradients, and optimizer
data---across a shard group; HSDP repeats shard groups & Full weights for the current wrapped
unit; other units remain sharded & Gather weights before each forward and
backward use; free them after each use and reduce-scatter gradients in backward
\\

\arrayrulecolor{black}
\rowcolor{black!5}
\multicolumn{4}{c}{\rule[-1.2ex]{0pt}{3.8ex}\textbf{Model and activation partitioning}} \\
\arrayrulecolor{black!22}
EP & Routed experts & The assigned expert shard; dense and shared weights
follow the other axes & Dispatch and combine in each routed layer and
microbatch; synchronize expert copies once per optimizer step \\\hline

PP & Transformer layers across physical and virtual stages & The layers
assigned to the local stages & Hidden activations and their gradients cross
stage boundaries for every microbatch \\\hline

TP & Individual tensors and matrix operations & One tensor or operator shard
& Activation or partial-result collectives inside each sharded layer
\\\hline

CP & Sequence positions & A local sequence slice; attention also needs keys
and values from the other slices & Ring attention passes key/value blocks
around the CP ranks during attention; Ulysses exchanges sequence and head
slices with all-to-all before and after attention \\
\arrayrulecolor{black}
\bottomrule
\end{tabularx}
\endgroup
\caption{\textbf{Parallel methods divide different work and communicate at
different times.} The rows are composable mechanisms, not mutually exclusive
systems. The FSDP/HSDP row describes the configuration used in this report's
primary comparison. TP is included as a standard point of comparison;
the Olmo stack described here does not use it. The Olmo MoE path uses the
Ulysses form of CP (Appendix~\ref{app-context-parallelism}).}
\label{tab-parallelism-toolbox}
\end{table}

\conceptlead{DDP and FSDP determine when compute weights move.}
We start with this choice because it is the data-parallel foundation beneath
EP and PP. FSDP remains a data-parallel method even though it shards stored
parameters: its ranks process different data while executing the same model.
EP and PP instead assign different experts or layers to different ranks.

In the FSDP configuration used by Section~\ref{sec-ddp-vs-fsdp}, each wrapped
unit is gathered before its forward
pass, freed afterward, gathered again before backward, and followed by
gradient reduce-scatter
\citep{zhao2023pytorchfsdp,pytorch2026fsdp2docs}. DDP instead keeps one \emph{static} copy of
the rank-local model partition at stable GPU locations across all microbatches
that contribute to an optimizer step, then synchronizes the accumulated
gradients. Here, \emph{static} describes placement, not value: the optimizer
still updates the weights after every successful step. This difference sets
both the memory cost and the communication frequency developed in the next
section. Appendix~\ref{app-fsdp-reshard-policy} discusses other FSDP reshard
policies.

\conceptlead{EP and PP make the DDP partition fit.}
Olmo's distributed optimizer shards FP32 main weights and optimizer states,
EP divides routed experts, and PP divides layers. Optional CP divides long
sequences when activation memory or attention work becomes limiting. Attention mixes token positions, so it is the layer CP must restructure.
The routed experts process each position independently and can treat CP
ranks as additional DP ranks. Olmo does not use TP. Every expert in our
configurations fits on one GPU, and TP's per-layer collectives would compete
with EP for each node's NVLink domain
(Section~\ref{sec-hardware-topology}).

The rest of Part~\ref{part-parallelization} follows this dependency. Section~\ref{sec-ddp-vs-fsdp}
first compares DDP with FSDP. Section~\ref{sec-expert-parallelism}
then explains how EP divides experts, and
Section~\ref{sec-pipeline-parallelism} explains how PP divides layers.
Section~\ref{sec-hardware-topology} finally maps their composition onto the
physical network.

Appendix~\ref{app-context-parallelism} covers the optional CP path in detail.
Olmo uses the Ulysses form of CP, which changes attention from a local sequence
slice with all heads to the full sequence for a subset of heads, then reverses
that layout after attention %
\citep{jacobs2023deepspeedulysses}. The MoE layers view the same ranks along two
axes, developed in Section~\ref{sec-expert-parallelism}: ranks along the
expert-sharding axis (EP-MP) hold different routed experts, while ranks along
the expert-replica axis (EP-DP) hold copies of the same experts and synchronize
their gradients. The appendix also explains how Olmo maps the dense
DP$\times$CP view and this MoE view onto the same ranks through MoE parallel
folding %
\citep{liu2025moeparallelfolding}.
Appendix~\ref{app-glossary-notation} collects the symbols and terms used across
these sections.

\FloatBarrier

\section{Distributed Data Parallel}
\label{sec-ddp-vs-fsdp}

\conceptlead{Sparse MoE changes the cost of parameter materialization.} In a dense model,
$P_{\mathrm{active}}\simeq P_{\mathrm{total}}$: nearly every parameter that is
materialized participates in useful computation for every token. Sparse MoE
breaks that coupling. Its expert parameter size grows with $P_{\mathrm{total}}$, while
the expert compute performed for one token follows $P_{\mathrm{active}}$.
On the FSDP path studied here, parameter materialization therefore
continues to follow the rank-local parameter size even as useful expert
work follows only the active experts, so the same gathering cost is spread
over less useful work. Olmo-core changes the foundation to DDP so
that the rank-local model partition, once it fits in memory, stays at a
stable location across the
accumulation window. A distributed optimizer, EP, and PP then reduce the
persistent training-state memory---compute weights, gradient buffers, FP32 main
weights, and optimizer states---that each rank must carry, so the partition
fits without gathering parameters in every microbatch.

This motivation is specific to sparse MoE configurations with a large gap between
total and active parameters and with a rank-local partition that fits after
optimizer, expert, and pipeline sharding. It is not a general argument against
FSDP. Throughout this section, \emph{FSDP} refers to FSDP2 with the ``full-reshard'' configuration;
Appendix~\ref{app-fsdp-reshard-policy} discusses FSDP configurations that keep
gathered parameters instead of freeing them. Dense
Olmo continues to use an FSDP/HSDP-oriented stack
\citep{teamolmo2025olmo3}. The Olmo-core DDP stack can also train dense models and
can be faster when its rank-local compute-weight partition fits. Within current Olmo-core,
it is also the path for fine-grained integrations such as native MXFP8. We
therefore treat DDP as a second training foundation, not as an MoE-only
mechanism or a universal replacement for FSDP.

\subsection{DDP Uses Static Compute-Weight Replicas}

\begin{figure}[!htb]
\centering
\includegraphics[width=0.92\linewidth]{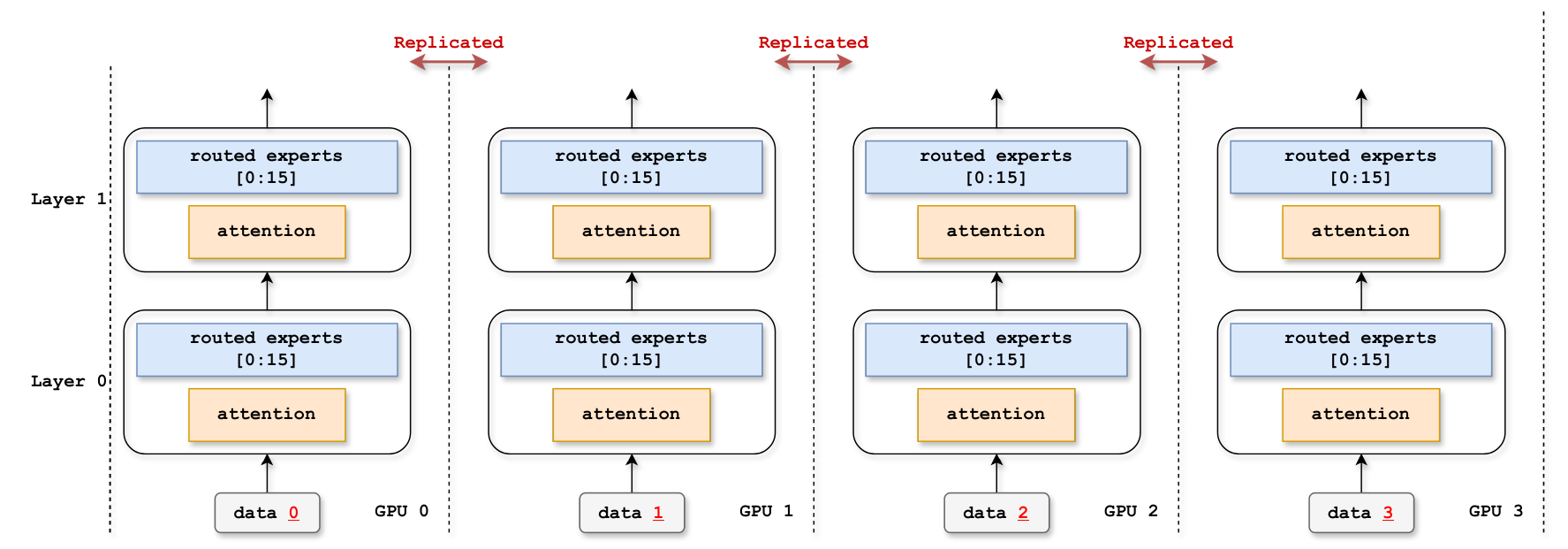}
\caption{\textbf{A four-GPU DDP example with no model-parallel sharding.}
With DP${}=4$, every GPU holds a static replica of the same attention layers and
complete routed-expert pool, experts $0$--$15$, while processing different
input data. DDP synchronizes the gradients of these replicated weights; it does
not itself shard the model shown here.}
\label{fig-ddp-topology-four-gpu}
\end{figure}

Start with replicated data parallelism and no model parallelism. Let $D$ denote
the DDP degree. Every rank holds the complete model
(Figure~\ref{fig-ddp-topology-four-gpu}), processes different
training examples, and accumulates gradients for corresponding parameters.
Before an optimizer update, a collective reduction produces the same
synchronized gradient for each replicated parameter, so all replicas apply the same update and stay identical. This is the standard replicated-data-parallel contract implemented by PyTorch DDP
\citep{li2020pytorchdistributed}.
As in Section~\ref{sec-parallelism-overview}, \emph{static} refers to
placement, not to persistent memory: the weights stay at stable GPU locations, and the optimizer still
updates their values after every successful step.

FSDP uses a different parameter-materialization policy for data parallelism.
Persistent parameter, gradient, and
optimizer-state allocations are sharded across the FSDP group. Before a wrapped unit---for
example, one Transformer block---runs its forward or backward computation,
FSDP gathers that unit's complete weights. After the computation consumes those
weights, FSDP frees the gathered copy and keeps only the local shard; backward also reduce-scatters gradients into
their persistent shards. FSDP thereby lowers persistent parameter, gradient,
and optimizer memory at the price of gathering and materializing weights around
each layer use. Both methods still process different input data. They differ in
which complete weights remain materialized between uses.
The original FSDP paper and the current FSDP2 documentation define this
mechanism and its reshard policies
\citep{zhao2023pytorchfsdp,pytorch2026fsdp2docs}.

Mixed-precision training separates the representation used by model
computation from the higher-precision values used by the optimizer
\citep{micikevicius2018mixedprecision}. The following terms distinguish the
three parameter-sized allocations used throughout this section.

\begin{definitionbox}[Compute Weights, Main Weights, and Optimizer States]
\textbf{Compute weights} are the representations consumed by forward and
backward computation. Ordinary layers use BF16 compute weights; selected MXFP8
linears use quantized data and scales, including orientation-specific weight
caches.

\textbf{Main weights} are the FP32 parameter values updated by the optimizer.
For storage accounting, we group them with the optimizer states because the
distributed optimizer stores and shards them together; they remain parameter
values, not optimizer states. The updated main weights are then used
to refresh the BF16 or MXFP8 compute weights.

\textbf{Optimizer states} are the per-parameter buffers that an optimizer keeps
between steps. In Olmo, parameters updated by AdamW keep two moment estimates %
\citep{kingma2015adam,loshchilov2019decoupled}, and matrices updated by Muon
keep one momentum buffer %
\citep{jordan2024muon,liu2025muon}; all of these states are stored in FP32.
\end{definitionbox}

These dtypes are training-recipe choices, not requirements imposed by DDP or
optimizer sharding. DeepSeek-V3 and Megatron-Core report storing both AdamW
moments in BF16 while retaining FP32 main weights and accumulated gradients
\citep{deepseekai2024deepseekv3,nvidia2026megatroncoremoe}. Olmo keeps both
AdamW moments in FP32 in the configurations described here; we have not
evaluated lower-precision moment storage in this stack.

Let $B$ be the tokens processed by one rank in an optimizer step, $m$ the
number of microbatches, $K$ the number of active experts, $d$ the model width,
and $D$ the DDP degree. Table~\ref{tab-ddp-step-cost-ledger} gives approximate
resource costs for the work introduced in
Section~\ref{sec-intro-dense-assumptions}. The table omits constant factors and hardware
efficiency.

\begin{table}[!htb]
\centering
\small
\begingroup
\renewcommand{\arraystretch}{1.06}
\renewcommand{\tabularxcolumn}[1]{m{#1}}
\setlength{\tabcolsep}{4pt}
\begin{tabularx}{\linewidth}{
  m{0.19\linewidth}
  m{0.15\linewidth}
  >{\raggedright\arraybackslash}m{0.23\linewidth}
  >{\raggedright\arraybackslash}X}
\arrayrulecolor{black}
\toprule
\rowcolor{ai2accent!7}
\multicolumn{4}{c}{\textbf{No model parallelism: every DP/DDP rank holds the whole model}} \\
\midrule
\multicolumn{1}{c}{\textbf{Training work}} &
\multicolumn{1}{c}{\makecell[c]{\textbf{When it}\\\textbf{happens}}} &
\multicolumn{1}{c}{\textbf{Approximate cost}} &
\multicolumn{1}{c}{\textbf{What changes for MoE}} \\
\midrule
\arrayrulecolor{black!22}
{\centering\bfseries Forward and backward model compute\par} &
{\centering Every microbatch\par} &
$\propto B P_{\mathrm{active}}$ for parameter-linear terms &
This is the useful work. It follows active, not total, parameters. \\
\hline
{\centering\bfseries Gradient storage and accumulation\par} &
{\centering Kept through the accumulation window; initialize once; add after each backward\par} &
Storage and zeroing $\propto P_{\mathrm{total}}$; additions recur for each
microbatch &
Buffer capacity follows total parameters. A rank's experts share one stacked
gradient tensor, so each accumulation can touch the whole tensor even when only
some experts received tokens. \\
\hline
{\centering\bfseries Gradient synchronization\par} &
{\centering Final microbatch, or every microbatch\par} &
Payload $\propto P_{\mathrm{total}}$; the exact factor depends on $D$ and the
collective algorithm &
Skipping it on the earlier microbatches (PyTorch's \texttt{no\_sync()}) leaves
one synchronization per optimizer step, but the payload still covers the full
gradient replica, including experts that received no tokens. \\
\hline
{\centering\bfseries Optimizer update\par} &
{\centering Once per optimizer step\par} &
Arithmetic and memory traffic $\propto P_{\mathrm{total}}$ &
An inactive expert still has a weight and optimizer states; in this unsharded
baseline every DP rank carries them. \\
\hline
{\centering\bfseries Updated-parameter distribution\par} &
{\centering After the optimizer step\par} &
Local only in unsharded DDP &
Every rank updates its complete replica, so there is no parameter collective.
Parameter movement appears only after choosing a sharded optimizer. \\
\hline
{\centering\bfseries Routing and expert work setup\par} &
{\centering Every MoE layer and microbatch\par} &
Activation movement $\propto BKd$ per routed layer, plus route metadata and
launches &
Top-$K$, counting, permutation, and combination already exist with all experts
local; the grouped expert GEMMs are useful model work. \\
\arrayrulecolor{black}
\bottomrule
\end{tabularx}
\endgroup
\caption{\textbf{Cost ledger before model parallelism.} Every DP/DDP rank
holds a complete model replica and an unsharded optimizer. These estimates show
which resources grow; they do not add up to elapsed time. Model-sharding choices
change the per-rank parameter costs and when they are paid.}
\label{tab-ddp-step-cost-ledger}
\end{table}

These proportionalities identify pressure points; they are not additive terms
in an elapsed-time model. Collective traffic also depends on the algorithm,
topology, and overlap with other work. For example, ring all-reduce moves
approximately $2(D-1)/D$ payload copies per rank, while aggregate network
traffic additionally grows with the number of ranks
\citep{patarasuk2009allreduce}.

The capacity-to-active gap is especially clear for the expert MLP. A bias-free
SwiGLU expert pool with model width $d$, expert intermediate width $h_e$, and $N$
routed experts stores approximately $3Ndh_e$ parameters, whereas one token
evaluates approximately $3Kdh_e$ of them. Parameter-sized work can therefore
grow with $N$ while useful expert compute grows with $K$. The next
subsection compares how often DDP and FSDP perform that parameter-sized work.

\subsection{FSDP Gathers per Microbatch; DDP Synchronizes per Global Step}

\begin{figure}[!htb]
\centering
\includegraphics[width=1.0\linewidth]{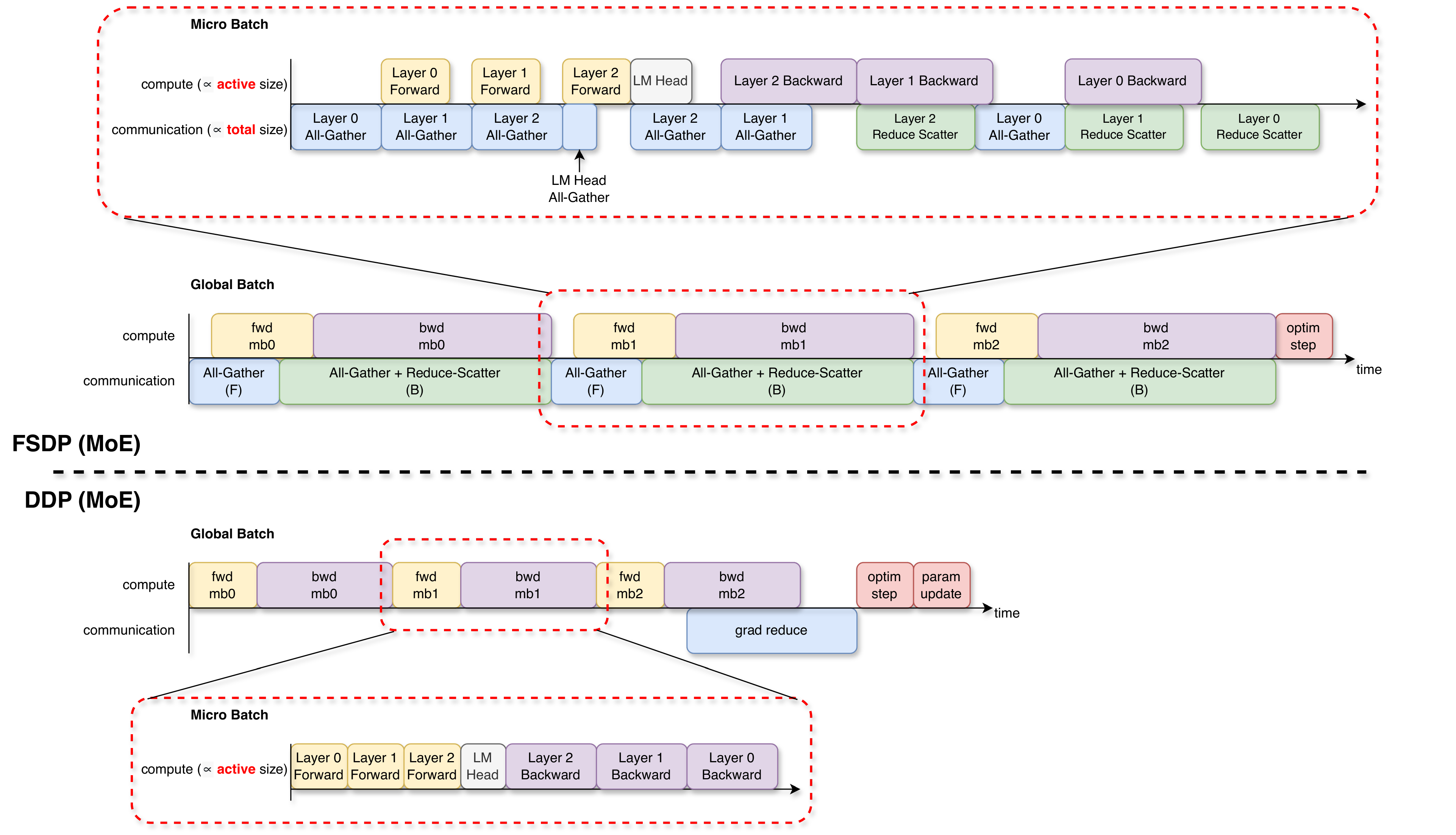}
\caption{\textbf{Weight materialization and gradient synchronization run on
different clocks.} Under the illustrated FSDP schedule, unsharded weights
are materialized around layer use in each microbatch. DDP instead uses static
replicas of its local weights throughout the accumulation window and
synchronizes accumulated gradients at the optimizer-step boundary. When EP is
enabled, token dispatch and combine still recur in every routed layer and
microbatch. The phase widths were manually reconstructed from one representative
profile and preserve its approximate timing proportions; the diagram is not an
aggregate benchmark.}
\label{fig-ddp-fsdp-two-clocks}
\end{figure}

An optimizer step is the complete accumulation window that produces one model
update. Suppose it contains $m$ microbatches of $b$ tokens per rank, so the
rank-local accumulated batch is $B=mb$. At fixed microbatch shape, useful model
work grows the same way under either data-parallel choice:
\begin{equation}
W_{\mathrm{useful}}(m) \propto mbP_{\mathrm{active}}.
\label{eq-useful-work-accumulation}
\end{equation}
The parameter-sized communication differs between the two alternatives
(Figure~\ref{fig-ddp-fsdp-two-clocks}):
\begin{itemize}[leftmargin=*,itemsep=3pt]
\item \textbf{With DDP,} gradient synchronization and, when using a distributed
optimizer, compute-weight reconstruction occur once per optimizer step. Their
payload scales with the rank-local parameter count,
$W_{\mathrm{DDP\ boundary}}\propto P_{\mathrm{local}}$.
\item \textbf{With FSDP,} weight gathering and freeing recur as each wrapped
unit is used in every microbatch. Across the accumulation window, this movement
scales as $W_{\mathrm{FSDP\ materialization}}\propto
mP_{\mathrm{local}}$.
\end{itemize}
The DDP and FSDP terms describe alternative designs; they are not paid
together.
Increasing $m$ spreads DDP's once-per-step synchronization over more useful
work,
whereas it repeats FSDP's weight movement. The optimizer update itself remains
once per optimizer step in both designs.

On the Olmo-core DDP path, non-final microbatches run under
\texttt{no\_sync()} %
\citep{li2020pytorchdistributed}.
Gradients accumulate in stable buffers, and the complete
window ends with one ordered sweep of bucketed all-reduces. When the distributed
optimizer is enabled, it then updates each rank's FP32 main-weight and
optimizer-state shards and
reconstructs the compute weights once. Replica synchronization therefore
occurs at the global-step boundary, even with optimizer sharding.

With FSDP, each wrapped unit must instead be materialized for its forward and
backward computation and freed after each use. Gradient accumulation
can delay the optimizer step, but it does not eliminate this repeated weight
movement. Retaining every gathered unit across the accumulation window instead
trades that movement for higher memory, as discussed in
Appendix~\ref{app-fsdp-reshard-policy}
\citep{pytorch2026fsdp2docs}.

DDP moves gradient reduction and optimizer reconstruction to the global-step boundary, but it
does not make all communication independent of $m$. Once EP is enabled,
dispatch and combine move routed token rows, and their backward adjoints, in
every routed layer and microbatch. Accumulating more microbatches spreads the
once-per-step gradient synchronization and optimizer update over more useful
work, but it does not reduce each microbatch's EP traffic or enlarge its
grouped-GEMM shapes.

\begin{figure}[!htb]
\centering
\begin{subfigure}[t]{0.475\linewidth}
  \centering
  \includegraphics[width=\linewidth]{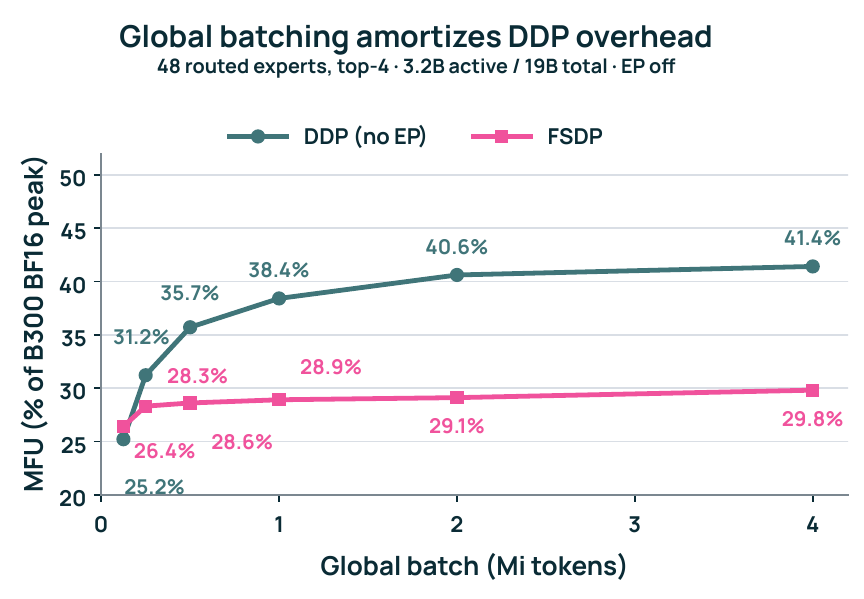}
  \caption{Accumulation amortizes DDP's per-step synchronization.}
  \label{fig-fsdp-ddp-global-batch-sweep}
\end{subfigure}\hspace{0.015\linewidth}%
\begin{subfigure}[t]{0.475\linewidth}
  \centering
  \includegraphics[width=\linewidth]{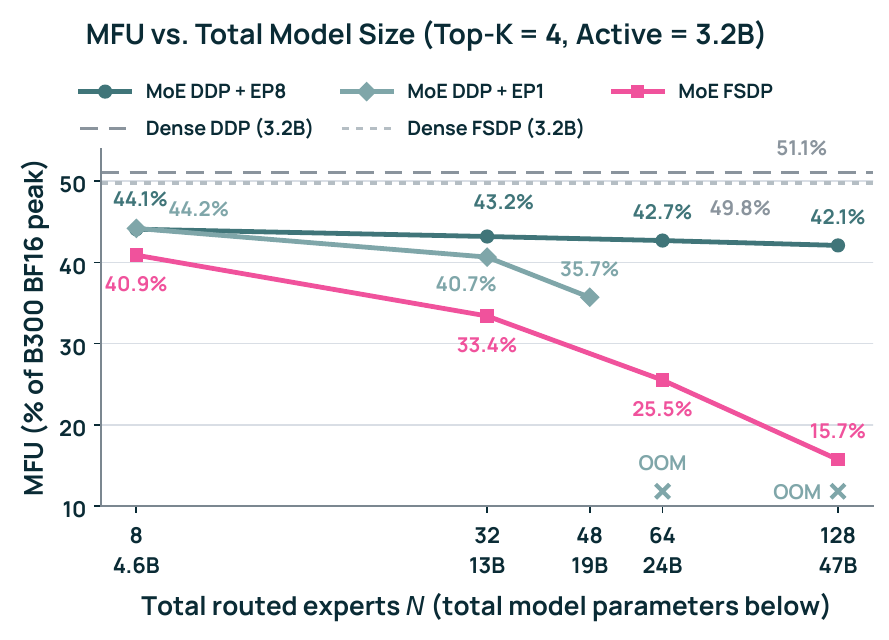}
  \caption{EP delays DDP's expert-parameter memory wall.}
  \label{fig-fsdp-ddp-capacity-sweep}
\end{subfigure}
\caption{\textbf{Global-batch and model-size sweeps expose different
limits.} Preliminary single-node measurements use eight B300 GPUs, an
eight-layer BF16 MoE model, top-4 random routing, no PP, and approximately 3.2B
active parameters. \textbf{Left:} with EP disabled and 48 routed experts (19B
total parameters), DDP useful-model FLOPs utilization (MFU) rises from 25.2\% to
41.4\% between
128-Ki and 4-Mi-token global batches ($1\ \mathrm{Ki}=2^{10}$ and
$1\ \mathrm{Mi}=2^{20}$ tokens), while FSDP rises from 26.4\%
to 29.8\%. \textbf{Right:} total parameters grow from 4.6B to 47B;
no-EP DDP runs out of memory at 64 and 128 experts, no-EP FSDP
continues to fit but loses useful-model MFU, and DDP+EP8 remains nearly flat.
Each point is one run, so the curves show preliminary scaling behavior rather
than run-to-run variance. They do not decompose peak memory or attribute the
trend to a specific communication phase.}
\label{fig-fsdp-ddp-sweeps}
\end{figure}

The preliminary single-node accumulation sweep in
Figure~\ref{fig-fsdp-ddp-global-batch-sweep} shows the expected global-batch
signature without mixing in EP. It fixes the model at 48 routed experts, approximately 3.2B
active and 19B total parameters, fixes the rank-microbatch shape, and changes
only the number of accumulated microbatches. Both paths run with EP disabled.

Figure~\ref{fig-fsdp-ddp-sweeps} places this accumulation result beside the
model-size sweep used in the next subsection. The panels test different
questions: the left varies how much useful work shares one synchronization
boundary, while the right varies how many inactive expert parameters the system
must store and manage.

These trends are consistent with the cost model: DDP spreads its once-per-step work over
more useful computation as the batch grows, while FSDP repeats its weight
gathering in every microbatch. Each point is a single run without a matched
communication profile, so the figure shows the trend but not its exact cause,
and it measures throughput, not training quality or time to a target loss.

\subsection{DDP Alone Hits a Memory Wall}

\conceptlead{Static replicas require a local partition that fits in memory.} DDP removes
repeated weight materialization only by keeping its compute weights and
selected training buffers present throughout the accumulation window. Peak
memory therefore combines persistent training state with execution working
memory, summarized in
Table~\ref{tab-ddp-peak-memory}.

\begin{table}[!htb]
\centering
\small
\begingroup
\renewcommand{\arraystretch}{1.14}
\renewcommand{\tabularxcolumn}[1]{m{#1}}
\setlength{\tabcolsep}{5pt}
\begin{tabularx}{\linewidth}{
  >{\centering\arraybackslash\bfseries}m{0.18\linewidth}
  >{\RaggedRight\arraybackslash}m{0.43\linewidth}
  >{\RaggedRight\arraybackslash}X}
\arrayrulecolor{black}
\toprule
\rowcolor{ai2accent!7}
\textbf{Memory class} & \textbf{What it includes} & \textbf{Why it matters} \\
\midrule
\arrayrulecolor{black!22}
\makecell[c]{Persistent\\training state} & Compute weights; gradient-accumulation buffers; FP32 main
weights; optimizer states and other long-lived optimizer buffers & Must remain available across
microbatches or update boundaries. DDP, the distributed optimizer, EP, and PP determine
which portions are replicated or sharded. \\
\hline
\makecell[c]{Execution\\working memory} & Activations; materialization, routing, and communication
buffers; kernel workspaces; allocator fragmentation; execution caches & Its
contents and peak demand depend on microbatch shape, recomputation, routing,
schedule, kernels, and runtime behavior. A preallocated fixed-address buffer
can still belong to this class. \\
\arrayrulecolor{black}
\bottomrule
\end{tabularx}
\endgroup
\caption{\textbf{Peak memory combines persistent training state and execution
working memory.} This distinction is based on role, not physical allocation
lifetime: a fixed-capacity routing buffer can be preallocated yet remain
execution working memory. It prevents parameter-byte accounting from being
mistaken for a complete peak-memory prediction.}
\label{tab-ddp-peak-memory}
\end{table}

FSDP has a direct advantage in persistent training-state memory. With enough
shard ranks, it can make each rank's long-lived parameter, gradient, and
optimizer-state contribution small, although the currently materialized unit,
activations, and temporary buffers still contribute to peak memory. DDP keeps
the compute-weight representation of its rank-local partition complete. That
partition, its gradient buffers, and the execution working set must all fit at
once.

\begin{figure}[!htb]
\centering
\includegraphics[width=0.96\linewidth]{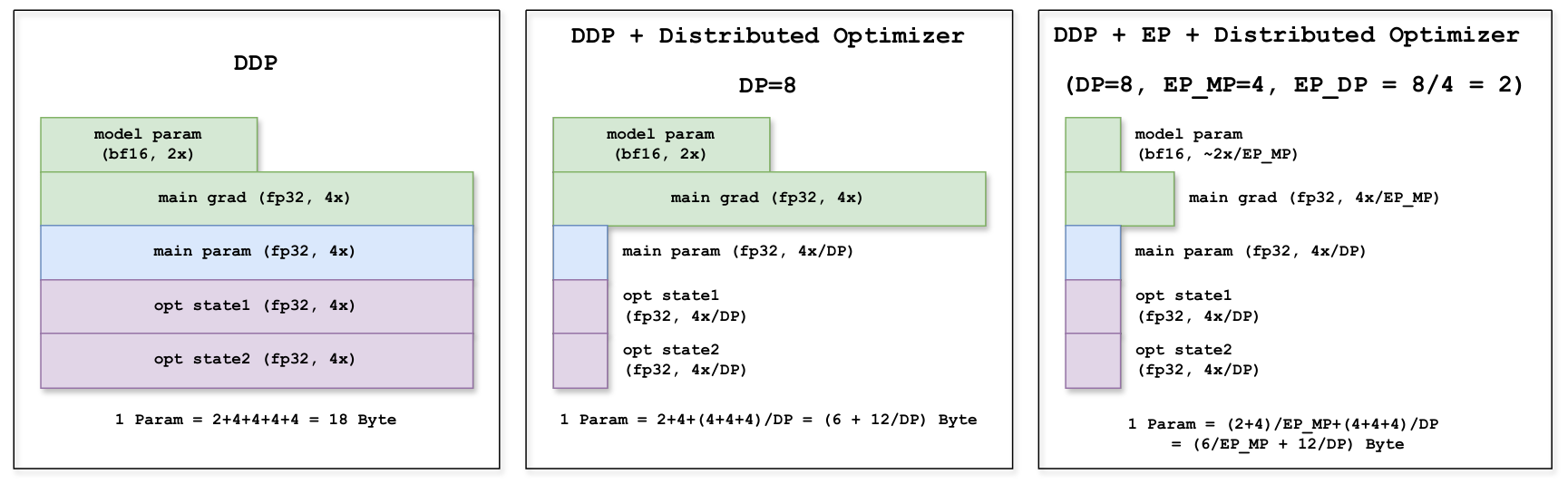}
\caption{\textbf{The distributed optimizer and EP reduce different parts of
DDP's persistent parameter-sized training state.} Under the illustrated BF16-execution, FP32-gradient,
and two-moment FP32 optimizer policy, naive DDP uses 18 bytes per parameter.
Sharding the FP32 main weights and optimizer states evenly over $D$ replica
ranks reduces this to $6+12/D$ bytes while compute weights and gradient
buffers remain replicated. Sharding routed-expert compute weights and gradients
over the expert-parallel model-sharding degree $M_E$ (EP-MP; defined in
Section~\ref{sec-expert-parallelism}) further reduces their per-rank memory contribution to
$6/M_E+12/D$ bytes when $D=M_ED_E$. Dense and shared parameters follow their
separate replica groups. This accounting excludes execution working memory,
including activations, routing and communication buffers, kernel workspaces,
execution caches, and allocator fragmentation, as well as uneven shards and
small auxiliary tensors that are not sharded.}
\label{fig-ddp-static-memory}
\end{figure}

Figure~\ref{fig-ddp-static-memory} isolates the parameter-sized persistent
training-state allocations directly changed by the distributed optimizer and
EP under one concrete policy.

The model-size sweep in Figure~\ref{fig-fsdp-ddp-capacity-sweep} makes the wall
visible while holding useful expert work
approximately fixed. Top-$K$ stays at 4 and the active model size near 3.2B
parameters as the routed-expert pool grows from 8 to 128 experts. DDP without
EP fits at 8, 32, and 48 experts, then runs out of memory at 64 and 128. FSDP
without EP continues to fit, but its measured MFU falls as total parameter size
grows. DDP+EP8, which also shards the routed-expert pool across eight ranks,
remains near 42--44\% MFU over its measured points.

In the right panel, DDP+EP8 stores one eighth of the routed-expert pool per
EP-MP shard; the dashed dense references were measured separately. Its measured MFU is nearly flat, but the per-rank expert shard
still grows as $N/8$.
This residual growth matters more as modern MoE designs increase $N$
while keeping $K$ small (Table~\ref{tab-intro-modern-moe-sparsity}): EP changes
the per-rank growth from $N$ to $N/M_E$, but it does not make memory independent
of total expert parameter size. Olmo therefore reduces
the rank-local partition with a distributed optimizer, EP, and PP.

\subsubsection{Distributed Optimizer Shards FP32 Main Weights and Optimizer States}

The first of these three mechanisms, the distributed optimizer, keeps model
execution unchanged while shrinking the memory for FP32 main weights and
optimizer states. Under the illustrated AdamW-like policy, one
parameter contributes a 2-byte BF16 compute weight, a 4-byte FP32
gradient-accumulation slot, a 4-byte FP32 main weight, and two 4-byte FP32
moments. Naive DDP therefore carries 18 bytes per parameter before execution
working memory. The distributed optimizer leaves the compute weight and
gradient buffer replicated within that parameter's replica group, but shards
eligible FP32 main weights and optimizer states over the group. The idealized
$6+12/D$ expression in Figure~\ref{fig-ddp-static-memory} assumes even
sharding. This optimizer-state partition follows the design family introduced
by ZeRO %
\citep{rajbhandari2020zero}.

\begin{figure}[!htb]
\centering
\includegraphics[width=\linewidth]{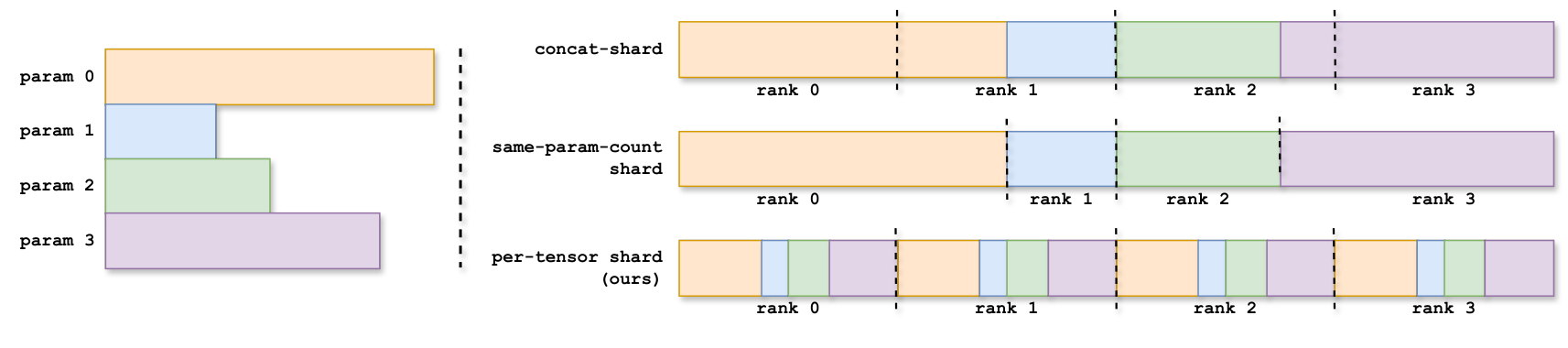}
\caption{\textbf{Optimizer-state layout choices.} Colors identify eligible
FP32 optimizer-state tensors associated with different model parameters; they do not
represent the replicated BF16 compute weights. Dashed lines mark the ranges
assigned to each rank in an optimizer replica group of size $D=4$. Concatenate-then-shard
balances the flattened optimizer state globally but requires explicit view and offset
metadata. The same-param-count baseline preserves whole tensors but can
severely imbalance bytes. Olmo instead divides each eligible main-weight and
optimizer-state tensor independently into $D$ equal contiguous shards.}
\label{fig-distributed-optimizer-layouts}
\end{figure}

Figure~\ref{fig-distributed-optimizer-layouts} compares the two useful balanced
layouts with a naive baseline. \emph{Concatenate-then-shard} flattens all
eligible optimizer-state tensors into one sequence and partitions that sequence across
ranks. It balances the aggregate optimizer state and does not require every tensor to be
individually divisible by the sharding degree. Its cost is bookkeeping: shard
boundaries may cut through tensors, so optimization, weight reconstruction, and
checkpointing must all preserve a global map from each tensor to its flat-buffer
views and offsets.\footnote{Megatron-LM's standard
\texttt{DistributedOptimizer} is a bucket-level instance of this layout. It
assigns equal contiguous bucket ranges to data-parallel ranks; because a range
may cut through a parameter, the implementation maps bucket slices back to
parameter slices %
\citep{nvidia2026megatronlm}.} The figure's
\emph{same-param-count} baseline assigns the same
number of complete parameter tensors to every rank. It does not equalize the
number of scalar elements. Because tensor sizes vary widely, equal tensor counts can
produce severely unequal byte counts and are the least useful of the three
layouts.

Olmo uses the per-tensor layout in the bottom row. Each eligible tensor is
flattened and divided independently across its optimizer replica group; its
FP32 main weight and optimizer states use the same rank-local slice. The
mapping is local to one tensor, so update, reconstruction, and checkpoint views
remain direct and easy to inspect. The trade-off is that the current policy
shards a tensor only when it is sufficiently large and its element count is
divisible by the replica-group size. Small or non-divisible optimizer-state tensors remain
replicated, as do scalar step counters. This fallback does not prevent training,
but it reduces the memory saved by the distributed optimizer.

In practice, this restriction is mild for the regular, mostly power-of-two
replica groups in our production configurations, including groups of
128 ranks: nearly all large model tensors are divisible by these group sizes.
Irregular sharding degrees such as 7 or 13 would cause more tensors to remain
replicated, but we have not encountered that case in our production runs. Muon
changes the number of optimizer history tensors, not this placement rule.

At the end of the accumulation window, DDP first produces reduced FP32
gradients. Each rank narrows the relevant gradient into the shard corresponding
to its FP32 main-weight shard, updates that shard and its optimizer states locally,
and then participates in coalesced all-gathers that reconstruct the updated
compute weights. Thus gradient reduction, sharded update, and
reconstruction happen once per optimizer step. The mechanism recovers much of
the optimizer-state memory that motivated FSDP without evicting or
rematerializing compute weights between microbatches, but it does not
remove the replicated weight and gradient floor.

\subsubsection{EP Shards Expert Weights and Gradients}

The second mechanism reduces the routed-expert memory that the distributed
optimizer leaves replicated. With EP-MP degree $M_E$, each rank stores roughly
$N/M_E$ routed experts and their gradient buffers rather than all $N$ experts.
Ranks holding copies of the same expert shard form an EP-DP group of size
$D_E$: they reduce that shard's gradients together, and the distributed
optimizer shards its FP32 main weights and optimizer states over those $D_E$ replicas.
Because the original data-parallel group has size $D=M_ED_E$, expert optimizer
state is still divided across the same total degree $D$ after the two stages of
sharding. EP's additional memory saving is therefore in compute weights and
gradient buffers; it is not another factor applied to optimizer state that was
already sharded over $D$. Section~\ref{sec-expert-parallelism} introduces the
EP-MP and EP-DP topology and the dispatch/combine traffic paid for this saving.

\subsubsection{PP Shards Layers and Their Persistent Training State}

EP divides the expert pool inside a layer, but every rank may still hold too
many layers, dense weights, shared experts, and activations. PP assigns
different layer ranges and their weights, gradient buffers, and optimizer
shards to different stages, reducing this persistent training-state memory in each
rank-local partition. It introduces a different
tax: activation transfers, pipeline bubbles, and multiple in-flight
microbatches. Section~\ref{sec-pipeline-parallelism} develops those schedules
and memory trade-offs. Together, the distributed optimizer, EP, and PP make
DDP practical at model sizes where ordinary replicated DDP is not.

\subsection{Olmo-Owned DDP}

The Olmo-core DDP training stack includes both replica synchronization and the
update path around it. PyTorch's standard
\texttt{DistributedDataParallel} wrapper is designed around one process group and ordinary
\texttt{Parameter.grad} handling %
\citep{li2020pytorchdistributed,pytorch2026ddpdocs}.
Olmo combines separate execution and update precision, parameter-specific
replica groups after EP, sharded optimizer states, derived MXFP8 representations,
and explicit completion after gradient accumulation or a pipeline schedule.
Each piece can be assembled from PyTorch mechanisms, but operating them as
one coordinated update path requires integration deep enough that Olmo
implements and controls the DDP path directly rather than layering it around
the standard wrapper.

\subsubsection{Hybrid Precision Control}

Execution precision and update precision serve different goals. BF16 and
MXFP8 reduce model execution and storage cost. Gradient accumulation sums
contributions across microbatches, gradient reduction combines ranks, and the
optimizer integrates updates over many steps. Mixed-precision training keeps
an FP32 master copy of the weights for updates and performs large sums in FP32 %
\citep{micikevicius2018mixedprecision}; we apply the same principle to
gradient accumulation and reduction. In the
Olmo-core DDP configuration described here, model execution
uses BF16 or selected MXFP8 paths, while normal parameter gradients accumulate
in dedicated FP32 bucket views, reduce in FP32, and feed FP32 main weights and
optimizer states.

This separation is more specific than generic mixed-precision execution. For
an ordinary BF16 parameter, autograd's leaf \texttt{AccumulateGrad} node first
produces a temporary BF16 \texttt{param.grad}. A post-accumulate hook
immediately adds that gradient into the parameter's FP32 accumulation buffer
and clears \texttt{param.grad}. Each FP32 accumulation buffer is a view into a
flat bucket built before training, and gradient reduction operates directly on
that bucket.\footnote{The wrapper constructs buckets by replica group, gradient-storage
dtype, communication dtype, and size limit. It binds each parameter's
\texttt{\_main\_grad\_fp32} to the corresponding slice of the bucket's flat
FP32 storage. Under the FP32-accumulation and FP32-reduction policy described here,
the all-reduce consumes that flat storage directly. This follows the standard
DDP bucket pattern; Olmo extends it to multiple replica groups and explicit
storage and communication dtypes.} The policy described here uses FP32
accumulation and FP32 reduction. After the optimizer updates
an FP32 main weight, the BF16 or MXFP8 compute-weight representation is
refreshed. MXFP8 weight stores use the optimizer-integrated path
described below.

\subsubsection{EP Requires Multiple DDP Groups}

EP makes a single process group insufficient for one model part. Dense
parameters are replicated across the dense data-parallel group. Ranks along
EP-MP hold different expert shards and must not average those shards together;
only EP-DP ranks holding copies of the same expert shard reduce its
  gradient. A single standard PyTorch DDP wrapper selects one process group for all of its
parameters, whereas the MoE model part contains both kinds of parameter.

For parameters handled by the bucketed DDP path, Olmo chooses a replica group
per parameter and forms buckets only among parameters with compatible
groups and storage and communication dtypes. Non-final microbatches accumulate
locally. Explicit finalization launches the remaining bucket reductions in a
stable order and preserves gradients from experts that appeared earlier in the
window but not in its final microbatch. Section~\ref{sec-expert-parallelism} derives the topology that
these groups implement.\footnote{The current implementation is named
\texttt{MultiGroupDistributedDataParallel}.}

\subsubsection{Native MXFP8 Integration}

An MXFP8 compute-weight cache is not independently updated. The FP32 main
weight is the source used to rebuild the derived, orientation-specific MXFP8
caches for forward and dgrad. After an update, the optimizer gathers any
sharded FP32 main weight, reconstructs the full weight, and refreshes
the caches coherently. The current path casts through a BF16 compute-weight
representation before quantization, so we describe the caches as
\emph{derived from} the FP32 main weight rather than as a direct FP32-to-MXFP8
conversion. Section~\ref{sec-mxfp8-training} covers formats, scales, and kernels, together with the
cache formats and refresh points.

\subsubsection{Collectives and Weight Reconstruction Stay Outside the Compiled Update}

Olmo keeps collective launches and compute-weight reconstruction outside the
compiled optimizer computation. This separates local shard updates from the
communication and refresh phases that surround them, making their order
explicit and keeping each phase visible in the step timeline. The same update
path coordinates gradient clipping with the optimizer step and performs
per-group initialization checks.

\subsection{DDP and FSDP Optimize Different Bottlenecks}

FSDP minimizes
long-lived unsharded parameter storage by materializing weights around use. Olmo-core DDP keeps
its rank-local compute-weight partition, which must fit in memory, materialized at stable locations across accumulation and moves
gradient synchronization and optimizer reconstruction to the global-step
boundary. The distributed optimizer lowers FP32 main-weight and
optimizer-state memory, EP divides the routed-expert pool, and PP divides
layers. Olmo's implementation integrates the precision, replica-group, and
derived-representation refresh rules needed by that composition.

Dense Olmo continues to use FSDP/HSDP. When useful
compute grows with materialized parameters or peak parameter, gradient, and
optimizer memory is the
primary constraint, FSDP or HSDP remains an effective foundation.
The current single-node sweeps establish two local scaling patterns in the
eight-layer model: accumulation amortizes more of DDP's step-boundary work, and
increasing expert parameter size raises pressure on an unsharded local partition.
They do not establish a universal sparsity crossover, optimization quality, or
time to a target loss.

The next section develops EP as the model-parallel mechanism that makes the
routed-expert partition fit. It explains which ranks hold different expert
shards, which ranks hold replicas of the same shard, and why saving weight and
gradient memory introduces dispatch and combine on the microbatch clock.

\FloatBarrier
\phantomsection\label{sec-ddp-vs-fsdp-end}%

\section{Expert Parallelism}
\label{sec-expert-parallelism}

\conceptlead{EP reduces routed-expert storage per rank.} DDP synchronizes model
replicas, but it does not reduce the size of each replica. Without model
parallelism, every DDP rank stores all $N$ routed experts even though each token
uses only $K$. Expert parallelism (EP) distributes the routed-expert pool
across the existing DP ranks. Each rank stores its assigned expert shard, and
selected token rows move to the ranks that store their experts. This reduces
per-rank expert weights and gradient buffers, but adds dispatch and combine
communication to every routed layer.

\conceptlead{Olmo-core introduces rowwise EP.} Conventional all-to-all EP
moves tokens in large contiguous blocks, but constructing those blocks
requires extra data reordering and can expose dynamic route sizes to the CPU.
Rowwise EP keeps routing on the GPU and moves each accepted row directly to
its destination, eliminating block packing and the associated host
synchronization from Olmo-core's EP path. NVSHMEM symmetric memory and
GPU-initiated communication make this direct placement possible.
Section~\ref{sec-hardware-topology} connects it to the underlying network.

\subsection{EP Shards Models That Do Not Fit}
\label{subsec-ep-shards-models}

\conceptlead{Persistent training-state memory grows with total
parameters.\footnotemark}\footnotetext{Relative to a dense model with the same total
parameter count, activating only $K$ of $N$ experts reduces per-token
computation but not the bytes required to store the complete parameter set and
its persistent training state.}
Top-$K$ routing limits how many experts compute for one token, but it does not
reduce how many expert weights a rank must store. Without model parallelism,
the rank-local model partition is the whole model, so ordinary DDP places all
$N$ routed experts on every one of its $D$ ranks. Increasing $N$ can therefore
increase total parameter size without increasing active expert work, yet still grow
the compute weights and gradient buffers until one replica no longer
fits. EP addresses this memory limit by partitioning the routed-expert pool,
as in established large-scale MoE systems
\citep{lepikhin2021gshard,rajbhandari2022deepspeedmoe}.

\conceptlead{EP uses the existing DP ranks.} EP shards expert weights, but it
does not require a separate set of ranks in addition to the data-parallel
ranks. In the base topology, the same $D$ ranks that form the dense
data-parallel group are organized into an \texttt{EP\_MP} rank axis of size
$M_E$ and an \texttt{EP\_DP} rank axis of size $D_E$. Let
$\mathcal{A}_{\mathrm{EP\_MP}}$ and $\mathcal{A}_{\mathrm{EP\_DP}}$ denote the
coordinate sets of these axes. The DP group is their Cartesian product:

\begin{equation}
    \mathcal{G}_{\mathrm{DP}}
    = \mathcal{A}_{\mathrm{EP\_MP}}
      \times \mathcal{A}_{\mathrm{EP\_DP}},
    \qquad D = M_E D_E.
    \label{eq-ep-cartesian-product}
\end{equation}

\noindent Each pair of coordinates identifies one rank in the DP group.
Dense parameters remain replicated across all $D$ ranks. Routed-expert
parameters are sharded across \texttt{EP\_MP} and replicated across
\texttt{EP\_DP}. Equation~\ref{eq-ep-cartesian-product} covers the case
without CP. Appendix~\ref{app-context-parallelism} derives the folded case,
where a CP degree $C$ gives $DC=M_E D_E$ within each pipeline stage.

\begin{figure}[!htb]
\centering
\includegraphics[width=0.95\linewidth]{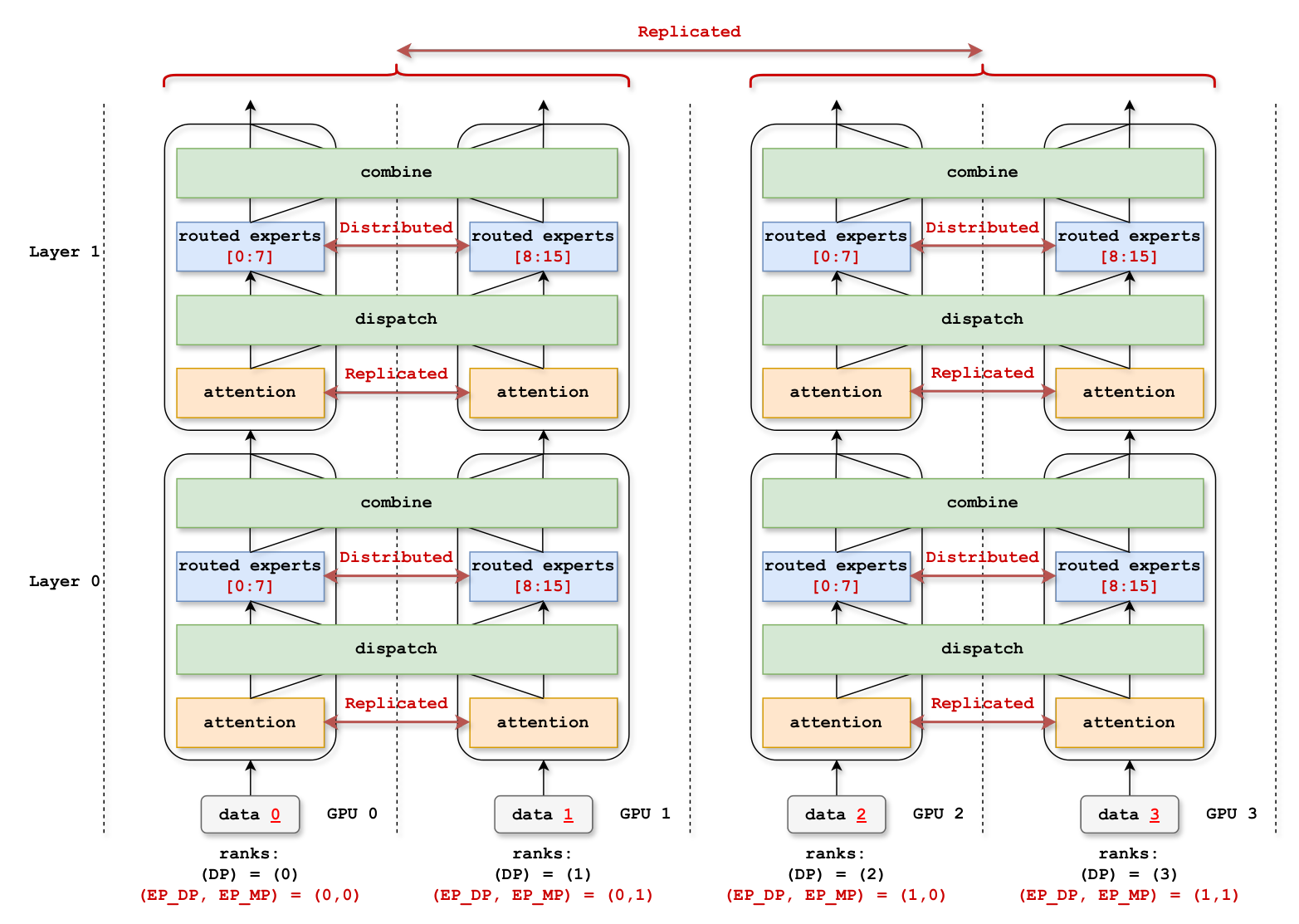}
\caption{\textbf{A four-GPU example with DP${}=4$, EP-MP${}=2$, and 16 total
experts.}
Starting from the fully replicated layout in
Figure~\ref{fig-ddp-topology-four-gpu}, the same four DP ranks form EP-MP groups
of size 2 and EP-DP groups of size 2. Each EP-MP group keeps attention
replicated and shards experts $0$--$7$ and $8$--$15$ across its two ranks.
Dispatch moves selected token rows to the appropriate expert rank, and combine
returns their contributions. Ranks holding corresponding shards in the two EP-MP groups
are EP-DP replicas and reduce gradients together.}
\label{fig-ep-topology-four-gpu}
\end{figure}

The four-GPU example in Figure~\ref{fig-ep-topology-four-gpu} shows this rank
organization. Begin with $D=4$ ranks and $N=16$ routed experts. Set
$M_E=2$ and $D_E=2$. Each two-rank EP-MP group collectively holds one complete
routed-expert set: one rank holds experts 0--7, and the other holds experts
8--15. There are two such EP-MP groups processing different data, so each
expert shard has two copies. Attention, the router, and other non-sharded
parameters remain replicated across the four dense-DP ranks.

\conceptlead{A route connects a token-source rank to an expert rank.} The
token-source rank holds the token before dispatch and receives its combined
contribution. The expert rank stores and executes the selected expert. These
are per-route roles, not additional parallel axes.

\begin{definitionbox}[EP Group Definitions]
\small
\begin{tabularx}{\linewidth}{@{}X@{\hspace{14pt}}X@{}}
\textbf{Expert-sharding group (\texttt{EP\_MP}).}
Ranks in this group hold different expert shards and exchange routed token rows
during dispatch and combine. Because their expert parameters differ, they do
not reduce expert gradients together.
&
\textbf{Expert-replica group (\texttt{EP\_DP}).}
Ranks in this group hold replicas of the same expert shard and reduce that
shard's gradients together.
\end{tabularx}
\end{definitionbox}

\conceptlead{EP reduces expert memory and gradient traffic.}
Figure~\ref{fig-ddp-static-memory} previewed this decomposition under the
report's BF16-execution and FP32-optimizer policy. Abstracting from that
specific precision policy, let $P_E$ denote the routed-expert parameter count
in the complete model. With an even expert assignment, each rank holds
approximately

\begin{equation}
    \text{expert weights per rank} \approx \frac{P_E}{M_E},
    \qquad
    \text{expert-gradient elements per rank} \approx \frac{P_E}{M_E}.
    \label{eq-ep-local-expert-cost}
\end{equation}

\noindent A rank synchronizes only its local expert shard over $D_E$
replicas, rather than reducing the full expert pool over all $D$ ranks. For a
fixed EP-DP group size and reduction algorithm, the communicated expert-gradient
volume therefore falls with $P_E/M_E$; the exact network bytes also depend on
$D_E$ and the collective algorithm. Dense, router, shared-expert, and other non-sharded
parameters do not receive this EP-MP reduction; they continue to follow the
dense DDP group. At fixed expert width, local expert weights and gradient
buffers still grow as $N/M_E$. Keeping these per-rank tensors bounded as $N$ grows therefore
requires increasing $M_E$.

\begin{figure}[!htb]
\centering
\includegraphics[width=\linewidth]{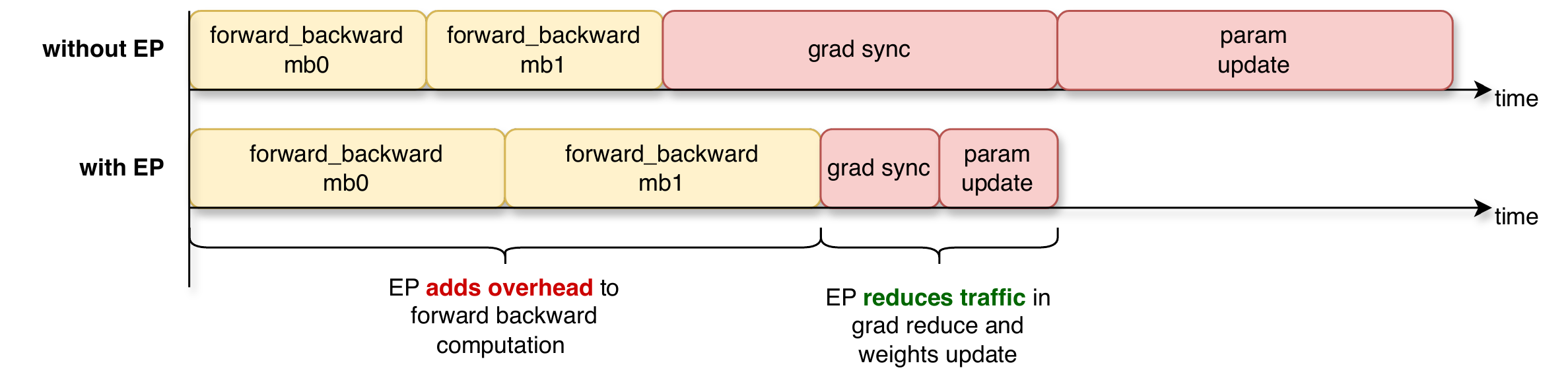}
\caption{\textbf{EP trades per-microbatch routing overhead for less
parameter-sized work at the optimizer-step boundary.} EP adds dispatch and
combine to forward and backward, but each rank synchronizes and updates only
its local routed-expert shard. With few accumulation microbatches and a large
routed-expert pool, the reduction in exposed gradient-synchronization and
parameter-update time can outweigh the added EP work and make end-to-end
training faster. The widths are conceptual and not measured timings. Dense,
router, shared-expert, and other non-sharded parameters do not receive the
EP-MP reduction; optimizer-state work may be sharded further over EP-DP.}
\label{fig-ep-reduce-traffic}
\end{figure}

\conceptlead{EP can reduce end-to-end step time when parameter-sized work
dominates.} Dispatch and combine add work to every routed layer and
microbatch, whereas gradient synchronization and parameter update occur once
per optimizer step. When the gradient-accumulation count is small and the
routed-expert state is large, reducing the rank-local expert shard from $P_E$
to approximately $P_E/M_E$ can reduce the exposed expert-gradient and update
phase enough to outweigh the added per-microbatch EP work. The exact optimizer
cost depends on its state placement: optimizer state for the local EP-MP shard
may be sharded further over EP-DP. Figure~\ref{fig-ep-reduce-traffic} therefore
shows a possible operating regime, not a topology-independent speedup.

\begin{minipage}{\linewidth}
One single-node eight-B300 run illustrates this operating regime. The run
used 32 routed experts, top-4 routing, fixed random routes, eight rank-local
microbatches per optimizer step, and 15 measured steps
(Table~\ref{tab-ep-memory-throughput-seed}).

\begin{center}
\small
\centering
\begin{tabular}{@{}lrr@{}}
\toprule
\textbf{Execution path} & \textbf{Tokens/s/GPU} & \textbf{Peak active memory} \\
\midrule
No EP & 53,512 & 129.3 GiB \\
Block all-to-all EP8 & 54,063 & 77.85 GiB \\
\bottomrule
\end{tabular}
\vspace{4pt}
\captionof{table}{\textbf{One intra-node EP feasibility point.} Conventional
block all-to-all EP8 (Section~\ref{subsec-ep-block-alltoall}) substantially
reduces peak active memory while preserving
similar end-to-end throughput at this operating point. Both runs completed
cleanly.}
\label{tab-ep-memory-throughput-seed}
\end{center}
\end{minipage}

In this run, EP reduced peak active memory by 51.45~GiB at similar
throughput. As a single end-to-end feasibility point, it does not show
whether reduced expert-gradient synchronization and optimizer-boundary work
caused the 1.0\% throughput difference, nor does it measure variance or
generalize beyond this intra-node configuration. Section~\ref{subsec-ep-block-alltoall} examines the costs of
the block all-to-all implementation used for this comparison.

\conceptlead{Per-rank routed activations stay roughly constant as the EP-MP degree grows.} Let
each rank originate $T$ tokens before routing. Without EP, the rank processes
approximately $TK$ routed rows. An EP-MP group of size $M_E$ contains $M_E T$
source tokens and distributes the resulting $M_E TK$ routes over $M_E$ expert
ranks. Under balanced routing, each rank therefore still processes
approximately $TK$ routed rows. Increasing $M_E$ reduces the number of local
experts and their persistent training state, but not the average routed rows
or capacity-sized communication buffers per rank. Once the local expert shard
is small, activations and other execution working memory set a floor, so
larger EP-MP degrees provide diminishing peak-memory savings.

\conceptlead{Dense and expert parameters use different reduction groups.} A
dense gradient must be reduced across all $D$ dense-DP replicas. An expert
gradient must instead be reduced across the $D_E$ ranks that hold copies of
that exact EP-MP shard. Reducing expert gradients across EP-MP would mix
different experts; reducing dense gradients only across EP-DP would leave
dense replicas out of agreement. The multi-group reducer introduced in
Section~\ref{sec-ddp-vs-fsdp} assigns each gradient bucket to its corresponding
replica group.

\begin{figure}[!htb]
\centering
\includegraphics[width=0.45\linewidth]{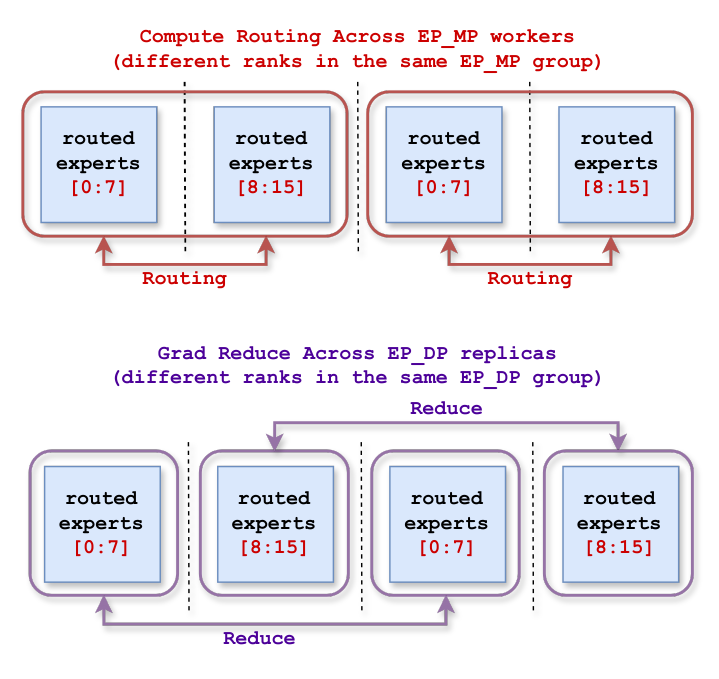}
\caption{\textbf{EP-MP routes tokens; EP-DP reduces gradients.} Within each
EP-MP group, different ranks hold different expert shards, so routed token rows
move between those ranks for expert computation. Within each EP-DP group, ranks
holding replicas of the same expert shard reduce that shard's gradients together:
experts $0$--$7$ reduce with their replicas, and experts $8$--$15$ reduce with
theirs. Token communication and gradient reduction therefore use different
rank groups.}
\label{fig-ep-mp-vs-ep-dp}
\end{figure}

Figure~\ref{fig-ep-mp-vs-ep-dp} shows both group types for the four-rank
example.

\conceptlead{EP adds token communication.} Before EP, every chosen expert is
local to the token-source rank. After the expert pool is sharded, a remote
route must dispatch its input row to the expert rank and combine the resulting
contribution back at the token-source rank. This communication occurs in every
routed layer and every microbatch, not only once at the global-step boundary.
The reduced expert memory and gradient traffic therefore come with additional
dispatch and combine communication. The next subsection compares these
execution paths.

\subsection{EP Adds Communication to Dispatch and Combine}
\label{subsec-ep-dispatch-combine}
\conceptlead{MoE dispatches and combines even when every expert is local.}
A dense MLP applies the same local computation to every token. It has no
router, expert-order permutation, or token communication. An MoE layer without
EP first chooses experts for each token, permutes the accepted routes into
expert-contiguous order, executes the grouped expert MLPs, and then unpermutes
and combines the expert outputs. The permutation is local dispatch; the final
unpermutation and weighted reduction are local combine. Grouped GEMM requires
this expert-contiguous layout even though no rows cross ranks.

\begin{figure}[!htb]
\centering
\includegraphics[width=\linewidth]{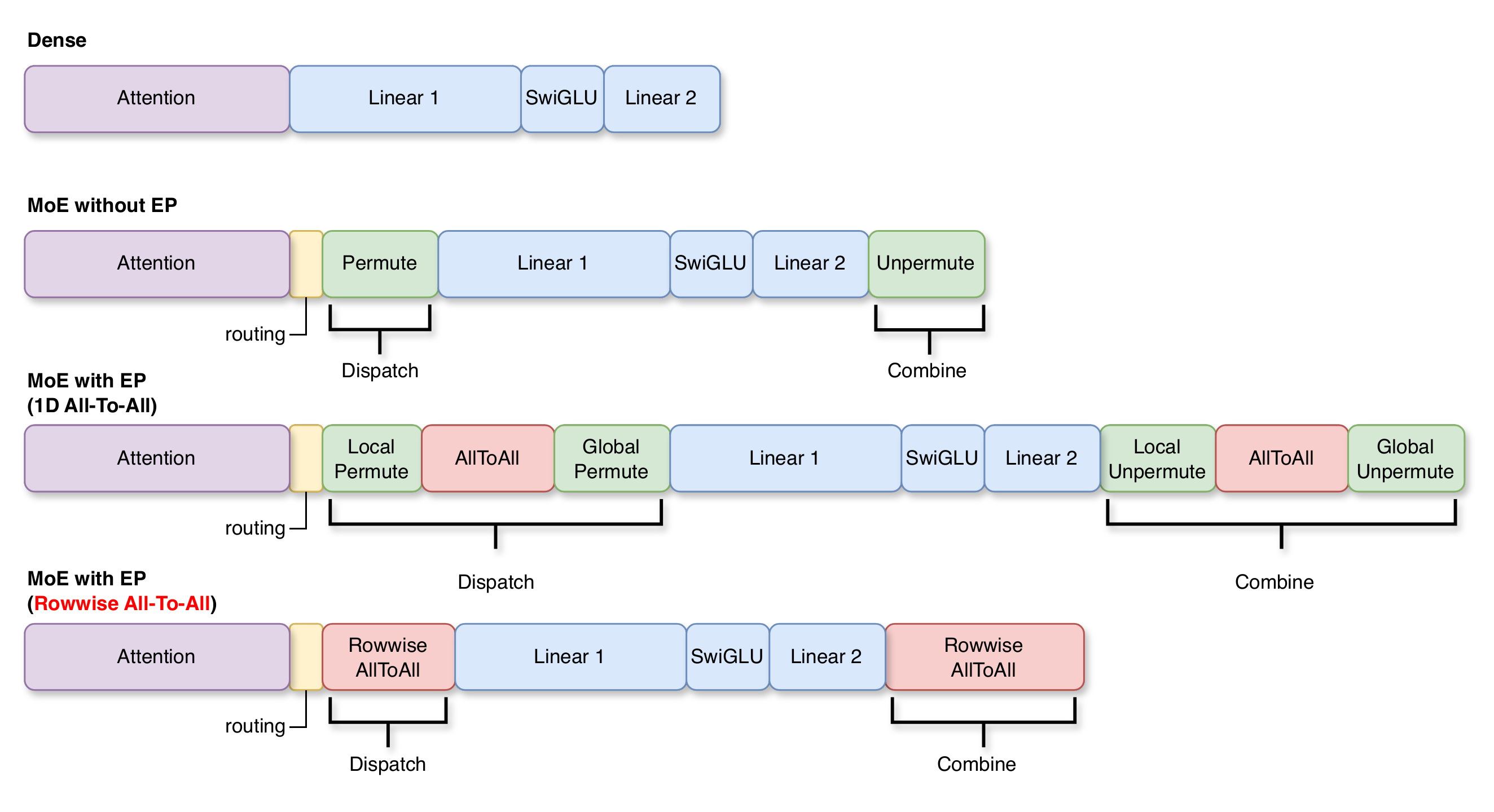}
\caption{\textbf{Routing overhead grows from dense execution to local MoE and
conventional all-to-all EP.} A dense MLP has no routing path. MoE without EP adds one local
permutation into expert order and one unpermutation back to token order.
Conventional all-to-all EP expands dispatch into permute--all-to-all--permute and combine
into unpermute--all-to-all--unpermute. Rowwise EP instead attempts to move
selected rows directly between token-source and expert ranks. The drawing is
not performance evidence.}
\label{fig-dense-vs-moe-execution}
\end{figure}

\conceptlead{EP extends dispatch and combine with an all-to-all exchange.} Conventional
all-to-all EP extends local dispatch to
permute--all-to-all--permute and local combine to
unpermute--all-to-all--unpermute. Figure~\ref{fig-dense-vs-moe-execution}
compares this operator path with dense and local-MoE execution.
Section~\ref{subsec-ep-block-alltoall} derives the intermediate layouts and
identifies the two additional peer-major/expert-major conversions beyond the
local-MoE path.

\conceptlead{Conventional all-to-all EP adds layout and synchronization work
beyond the transfer itself.} Relative to
local MoE, it adds a rank exchange, up to two additional layout
passes, peer-packed buffers, launches, and synchronization around the useful
expert GEMMs. These EP-specific costs can remain on the critical path even when
network bandwidth is sufficient;
Section~\ref{subsec-ep-block-alltoall} traces them to concrete operations.

Comparing these execution paths fairly requires giving every path the same
inputs, routes, and expert work; otherwise a routing or workload difference
can be mistaken for an execution-path improvement. The profiled comparison in
Section~\ref{subsec-rowwise-ep-motivation} matches these factors. End-to-end
training speed is a separate question, because EP also changes the
gradient-synchronization and optimizer work at the step boundary.

\subsection{Packed All-to-All Requires Peer-Contiguous Blocks}
\label{subsec-ep-block-alltoall}
\begin{figure}[!htb]
\centering
\includegraphics[width=\linewidth]{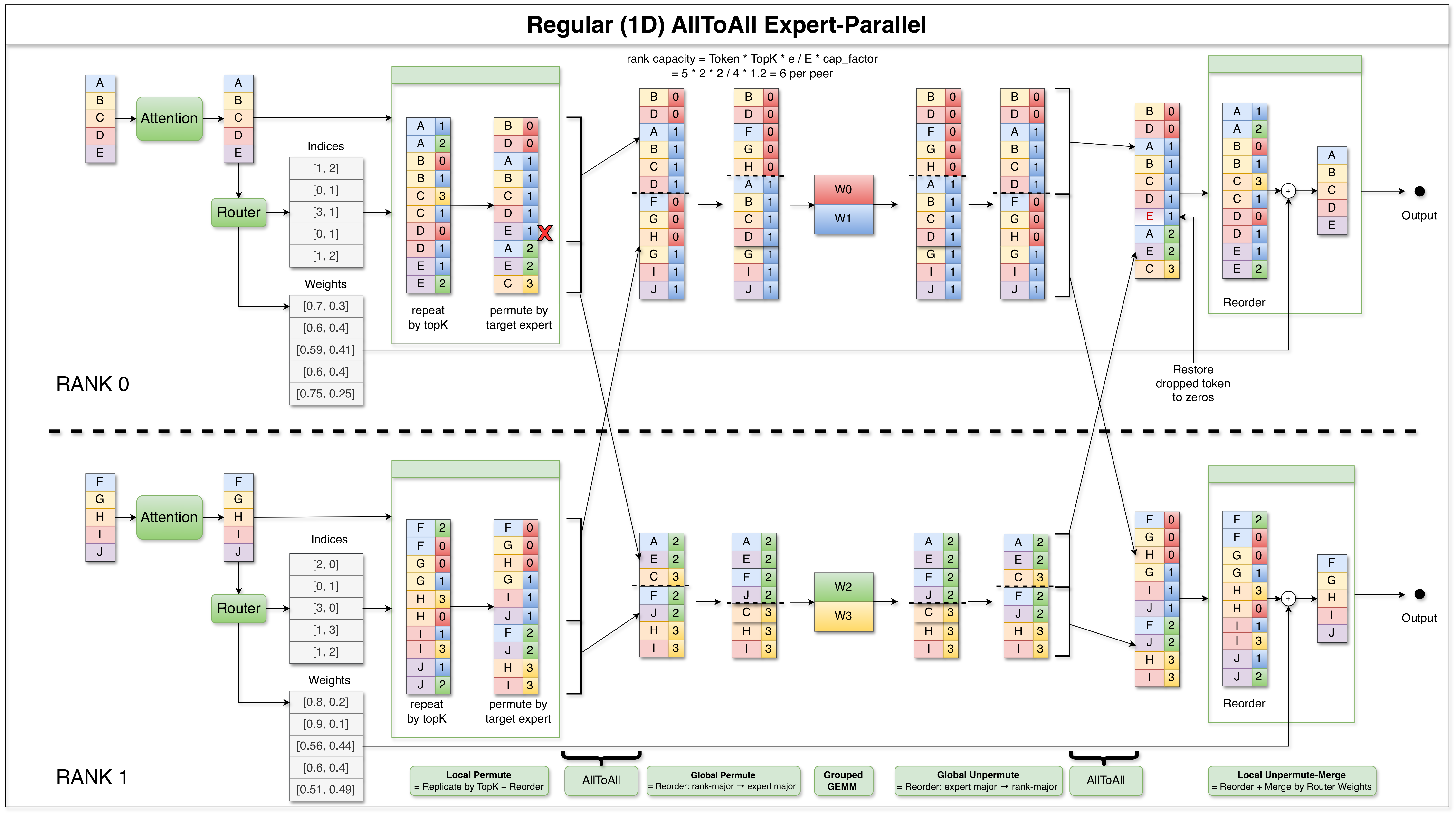}
\caption{\textbf{Block all-to-all EP exposes the complete packing path.}
In this illustrative two-rank, top-2 example, local dispatch replicates token
rows and packs routes by destination rank. The first all-to-all exchanges
peer-contiguous blocks, and a global permutation converts the received
rank-major layout into local-expert order for grouped GEMM. Combine reverses
the global layout conversion and all-to-all, restores token order, fills a
capacity-dropped route with zero, and merges route contributions using the
router weights. The drawing explains data layout and movement, not measured
performance.}
\label{fig-ep-1d-alltoall-detail}
\end{figure}

\conceptlead{Block all-to-all moves one contiguous block per peer.}
We call the conventional path \textbf{block all-to-all EP}: its transport
moves one contiguous block of rows per peer rank, never individual token
rows. The path expresses dispatch with PyTorch variable-split
\texttt{all\_to\_all\_single} on an NCCL process group
\citep{pytorch2026distributed,nvidia2026ncclcollectives}. Each rank supplies one
flat send buffer partitioned into a contiguous slice per destination and
receives a contiguous slice per source. The call cannot scatter arbitrary token
rows directly into arbitrary remote expert slots. The sender must therefore
pack all accepted routes for destination rank 0, then destination rank 1, and
so on. This peer-block packing is a common expert-parallel implementation
pattern %
\citep{hwang2023tutel,nvidia2026megatroncoremoe}.
Figure~\ref{fig-ep-1d-alltoall-detail} traces the complete packing path for a
two-rank example.

\par
\conceptlead{Expert order requires a second layout change.} After the transfer,
the receive buffer is grouped by source rank, not by the local expert that will
process each row. A second permutation places those rows into
expert-contiguous intervals before grouped GEMM. Combine reverses the layout
changes: it converts expert order into source-rank blocks, returns those blocks
through the reverse all-to-all, and then restores token order and applies the
router weights. The outer token-to-expert and expert-to-token conversions are
already present in local MoE. With multiple local experts, packed peer exchange
can require two additional peer-major/expert-major conversions, for four passes
total; with one local expert, the extra pair is skipped.

\conceptlead{Dynamic peer-block lengths create a host dependency in the current
path.} Because routing changes the peer-block lengths every microbatch,
Olmo-core's implementation converts GPU-produced peer counts into host split
lists before launching the variable-split exchange; when we emphasize this
host dependency, we call it the \emph{synchronized} path.
Section~\ref{sec-sync-d2h} explains how this device-to-host dependency reduces
submission headroom and how independent work can cover, but not remove, the
wait.

\conceptlead{Faster links shrink only the all-to-all itself.} The
peer/expert layout conversions, peer-packed buffers, and this
implementation's host dependency remain. Network bandwidth alone therefore
cannot explain the cost of the complete block all-to-all path.

\subsection{Rowwise EP Removes Block Packing}
\label{subsec-rowwise-ep-motivation}
\conceptlead{Rowwise EP transfers each row directly to its expert slot.}
\textbf{Rowwise EP} replaces block packing with direct row placement: each
accepted token row is transferred on its own, straight into the destination
buffer row where its selected expert expects it. For $T$
local tokens and top-$K$ routing, Olmo-core constructs two GPU tensors of shape
$[T,K]$. One gives the destination EP-MP rank for each accepted route; the
other gives the row in that rank's expert-capacity buffer. Negative entries
represent invalid or capacity-dropped routes;
Section~\ref{sec-routing-rank-capacity} defines the capacity policy and its
drop metrics, and in stable training the routing distribution stays nearly
dropless. A dispatch kernel uses these
two indices to move $x_t$ directly to its destination row. The communication
output is already expert-contiguous on each destination rank, so grouped expert
computation does not require a packed peer buffer or a second permutation.

The destination row is deterministic. The route-map builder combines the
destination expert, the number of earlier accepted routes for that expert, and
the offsets of the preceding local experts. It also applies the capacity
decision before communication. The map therefore records both the remote
destination rank and the precise row consumed by the selected expert. Route
counts and overflow metadata remain on
the GPU; they are not converted into host split lists for the data movement.
Direct placement presupposes that a kernel on the token-source rank can write
a row that lives on the expert rank.
Section~\ref{subsec-ep-symmetric-memory} introduces the symmetric-memory and
GPU-initiated communication (NVSHMEM/RDMA) mechanisms that make remote rows
addressable.

\conceptlead{Rowwise EP adds indexing and ordering costs.} Rowwise movement
replaces a small number of block transfers with many indexed row transfers and
adds route-map construction, fixed-capacity storage, and explicit
write-ordering between communication and compute. It is beneficial only when
those costs are lower than the permutations,
temporary buffers, and host stalls of the block all-to-all path. Its end-to-end advantage
depends on expert shape, EP degree, route imbalance, and network topology,
which the current measurements only partly cover.

\begin{figure}[!htb]
\centering
\includegraphics[width=\linewidth]{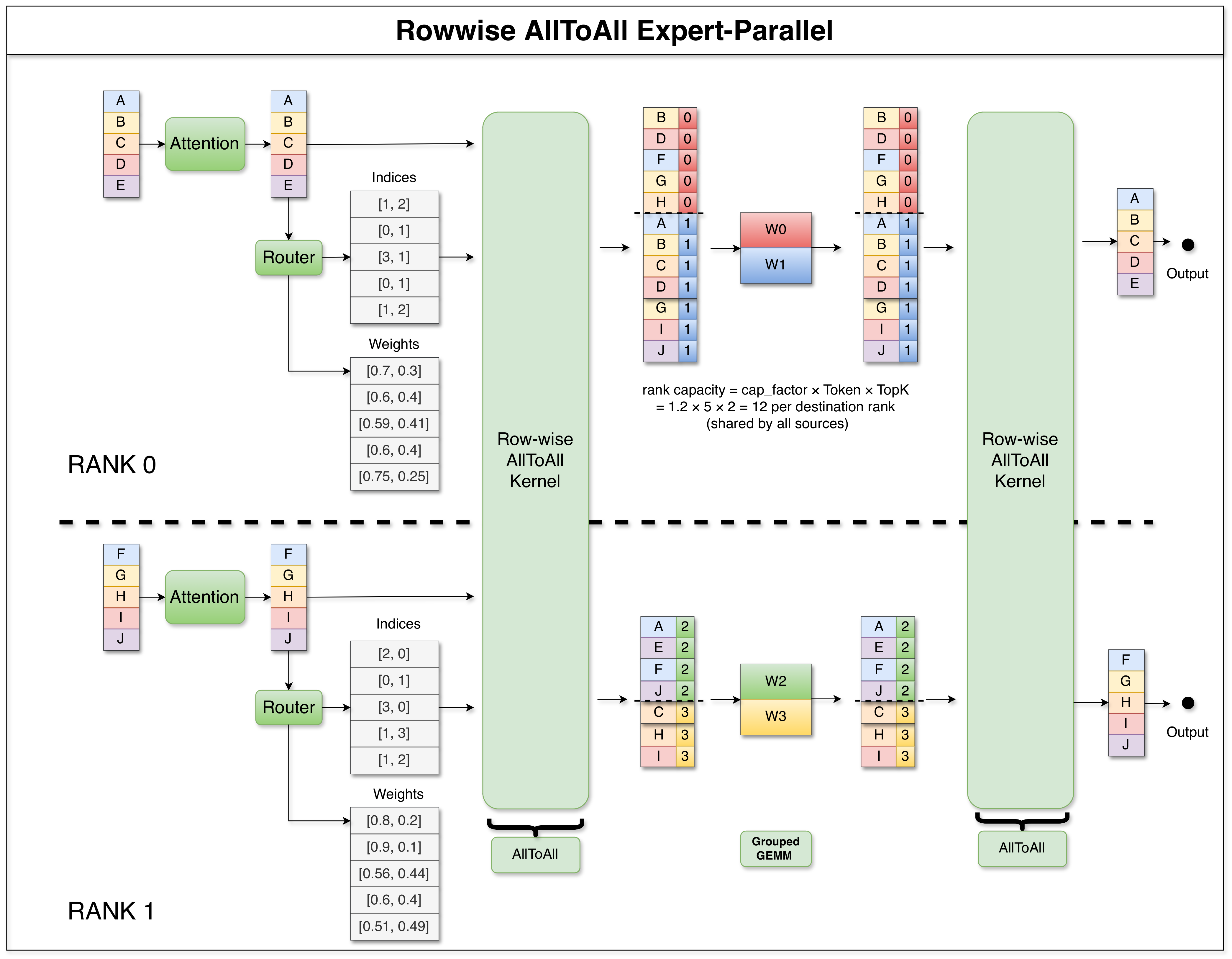}
\caption{\textbf{Rowwise EP targets remote expert rows without block
packing.} In the same illustrative two-rank, top-2 example as
Figure~\ref{fig-ep-1d-alltoall-detail}, GPU-resident route indices drive a
rowwise dispatch kernel that places accepted rows directly into fixed-capacity
rank- and expert-major storage. Grouped GEMM consumes that layout without a
second permutation. Rowwise combine returns expert outputs directly to their
source-token positions for weighted merging, without materializing a
peer-packed return buffer. The ``AllToAll'' labels inside the drawing denote
the logical rank exchange; the implementation uses indexed one-sided row
movement rather than an NCCL all-to-all collective. As in
Figure~\ref{fig-ep-1d-alltoall-detail}, token E's route to expert 1 does not
fit in rank 0's 12-row capacity, so it is dropped and combine treats it as
zero. The drawing explains data
layout and movement, not measured performance.}
\label{fig-ep-rowwise-detail}
\end{figure}

Figures~\ref{fig-ep-1d-alltoall-detail} and
\ref{fig-ep-rowwise-detail} make the layout difference explicit. The
block all-to-all path packs accepted routes into contiguous
destination-rank blocks, exchanges those blocks, and then reorders the
received rows into local-expert order. Combine reverses both layout
conversions. Rowwise EP instead uses the GPU route map to place each accepted
row directly in its destination expert-capacity slot and reverses that indexed
movement during combine. It removes packed peer buffers and up to four explicit
layout conversions; it does not remove routing, communication, grouped expert
computation, or the synchronization that orders remote writes between ranks.

Table~\ref{tab-ep-sync-rowwise-profile-snapshot} groups the displayed ranges
from one inspected rank-0 MoE forward instance in each of two
configuration-matched traces, each captured on one node with eight B300 GPUs.
Both use an eight-layer BF16 model with 128 routed experts, top-8 routing, no
shared expert, EP8, and four sequences per rank microbatch. The traces were
captured on different nodes, so the comparison is mechanism evidence rather
than a controlled performance result.

\begin{table}[!htb]
\centering
\small
\renewcommand{\arraystretch}{1.14}
\setlength{\tabcolsep}{5pt}
\begin{tabularx}{\linewidth}{@{}
  >{\bfseries\raggedright\arraybackslash}p{0.09\linewidth}
  >{\raggedright\arraybackslash}X
  >{\raggedleft\arraybackslash}p{0.075\linewidth}
  @{\hspace{14pt}}
  >{\raggedleft\arraybackslash}p{0.075\linewidth}
  >{\raggedleft\arraybackslash}X@{}}
\toprule
Phase & \multicolumn{2}{c}{\textbf{Block all-to-all EP}} &
  \multicolumn{2}{c}{\textbf{Rowwise EP}} \\
\cmidrule(lr){2-3}\cmidrule(lr){4-5}
& Operation & ms & ms & Operation \\
\midrule
\multirow{3}{*}{Dispatch}
  & Local permute & 0.351
  & \multirow{3}{*}{\textbf{2.298}}
  & \multirow{3}{=}{Route-map build + row dispatch} \\
  & All-to-all & 1.960 & & \\
  & Global permute & 0.532 & & \\
\midrule
Compute
  & Grouped GEMM & \multicolumn{2}{c}{3.630} & \\
\midrule
\multirow{3}{*}{Combine}
  & Global unpermute & 0.712
  & \multirow{3}{*}{\textbf{2.423}}
  & \multirow{3}{=}{Row gather + weighted reduce} \\
  & All-to-all & 1.970 & & \\
  & Local unpermute + merge & 0.391 & & \\
\midrule
\rowcolor{white}
\multicolumn{2}{@{}l}{\textbf{Dispatch + combine}}
  & \textbf{5.916} & \textbf{4.721} & \\
\bottomrule
\end{tabularx}
\caption{\textbf{A single profiled forward example localizes the rowwise
saving.}
The rowwise path is 20.2\% shorter across the displayed non-expert ranges
while the displayed routed-expert computation is identical.
Durations are manually transcribed from one Nsight Systems instance per path,
not repetition means or an additive decomposition of complete step time.}
\label{tab-ep-sync-rowwise-profile-snapshot}
\end{table}

\begin{figure}[!htb]
\centering
\includegraphics[width=\linewidth]{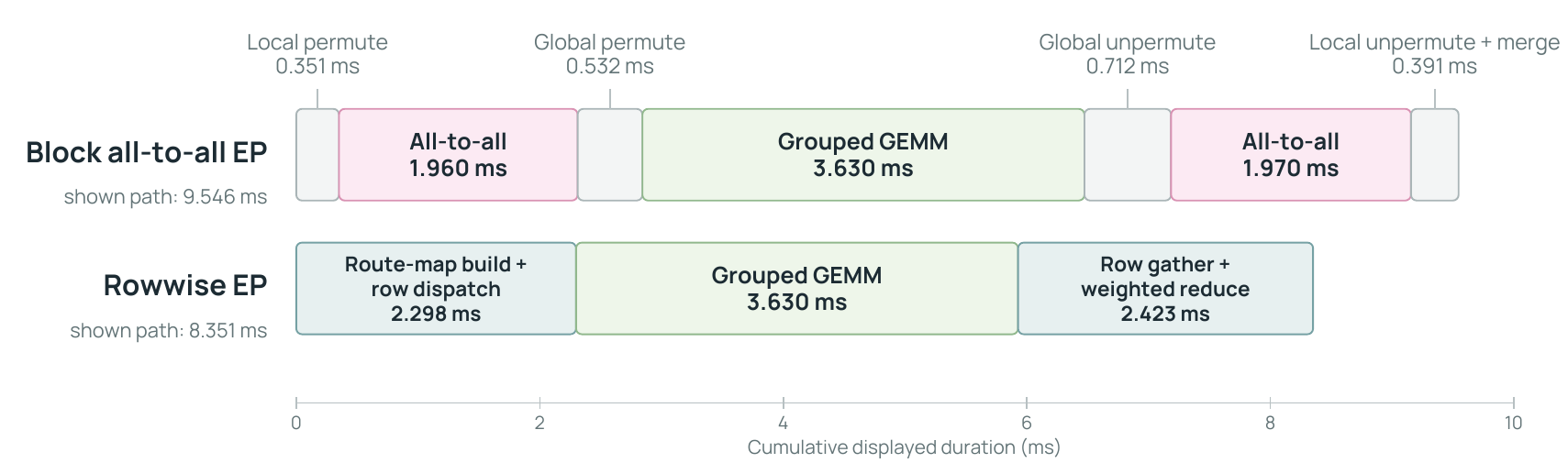}
\caption{\textbf{Rowwise EP removes the block-packing stages around expert
computation.} Block all-to-all EP exposes separate layout and collective
stages in dispatch and combine. Each rowwise phase aggregates two displayed
subranges: route-map construction plus row dispatch, or row gather plus
weighted reduction. Horizontal width shows cumulative displayed duration
within each trace, not temporal alignment between traces. The
traces were captured on different nodes, so this is mechanism evidence rather
than a controlled performance comparison.}
\label{fig-ep-profile-path-anatomy}
\end{figure}

Figure~\ref{fig-ep-profile-path-anatomy} shows the same example as two
cumulative-duration paths: three explicit block-path stages on each side of
expert computation versus one aggregated rowwise phase.

This example is consistent with the intended mechanism: rowwise replaces four
explicit layout transformations and two peer-block collectives with a
GPU-resident route map and indexed row movement, with device route-map
construction included in the rowwise total. The displayed ranges exclude the
block path's host split-list wait (Section~\ref{sec-sync-d2h}). This instance
identifies the affected operations but does not establish the size of the
saving across workloads and networks.

\par\noindent\begin{minipage}{\linewidth}
\conceptlead{Direct placement also lets communication use the MXFP8 representation consumed by grouped GEMM.}
Olmo can quantize rows during dispatch, let expert grouped GEMM consume the
received qdata (quantized data) and scales directly, and return quantized
outputs before local reduction. Communication and expert compute thus share
one representation without an intervening BF16 materialization.
Section~\ref{sec-mxfp8-recipe} follows the integrated forward and backward path.
\end{minipage}\par

\needspace{6\baselineskip}
\subsection{Symmetric Memory Addresses Remote Rows}
\label{subsec-ep-symmetric-memory}

\begin{figure}[!htb]
\centering
\includegraphics[width=0.36\linewidth]{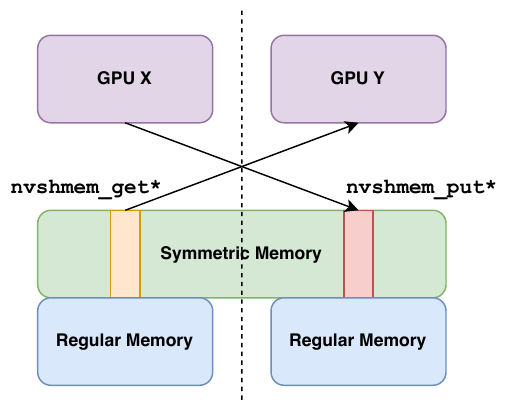}
\caption{\textbf{Symmetric memory makes matching offsets remotely
addressable.} Participating EP-MP ranks create corresponding heap regions.
A local offset and a peer-rank index therefore identify the matching remote
location, allowing a GPU kernel to address an expert-capacity row directly.
The allocation establishes addressability; ordering and buffer-reuse rules
still determine when a row may be read or overwritten.}
\label{fig-ep-symmetric-memory}
\end{figure}

\conceptlead{Direct placement requires a remote address.} Ordinary device
pointers name local memory. Symmetric memory gives the participating ranks
corresponding regions in a symmetric heap, so a local offset together with a
peer-rank index names the matching remote location
(Figure~\ref{fig-ep-symmetric-memory}). In the current Olmo-core
setup, each EP-MP process group forms the NVSHMEM bootstrap world used by its
rowwise kernels. All ranks in that group allocate the same set of buffers,
with matching shapes and allocation order %
\citep{nvidiaNVSHMEMUsing}.

Olmo-core uses NVSHMEM because it permits GPU-initiated one-sided operations.
A CUDA kernel can issue a \texttt{PUT} to a remote expert row or a
\texttt{GET} from one without first constructing a host-managed peer block.
This does not mean that NVSHMEM is universally faster than NCCL. It provides
the addressability and fine-grained movement required by this particular
rowwise design %
\citep{nvidiaNVSHMEMUsing,nvidia2026nvshmemmemory}.

\conceptlead{Safe buffer reuse is a separate problem from addressability.} A
symmetric pointer may remain valid while the value at that address is still
needed by an earlier microbatch. Every rank must agree on buffer capacity and
initialization before communication begins, and the runtime must prevent a
later operation from overwriting a row until its final forward or backward
consumer has completed. Matching allocations alone cannot prevent premature
reuse; the runtime also needs explicit ordering and buffer-reuse rules.

\subsection{Routes Use PUT and GET}
\label{subsec-ep-put-get}

\conceptlead{MoE operations and transport directions are distinct.}
Dispatch and combine describe the MoE computation; \texttt{PUT} and
\texttt{GET} describe the direction of the one-sided transfer. Table~\ref{tab-ep-put-get}
shows the mapping used by the rowwise path.

\begin{table}[!htb]
\centering
\small
\renewcommand{\arraystretch}{1.22}
\setlength{\tabcolsep}{7pt}
\begin{tabularx}{0.92\linewidth}{@{}>{\bfseries}l >{\centering\arraybackslash}m{0.13\linewidth} X@{}}
\toprule
MoE operation & Transfer & Row movement \\
\midrule
Dispatch forward & \texttt{PUT} &
The token-source rank pushes $x_t$ to the capacity row assigned to the selected
expert. \\
\arrayrulecolor{black!22}\hline\arrayrulecolor{black}
Combine forward & \texttt{GET} &
The token-source rank pulls the selected expert outputs and performs the
router-weighted reduction locally. \\
\arrayrulecolor{black!22}\hline\arrayrulecolor{black}
Combine backward & \texttt{PUT} &
The token-source rank pushes each router-weighted output gradient to the
corresponding expert-output gradient row. \\
\arrayrulecolor{black!22}\hline\arrayrulecolor{black}
Dispatch backward & \texttt{GET} &
The token-source rank pulls the route gradients and sums them into the gradient
of the original token row. \\
\bottomrule
\end{tabularx}
\caption{\textbf{Forward and backward use adjoint row movements.} The saved
route map is shared by all four operations. Backward reverses each forward
movement, so dispatch and combine swap transport directions in the
backward pass.}
\label{tab-ep-put-get}
\end{table}

\begin{figure}[!htb]
\centering
\includegraphics[width=0.70\linewidth]{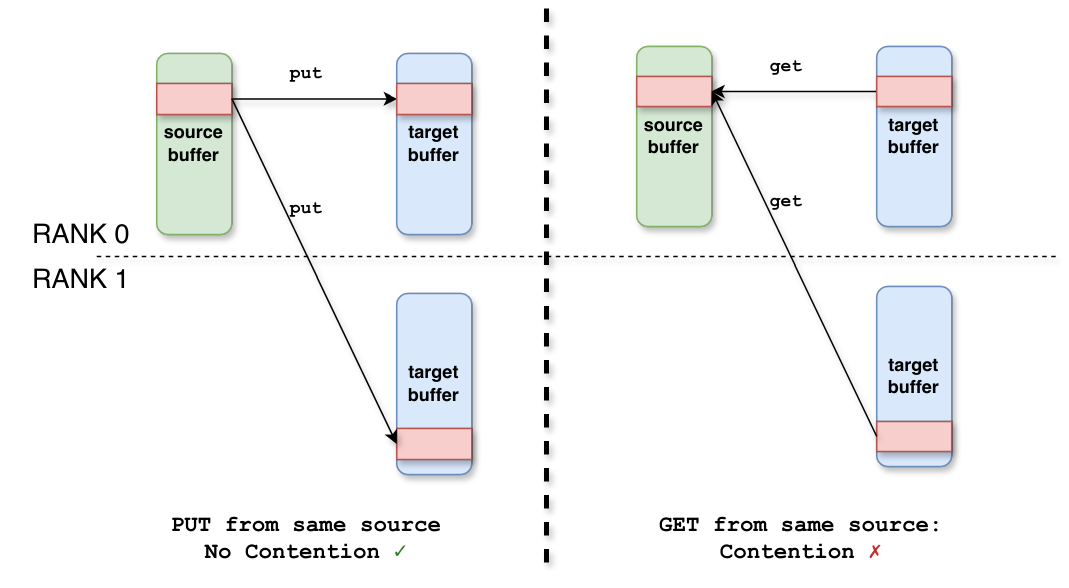}
\caption{\textbf{PUT fan-out writes distinct destinations; GET fan-out
converges on one source.} On the left, one rank issues PUTs from a local source
row into separate remote target rows. On the right, multiple target ranks
issue GETs against the same remote source row. The arrows show which rank
issues the one-sided operation toward the remote address. The check and cross
refer only to same-source access contention: PUT kernels still share memory
bandwidth, streaming multiprocessors (SMs), and fabric resources with other
work.}
\label{fig-ep-get-put-contention}
\end{figure}

\conceptlead{Fan-out favors source-initiated PUT.} A top-$K$ token row may need
to reach several expert ranks. With PUT, the token-source rank reads its local
row and writes each route into a distinct remote capacity slot. Expressing the
same fan-out as destination-initiated GETs makes several peers read the same
remote source row, concentrating access pressure at that source
(Figure~\ref{fig-ep-get-put-contention}).

Forward dispatch writes directly into expert-contiguous capacity storage. The
expert rank then runs grouped expert computation over the valid row
intervals. Forward combine gathers the $K$ selected expert rows for each local
token, multiplies them by their router weights, and sums them into token order.
The custom autograd functions save the route map so backward applies the exact
transpose movements in Table~\ref{tab-ep-put-get}. The router-weight gradient
uses the gathered expert outputs and the token-output gradient from the same
accepted routes.

Invalid routes remain explicit throughout this path. Rowwise \texttt{PUT}
kernels skip negative destinations; combine gathers zero for invalid entries
before the local reduction. Capacity handling is therefore part of the route
map rather than an implicit property of unwritten memory.

\subsection{Buffer Reuse and Ordering Are Explicit in Rowwise EP}
\label{subsec-rowwise-ep-buffer-reuse}

\conceptlead{A microbatch fills only a prefix of each capacity buffer.} Rowwise
communication reserves capacity-sized destination buffers, but a microbatch
usually writes only a dense prefix of each buffer: the local experts' rows are
packed back to back from row 0, and one unwritten tail follows the last expert.
Tail rows may contain values from an earlier use. Expert forward and
backward therefore receive the real per-expert row offsets, and grouped GEMM
and the expert weight-gradient GEMM (Wgrad) skip the unwritten tail. Clearing the whole capacity buffer is
useful as a diagnostic, but it adds unnecessary memory traffic to the normal
path.

\begin{figure}[!htb]
\centering
\includegraphics[width=\linewidth]{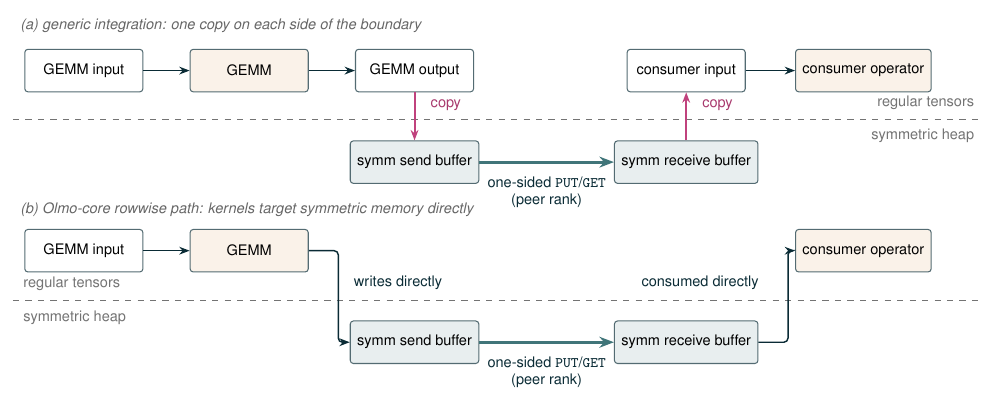}
\caption{\textbf{Crossing the regular\textendash{}symmetric boundary: staging copies
versus direct access.} The remote endpoint of a one-sided transfer must
live in the symmetric heap, which is allocated separately from regular
PyTorch tensors; inter-node GPU-initiated transfers require the local
endpoint to be symmetric as well. (a) A generic integration pays one copy
on each side of the boundary. (b) The Olmo-core rowwise path removes these
copies where kernels can target symmetric memory directly: the expert
grouped GEMM reads the dispatch buffer in place, the down projection writes
its output directly into the symmetric combine buffer, and MXFP8 dispatch
fuses quantization into the \texttt{PUT} kernel. Conceptual illustration
with no measured quantities.}
\label{fig-ep-symm-boundary}
\end{figure}

\conceptlead{Direct access avoids staging copies but extends buffer use into backward.}
Grouped GEMM reads the symmetric dispatch buffer in place instead of copying
its contents into a regular tensor. Its saved Wgrad input therefore remains
in the same buffer that a later dispatch could overwrite.
Figure~\ref{fig-ep-symm-boundary} shows where direct access removes copies;
the resulting buffer-reuse constraints depend on how long each consumer needs
the data.

\begin{figure}[!htb]
\centering
\includegraphics[width=\linewidth]{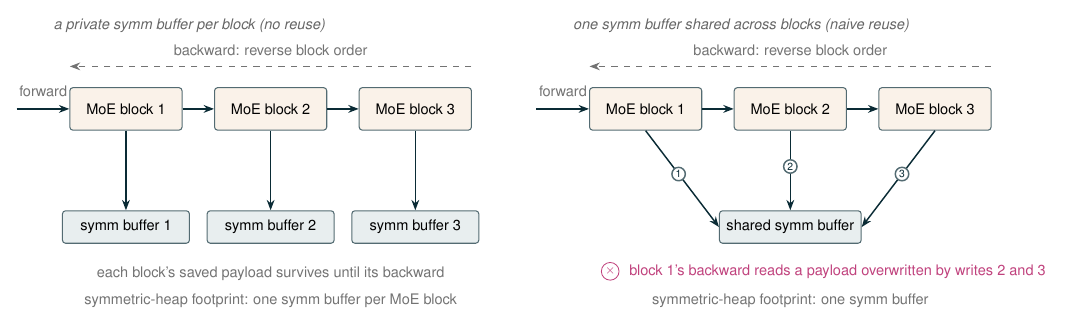}
\caption{\textbf{A shared symmetric buffer saves memory but can overwrite saved
payloads.} With a private
symmetric buffer per block, each saved payload survives untouched until its
block's backward, but resident symmetric memory grows with the number of
local blocks; one shared buffer keeps the footprint constant, but forward
visits blocks in order $1,2,3$ while backward returns in order $3,2,1$, so
block 1's backward reads a payload that later blocks have overwritten.
Scratch buffers, which no later operation reads, safely take the shared
path; payloads saved for backward keep per-block slots whose reuse is
controlled by the lease system of Figure~\ref{fig-ep-symm-buffer-lease}, and
activation recomputation turns a saved payload back into scratch, restoring
the shared option. Conceptual illustration with no measured quantities.}
\label{fig-ep-symm-block-reuse}
\end{figure}

\conceptlead{Reusing buffers across blocks saves memory but can overwrite saved
payloads.} Symmetric buffers are allocated collectively and stay resident,
so giving every MoE block private symmetric buffers makes resident symmetric
memory grow with the number of local blocks. Sharing buffers across blocks
removes that growth, but forward and backward visit blocks in opposite
orders: block $\ell{+}1$'s dispatch would overwrite the payload that block
$\ell$ saved, long before block $\ell$'s backward reads it
(Figure~\ref{fig-ep-symm-block-reuse}). Whether a buffer faces this hazard
is decided by its immediate consumer. The dispatch buffer feeds the
up-projection grouped GEMM, which saves its input for Wgrad, so the buffer
must survive until backward. The expert-output buffer is only read by the
combine transfer, and backward re-routes gradients with the saved route map
(the router-weight gradient uses a separate gather capture on the
token-source rank), so it is scratch and may be reused as soon as the
transfer completes.

\begin{figure}[!htb]
\centering
\includegraphics[width=\linewidth]{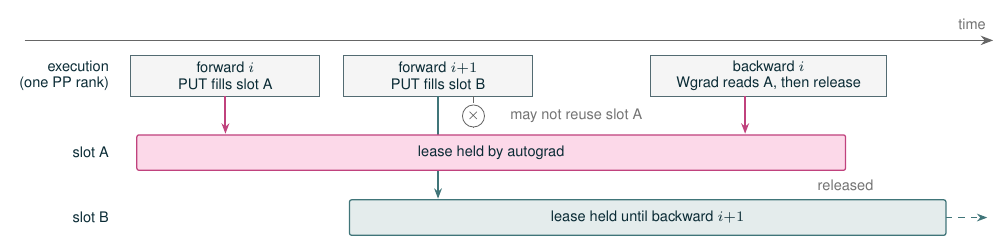}
\caption{\textbf{A lease pins the dispatch buffer until its backward
consumer finishes.} The routed
grouped GEMM saves the dispatch buffer as its Wgrad input, so an
autograd-held lease pins the slot until the backward that consumes it
releases it, and a later forward acquires a different slot from a pool sized
before training starts. Conceptual illustration with no measured
quantities.}
\label{fig-ep-symm-buffer-lease}
\end{figure}

\conceptlead{Autograd determines when a buffer may be reused.} Under pipeline
parallelism, several later forwards may execute before the Wgrad that reads a
block's saved dispatch payload.
Remote writes through custom communication kernels are not protected by
PyTorch's ordinary tensor version checks, so Olmo-core attaches autograd-held
leases to symmetric payloads that remain visible to backward. Each lease
prevents reuse until the corresponding autograd node releases it after its
final consumer. When
activation checkpointing recomputes the payload for backward, the original
forward can instead use shared scratch storage.
Figure~\ref{fig-ep-symm-buffer-lease} shows how a lease keeps a dispatched
payload available for its backward consumer.

\conceptlead{One-sided communication requires explicit ordering.} Expert compute
must not read a destination row before all required dispatch writes are
visible. A buffer must not be overwritten while a peer is still reading it,
and all ranks participating in a barrier or collective kernel must enter a
compatible operation order. The rowwise path uses stream ordering and
NVSHMEM completion operations around these phase boundaries. Removing the host
split-size wait does not remove distributed synchronization; it moves the
required agreement and completion into the GPU execution path
\citep{nvidia2026nvshmemmemory}.

\conceptlead{MXFP8 adds a reuse constraint between quantized activations and gradients.}
As on the BF16 path, Wgrad needs the dispatched input. A BF16 route carries
one row; an MXFP8 route carries qdata and scale rows that follow the same
destination indices and validity mask. Backward must consume both saved
components before the slot is reused for quantized gradient rows.
This shared-buffer constraint is one reason Olmo-core fuses dispatch, expert
computation, and combine into a single autograd node on the MXFP8 path
(Section~\ref{sec-mxfp8-recipe}).

\subsection{DeepEP v2 Provides an Optional EP Backend}
\label{subsec-ep-deepep-v2}

\conceptlead{DeepEP v2 is an effective inter-node alternative.} Olmo-core
can select DeepEP v2 for dispatch and combine while retaining its router,
capacity policy, expert kernels, autograd integration, and optimizer
\citep{zhao2025deepep}. DeepEP v2 is block-oriented like the path in
Section~\ref{subsec-ep-block-alltoall}: it exchanges per-peer batches of rows
rather than individual rows. The difference is that packing, transfer, and
unpacking happen inside its dispatch and combine calls with GPU-resident
metadata, so the model-facing path keeps no host split lists.

\conceptlead{Block transfers reduce the NIC request rate.}
The block orientation matters most across nodes. Within a node, NVLink
transfer time is governed mainly by payload bytes, so rowwise EP's many
small row transfers cost little beyond the bytes they move. Across nodes,
every transfer becomes a request that the network interface controller
(NIC) must process, and a
pattern that issues one request per route can saturate the NIC's message
rate before it saturates link bandwidth
(Section~\ref{sec-hardware-topology} describes the GPU-to-NIC submission
path). Aggregating rows into a few large per-peer transfers keeps the
request count low. Our observations are consistent with this reasoning, though we have not
measured the attribution. DeepEP v2 has performed well in our current
inter-node experiments, which makes it useful when an EP-MP group spans nodes
(on our eight-GPU nodes, EP-MP degrees greater than eight); rowwise EP remains
the production intra-node path.

\conceptlead{The trade-off is integration depth.} Olmo-core does not control
DeepEP's dispatch/combine kernels and public interface, so it cannot specialize
every communication boundary as freely as it can in the rowwise backend. In
the current MXFP8 integration (Section~\ref{sec-mxfp8-training}), forward
dispatch transports E4M3 data with E8M0 scales and the expert GEMMs use
MXFP8. Forward combine, cached backward
dispatch, and backward combine remain BF16. Extending MXFP8 across those
remaining communication legs would require new DeepEP kernels and interfaces,
not only an Olmo-core configuration change. We therefore use DeepEP v2 as an
inter-node option and rowwise EP where its closer integration with Olmo-core's
communication, precision, and autograd implementation is useful.

\subsection{Overlap Is Not Free: Resource Contention}
\label{subsec-ep-overlap}

A \textbf{wave schedule} splits a microbatch's routed work into waves and
attempts to dispatch wave $i+1$ and combine wave $i-1$ while the experts
compute wave $i$. The idea comes from DeepSeek's inference systems, where
Mega MoE in DeepGEMM fuses EP dispatch, the expert MLP, and EP combine into
one persistent megakernel that overlaps communication with tensor-core
computation %
\citep{deepseekai2026deepseekv4,deepseekai2026deepgemm}. We
adopted the schedule for training but not the megakernel: the Olmo-core wave
path launches ordinary dispatch, grouped-GEMM, and combine kernels on
separate streams, so the overlapped stages compete for the same GPU
resources instead of sharing one kernel's internal schedule.
Section~\ref{subsec-megamoe-lesson} describes MegaMoE, our own BF16
megakernel prototype in the style of DeepGEMM's Mega MoE, and why we removed
it. Wave execution
can hide exposed communication, but it also
turns one large expert batch into several smaller grouped GEMMs. A complete comparison needs three
schedules: no-wave phase-sequential execution, wave-sequential execution, and
wave-overlapped execution. Wave-sequential execution is the control that
separates the cost of smaller GEMMs from the cost of concurrent kernels. Our
archived experiments lack it, so they show whether the complete wave schedule
was faster but cannot split the difference between the two causes.

Let $T_{\mathrm{comm}}$ and $T_{\mathrm{gemm}}$ be the standalone times for
communication and expert computation. Sequential execution costs
$T_{\mathrm{comm}}+T_{\mathrm{gemm}}$. After overlap, the same operations may
stretch to $\widetilde{T}_{\mathrm{comm}}$ and
$\widetilde{T}_{\mathrm{gemm}}$ as they contend for GPU and communication
resources, so the overlapped schedule takes
$T_{\mathrm{overlap}} \ge \max(\widetilde{T}_{\mathrm{comm}},
\widetilde{T}_{\mathrm{gemm}})$. Overlap improves elapsed time only if

\begin{equation}
    T_{\mathrm{overlap}}
    < T_{\mathrm{comm}} + T_{\mathrm{gemm}}.
    \label{eq-ep-overlap-condition}
\end{equation}

\begin{figure}[!htb]
    \centering
    \includegraphics[width=\linewidth]{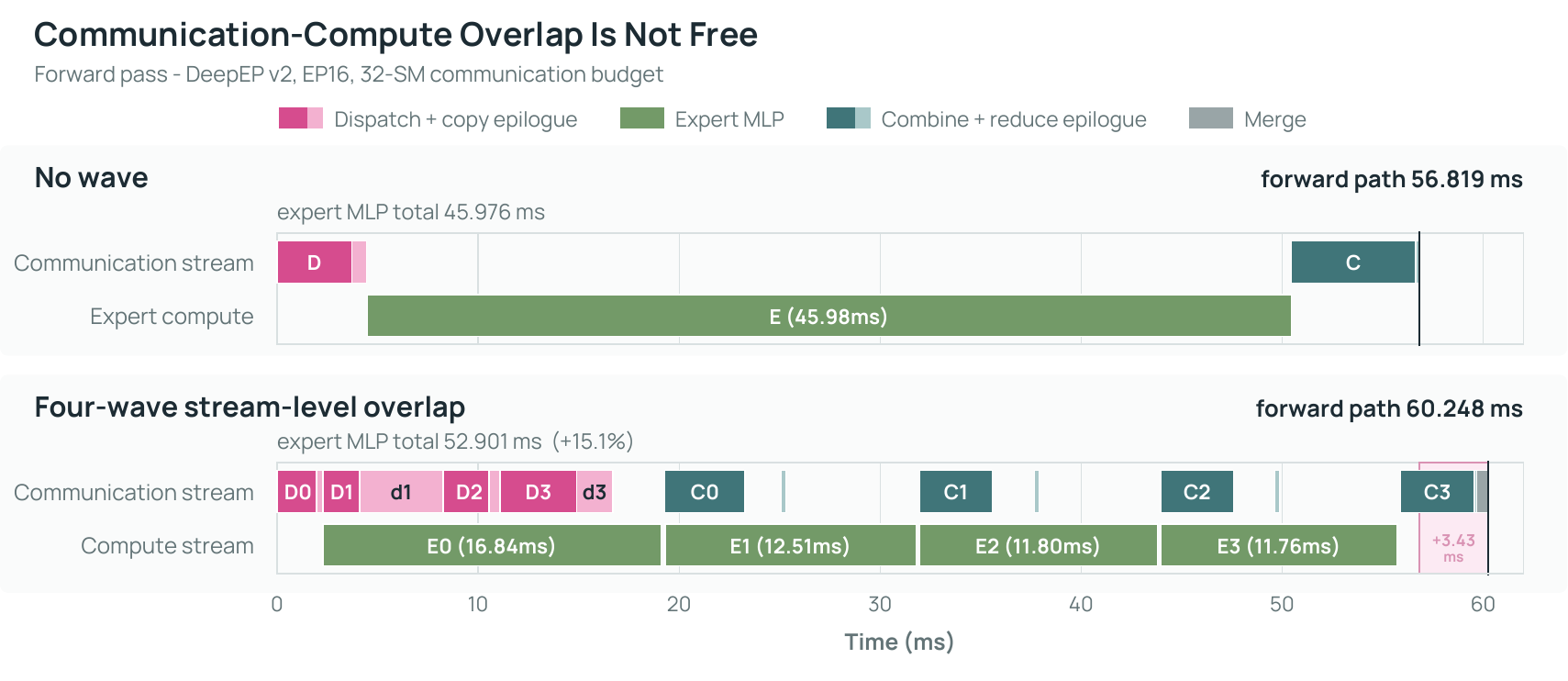}
    \caption{\textbf{Communication\textendash{}compute overlap lengthens this measured
    forward path.} This two-node DeepEP v2, EP-MP${}=16$, 128-expert,
    top-$8$ BF16 benchmark uses 16,384 tokens per rank, $d=h=8{,}192$, and a
    32-SM communication budget. Both panels align time to the first dispatch.
    The four-wave path overlaps communication-stream work with expert
    computation, but its displayed forward interval is 60.248~ms versus
    56.819~ms without waves, a 6.0\% increase. Its four expert-compute
    intervals total 52.901~ms versus 45.976~ms for the single no-wave
    interval, a 15.1\% increase. Durations are manually transcribed from one
    rank and one measured iteration in the two node-0 Nsight traces. The
    timeline establishes concurrent execution and component stretch for this
    instance; it does not identify the contended resource or separate wave
    granularity from concurrency because no wave-sequential control was
    archived.}
    \label{fig-ep-wave-forward-timeline-measured}
\end{figure}

Figure~\ref{fig-ep-wave-forward-timeline-measured} makes the no-wave and
wave-overlapped order concrete with one measured forward instance at a 32-SM
communication budget; the missing
wave-sequential control cannot be inferred from it.

\conceptlead{Expert computation takes longer in the wave trace.}
Figure~\ref{fig-ep-wave-forward-timeline-measured} shows real
communication--compute concurrency, yet the displayed wave forward interval is
3.429~ms longer and its expert-compute intervals grow by 15.1\% in aggregate.
Across the full profiles, the three dominant grouped-GEMM kernel families accumulated 4.63~s in
the no-wave trace and 5.74~s in the wave trace. The wave schedule both split
the expert work into four smaller grouped GEMMs and ran DeepEP kernels beside
them, so these profiles cannot assign the additional time uniquely to GEMM
granularity or concurrent resource contention.

\conceptlead{Concurrent rowwise communication slows expert GEMM.} In
a separate two-node, 16-GPU diagnostic with EP-MP${}=16$, 16,384 tokens per rank, top-$8$
routing, and $d=h=6{,}144$, we launched either a rowwise NVSHMEM dispatch or
combine transfer concurrently with one grouped expert workload while
sweeping the communication launch width from 1 to 128 blocks.
Table~\ref{tab-ep-rowwise-interference} reports the GEMM stretch and the
makespan, the total elapsed time of the concurrent pair. Concurrent
communication slowed the grouped GEMM by roughly a quarter at every launch
width, and shrinking the communication launch to a single block did not
avoid the GEMM slowdown; it stretched communication by an order of magnitude.

\begin{table}[!htb]
\centering
\small
\begin{tabular}{@{}lrr@{}}
\toprule
\textbf{Concurrent pair} & \textbf{GEMM time (ms)} & \textbf{GEMM slowdown} \\
\midrule
Grouped GEMM alone & 4.645--4.679 & --- \\
GEMM + rowwise dispatch & 5.736--5.874 & 22.9--26.0\% \\
GEMM + rowwise combine & 5.724--5.907 & 23.0--27.0\% \\
\midrule
\textbf{Pair schedule (128-block launch)} & \textbf{Makespan (ms)} & \\
\midrule
Dispatch then GEMM, sequential & 7.768 & \\
Dispatch and GEMM, overlapped & 5.833 & \\
Combine then GEMM, sequential & 7.560 & \\
Combine and GEMM, overlapped & 5.868 & \\
\bottomrule
\end{tabular}
\caption{\textbf{Concurrent rowwise communication slows the grouped GEMM,
yet a wide overlapped launch still shortens the pair.} These measurements come from a
separate two-node rowwise NVSHMEM diagnostic ($d=h=6{,}144$), not from the
DeepEP v2 run of Figure~\ref{fig-ep-wave-forward-timeline-measured}. GEMM-time ranges span
the 1--128-block communication-launch sweep. Restricting communication to
one block stretched dispatch and combine to 40.754 and 41.706~ms. Values are
max-rank medians over five measured iterations with independent
communication and compute buffers; the diagnostic isolates resource
interference and does not reproduce the dependencies of an end-to-end MoE
layer.}
\label{tab-ep-rowwise-interference}
\end{table}

GEMM stretch alone, however, does not decide the outcome. At the 128-block
launch, overlap still shortened the pair's makespan by 22--25\% relative to
sequential execution: the wide launch finished communication soon enough to
hide behind useful work despite the contention tax, while the one-block launch
did not.

\begin{figure}[!htb]
    \centering
    \includegraphics[width=\linewidth]{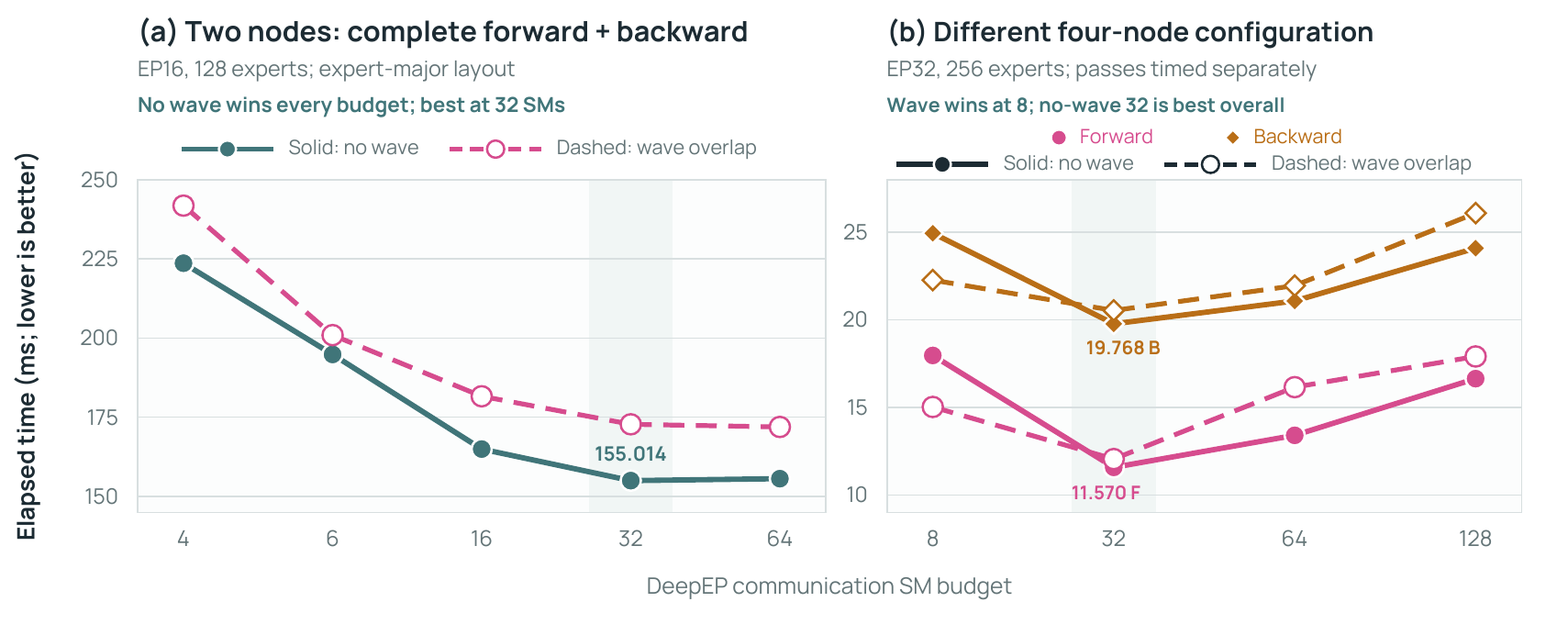}
    \caption{\textbf{No-wave execution is fastest after tuning the communication-SM
    budget in both sweeps.} The panels show absolute elapsed time with independent vertical
    scales; lower is better. (a) On the two-node EP-MP${}=16$, 128-expert,
    expert-major BF16 configuration, no-wave was faster at every communication-SM
    budget and reached its minimum at 32 SMs. (b) On a different four-node
    EP-MP${}=32$, 256-expert, source-token-major BF16 configuration, overlap beat no-wave
    at the communication-constrained eight-SM point. Nevertheless, no-wave at
    32 SMs remained the best measured forward and backward operating point.
    Panel (a) reports max-rank medians over five iterations after 10 warm-ups
    with inter-iteration synchronization; panel (b) reports max-rank medians
    over five iterations after three warm-ups without that synchronization.
    Both are Nsight-instrumented standalone development runs.}
    \label{fig-ep-overlap-operating-points}
\end{figure}

\conceptlead{The complete wave candidate did not beat its no-wave baseline.}
A two-node DeepEP v2 sweep measured BF16 forward and backward
for EP-MP${}=16$, 128 experts, top-$8$, 16,384 tokens per rank, and
$d=h=8{,}192$. The four-wave path used an expert-major static layout and
overlapped dispatch, expert compute, and combine. Figure~\ref{fig-ep-overlap-operating-points}a
shows that it remained slower than no-wave execution at every matched
communication-SM setting.

The sweep also shows why lowering the communication SM count is not
sufficient: at four and six SMs communication
became the bottleneck, while larger budgets still did not recover the no-wave
time.

\conceptlead{Overlap helps at eight SMs, but no-wave at 32 SMs is faster.}
A later four-node, 32-GPU source-token-major experiment is shown in
Figure~\ref{fig-ep-overlap-operating-points}b. At the
eight-SM budget, where communication dominates, four-wave execution beat
no-wave for both forward and backward. At 32 SMs, no-wave was faster again
and was the best measured point for both passes. Overlap can therefore help
a communication-constrained configuration without improving the best
configuration of the sweep.

DeepSeek-V3, Megatron-Core, DeepEP, COMET, and Lancet report overlap designs
that win in their settings
\citep{deepseekai2024deepseekv3,nvidia2026megatroncoremoe,zhao2025deepep,zhang2025comet,jiang2024lancet}.
Our measurements are standalone
development benchmarks with five within-run timings rather than repeated
training runs, and the earlier DeepEP wave layout can change communication
traffic relative to no-wave execution. On this evidence, we retain
phase-sequential execution as the Olmo-core reference. Replacing it with a
wave backend would require a speedup over the best no-wave setting for the
same shape and topology. Concurrent kernels or a lower nominal communication
resource budget alone do not establish that speedup.

EP makes a larger routed-expert pool fit by sharding expert weights and
gradient buffers across the existing DP ranks, and it reduces each rank's expert-gradient-reduction
payload. The cost is dispatch and combine in every routed layer and
microbatch. Rowwise EP removes contiguous-block packing and host split lists by
addressing remote expert rows directly, but it is worthwhile only when route
construction, symmetric-buffer management, one-sided communication, and any
resource contention cost less than the work they replace.

\FloatBarrier

\section{Pipeline Parallelism}
\label{sec-pipeline-parallelism}

\conceptlead{PP distributes the layers that EP leaves replicated.} Expert parallelism (EP) divides the routed experts inside every MoE
layer, but it does not divide the stack of layers, dense weights, shared
experts, or the activation schedule. Pipeline parallelism (PP) addresses that
remaining depth-wise memory problem by assigning consecutive Transformer
blocks to different stages. A stage keeps its assigned parameters, gradients,
and optimizer state; complete hidden activations, rather than selected token
rows, cross stage boundaries
\citep{huang2019gpipe,narayanan2019pipedream}.

For PP degree $P$ and EP-MP degree $M_E$, a balanced repeated
stack gives the rough rank-local parameter count
\[
P_{\mathrm{local}}
\approx \frac{1}{P}\left(P_{\mathrm{dense}}+
\frac{P_E}{M_E}\right).
\]
This is an estimate of persistent training state, not a peak-memory formula.
The first pipeline stage also stores the token embeddings, while the last
stores the language-model head and loss. Unequal block types, communication buffers, and
the number of simultaneously live activations further prevent peak memory from
scaling exactly as $1/P$.

PP does not change which ranks store each parameter or which group reduces
each gradient within a stage. Dense
parameters still reduce over dense data-parallel groups, routed experts still
shard over EP-MP, and expert replicas still reduce over EP-DP. If one stage
contains $D=M_E D_E$ ranks, then a $P$-stage job contains $W=P M_E D_E$ ranks
before context parallelism. PP adds a group connecting corresponding stage
ranks; it does not replace the placement and reduction rules established in
Sections~\ref{sec-ddp-vs-fsdp} and~\ref{sec-expert-parallelism}.

\conceptlead{PP trades schedule time for depth-wise memory.}
A single microbatch advances stage by stage and leaves the other stages
idle. Multiple microbatches fill more of the pipeline, but they also add
stage-boundary transfers, scheduler and launch overhead, and forward values
that remain live until backward. We therefore use PP when depth-wise sharding
is needed to fit the model, then tune the schedule to control the added bubbles,
communication, and activation memory.

\subsection{Olmo Uses the 1F1B Family}

\conceptlead{Interleaved-1F1B is compile-friendly because it keeps backward
whole.} The selected pipeline schedule uses the
\textbf{interleaved one-forward-one-backward (1F1B)} family
\citep{narayanan2019pipedream,narayanan2021megatron}. The scheduler decides
when each stage runs, but each stage still executes one ordinary, unsplit
backward. The compiled model path therefore does not need to expose separate
input-gradient and weight-gradient programs to the scheduler. This preserves
the existing \texttt{torch.compile} and TorchInductor contract
\citep{ansel2024pytorch2} while gaining
the activation-memory benefit of 1F1B scheduling.

Other schedules can reduce the symbolic bubble rate, with different memory
and execution requirements.
DualPipe uses opposing pipeline directions,
but its published comparison requires twice the per-device parameter storage
of 1F1B at the same PP degree
\citep{deepseekai2024deepseekv3,deepseekai2025dualpipe}. Zero-bubble schedules
instead separate input-gradient work from deferrable weight-gradient work
\citep{qi2023zerobubble,huang2024torchtitanzb}. Olmo's compiled stage
backward computes both in one unsplit pass, so the scheduler cannot place
them separately; splitting it would require coordinating compiled operators,
retained operands, recomputation, gradient accumulation, and the scheduler.

\subsection{Virtual-Stage Placement Controls Activation Retention}

In Olmo-core's custom interleaved schedule, $P$ physical PP ranks hold $2P$
contiguous \textbf{virtual stages}, two per rank. Splitting each rank's work into two shorter chunks creates
more opportunities to fill a bubble, at the cost of more virtual-stage
boundaries and activation handoffs. Placement also determines how long each
chunk's forward values remain live before its backward executes.

Interleaved-1F1B uses \textbf{loop placement}: physical rank $r$ holds virtual
stages $r$
and $r+P$ %
\citep{narayanan2021megatron}. Early pipeline stages can then retain
more outstanding activations than late stages, concentrating peak memory on a
few ranks. \textbf{1F1B-V}, experimental in Olmo-core, instead assigns stages
$r$ and $2P-1-r$ to rank
$r$ %
\citep{qi2024pipeline}. Pairing an early, long-lived chunk with a late,
short-lived chunk can equalize activation memory, but it changes the bubble and
communication pattern. It is a memory-balancing option, not an assumed
throughput improvement. Both placements preserve unsplit backward.

\begin{figure}[!htb]
\centering
\begin{subfigure}{\linewidth}
  \centering
  \includegraphics[width=\linewidth]{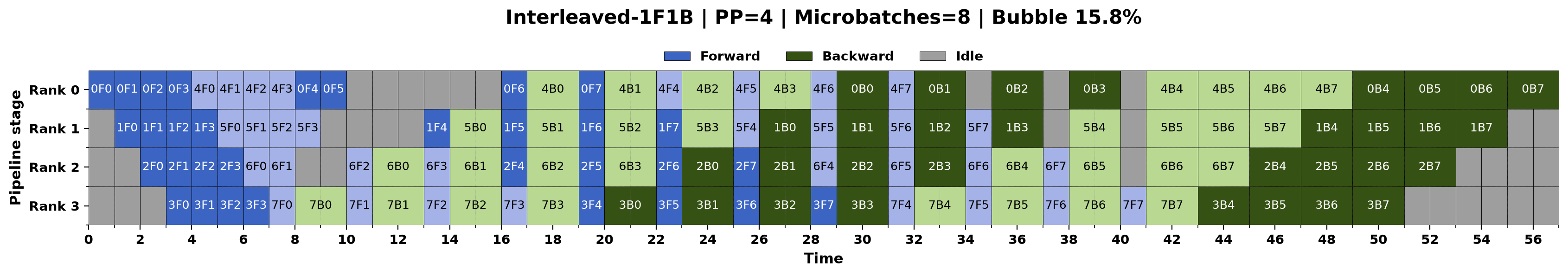}
  \caption{Interleaved-1F1B.}
\end{subfigure}

\vspace{0.35em}

\begin{subfigure}{\linewidth}
  \centering
  \includegraphics[width=\linewidth]{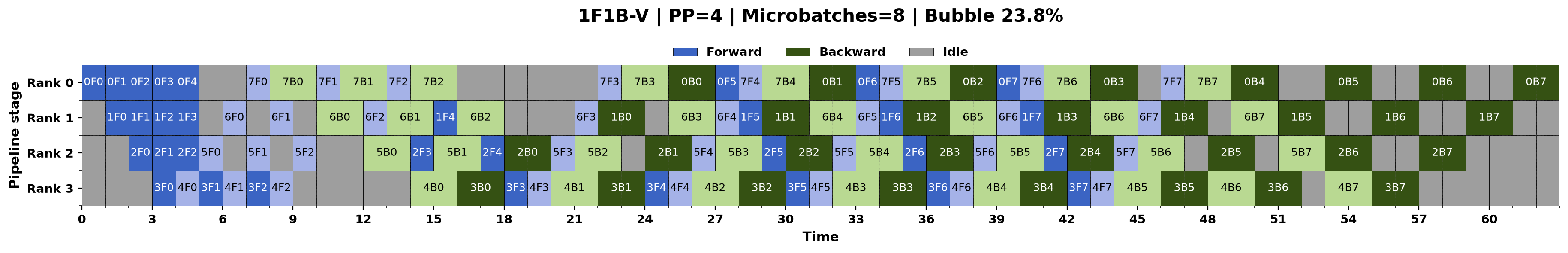}
  \caption{1F1B-V placement.}
\end{subfigure}
\caption{\textbf{Interleaved-1F1B and 1F1B-V schedules with PP${}=4$ and eight
microbatches.} A label such as \texttt{4B2} denotes backward for virtual stage
4 and microbatch 2; the lighter forward/backward shades identify stages 4--7,
the second virtual stage stored by each rank. Gray cells are empty schedule
slots. The bubble percentages are empty-slot fractions in the symbolic
schedule, not measured GPU idle time.}
\label{fig-pp-interleaved-v-schedules}
\end{figure}

Figure~\ref{fig-pp-interleaved-v-schedules} compares the two placements for
four pipeline ranks, eight microbatches, and eight virtual stages in total.

The symbolic PP${}=4$ example makes the trade concrete
(Table~\ref{tab-pp-activation-memory-example}). With eight microbatches, loop
placement concentrates unmatched forwards on the early ranks---up to eleven
on rank~0---while V placement holds six on every rank. These are schedule
high-water marks, not gigabytes: stages can retain different payload sizes, and
recomputation can change which values survive. Placement controls both the maximum and the skew of retained forward
activations.

\begin{table}[!htb]
\centering
\small
\begingroup
\renewcommand{\arraystretch}{1.08}
\setlength{\tabcolsep}{5pt}
\renewcommand{\tabularxcolumn}[1]{m{#1}}
\begin{tabularx}{\linewidth}{
  >{\raggedright\arraybackslash}m{0.22\linewidth}
  >{\centering\arraybackslash}m{0.10\linewidth}
  >{\centering\arraybackslash}X
  >{\centering\arraybackslash}X
  >{\centering\arraybackslash}X}
\arrayrulecolor{black}
\toprule
\rowcolor{black!5}
\multicolumn{2}{c}{\textbf{Assumed activation copy: 10\,GB}} &
\multicolumn{2}{c}{\textbf{Assumed persistent training state: 12\,GB}} &
\textbf{Budget: 80\,GB} \\
\midrule
\rowcolor{ai2accent!7}
\textbf{Schedule} &
\textbf{Rank} &
\makecell[c]{\textbf{Peak activation}\\\textbf{copies}} &
\makecell[c]{\textbf{Illustrative peak}\\\textbf{memory (GB)}} &
\makecell[c]{\textbf{Fits assumed}\\\textbf{budget?}} \\
\midrule
\arrayrulecolor{black!22}
\multirow{4}{*}{Interleaved-1F1B}
  & 0 & 11 & 122 & \textcolor{red}{\(\boldsymbol{\times}\)} \\
  & 1 & 9  & 102 & \textcolor{red}{\(\boldsymbol{\times}\)} \\
  & 2 & 7  & 82  & \textcolor{red}{\(\boldsymbol{\times}\)} \\
  & 3 & 5  & 62  & \textcolor{green!50!black}{\(\checkmark\)} \\
\arrayrulecolor{black}
\midrule
\arrayrulecolor{black!22}
\multirow{4}{*}{1F1B-V}
  & 0 & 6 & 72 & \textcolor{green!50!black}{\(\checkmark\)} \\
  & 1 & 6 & 72 & \textcolor{green!50!black}{\(\checkmark\)} \\
  & 2 & 6 & 72 & \textcolor{green!50!black}{\(\checkmark\)} \\
  & 3 & 6 & 72 & \textcolor{green!50!black}{\(\checkmark\)} \\
\arrayrulecolor{black}
\bottomrule
\end{tabularx}
\endgroup
\caption{\textbf{Illustrative peak-memory consequence of activation-retention
imbalance.} This is a worked example, not a measurement. It applies the stated
10-GB-per-copy and 12-GB persistent-training-state assumptions to the symbolic PP${}=4$,
eight-microbatch high-water marks using
$12+10\times\text{copies}$~GB. Real stages can retain unequal payloads, and
recomputation changes what must remain live.}
\label{tab-pp-activation-memory-example}
\end{table}

Placement also decides which physical rank computes the loss. The final
virtual stage computes the language-model loss and begins backward, and its
physical rank follows the placement rather than a rank-number convention.
Loss and metric handling must therefore follow stage identity, not the
highest-numbered PP rank.

\subsection{P2P Overlap}
Placement fixes which stage boundaries are remote; the remaining cost is
moving activations across them.
Adjacent virtual stages on one physical rank use a local tensor handoff. Remote
boundaries use NCCL point-to-point (P2P) communication: forward sends hidden
activations to the next stage, and backward sends their activation gradients
in the reverse direction. For a microbatch activation of shape $[b,s,d]$, each
remote boundary moves $O(bsd)$ values in each direction per microbatch. This
traffic scales with activation shape and remote-boundary count, not total model
parameters. The boundary payload remains BF16 even when expert GEMMs use
MXFP8 (Section~\ref{sec-mxfp8-training}), so each boundary value costs two
bytes in each direction.

\begin{figure}[!htb]
\centering
\includegraphics[width=0.7\linewidth]{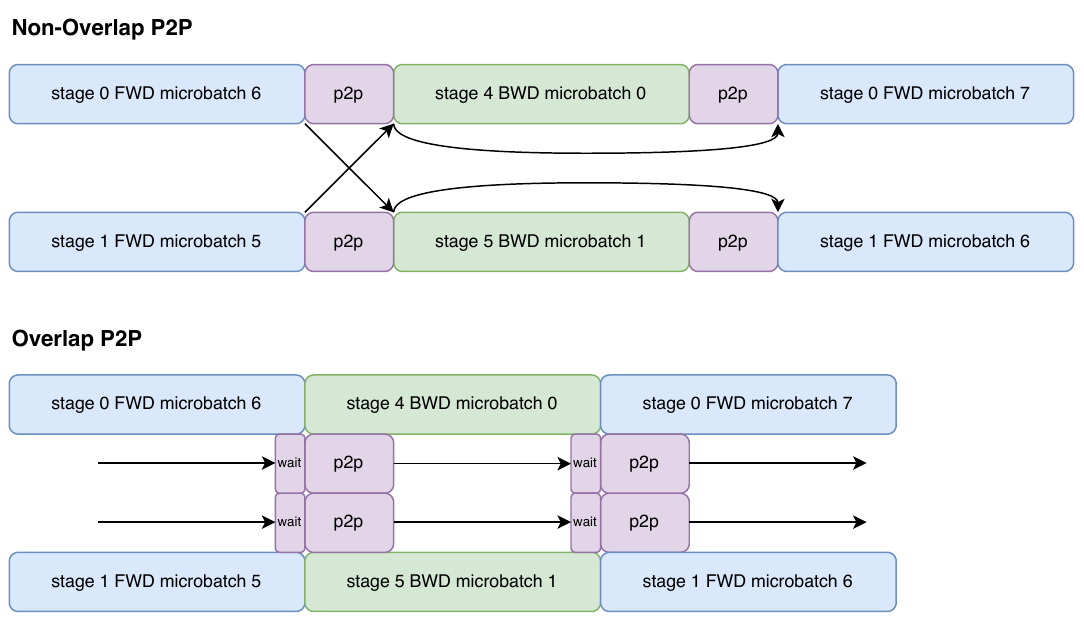}
\caption{\textbf{Interleaved-1F1B can overlap P2P communication with
independent compute.}  Without overlap, each stage-boundary transfer is
serialized between compute actions.  In the steady-state Interleaved-1F1B
pattern shown here, both ranks can post their transfers, execute an unrelated
virtual-stage action, and defer each wait until the transferred activation or
activation gradient is needed.  The overlap hides only the portion of P2P time
that fits inside this independent compute window.}
\label{fig-pp-p2p-overlap}
\end{figure}

\conceptlead{P2P overlap requires independent scheduled work.}  An
asynchronous transfer creates overlap only when a rank can execute something
that does not consume the tensor in flight.  In steady-state
Interleaved-1F1B, the peer's next scheduled action can operate on another
virtual stage or microbatch.  Olmo can launch the transfer, execute one
independent forward or backward action, and wait only when the actual consumer
is reached (Figure~\ref{fig-pp-p2p-overlap}). This can reduce the exposed
transfer wait; it does not remove the
communication or the final dependency
wait.

A 1F1B-V handoff cannot be hidden when the receiving rank's next scheduled
action consumes the activation or gradient just produced by its peer.
Posting that transfer asynchronously
cannot start the dependent kernel earlier: the receiving rank must wait for the tensor at
the same schedule step.  Other V transfers can still overlap, but hiding this
handoff requires changing the schedule and accepting a different bubble or
memory trade-off.

\begin{figure}[!htb]
\centering
\includegraphics[width=\linewidth]{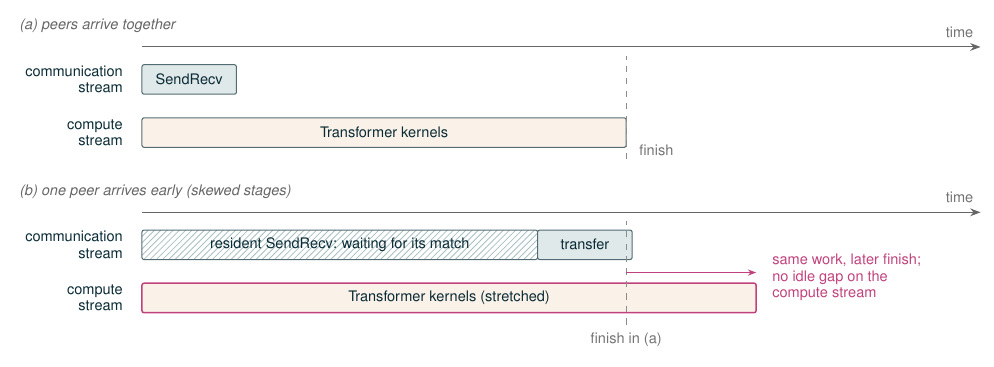}
\caption{\textbf{A resident P2P wait can slow the compute meant to hide it.}
With matched arrival (a), the send/receive kernel transfers and exits, and
the overlapped Transformer kernels run at full speed. With skewed arrival
(b), the early peer's resident send/receive kernel occupies streaming
multiprocessors while it waits for its match, so the overlapped compute
stretches even though no idle gap appears on the compute stream. Conceptual
illustration with no measured quantities.}
\label{fig-pp-p2p-wait-interference}
\end{figure}

\conceptlead{Overlapped P2P still competes for shared resources.} The default
two-sided NCCL path
requires matching send and receive participation. When stage runtimes are
skewed, an early peer can leave a resident NCCL send/receive kernel waiting for
its match. That kernel can contend with the Transformer kernel intended to hide
the transfer, so overlapped compute may run more slowly even when the wait is
not visible as an idle gap (Figure~\ref{fig-pp-p2p-wait-interference}).
Capping the resident kernel's thread-block count shrinks its footprint but
does not remove the rendezvous, and a smaller communication footprint can
lengthen the transfer. As in Section~\ref{subsec-ep-overlap}, the net time
saved depends on both the exposed wait and the slowdown of concurrent
computation.

This contention motivated a one-sided alternative for the pipeline boundary.
Prototype NCCL remote memory access (RMA) backends replace the matched send and receive with a
producer-side put and a consumer-side wait placed at the latest consume
point, so no resident kernel waits for a peer. The prototypes currently work
only within a node, while PP boundaries usually span nodes, so the two-sided
path remains the default.

\subsection{PP Delays Reuse of Saved EP Payloads}

\conceptlead{An EP buffer becomes reusable only after its pipeline backward
finishes.} Section~\ref{subsec-rowwise-ep-buffer-reuse} introduced the hazard
and the lease that guards against it: a dispatched payload saved for Wgrad
must not be overwritten before its backward consumer runs. Without PP, a
forward's temporary EP buffers can
often be reclaimed by the following backward. PP interleaves several
microbatches, so a later forward may arrive while an earlier forward still
needs its saved EP payload for weight-gradient computation. Buffer reuse must
therefore follow the pipeline schedule and the point at which each saved
value's backward consumer finishes, not simply the order of forward calls.
The pipeline schedule determines how many slots must exist at once.

Let
$F_s(t)$ and $B_s(t)$ count completed forwards and backwards for stage $s$ by
schedule time $t$. The \textbf{outstanding-forward high-water}
\[
H_s=\max_t\bigl(F_s(t)-B_s(t)\bigr)
\]
predicts how many stage-$s$ payloads can be live concurrently. It sizes a
buffer pool, but it does not determine when any particular slot is reusable.
A slot becomes reusable only after the matching backward has consumed the
saved value.

We encountered this hazard in a PP--EP correctness failure. Rowwise expert
execution saved a self-managed symmetric-memory dispatch payload needed later
for expert weight gradients. A later pipeline forward could reuse that storage
before the matching backward consumed it. The corrupted values could remain
finite in BF16, so loss finiteness alone did not rule out a silent wrong
gradient; the MXFP8 form of the same bug instead surfaced as an infinite
gradient norm from stale scales. The repair uses two complementary rules: the
schedule high-water
determines how many slots to reserve, while the saved autograd value retains a
lease on its particular slot until its backward consumer finishes
(Figure~\ref{fig-ep-symm-buffer-lease}).

Two schedule features change what the lease rule must cover. Recomputation
(Section~\ref{sec-recompute}) changes whether a value must survive from the
original forward, and two-batch overlap (TBO,
Section~\ref{subsec-tbo-lesson}) can multiply simultaneous acquisitions
within one scheduled forward.

PP peak memory therefore includes not only stage parameters but also the
schedule-dependent high-water of ordinary activations, communication buffers,
route metadata, and rowwise payload leases.
Appendix~\ref{app-context-parallelism}
describes how context parallelism changes the activation shape entering this
same reuse rule.

\subsection{Runtime Determines Pipeline Throughput}

Once a configuration fits, the microbatch schedule and stage runtimes determine
its throughput. Let $m$ be the number of pipeline microbatches in one
optimizer step. Increasing
$m$ amortizes warmup and cooldown because more work occupies the steady-state
portion of the pipeline. It does not make those bubbles disappear, and it is
not equivalent to increasing the microbatch size. At fixed global batch,
increasing $m$ makes each microbatch smaller, which can reduce attention and
GEMM efficiency and increase launch and P2P overhead per token.
The useful operating point therefore balances pipeline occupancy against the
efficiency of each stage invocation.

\conceptlead{One straggling stage throttles the complete pipeline.} Equal layer
counts are only a starting point because embeddings, the language-model head,
dense versus routed blocks, shared experts, recomputation, and routing
imbalance all change stage time. In PP recipes where the last virtual stage's
language-model head and loss would make it the straggler, we leave that stage
one Transformer block lighter. For example, some nominal 32-block designs use
31 blocks while retaining split points from the 32-block partition, so the
missing block comes from the final stage. This is a profile-guided recipe
choice, not an automatic scheduler policy. We balance stages using measured
time and memory rather than layer count alone.

\conceptlead{Symbolic timelines omit execution costs.} Generated
timelines reveal action order, empty slots, activation high-water, and possible
P2P overlap windows. They do not include kernel duration, transfer latency, CPU
launch gaps, GPU idle time, or contention between concurrent operations. A
smaller symbolic bubble can still lose to smaller microbatch kernels, an
imbalanced stage, or communication interference. Pipeline throughput must
therefore be measured at runtime for the complete configuration rather than
inferred from the schedule diagram alone.

The production operating points in Section~\ref{sec-production-throughput}
report measured throughput for the PP4 and PP8 configurations used at scale.

\section{Mapping Parallelism to Hardware}
\label{sec-hardware-topology}

\conceptlead{Rank placement constrains the physical communication path.} Every parallelism
axis in the preceding sections was defined by a process group: a list of ranks
that must communicate. The list defines the participants but does not predict
communication cost. Two groups of the same size may lie inside one NVLink
domain or span an InfiniBand fabric. They may also communicate at different
frequencies: once per optimizer step or at every routed layer. Parallelism
determines what moves and how often; rank placement determines the available
physical paths. Together, they affect latency, usable bandwidth, interference,
and end-to-end cost.

\begin{figure}[!htb]
\centering
\includegraphics[width=0.80\linewidth]{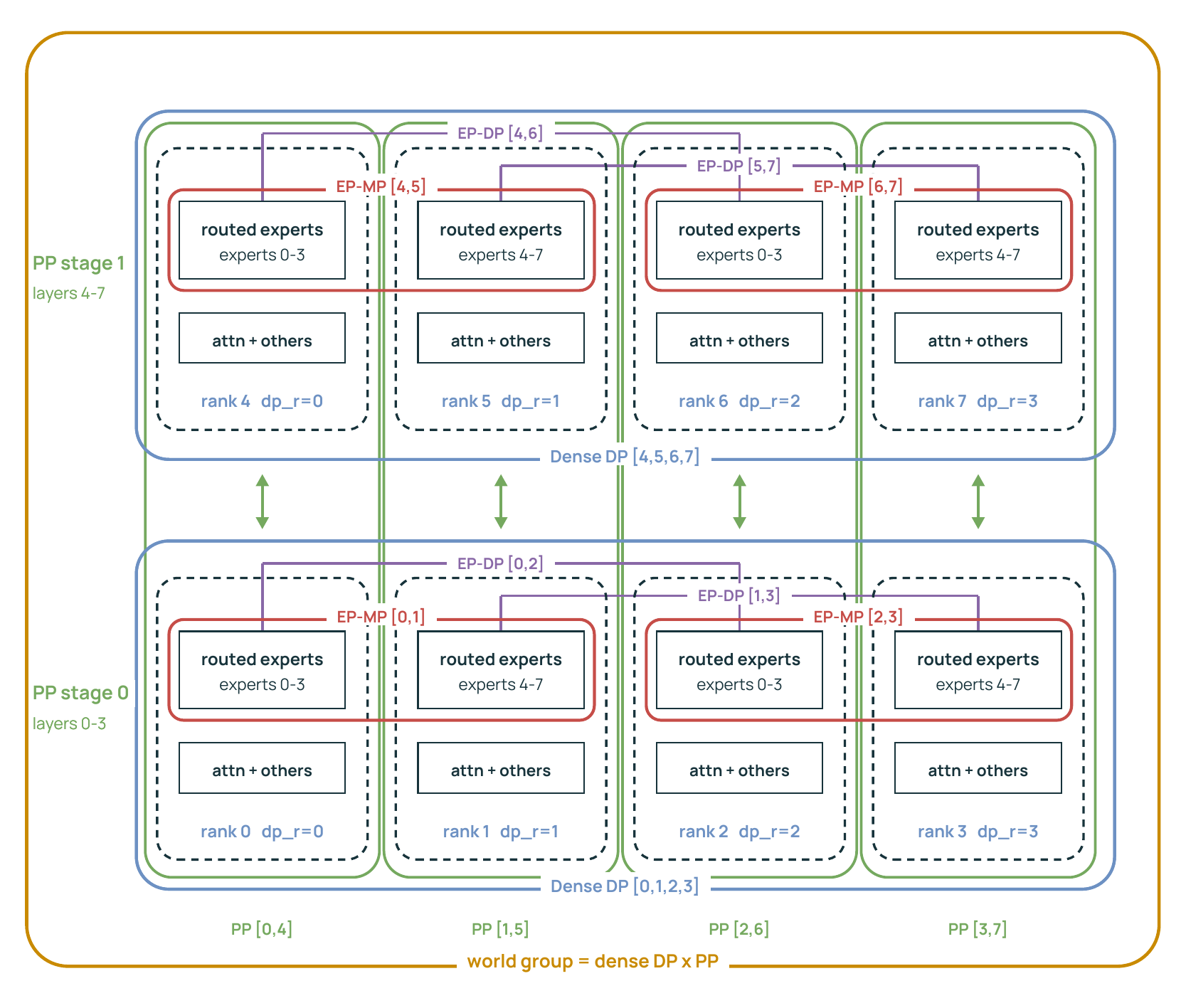}
\caption{\textbf{The same ranks participate in several logical topologies.}
This illustrative eight-rank example uses PP${}=2$, DP${}=4$, EP-MP${}=2$, and
EP-DP${}=2$. Each PP stage stores one layer range; EP-MP pairs form a complete
expert set; EP-DP pairs connect replicas of an expert shard; and PP groups join
matching data-parallel coordinates across stages. Context parallelism is
omitted. The diagram establishes group membership, not the hardware path or a
measured production placement.}
\label{fig-combined-ddp-ep-pp-topology}
\end{figure}

Figure~\ref{fig-combined-ddp-ep-pp-topology} recalls the logical topology from
the preceding sections. PP divides layer ranges. Within
each stage, EP-MP ranks hold different expert shards,
EP-DP ranks hold replicas of the same shard, and dense
parameters reduce over the full data-parallel group. PP groups connect matching
coordinates across layer stages. The diagram leaves physical placement
unspecified: the same logical edge may stay inside a node or cross the network.

\subsection{Rank Order Turns a Mesh into Placement}

\conceptlead{The mesh builder arranges rank numbers.} Olmo-core constructs its
device mesh without consulting a switch table or adapter inventory.
It orders the dense mesh as PP, then data parallelism
(DP), then context parallelism (CP), so CP varies fastest. MoE layers take a
second view of the same per-stage ranks, refolding the combined DP--CP pool
into EP-DP and EP-MP with $DC = D_E M_E$ for CP degree $C$
(Appendix~\ref{app-context-parallelism}). Ranks are laid out in row-major
order, so whichever dimension sits innermost occupies contiguous global ranks:
CP in the dense view, EP-MP in the MoE view. A PP group varies the outermost
dimension and strides by the width of one pipeline stage.

For the eight-rank example in Figure~\ref{fig-combined-ddp-ep-pp-topology}, the
resulting groups are
\[
\begin{aligned}
\text{EP-MP:}&\quad [0,1], [2,3], [4,5], [6,7],\\
\text{EP-DP:}&\quad [0,2], [1,3], [4,6], [5,7],\\
\text{PP:}&\quad [0,4], [1,5], [2,6], [3,7].
\end{aligned}
\]
With eight contiguous local ranks per node, EP-MP occupies the innermost,
adjacent ranks, so at the production degree of eight an EP-MP group fills
exactly one NVL8 domain---eight GPUs on one NVLink fabric
(Section~\ref{subsec-tested-nvl8}). PP groups and the replica groups, which
usually span several nodes, stride across those local blocks. This layout
keeps frequent, irregular EP traffic on NVLink. An EP-MP
group wider than eight has to leave the node.

This mapping follows from mesh dimension order and launcher rank order, with
no topology optimizer. Reordering the hostfile, changing the number of local
ranks, or choosing a stage width that does not divide the node can change the
physical mapping even when the parallelism degrees stay the same. Rank order
therefore belongs in the recorded performance configuration.

\subsection{GPUDirect RDMA Moves Payload; IBGDA Submits Work}

\conceptlead{Payload movement and work submission have separate costs.} Inside one NVL8 node, a GPU reaches peer GPU memory
through the NVLink/NVSwitch fabric. Across nodes the path is longer:
\[
\text{GPU memory}
\rightarrow \text{PCIe}
\rightarrow \text{local HCA}
\rightarrow \text{InfiniBand}
\rightarrow \text{remote HCA}
\rightarrow \text{PCIe}
\rightarrow \text{remote GPU memory}.
\]
The host channel adapter (HCA) is InfiniBand's name for the network adapter---the NIC of
Section~\ref{subsec-ep-deepep-v2}---and it connects to the GPU through PCIe, which is why PCIe appears at both ends of every cross-node
transfer. The nominal fabric rate is only one limit: GPU--HCA locality,
shared PCIe bridges, rail selection (which of the node's per-GPU adapters
carries the transfer, Section~\ref{subsec-tested-nvl8}), message size, and
concurrent traffic can
each become the limiting segment. The chain above is conceptual, since
NCCL or NVSHMEM may select another local adapter or relay path, and only
runtime diagnostics establish the route actually used.

\textbf{GPUDirect RDMA} lets
an HCA read and write GPU memory directly, removing a host-memory bounce buffer
from the payload path %
\citep{nvidia2026gpudirectrdma}; a host thread or proxy
may still be the one preparing the work. \textbf{InfiniBand GPUDirect Async (IBGDA)}, as
used by NVSHMEM, can move that preparation onto the GPU for one-sided operations
\citep{nvidiaNVSHMEMUsing,markthub2022nvshmemibgda}. Symmetric memory
(Section~\ref{subsec-ep-symmetric-memory}) already makes remote expert rows
addressable; the GPU-driven IBGDA path lets the rowwise EP kernel submit the
inter-node transfer itself, with no CPU proxy preparing the work.

The two mechanisms are independent: a direct payload path removes staging
copies, and GPU-side submission removes a host dependency from the critical
path. Both still require completion ordering and memory registration, and
rowwise EP transfers also depend on the lifetime rules of the remotely visible
symmetric buffers.
Section~\ref{sec-sync-d2h} follows the host-synchronization consequence in
detail.

\subsection{Communication Clock Determines Placement Priority}

Table~\ref{tab-parallelism-hardware-clock} combines what moves with when it
moves. A single DDP collective can carry more bytes than a single EP dispatch,
but it runs once per optimizer step and several accumulation microbatches
share its cost. EP and PP communication recurs inside every microbatch, and
full-reshard FSDP repeats parameter materialization at every wrapped execution
boundary. Placement must account for this frequency and critical-path
sensitivity as well as byte count.

\begin{table}[!htb]
\centering
\small
\renewcommand{\arraystretch}{1.16}
\setlength{\tabcolsep}{5pt}
\renewcommand{\tabularxcolumn}[1]{m{#1}}
\begin{tabularx}{\linewidth}{@{}>{\RaggedRight\arraybackslash}m{0.16\linewidth}
>{\RaggedRight\arraybackslash}m{0.40\linewidth}
>{\RaggedRight\arraybackslash}X@{}}
\toprule
\textbf{Axis} & \textbf{Traffic and clock} & \textbf{Placement consequence} \\
\midrule
\arrayrulecolor{black!22}
DDP / optimizer & Parameter-sized gradient and updated-parameter buckets, once
per optimizer step & Bandwidth and rail balance matter; accumulation
microbatches amortize this boundary cost. \\
\midrule
Full-reshard FSDP & Parameter shards materialized into full wrapped weights,
around each wrapped microbatch execution & Repetition makes remote placement
costly even when each collective is regular. \\
\midrule
EP-MP & Routed token rows and returned expert outputs, at every routed layer in
forward and backward for every microbatch & Fast locality, low latency, route
balance, and small-message behavior all matter. \\
\midrule
PP & Hidden activations and their gradients, at every remote stage boundary for
every microbatch & Physical adjacency and independent-work windows determine
the exposed P2P cost. \\
\arrayrulecolor{black}
\bottomrule
\end{tabularx}
\caption{\textbf{Placement pressure follows both payload and recurrence
clock.} The table describes cost structure, not a universal placement ranking.
Actual priorities depend on message shape, degree, model balance, transport,
and the measured hardware path.}
\label{tab-parallelism-hardware-clock}
\end{table}

The axes also use different communication libraries. Dense DDP and EP-DP reduction use NCCL
collectives; PP uses NCCL point-to-point communication; block all-to-all EP uses an
NCCL all-to-all collective; and the rowwise EP-MP path uses NVSHMEM
symmetric memory, because it benefits from indexed one-sided movement among
the ranks of an EP-MP group.

\subsection{The Tested System Has Eight GPUs per NVLink Domain}
\label{subsec-tested-nvl8}

The audited cluster used for the main training experiments has eight NVIDIA B300 SXM6 GPUs per node
in a single NVLink domain, with every GPU pair reported as NV18 (eighteen
NVLink links per GPU). Each node
reaches the fabric over eight active 800-Gb/s-class XDR InfiniBand rails, one
aligned to each GPU across a shared PCIe bridge (PXB)
\citep{nvidia2026hgxb300,nvidia2026quantumx800}.
This is an NVL8 system:
any parallel group wider than eight leaves the NVLink domain and crosses the
fabric.

These specifications describe hardware capabilities, not achieved bandwidth.
``One local rail per GPU'' describes a favorable
GPU--HCA pairing in the physical topology; which rail a given NCCL or NVSHMEM
transfer actually used is a separate question. Enabling an IBGDA environment
option is likewise a request rather than evidence that the effective handler
mode was fully GPU-side. The rail and IBGDA statements in this report
therefore describe the configured paths, taken from the audited inventory
and requested configuration rather than per-transfer counters.

The distinction matters most when comparing training systems. An NVL72 domain
keeps much larger model-parallel groups on NVLink, so results measured there and our NVL8
measurements are not an apples-to-apples topology comparison. The comparison
can also run the other way: the unusually well-provisioned XDR fabric on these
nodes can flatter an inter-node method that would fare worse on a cluster with
fewer or slower rails. Comparing systems therefore requires the end-to-end
path available to each rank group, not just the GPU and fabric names.

\subsection{Topology Changes the Operating Point}

Increasing EP-MP degree from eight to sixteen halves each rank's expert
shard, but also moves EP from an intra-node NVLink exchange to an inter-node
operation, and it can change which backend is worth using: many small row
transfers can become limited by the HCA's message rate, favoring the
block-oriented DeepEP v2 backend
(Section~\ref{subsec-ep-deepep-v2}). Where the crossover falls is an empirical property of
model shape and fabric rather than a universal degree threshold.

Adding accumulation microbatches
spreads DDP gradient and optimizer-boundary traffic across more useful token
work, while EP dispatch/combine and PP activation traffic recur for every
microbatch regardless. One batch-size change can therefore relieve one
communication axis and, at the same time, make another communicate more often.

Separate communicators or CUDA streams still use the same hardware.
Concurrent communication and GEMM can
still contend for SM scheduling, cache, memory bandwidth, PCIe, HCAs, and
fabric links. Section~\ref{subsec-ep-overlap} shows this directly in one
two-node EP-MP${}=16$ diagnostic, where concurrent rowwise communication slowed
the grouped expert work by roughly 23--27\% on the tested shape. The magnitude
is specific to that shape. As Section~\ref{subsec-ep-overlap} shows, the
benefit of hidden communication must be evaluated together with the slowdown
of concurrent kernels.

Parallelism defines which values communicate and how often; mesh dimension
order and launcher rank order turn those logical groups into physical paths.
Our placement gives frequent, latency-sensitive traffic priority within the
NVLink domain, subject to the launcher producing the intended mapping.
Changing degree, batching, backend, or overlap policy can change the limiting
resource, so their combined effect is specific to each model and
hardware configuration.

\FloatBarrier

\clearpage
\part{Efficient Training}
\label{part-efficient-training}
\section{Synchronization-Free Training}
\label{sec-sync-free-training}

\begin{figure}[!htb]
\centering
\includegraphics[width=\linewidth]{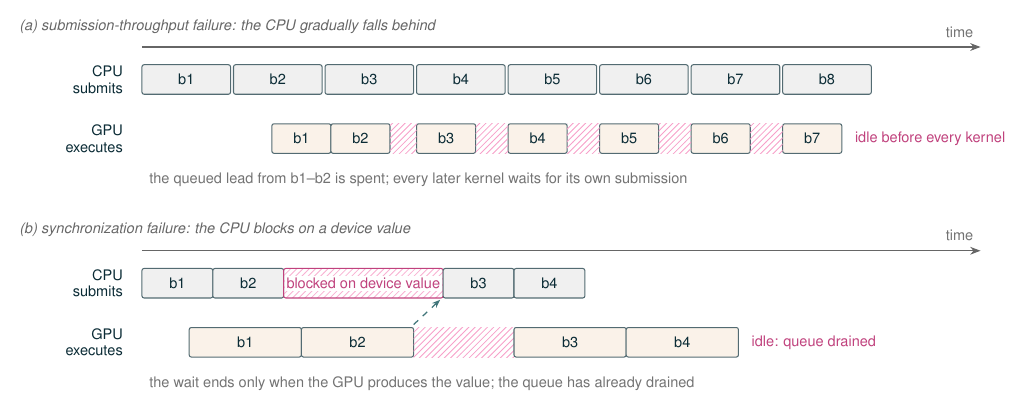}
\caption{\textbf{GPU idle intervals caused by slow submission and host waits.}
(a) When submitting a block costs the CPU more time than executing it costs
the GPU, the queued lead is spent and every later kernel waits for its own
submission; the GPU idles a little before every block. (b) When the CPU
blocks on a device value, submission stops entirely: the GPU finishes the
queued work, sits idle until the value is produced and the host resumes, and
the host must then rebuild its lead. Conceptual illustration with no measured
quantities.}
\label{fig-host-failure-modes}
\end{figure}

The GPU executes each training step, but the CPU prepares and submits nearly
all of that work. A CUDA kernel launch normally returns without waiting for
the kernel to finish, allowing the CPU to submit later operations while the GPU executes
earlier ones %
\citep{nvidiaCUDAAsyncExecution,pytorchCUDASemantics}. This
asynchronous execution is essential for high utilization: the GPU should find
its next operation already waiting when the current operation finishes.
The GPU can run out of queued work if the CPU submits operations too slowly
or blocks while waiting for a device-produced value
(Figure~\ref{fig-host-failure-modes}). Slow submission calls for less host
work or more device work per launch. A device-to-host wait instead calls for
an interface that keeps device results off the host's steady-state path;
initialization, compilation and warmup, and buffer allocation may still
synchronize.

MoE training is especially sensitive to both problems. Routing introduces
many short metadata operations around the expert GEMMs, and the number of
tokens assigned to each rank and expert is data dependent. A direct
implementation may therefore combine small GPU work units with host-visible
counts, variable allocations, and variable communication schedules. These
costs are part of the systems tax introduced in
Section~\ref{sec-intro-dense-assumptions}; they are not counted
by the model-FLOP estimate.

In this section, \textbf{synchronization-free} has a narrow meaning: the
steady-state training path contains no avoidable host-blocking synchronization
for routing, allocation, communication, or expert scheduling. It does not mean
that execution has no ordering. CUDA stream dependencies, collective
completion, autograd dependencies, and optimizer-step boundaries are still
required. In the MoE path examined here, the all-to-all and grouped GEMM
required host-visible counts, which cost one host wait per routed block.
Section~\ref{sec-sync-device-scheduling}
describes how Olmo removes that wait without introducing a new one.

\subsection{The CPU Must Stay Ahead of the GPU}
\label{sec-sync-cpu-ahead}

Consider a repeated region such as one transformer block. Let
$T_{\mathrm{submit}}$ be the CPU time required to submit the region and
$T_{\mathrm{device}}$ be the time required by its GPU work. If
$T_{\mathrm{submit}} < T_{\mathrm{device}}$, the CPU can usually move ahead
while the GPU executes. If $T_{\mathrm{submit}} \geq T_{\mathrm{device}}$,
the GPU eventually reaches the end of the submitted queue and waits for the
host. Dependencies and concurrent streams make this comparison a diagnostic,
not a complete performance model, but it identifies the relevant imbalance.

\begin{figure}[!htb]
\centering
\includegraphics[width=\linewidth]{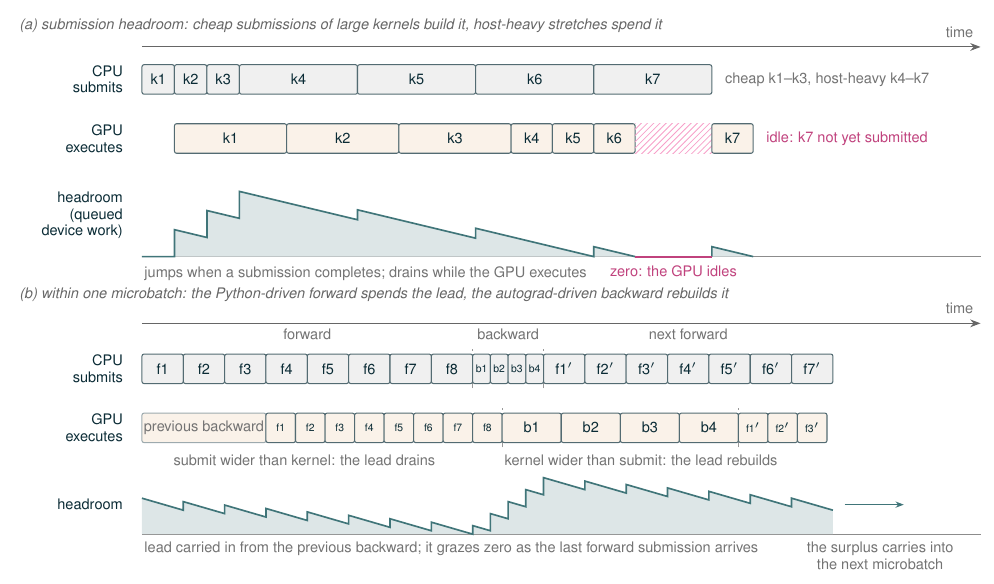}
\caption{\textbf{Submission headroom is the queued, not-yet-executed device
work.} Both curves are computed exactly from the drawn boxes: headroom jumps
up by a kernel's device duration when the CPU completes that kernel's
submission and drains at unit rate while the GPU executes, so the GPU is
idle exactly where the curve sits at zero. (a) Cheap submissions of large
kernels (k1--k3) build the lead; a host-heavy stretch of small kernels
(k4--k7) spends it faster than it is replaced, and after the lead reaches
zero the GPU waits for k7's submission. (b) Within one microbatch, each
Python-driven forward submission costs the CPU more time than its kernel
costs the GPU, so the lead carried in from the previous backward drains
layer by layer and grazes zero just as the last forward submission arrives.
The autograd-driven backward inverts the imbalance, and while the GPU
executes it, the CPU is already submitting the next microbatch's forward:
the rebuilt surplus is the lead that forward will spend. Each CPU box's
width is the full host cost of its submission, including Python and
dispatcher time. Conceptual illustration with no measured quantities.}
\label{fig-sync-headroom}
\end{figure}

\begin{figure}[!htb]
\centering
\includegraphics[width=\linewidth]{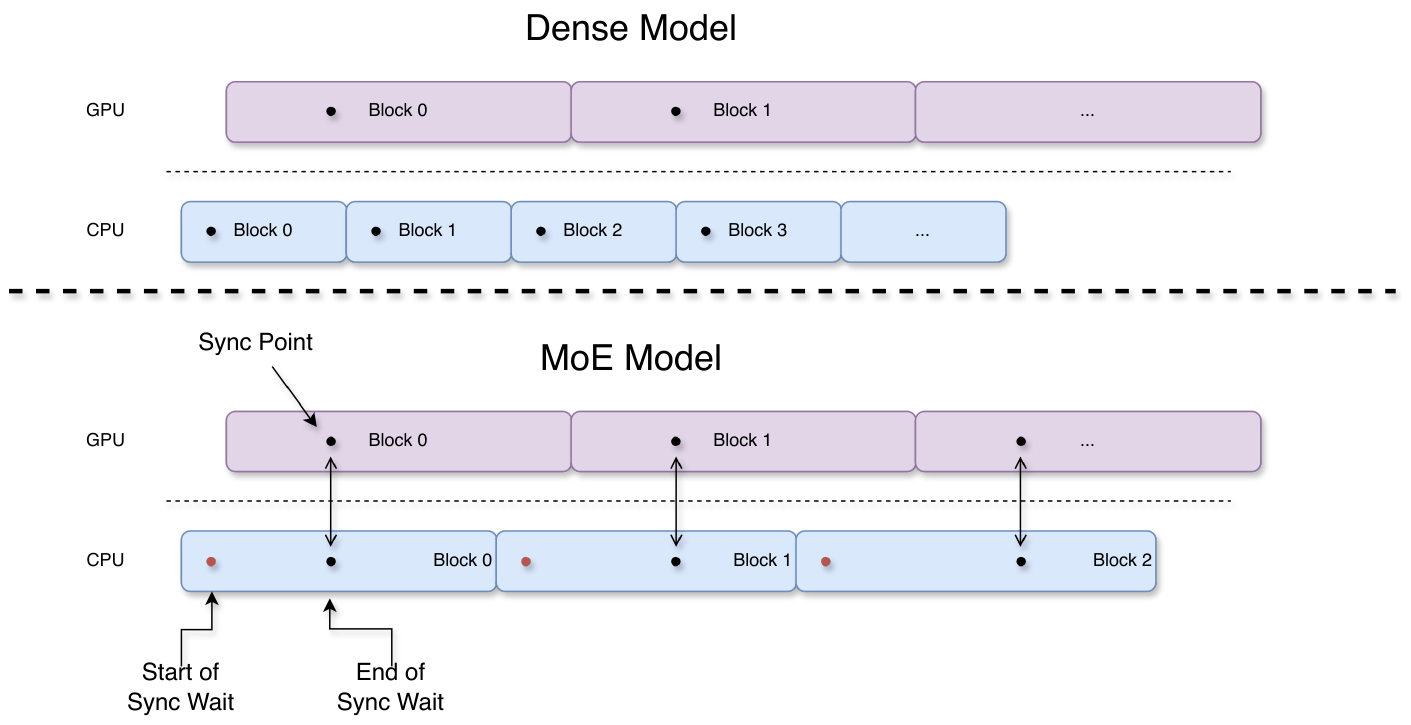}
\caption{\textbf{A per-block synchronization point stops the CPU from running
ahead.} In the dense timeline (top), the CPU submits blocks well in advance of
their GPU start times, so several blocks of device work stay queued and the GPU
always finds its next operation waiting. In the routed timeline (bottom), every
block contains a synchronization point: the CPU begins a sync wait, blocks until
the GPU reaches the matching point, and only then submits the remainder of the
block. Submission headroom cannot accumulate beyond one block, so the two
processors advance in lock step and any GPU stall is exposed. The timeline is
conceptual and does not encode measured duration.
Section~\ref{sec-sync-d2h} examines the synchronizing operations themselves.}
\label{fig-sync-cpu-ahead-timeline}
\end{figure}

We call the amount of submitted but not-yet-completed GPU work the
\textbf{submission headroom}. Headroom is a function of wall-clock time: it
jumps up by a kernel's device
duration when the CPU completes that kernel's submission, drains at unit
rate while the GPU executes, and the GPU is idle exactly while it is zero.
Large GEMMs, scaled dot-product attention, fused
kernels, compiled regions, and C++ execution can build headroom because they
submit substantial device work with relatively little host work. Python call
stacks, dispatcher and compilation-cache lookups, metadata bookkeeping, and
many short launches consume it. Submission headroom is an interpretation of
a CPU/GPU timeline, not a CUDA metric: the CPU's lead over GPU execution
shrinks to zero when the device runs out of queued work.
Figure~\ref{fig-sync-headroom}a draws headroom as this quantity,
and Table~\ref{tab-sync-headroom-ops} summarizes which common operations
build or consume it. Figure~\ref{fig-sync-cpu-ahead-timeline} then contrasts
a dense block
sequence that sustains that lead with a routed one in which a per-block
synchronization point removes it.

Backward passes usually rebuild headroom. A layer's backward launches roughly
twice the GPU work of its forward---the input-gradient and weight-gradient
GEMMs (Dgrad and Wgrad, Section~\ref{sec-grouped-gemm})---while the autograd
engine submits that work with less per-launch overhead than the Python-driven
forward. The lead earned during one microbatch's backward then carries into
the next microbatch: the first forward blocks can still run with positive
headroom, while each submission-bound block spends part of the lead and
leaves less for the blocks behind it. Figure~\ref{fig-sync-headroom}b shows
this drain-and-rebuild cycle within one microbatch.

\begin{table}[!htb]
\centering
\small
\renewcommand{\arraystretch}{1.14}
\renewcommand{\tabularxcolumn}[1]{m{#1}}
\begin{tabularx}{\linewidth}{@{}
  >{\raggedright\arraybackslash}m{0.28\linewidth}
  >{\raggedright\arraybackslash}m{0.15\linewidth}
  >{\raggedright\arraybackslash}X@{}}
\toprule
\textbf{Operations} & \textbf{Effect} & \textbf{Why} \\
\midrule
Large dense and grouped GEMMs & Builds & Much device work per launch \\
\arrayrulecolor{black!22}\hline\arrayrulecolor{black}
Scaled dot-product attention & Builds & Same \\
\arrayrulecolor{black!22}\hline\arrayrulecolor{black}
Fused kernels and compiled regions & Builds & Many operations submitted as
one launch \\
\arrayrulecolor{black!22}\hline\arrayrulecolor{black}
Autograd-engine backward & Builds & Roughly twice forward's device work,
submitted from C++ with low per-launch overhead \\
\arrayrulecolor{black!22}\hline\arrayrulecolor{black}
Python call stacks and dispatcher overhead & Consumes & Host time passes
while little or no device work is queued \\
\arrayrulecolor{black!22}\hline\arrayrulecolor{black}
Compilation-cache lookups and metadata bookkeeping & Consumes & Same \\
\arrayrulecolor{black!22}\hline\arrayrulecolor{black}
Routing metadata and other short kernels & Consumes & Little device work per
launch \\
\arrayrulecolor{black!22}\hline\arrayrulecolor{black}
Device-to-host synchronization & Stops submission & The host blocks until
the device produces the value; the queue drains and the lead must be rebuilt
(Section~\ref{sec-sync-d2h}) \\
\bottomrule
\end{tabularx}
\caption{\textbf{Common operations build, consume, or stop submission
headroom.} The classification is the interpretive lens of
Figure~\ref{fig-sync-headroom}, not a CUDA metric; whether a specific
timeline stays healthy depends on the mixture and on stream dependencies.}
\label{tab-sync-headroom-ops}
\end{table}

The distance between an API launch on the CPU and the corresponding GPU start
is useful, but it must be interpreted with stream dependencies. A long distance
can mean that useful work is already queued, or that the operation is waiting
on another stream. The decisive observation is whether the GPU has runnable
work when the preceding operation completes. GPU utilization and power are too
coarse to establish this cause by themselves.

Captured profiles of the eight-layer development model show both timeline
shapes, one per expert-parallel path
(Figure~\ref{fig-sync-profile-pair} in
Appendix~\ref{app-sync-profile-pair}). In the synchronized trace, a host wait
inside every routed block keeps the
CPU and GPU spans of each forward block aligned, and the kernel rows show
idle gaps at the block boundaries. In the rowwise trace, the CPU submits an
entire microbatch while the GPU is still executing earlier work, and the GPU
then executes the submitted blocks back to back.

\subsection{Small Work Units Expose Submission Overhead}
\label{sec-sync-small-work}

For a fixed operator sequence, a simple host-time approximation is

\begin{equation}
T_{\mathrm{submit}}
  \approx N_{\mathrm{launch}}\,t_{\mathrm{launch}}
  + T_{\mathrm{framework}}
  + T_{\mathrm{bookkeeping}},
\label{eq-host-submission-cost}
\end{equation}

whereas the GPU time of its GEMMs and attention generally grows with the
number of tokens. Reducing the microbatch can therefore shorten device
execution without reducing host work in proportion. This is one reason that
a kernel-level speedup may disappear when the
same kernels are invoked through an eager training stack.

Routed blocks add host submission work to the transformer operations shared
with a dense model: each block constructs routing metadata, counts expert
assignments, attaches router losses, permutes or maps rows, launches expert
communication, and schedules grouped GEMMs. Expert parallelism can also divide
the local tokens among many experts, making each GEMM shorter. The resulting
sequence of short GPU operations leaves less time for the CPU to prepare
the next group of launches, even when each operation is cheap in isolation.

The same reasoning applies to very fast normalization, elementwise, and
reduction kernels. Their device time can be shorter than the framework and
launch path around them. The appropriate remedies are then larger work units,
fusion, compilation, or a lower-overhead C++ interface. These remedies reduce
ordinary submission cost; they do not remove an explicit device-to-host wait.
The balance between submission and execution sets the microbatch-size
crossover discussed next.

\subsection{Critical Microbatch Size}
\label{sec-critical-microbatch-size}

This host--device balance defines a \textbf{systems critical microbatch size}.
It is unrelated to the optimization-theoretic critical batch size discussed
in Section~\ref{sec-lr-batch-scaling}.  Let $b$ denote the rank-local amount
of work in one microbatch---for example, the number of sequences at fixed
sequence length---and compare the CPU submission time
$T_{\mathrm{submit}}(b)$ with the corresponding GPU execution time
$T_{\mathrm{device}}(b)$.  The crossover occurs near

\begin{equation}
T_{\mathrm{device}}(b_{\mathrm{crit}})
  \approx T_{\mathrm{submit}}(b_{\mathrm{crit}}).
\label{eq-critical-microbatch-size}
\end{equation}

Below this point, the GPU completes the submitted work faster than the CPU can
prepare and launch the next work, so the GPU can run out of queued operations
and become idle.  At or above the crossover, each submitted region contains
enough device work for the CPU to stay ahead.  This threshold is not a model
constant: it depends on the operator mix, fusion and compilation boundaries,
software stack, CPU, GPU, and sequence length.

\begin{figure}[!htb]
  \centering
  \includegraphics[width=\linewidth]{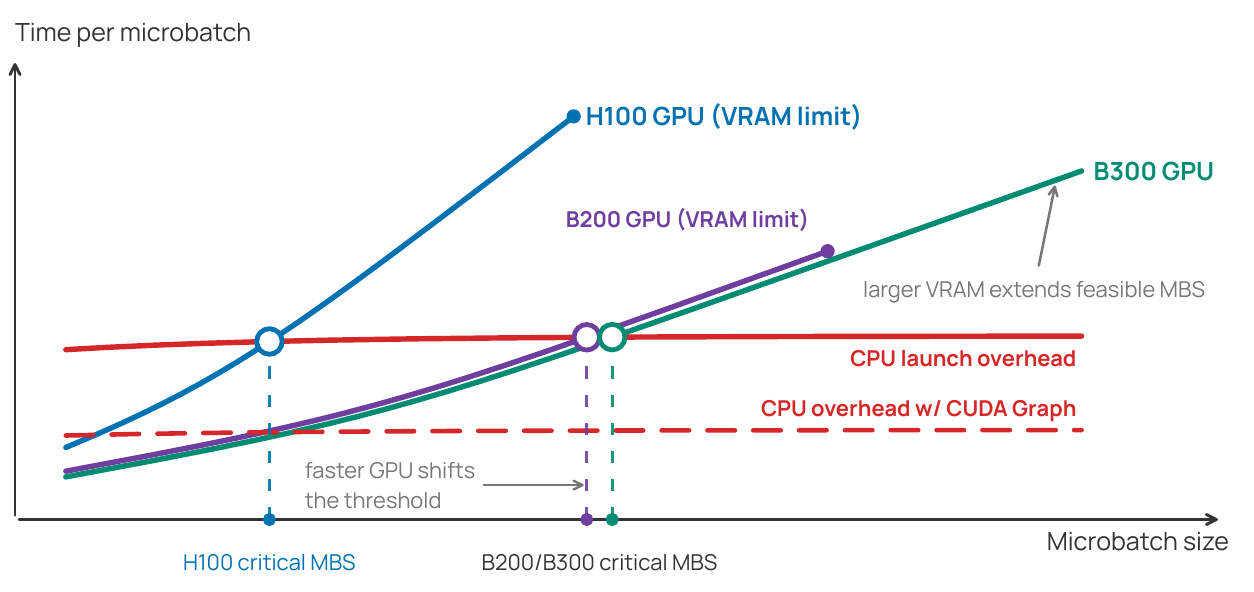}
  \caption{\textbf{A faster GPU can move the critical microbatch size to the
  right.}  The solid red curve represents CPU submission time, while the blue,
  purple, and green curves represent H100, B200, and B300 execution time.
  Below a crossover, the device can consume work faster than the host supplies
  it; faster execution with an unchanged host path therefore requires a larger
  microbatch to restore submission headroom.  Additional GPU memory can extend
  the feasible microbatch range, while reducing submission overhead (dashed
  red; CUDA Graph replay is one way to reduce this overhead) can move the crossover to the
  left.  The curves,
  intersections, and memory limits are qualitative annotations rather than
  measurements; no numerical critical size is implied.}
  \label{fig-critical-microbatch-size}
\end{figure}

The crossover must be identified from the CPU/GPU profile of the selected
configuration. Figure~\ref{fig-critical-microbatch-size} illustrates how it
can move; it does not estimate a critical size for a production run.

This trend matters as accelerators improve.  Moving from Hopper to Blackwell
can shorten the device side of the same operator sequence without reducing
Python, dispatcher, routing-bookkeeping, or launch work by the same factor.
The crossover can therefore shift to a larger microbatch.  A configuration
that kept an H100 continuously supplied may become host-submission limited on
a faster GPU even though every individual GPU kernel improved.

Memory-saving techniques can keep the GPU supplied by making a larger
microbatch feasible: lower-precision activation
storage and activation recomputation reduce the per-token memory cost, as
discussed in Sections~\ref{sec-mxfp8-training} and \ref{sec-recompute}.
Cheaper submission instead moves $b_{\mathrm{crit}}$ to the left: fusion,
compiled regions, and C++ execution reduce repeated host work in the
production path, and CUDA Graph replay provides a low-host-overhead reference
\citep{nvidiaCUDAGraphs,pytorchCUDASemantics}. Larger microbatches consume more activation memory,
while static PyTorch capture/replay imposes address, shape, and buffer-retention constraints;
their value depends on whether CPU submission or GPU execution limits the
measured step at the selected microbatch size.

\subsection{Device-to-Host Synchronization Drains the Queue}
\label{sec-sync-d2h}

A \textbf{device-to-host (D2H) synchronization} occurs when host execution cannot
continue until a value computed on the GPU is available to the CPU. The CPU
stops submitting later work while the GPU completes all dependencies needed to
produce that value. The wait consumes submission headroom, but the GPU becomes
idle only if it outlasts useful independent work that was already submitted.
After the value arrives, the host must rebuild its lead over the device.

Block all-to-all EP (Section~\ref{subsec-ep-block-alltoall}) provides
a concrete example. Routing produces the
number of tokens sent to each destination on the GPU. In Olmo-core's
synchronized path, ranks exchange token counts on the GPU, then copy the send,
receive, and local-expert counts to CPU memory and wait. The send and receive
counts become the Python split lists used to launch the data exchange. A fully
dynamic receive buffer has the same problem when its allocation shape is
derived from device counts. Some legacy grouped-GEMM
backends\footnote{In Olmo-core, the legacy grouped-GEMM fallback used
before the PyTorch device-offset fast path takes host-side split sizes.}
likewise require
host-visible expert sizes, although that dependency can occur without EP.
The simplified pseudocode shows one host copy and wait serving both host
dependencies:

\begin{lstlisting}[language=Python,escapeinside={(*@}{@*)}]
peer_sizes_gpu = count_routes_per_peer(routes)
expert_sizes_gpu = count_routes_per_expert(routes)
peer_sizes, expert_sizes = copy_to_cpu_and_wait( (*@\textcolor{red}{\bfseries\# host sync}@*)
    peer_sizes_gpu, expert_sizes_gpu)
(*@\textcolor{green!50!black}{\bfseries\# All-to-all requires host-side split sizes.}@*)
all_to_all(x, split_sizes=peer_sizes)
(*@\textcolor{green!50!black}{\bfseries\# Legacy grouped GEMM also requires host-side expert sizes.}@*)
grouped_gemm(x, expert_sizes)
\end{lstlisting}

Both dependencies arise when an interface turns device-produced counts into host
inputs. A receive buffer allocated from the current route count creates the
same dependency even if the transfer itself accepts device metadata.
Section~\ref{sec-sync-device-scheduling} follows the changes to communication,
allocation, and grouped-GEMM scheduling that keep these counts on the GPU.

\begin{figure}[!htb]
\centering
\includegraphics[width=\linewidth]{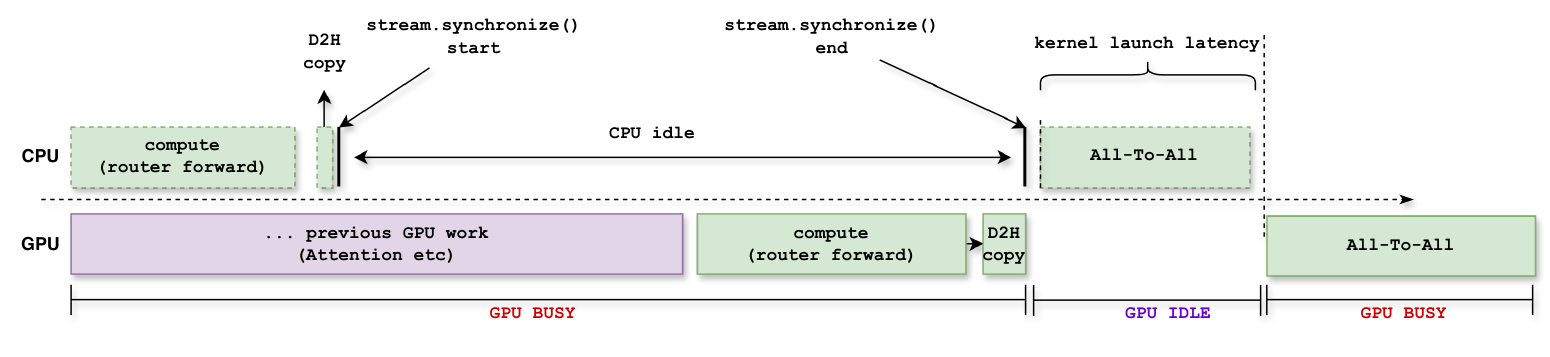}
\caption{\textbf{Host-visible split sizes delay the all-to-all launch.} The GPU
produces route counts, but the synchronized variable-split path must expose them to
the CPU before the collective can be launched. If the queue drains at this
boundary, the GPU becomes idle until the host receives the counts and submits
the all-to-all. The timeline is conceptual and does not encode measured
duration.}
\label{fig-d2h-sync-a2a}
\end{figure}

\begin{figure}[!htb]
\centering
\includegraphics[width=\linewidth]{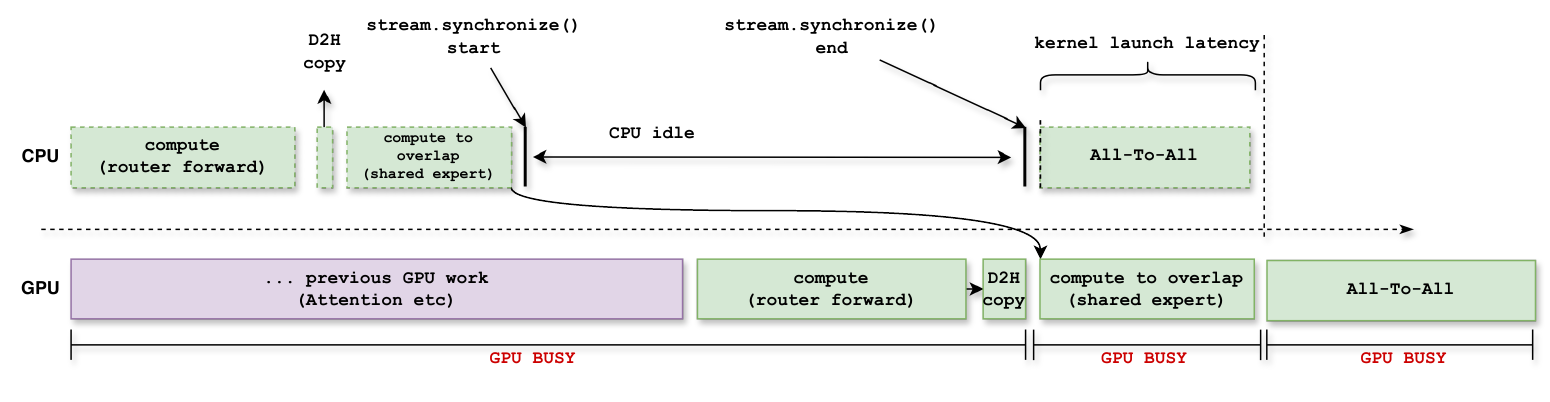}
\caption{\textbf{Independent compute can cover a host wait without removing
it.} The CPU submits shared-expert work before blocking on GPU-produced route
counts, so the GPU remains busy while the host waits. The all-to-all still
cannot be submitted until the counts are host-visible. This is latency
coverage, not communication--compute overlap; the timeline is conceptual and
does not encode measured duration.}
\label{fig-d2h-sync-a2a-overlap}
\end{figure}

Independent work submitted before the host wait can occupy some or all of this
interval. In Figure~\ref{fig-d2h-sync-a2a-overlap}, shared-expert compute runs while the CPU waits
for the route counts. This scheduling hides the GPU bubble but does not remove
the synchronization or allow the all-to-all to launch before its split sizes
reach the host.

Dynamic routing itself does not require a host synchronization. The
synchronization is introduced by an interface that asks the CPU to interpret a
device result. Table~\ref{tab-host-sync-patterns} separates definite
host-visible operations from operations that are often blamed incorrectly.

\begin{table}[!htb]
\centering
\small
\renewcommand{\arraystretch}{1.18}
\setlength{\tabcolsep}{5pt}
\begin{tabularx}{\linewidth}{>{\raggedright\arraybackslash}p{0.25\linewidth}
                              >{\raggedright\arraybackslash}p{0.25\linewidth}
                              X}
\toprule
\textbf{Pattern} & \textbf{Host wait} & \textbf{Reason} \\
\midrule
\texttt{x.item()} & Yes & Materializes a CUDA scalar as a Python value. \\
\arrayrulecolor{black!22}\hline\arrayrulecolor{black}
\texttt{x.cpu()} followed by host use & Yes & The host-visible copy must wait
for the required source data. \\
\arrayrulecolor{black!22}\hline\arrayrulecolor{black}
\texttt{torch.nonzero(x)} on CUDA & Yes & PyTorch documents a host--device
synchronization for this dynamic-output operator. \\
\arrayrulecolor{black!22}\hline\arrayrulecolor{black}
\texttt{torch.empty([int(x)])} & Yes, before allocation & The conversion of
the CUDA scalar to a Python integer synchronizes; \texttt{empty} itself is not
the cause. \\
\arrayrulecolor{black!22}\hline\arrayrulecolor{black}
Device indexing, gather, or scatter & Not inherently & These remain GPU
operations when indices, shapes, and results stay on the device. \\
\bottomrule
\end{tabularx}
\caption{\textbf{Host synchronization depends on how a GPU result is consumed.}}
\label{tab-host-sync-patterns}
\end{table}

PyTorch explicitly documents the synchronization behavior of
\texttt{nonzero()} on CUDA %
\citep{pytorchNonzero}. Scalar extraction and
host-visible copies have the same fundamental dependency: Python cannot use a
value that the device has not produced. This does not imply that every dynamic
operator or tensor allocation synchronizes. Whether a hot-path call blocks
depends on where its shape and control data reside.

CUDA stream events illustrate the distinction between required device ordering
and host blocking. A stream can wait on an event while the CPU continues to
submit independent work. Calling a host synchronization routine waits on the
CPU and therefore changes the submission path. The goal is to retain necessary
device ordering without making the host participate in data-dependent
scheduling.

\subsection{How to Become Synchronization-Free}
\label{sec-sync-device-scheduling}

Device-resident route metadata and device-side grouped-GEMM offsets remove
the host-visible counts that the all-to-all and grouped GEMM required, and
with them the per-block host wait, from Olmo's rowwise MoE path.
Capacity-shaped storage prevents route counts from becoming host allocation
decisions. The selected routes remain dynamic; buffer capacities are fixed
before the training loop.

The rowwise mechanics are those of
Sections~\ref{subsec-rowwise-ep-motivation}
and~\ref{subsec-rowwise-ep-buffer-reuse}; this subsection isolates their
host-facing consequence. Table~\ref{tab-sync-device-envelope} lists the
host dependency removed by each mechanism and its associated cost. The first two rows remove
the all-to-all dependency; the third removes the grouped-GEMM dependency.
Each removal has a focused illustration: Figures~\ref{fig-ep-1d-alltoall-detail}
and~\ref{fig-ep-rowwise-detail} contrast the synchronized block all-to-all
with the rowwise path whose route metadata never leaves the GPU, and
Figures~\ref{fig-grouped-gemm-host} and~\ref{fig-grouped-gemm-device}
contrast the host-loop grouped GEMM with the device-scheduled interface that
consumes offsets where they are produced.

\begin{table}[!htb]
\centering
\small
\renewcommand{\arraystretch}{1.16}
\setlength{\tabcolsep}{5pt}
\renewcommand{\tabularxcolumn}[1]{m{#1}}
\begin{tabularx}{\linewidth}{@{}>{\RaggedRight\arraybackslash}m{0.27\linewidth}
>{\RaggedRight\arraybackslash}m{0.32\linewidth}
>{\RaggedRight\arraybackslash}X@{}}
\toprule
\textbf{Mechanism} & \textbf{Host dependency removed} &
\textbf{Cost accepted} \\
\midrule
\arrayrulecolor{black!22}
Device-resident route metadata: route maps, counts, and offsets stay CUDA
tensors & Python split lists interpreted per microbatch & Metadata kernels in
every routed layer \\
\midrule
Capacity-shaped buffers with a valid-row prefix & Data-dependent allocation
shapes & Reserved capacity memory; possible route drops \\
\midrule
Device-side cumulative offsets for grouped GEMM & \texttt{item()} per expert
count and a host GEMM loop & Backend must accept device offsets \\
\arrayrulecolor{black}
\bottomrule
\end{tabularx}
\caption{\textbf{Device-side metadata and fixed-capacity storage remove
host dependencies.} The mechanisms retain dynamic routing while keeping
data-dependent counts and allocation decisions off the host path.}
\label{tab-sync-device-envelope}
\end{table}

\begin{keyidea}[Dynamic Routes, Fixed Buffer Capacities]
Keep route counts dynamic on the GPU; fix buffer capacities before the
training loop.
\end{keyidea}

\paragraph{Device-resident route metadata.}
The router produces route maps, destination ranks and rows, per-rank counts,
and expert-group offsets as CUDA tensors. The rowwise dispatch and combine
kernels consume those tensors directly
(Figure~\ref{fig-ep-rowwise-detail}). Olmo currently uses NVSHMEM because it
supports GPU-initiated one-sided PUT and GET operations over symmetric memory
\citep{nvidiaNVSHMEMUsing}. The host still launches kernels and establishes the
communication resources; it does not retrieve a variable split list for each
microbatch. Per-rank count exchange, stream waits, and NVSHMEM barriers still
provide required device ordering; their route-dependent values remain CUDA
tensors and are not consumed by the host. The count exchange is one
all-gather over the EP-MP group. This is the rowwise NVSHMEM path
analyzed in this section.

\paragraph{Capacity-shaped storage.}
Dispatch, combine, scale, and metadata buffers are allocated to a configured
capacity. A route map identifies which rows are valid in the current
microbatch. This separates a static upper bound, which the allocator needs,
from the dynamic valid-row count, which remains on the device. The distinction
matters because unwritten capacity rows may contain stale data; expert compute
and combine therefore consume only the rows marked valid for the current
microbatch.

\paragraph{Device-side grouped-GEMM scheduling.}
Current Olmo expert compute uses a grouped-GEMM interface with a CUDA
\texttt{int32} cumulative-offset tensor
(Figure~\ref{fig-grouped-gemm-device}). The backend, rather than a Python
loop, identifies the row range assigned to each local expert. This removes the
need to call \texttt{item()} on each expert count and also reduces the number
of GEMM launches. A CPU-loop implementation would instead copy offsets to the
host and serialize group launches, recreating the synchronization and launch
gaps that the device-offset interface removes.

\paragraph{Capacity trades memory for the removed stall.}
A fixed envelope can reserve more memory than the valid routes require and can
drop routes if the configured capacity is exceeded. Current grouped GEMMs use
device offsets and stop at the valid expert rows
(Section~\ref{sec-grouped-gemm}), but capacity tails still
occupy buffer memory and can be touched by non-GEMM elementwise work; a
tail-aware SwiGLU kernel that applies the activation only to the valid prefix
avoids touching the tails on forward-only paths. Other
implementations may also communicate padded rows. Useful-token FLOPs must
exclude those rows, while physical memory, transferred bytes, and executed
kernel work must include them. Capacity overhead is therefore the cost of removing the
host synchronization.

\paragraph{Reducing remaining submission overhead.}
Fusion, compiled regions, and C++ execution reduce the number and cost of
ordinary launches after dynamic metadata has been kept on the device. These
techniques solve a different problem from D2H synchronization. A faster Python
or C++ call stack cannot repair an API that fundamentally requires a GPU value
in Python. In the current implementation, the custom symmetric-memory
transport remains a deliberate C++/CUDA boundary of the compiled block;
compilation reduces surrounding work but does not capture the complete rowwise
path as one graph.

CUDA Graph replay could remove much of this remaining submission cost, but
its stable-address and fixed-shape requirements conflict with live pipeline
microbatches, and the production path is not graph captured.
Section~\ref{subsec-static-cudagraph-lesson} analyzes that trade and uses
replay only as a diagnostic ceiling that separates device service time from
eager submission cost.

Host-wait behavior depends on the running configuration.
Appendix~\ref{app-sync-verification} records the verification checklist and
a graph-replay microbenchmark that isolates the submission tax on short
expert operations. In the captured rowwise timeline of
Appendix~\ref{app-sync-profile-pair}, no host wait interrupts submission for
the eight-layer BF16 EP8 development model with a shared expert. This is a
single-rank captured instance of the mechanism, not a guarantee for every
configuration or execution phase; initialization, compilation, warmup, and
buffer allocation remain outside the steady-state scope.

Capacity memory and padding remain costs of this design. Once the
count-dependent wait is removed, ordinary submission overhead still
determines whether the CPU can keep the GPU supplied.

\section{Grouped GEMM}
\label{sec-grouped-gemm}

Routing turns one expert layer into a ragged batch of matrix
multiplications. After dispatch, each rank holds token rows ordered by local
expert and a device tensor containing the number of rows assigned to every
expert.  \textbf{Grouped general matrix multiplication (grouped GEMM)}
executes these unequal row counts without padding every expert to a
fixed capacity. A device-scheduled implementation also avoids returning the
counts to the CPU.

\subsection{Routing Creates Uneven GEMMs}

Let $M_e$ be the number of rows assigned to local expert $e$, $d$ the model
dimension, and $h$ the expert intermediate dimension.  A SwiGLU expert applies

\begin{equation}
\begin{aligned}
U_e &= X_e W_{\mathrm{ug},e}^{\mathsf T},
& X_e &\in \mathbb{R}^{M_e\times d},
& W_{\mathrm{ug},e} &\in \mathbb{R}^{2h\times d}, \\
Y_e &= \operatorname{SwiGLU}(U_e)W_{\mathrm{down},e},
& Y_e &\in \mathbb{R}^{M_e\times d},
& W_{\mathrm{down},e} &\in \mathbb{R}^{h\times d}.
\end{aligned}
\label{eq-grouped-expert-mlp}
\end{equation}

Within one routed layer, every expert has the same weight shapes; routing
changes only $M_e$.  The rank therefore receives
$M=\sum_e M_e$ useful rows but must execute a collection of GEMMs whose first
dimension varies by expert.  A single dense GEMM cannot select a different
weight matrix for each segment, while launching every expert independently
creates a long sequence of small operations.

\subsection{Padding, Host Scheduling, and Device Scheduling}

\begin{figure}[!htb]
  \centering
  \includegraphics[width=0.96\linewidth]{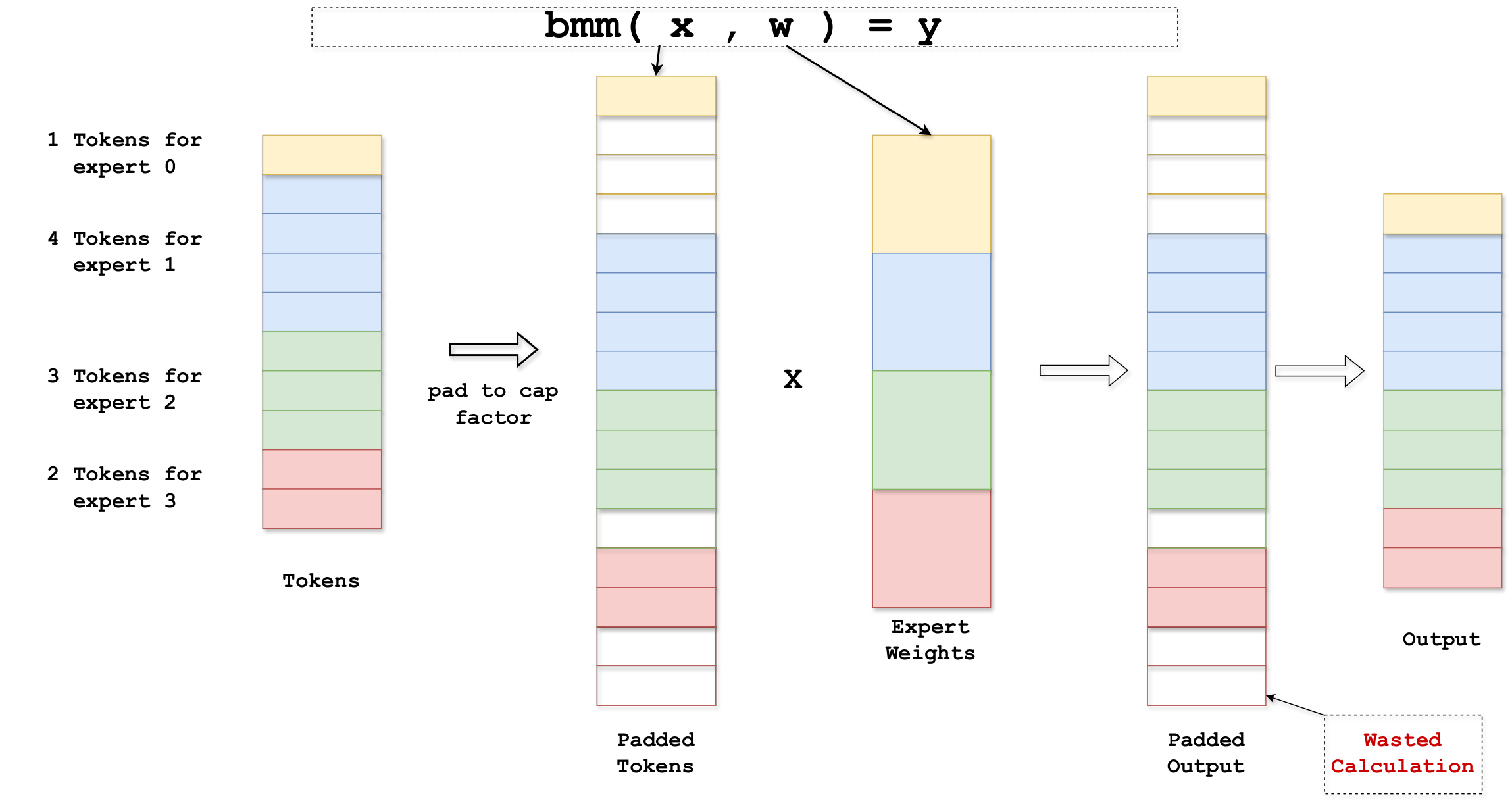}
  \caption{\textbf{Padded BMM regularizes the ragged row dimension.}
  Every expert is padded to capacity $C$, enabling one regular batched
  operation.  Empty rows nevertheless consume matrix-multiply work, and their
  outputs are discarded.  Expert weight shapes are fixed; only routed row
  counts differ.}
  \label{fig-grouped-gemm-bmm}
\end{figure}

\begin{figure}[!htb]
  \centering
  \includegraphics[width=\linewidth]{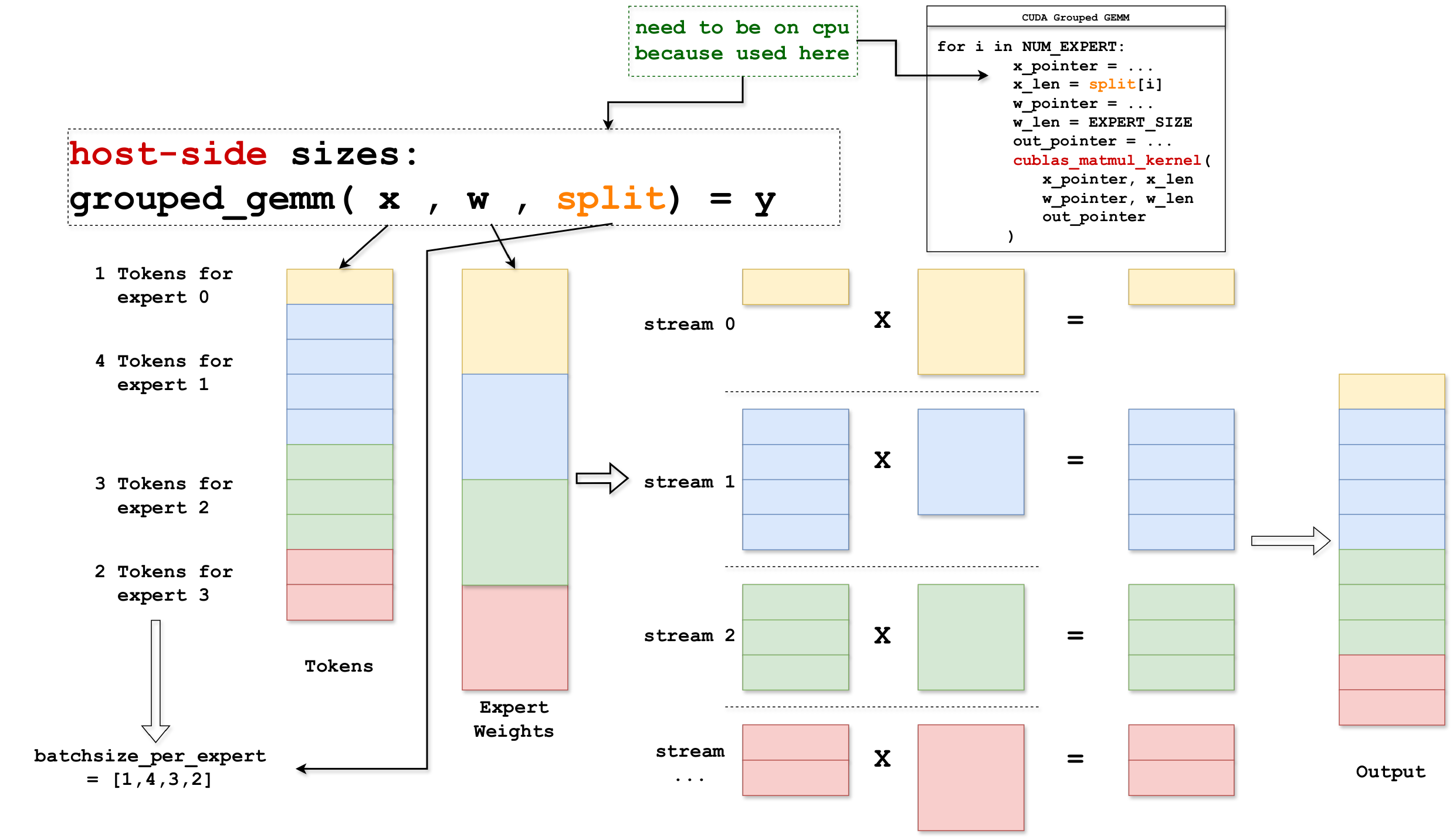}
  \caption{\textbf{A host loop executes exact-size expert GEMMs.}
  It avoids padded arithmetic, but GPU-produced row counts must become
  host-visible before the runtime can submit the per-expert operations.  The
  separate streams in this conceptual drawing are not required by the method;
  the legacy Olmo fallback submits its GEMMs on the current stream.}
  \label{fig-grouped-gemm-host}
\end{figure}

\begin{figure}[!htb]
  \centering
  \includegraphics[width=0.6\linewidth]{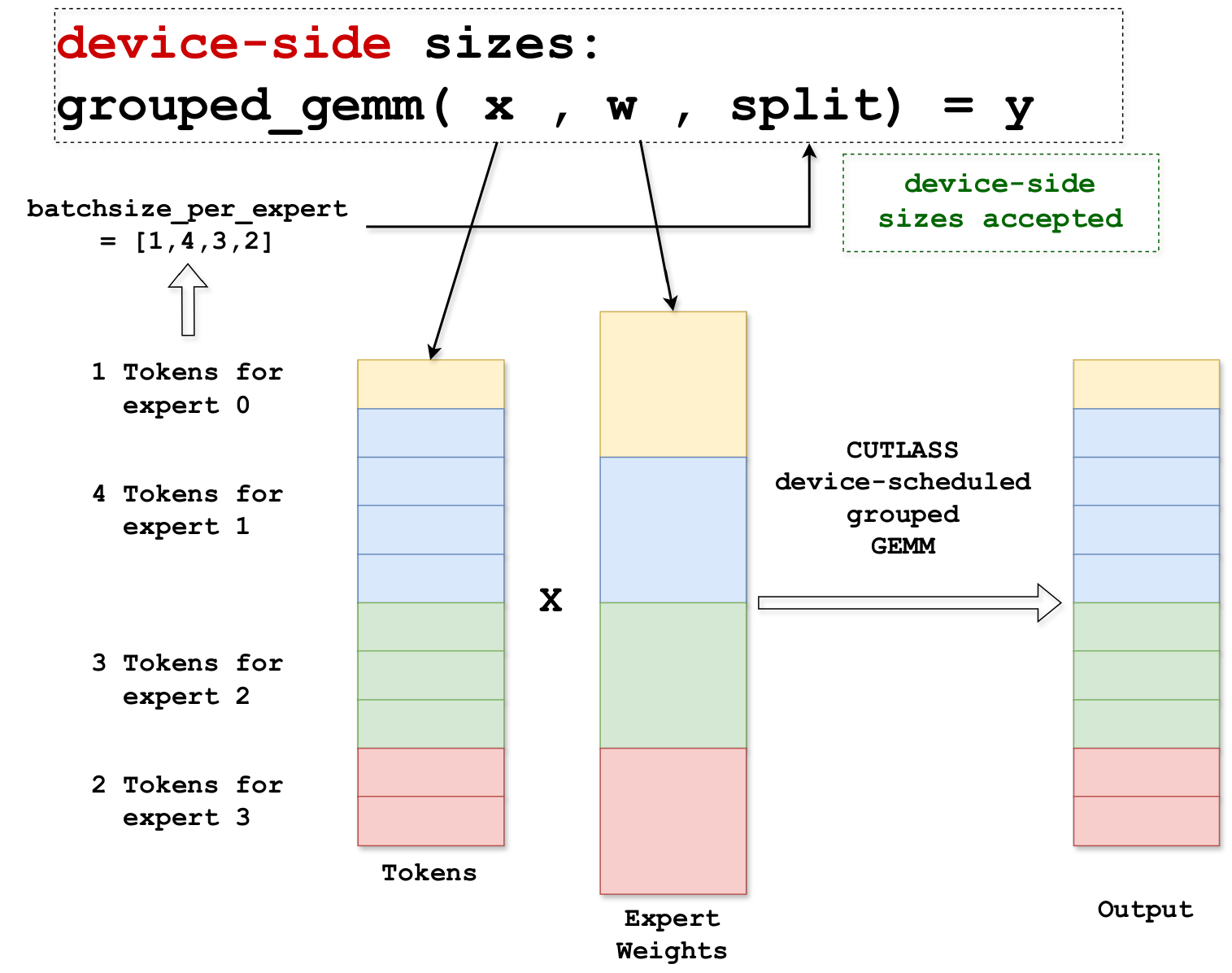}
  \caption{\textbf{Device-scheduled grouped GEMM executes exact row segments
  without a host round trip.}
  The drawing shows the device row counts; Olmo converts them on the GPU to
  cumulative offsets ($[1,5,8,10]$ here).  GPU-side preparation
  then forms the grouped problem list before CUTLASS schedules the compute.
  The physical buffers retain fixed shapes while the offsets restrict work to
  the rows assigned to each expert.}
  \label{fig-grouped-gemm-device}
\end{figure}

Figures~\ref{fig-grouped-gemm-bmm}, \ref{fig-grouped-gemm-host},
and~\ref{fig-grouped-gemm-device} compare three ways to execute the same
expert projections.  They differ in whether they pad the ragged row dimension,
make its sizes visible to the host, or schedule the variable-size problems
directly from device metadata.

\paragraph{Padded BMM.}
The simplest strategy pads every expert to a common capacity $C$ and forms an
$N_{\mathrm{local}}\times C\times d$ activation tensor.  \textbf{Batched matrix
multiplication (BMM)} then launches one regular operation, but it executes
$N_{\mathrm{local}}C$ rows even though only $M$ rows contain routes.  Its
padding fraction is%
\nopagebreak
\begin{equation}
\rho_{\mathrm{pad}}
  = 1 - \frac{\sum_e M_e}{N_{\mathrm{local}}C}.
\label{eq-bmm-padding-fraction}
\end{equation}

The cost rises with reserved capacity and, once experts overflow their
reserved rows, with load imbalance.  Empty rows
consume the same matrix-multiply work as routed rows.

\paragraph{Host-scheduled variable GEMMs.}
An exact-size alternative makes $(M_1,\ldots,M_{N_{\mathrm{local}}})$ visible to the CPU and
launches the corresponding GEMMs from a host loop.  This avoids padded
arithmetic, but it introduces a device-to-host dependency before expert compute
can be submitted and replaces the single grouped kernel with a sequence of smaller
launches.

\paragraph{Device-scheduled grouped GEMM.}
A device-scheduled backend keeps the cumulative expert offsets on the GPU.  A
GPU preparation kernel derives the per-expert pointers, strides, and matrix
dimensions; one CUTLASS grouped kernel then schedules
tiles across that problem list %
\citep{nvidiaCUTLASSGroupedScheduler}.  The
current Olmo BF16 path reaches this mechanism through PyTorch grouped matrix
multiplication on its CUDA fast path. MegaBlocks instead expresses expert
compute as block-sparse matrix multiplication over a block-compressed layout.
This avoids padding every expert to a common capacity, although its published
implementation still pads each expert batch to a multiple of 128 rows for
block alignment. It is a related, but distinct, execution strategy
\citep{gale2023megablocks}.

\subsection{Device-Side Offsets Avoid a Host Synchronization}

The route histogram is produced on the GPU.  If a GEMM interface requires a
Python list or CPU tensor of split sizes, the CPU must wait for that histogram
before it can submit the expert work.  As Section~\ref{sec-sync-d2h} explains,
this device-to-host dependency drains the queued GPU work and can expose a
submission bubble after the synchronization.

Olmo instead computes cumulative expert offsets on the device and passes them
to grouped matrix multiplication.  The preparation and scheduling kernels can
therefore be queued without the CPU interpreting any route count.
The selected backend must support device offsets. A fallback that copies
them to the CPU recreates the
synchronization even when the call site still says \texttt{grouped\_mm}.
Appendix~\ref{app-device-grouped-gemm-availability} records the
architecture-specific availability of the current device-offset fast path.

\subsection{Padded Shape, Unpadded Compute}
Static physical shapes are a central advantage of the rowwise path
(Section~\ref{subsec-rowwise-ep-motivation}).  Rowwise
EP reserves capacity-shaped communication buffers, so their addresses and
dimensions do not change with every routing decision.  The device offsets
separately describe the variable amount of work for each expert.  Dispatch
writes the received routes directly into expert-contiguous segments, and the
cumulative offsets are computed from the real $M_e$ values.  The grouped GEMM
consumes only those segments; unused capacity remains outside the final
offset.  Changing route counts therefore does not require new allocations or
new tensor shapes at the compiled boundary, yet the unused tails are not
executed.

This layout also removes a local permutation between communication and expert
compute.  The up/gate projection reads the dispatch buffer directly, and the
down projection can write directly into the buffer consumed by combine---the
direct-access pattern of Figure~\ref{fig-ep-symm-boundary}b.  In
backward, the buffer-aware grouped GEMM wrapper can write the input gradient
back into the designated dispatch-gradient storage.  Thus, the rowwise path
combines fixed-shape allocation with unpadded compute and avoids an additional
pack--unpack pair at the grouped-GEMM boundary.
Appendix~\ref{app-kernel-interface-reference} documents the explicit
output-buffer interfaces.

\needspace{8\baselineskip}
\subsection{Forward, Dgrad, and Wgrad Are Different Problems}

For a grouped linear projection $Y_e=X_eW_e$, let
$X_e\in\mathbb{R}^{M_e\times d_{\mathrm{in}}}$ and
$W_e\in\mathbb{R}^{d_{\mathrm{in}}\times d_{\mathrm{out}}}$.  Backward
requires the input gradient (\textbf{Dgrad}) and the weight gradient
(\textbf{Wgrad}):

\begin{equation}
\nabla X_e = \nabla Y_e W_e^{\mathsf T},
\qquad
\nabla W_e = X_e^{\mathsf T}\nabla Y_e.
\label{eq-grouped-linear-backward}
\end{equation}

\begin{table}[!htb]
  \centering
  \small
  \setlength{\tabcolsep}{3.2pt}
  \renewcommand{\arraystretch}{1.22}
  \begin{tabularx}{\linewidth}{@{}lcccc>{\centering\arraybackslash}X@{}}
    \toprule
    \textbf{Phase} & \textbf{Left operand} & \textbf{Right operand} &
    \textbf{Contract} & \textbf{Result} & \textbf{Role of $M_e$} \\
    \midrule
    Forward & $X_e\,[M_e\!\times\!d_{\mathrm{in}}]$ &
      $W_e\,[d_{\mathrm{in}}\!\times\!d_{\mathrm{out}}]$ &
      $d_{\mathrm{in}}$ & $Y_e\,[M_e\!\times\!d_{\mathrm{out}}]$ &
      Output rows \\
    Dgrad & $\nabla Y_e\,[M_e\!\times\!d_{\mathrm{out}}]$ &
      $W_e^{\mathsf T}\,[d_{\mathrm{out}}\!\times\!d_{\mathrm{in}}]$ &
      $d_{\mathrm{out}}$ & $\nabla X_e\,[M_e\!\times\!d_{\mathrm{in}}]$ &
      Output rows \\
    Wgrad & $X_e^{\mathsf T}\,[d_{\mathrm{in}}\!\times\!M_e]$ &
      $\nabla Y_e\,[M_e\!\times\!d_{\mathrm{out}}]$ &
      $M_e$ & $\nabla W_e\,[d_{\mathrm{in}}\!\times\!d_{\mathrm{out}}]$ &
      Contraction length \\
    \bottomrule
  \end{tabularx}
  \caption{\textbf{The route-dependent dimension changes roles during
  training.}  Forward and input-gradient computation (Dgrad) place $M_e$ in
  the output-row dimension.  Weight-gradient computation (Wgrad) contracts
  over $M_e$.  Implementations may evaluate a transpose-equivalent Wgrad to
  satisfy operand-layout constraints.}
  \label{tab-grouped-gemm-training-shapes}
\end{table}

The algebra is the same as for a dense linear layer.  The difference is that
a dense layer has one common token dimension, whereas a routed layer has a
different $M_e$ for every expert (Table~\ref{tab-grouped-gemm-training-shapes}).
Forward and Dgrad therefore contain
different numbers of output rows per expert, while Wgrad contains different
contraction lengths.  For the up/gate projection,
$(d_{\mathrm{in}},d_{\mathrm{out}})=(d,2h)$; for the down projection, it is
$(h,d)$.  These differences change operand layout, tile geometry, and output
buffer opportunities.  A forward-only kernel result therefore cannot predict
the complete training benefit.

Precision makes the distinction sharper.  Olmo's routed MXFP8 path uses
scaled grouped GEMM for forward and Dgrad, while its current Wgrad path uses
BF16 grouped GEMM after reconstructing the required operands.  Section~\ref{sec-mxfp8-challenges}
explains why the natural MXFP8 Wgrad contraction is not simply another copy of
the forward operation.

\subsection{Rows per Expert Determine the Throughput Crossover}
\label{subsec-grouped-gemm-crossover}
Grouped scheduling removes per-expert host launches and avoids BMM's padded
rows, but it does not turn many small expert matrices into one large dense
matrix.  Its useful geometry is set by the vector
$(M_1,\ldots,M_{N_{\mathrm{local}}})$, not only by the sum $M$.  The mean
$\overline{M}=M/N_{\mathrm{local}}$ indicates how much work is available per
expert, while the spread of the $M_e$ values determines BMM padding and how
evenly the grouped scheduler can distribute tiles.
Small $M_e$ also means less weight reuse, which can make memory bandwidth
the limit even before scheduling overhead. Appendix~\ref{app-expert-roofline}
derives this idealized bound; the comparisons below measure actual kernel
performance.

We measure the costs of fixed capacity and of splitting one projection
into expert-sized problems, then vary the expert-load histogram at fixed
work. A separate stress diagnostic extends beyond the training range to
investigate the decline at large shapes.
We omit the host loop from the kernel-speed comparison because CUDA events
do not capture the preceding host wait.

\paragraph{The cost of fixed capacity.}
The legacy BMM path sets its per-expert capacity from the
ideal mean load $\overline{M}$ and a capacity factor $c$.  Olmo applies
eight-row alignment to each source rank's share of that load, so with
$M_E$ EP ranks

\begin{equation}
C = M_E\cdot 8\left\lceil\frac{\lfloor c\,\overline{M}/M_E\rfloor}{8}\right\rceil.
\label{eq-grouped-bmm-capacity}
\end{equation}

For $c=1.25$, the ideal capacity already contains 25\% more rows than
useful work, or a 20\% padding fraction; eight-row alignment can increase that
gap further.\footnote{The sweep below uses $C=\lceil1.25\overline{M}\rceil$
exactly, isolating the capacity factor itself; the eight-row alignment of
Equation~\ref{eq-grouped-bmm-capacity} would add a second, shape-dependent
padding term on top of the factor.}
This is different from choosing $C=\max_e M_e$ after observing the routes.
Grouped GEMM instead uses the real cumulative counts and excludes every unused
capacity row.

\begin{figure}[!htb]
  \centering
  \includegraphics[width=\linewidth]{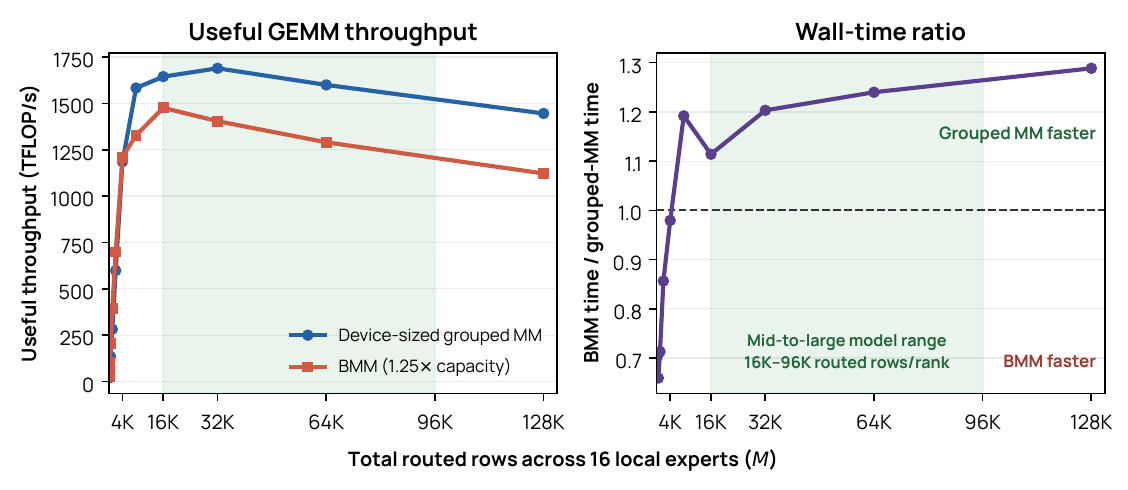}
  \caption{\textbf{Grouped GEMM and capacity-padded BMM across routed-row counts.}
  Grouped GEMM is faster throughout the shaded mid-to-large-model range.
  The experiment isolates one BF16 square expert projection on one
  B300 with $N_{\mathrm{local}}=16$ and $d=h=4096$.  The horizontal axis is total routed rows
  across the 16 local experts; routed rows count expert rows, not unique
  input tokens, since each token contributes one row to each of its top-$K$
  experts, and balanced routing gives $M/16$ valid rows per expert.
  The green band marks the representative 16K--96K rank-local operating
  range.  Its endpoints summarize variation in microbatch and routing choices
  rather than one fixed microbatch size, sequence length, top-$K$, or
  local-expert count; Appendix
  Table~\ref{tab-common-grouped-gemm-shapes} gives concrete configurations.  BMM
  reserves exactly 1.25$\times$ capacity per expert, while grouped GEMM executes
  only valid rows.  \textbf{Left:} useful
  throughput counts only valid-row FLOPs.  \textbf{Right:} values above one
  indicate that grouped GEMM is faster.  Each point is the mean of 20 queued
  CUDA-event samples after 10 warm-ups; all timed operations are submitted
  behind one device-side gate and released only after the host has queued the
  complete batch, so small shapes are not dominated by Python launch gaps.}
  \label{fig-grouped-gemm-capacity-sweep}
\end{figure}

\paragraph{The crossover sits at a few hundred rows per expert.}
On one B300 with 16 local experts and $d=h=4096$,
Figure~\ref{fig-grouped-gemm-capacity-sweep} compares one BF16 exact-row
grouped expert projection with
BMM at 1.25$\times$ per-expert capacity while sweeping the total routed-row
count $M$ from 64 rows to 128K.
BMM executes 25\% more rows at every
point, but its regular kernel is faster through $M=4\mathrm{K}$.  The crossover lies
between $M=4\mathrm{K}$ and $M=8\mathrm{K}$, corresponding to 256--512 routed
rows per expert, and
grouped GEMM is faster at every measured point above it, although its
advantage remains shape dependent. Avoiding padding is insufficient below
the crossover, while at $M=128\mathrm{K}$ BMM takes 28.9\% longer, more
than its 25\% of padded work. The common mid-to-large-model operating range of
16K--96K routed rows per rank, shaded in the figure, lies entirely above the
crossover.  Smaller models use narrower expert projections and often place
more rows on each rank, so their crossover must be measured separately;
Appendix Figure~\ref{fig-grouped-gemm-capacity-sweep-dh2048} repeats the
experiment with $d=h=2048$, representative of that regime, and extends the
routed-row sweep to 256K.

\paragraph{The cost of decomposing one projection.}
To isolate decomposition, we begin with one BF16 weight
$W_{\mathrm{dense}}\in\mathbb{R}^{8192\times81920}$ and partition its output
columns into $N\in\{4,8\}$ expert matrices.  For
$X\in\mathbb{R}^{T\times8192}$,

\begin{equation}
W_{\mathrm{dense}} = [W_1\;W_2\;\cdots\;W_N],
\qquad
XW_{\mathrm{dense}} = [XW_1\;XW_2\;\cdots\;XW_N].
\label{eq-grouped-dense-matched-projection}
\end{equation}

All three layouts therefore execute the same
$2T\times8192\times81920$ useful FLOPs, contain the same BF16 weight elements,
and produce the same output elements from the exact same BF16 values.  The
eight-expert case isolates the cost of dividing the same projection into
more, narrower GEMMs, including the additional input reads.  It is a
projection benchmark, not a complete dense-versus-MoE model
comparison.\footnote{To ground $T$: in a balanced EP32, top-$4$, 128-expert
configuration, each rank contributes $T$ input tokens and holds four
experts, so each local expert receives $T$ rows.  The eight-expert curve is
a controlled partition of the same projection, not a second end-to-end
routing configuration.} Splitting a microbatch into waves
(Section~\ref{subsec-ep-overlap}) reduces $T$ and lowers GEMM efficiency
further, one ingredient of the wave-scheduling result in
Section~\ref{subsec-wave-overlap-lesson}. The compared paths also use the
same seeded operand distribution, since operand values can affect throughput
(Section~\ref{sec-gemm-operand-values}).

\begin{figure}[!htb]
  \centering
  \includegraphics[width=0.98\linewidth]{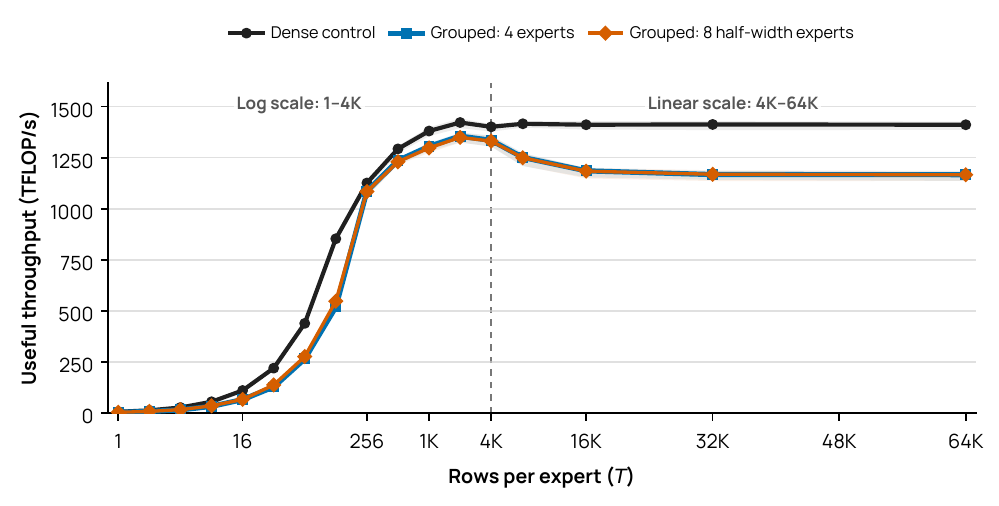}
  \caption{\textbf{Splitting one projection into four or eight expert GEMMs.}
  The four-expert path executes four
  $[T,8192]\mathbin{@}[8192,20480]$ GEMMs; the eight-expert path executes eight
  $[T,8192]\mathbin{@}[8192,10240]$ GEMMs.  Both are exact partitions of the
  same $[T,8192]\mathbin{@}[8192,81920]$ dense projection and therefore match
  its useful FLOPs, weights, and outputs.  Grouped inputs repeat the dense input
  four or eight times.  Curves show means across eight paired trials, one on
  each of eight B300 GPUs, and light bands show the full eight-GPU ranges;
  each GPU measures the dense control with both grouped partitions, and its
  two dense measurements are averaged before the dense curve is summarized.
  The horizontal axis is logarithmic through $T=4096$ and linear thereafter;
  the dashed divider marks the scale change, and no values are omitted.  The
  final two points, $T=32{,}768$ and $65{,}536$, are oversized diagnostics
  rather than representative training shapes.  Only the projection is timed;
  Appendix~\ref{app-grouped-dense-matched-projection} records the complete
  timing protocol and representative ratios.}
  \label{fig-grouped-dense-matched-projection}
\end{figure}

Figure~\ref{fig-grouped-dense-matched-projection} shows the absolute useful
throughput of Equation~\ref{eq-grouped-dense-matched-projection}.  At very
small $T$, decomposing the projection is expensive: both grouped layouts
deliver only about 58--65\% of dense throughput through $T=128$.  Both improve
sharply at $T=256$, reaching about 96\%, and remain at 94--96\% through
$T=4096$.
Doubling the number of expert GEMMs while halving their width changes
throughput only slightly in this experiment.  The larger gap is between either
grouped layout and the single dense projection.  At still larger $T$, both
grouped layouts drift to 83--88\% of dense across $T=8192$--$65{,}536$,
settling at 32K and 64K in the power-limited regime examined below.

\paragraph{Very large grouped GEMMs run at a power-limited clock.}
The large-shape decline motivates a separate stress diagnostic extending to
32K rows per local expert. Its largest shapes are deliberately oversized
diagnostics rather than representative training points. Across that range,
grouped GEMM throughput declines together with the SM clock while useful
throughput per MHz stays nearly constant. Nsight timelines from a separate
profiling run show the same
CUTLASS grouped-GEMM kernel template at $M_e=2\mathrm{K}$ and $32\mathrm{K}$,
and the software power-cap limiter is active in effectively every telemetry
sample near the board limit. The equal-FLOP dense control holds nearly flat
throughput at a higher clock, and the capacity-padded BMM control, also nearly
flat, overtakes grouped GEMM at 16K and 32K rows per expert. Together, these observations indicate a lower
power-limited operating point rather than a kernel-selection change or a
collapse in per-clock efficiency.

Clocks and power help interpret the extreme-shape decline; operand values
can also affect throughput (Section~\ref{sec-gemm-operand-values}).
Appendix~\ref{app-grouped-gemm-large-m-power} records the diagnostic protocol
and complete curves.

\paragraph{Sensitivity to the load histogram.}
Even with total rows and arithmetic fixed, the shape of the expert-load
histogram changes grouped GEMM time.
Let $M_e$ be the rows assigned to local expert $e$,
$M=\sum_e M_e$, and $p_e=M_e/M$.  We summarize concentration with the
normalized entropy deficit

\begin{equation}
I_H
  \defeq 1-\frac{H(\boldsymbol{p})}{\log N_{\mathrm{local}}}
  = 1+\frac{\sum_e p_e\log p_e}{\log N_{\mathrm{local}}}.
\label{eq-grouped-gemm-entropy-imbalance}
\end{equation}

$I_H=0$ denotes uniform load and $I_H$ approaches one as the routes
concentrate on one expert.  Unlike a maximum-to-mean ratio, this metric uses
every expert count.

\begin{figure}[!htb]
  \centering
  \includegraphics[width=\linewidth]{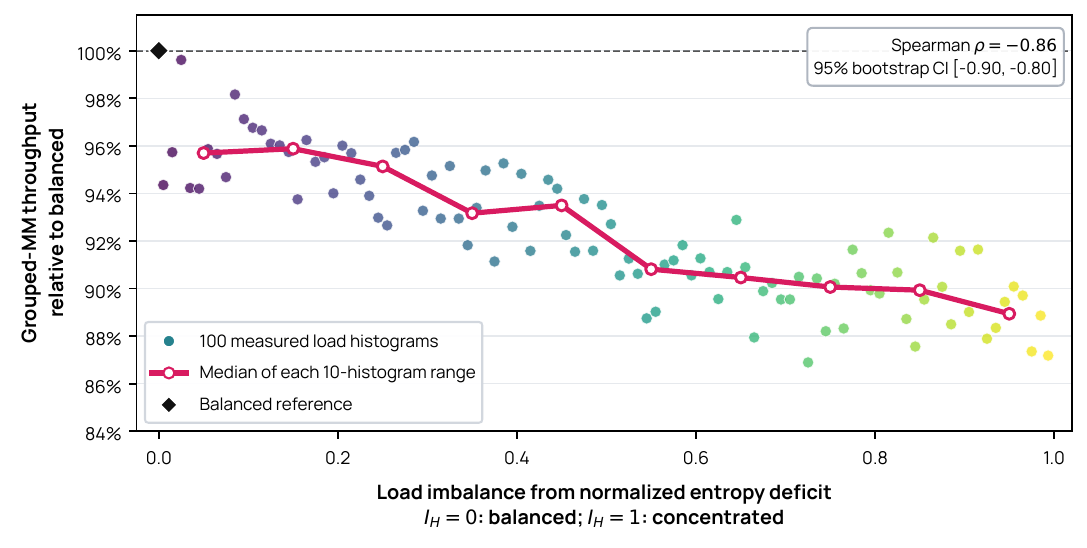}
  \caption{\textbf{Grouped GEMM throughput generally falls as expert loads
  concentrate.}  The experiment isolates one BF16 square expert projection
  on one B300 with $N_{\mathrm{local}}=16$, $M=65{,}536$ routed rows, and
  $d=h=4096$.  We draw symmetric-Dirichlet proposals, convert them to exact
  eight-row-aligned counts with at least one tile per expert, and select ten
  distinct histograms from each of ten equal-width $I_H$ ranges.  Every dot
  therefore has the same total rows, operands, dimensions, and useful FLOPs.
  Each dot averages four fixed expert-index permutations in each of three
  fresh processes.  Each of those 12 conditions averages 20 queued CUDA-event
  samples after 10 warm-ups; execution order is randomized and balanced
  controls bracket every ten cases.  The magenta curve connects the median of
  each ten-histogram range.  The correlation and bootstrap interval are
  computed over the 100 histogram means.  This kernel-only test excludes
  dispatch, communication, and cross-rank waiting.}
  \label{fig-grouped-gemm-entropy-sweep}
\end{figure}

Figure~\ref{fig-grouped-gemm-entropy-sweep} evaluates 100 distinct load
histograms spanning the entropy range.
Across these fixed-work cases, entropy-defined imbalance
has a Spearman correlation of $-0.86$ with grouped GEMM throughput; a
histogram-level bootstrap gives a 95\% interval of $[-0.90,-0.80]$.  The
median normalized throughput falls from approximately 96\% in the
least-concentrated entropy range to 89\% in the most-concentrated range.
Individual points are not monotonic: histograms with similar entropy can
still present different ordered collections of matrix problems to the
kernel.  The result supports the systems-level trend---greater route
concentration usually lowers grouped GEMM throughput at fixed arithmetic---not
the stronger claim that entropy alone determines execution time.  Appendix
Figure~\ref{fig-grouped-gemm-histogram-shapes} gives a complementary example
in which two histograms share the same maximum load but have different
throughput.

\begin{sectiontakeaway}
Grouped GEMM executes route-dependent expert shapes without turning reserved
capacity into matrix-multiply work; the rowwise layout feeds it directly from
fixed-shape buffers, and device offsets restrict computation to real routes.
In our B300 sweep, which stops at 8K rows per expert, grouped GEMM becomes
faster than capacity-padded BMM once enough rows reach each expert. BMM's regular kernels are faster below
the crossover, and concentrated expert loads still reduce throughput above
it. Even at exactly matched FLOPs,
decomposing one dense GEMM into expert-sized problems changes efficiency,
most at very small shapes and again in the power-limited large-shape regime.
\end{sectiontakeaway}

\FloatBarrier

\section{Routing, Capacity, and Load Balancing}
\label{sec-routing-load-balancing}

\conceptlead{A routing decision is both a model choice and a distributed-work
request.}  For each token, the router produces soft scores and selects $K$
experts, creating $K$ requested routes.  Capacity then determines which
requests are kept.  Dispatch moves the kept rows to the expert-owning ranks;
on the rowwise real-count path, grouped GEMM
executes their densely packed expert segments, and combine returns their
weighted contributions.  Requested and kept routes coincide when no route is
dropped, but they are different objects whenever capacity overflows.

Load-balancing mechanisms influence future requests, while capacity limits
how many of the current requests are admitted. The resulting workload has
both an expert-level shape and a
destination-rank total.  The former controls expert training signal and
grouped GEMM geometry, while the latter controls communication, buffer use,
overflow, and straggling.  After establishing those systems consequences, we
compare the scopes and mechanisms used to balance routes and then analyze a
failure of the standard Switch-style auxiliary objective
\citep{fedus2022switch}: its scalar can fall
while requested routing becomes less balanced.

\begin{keyidea}[Requested Routes and Kept Routes]
The current \textbf{load-balancing loss (LBL)} and post-batch bias controller
(Section~\ref{sec-lbl-aux-versus-loss-free}) use requested top-$K$ assignments.  Capacity and dispatch use kept
assignments, and the rowwise path also uses their real counts for grouped
GEMM.  When routes are dropped, neither histogram is a substitute for the
other.
\end{keyidea}

\subsection{From \texorpdfstring{Top-$K$}{Top-K} Selection to the Executed Workload}
\label{sec-routing-requested-kept-work}

Consider one routed layer with $N$ global experts and $M_E$ expert-parallel
ranks.  Let source rank $s$ hold $T_s$ tokens, and let $S_{s,t}$ be the $K$
experts selected for token $t$.  One pair $(t,i)$ with $i\in S_{s,t}$ is a
\textbf{route}.  The number of routes requested from source rank $s$ to expert
$i$ is
\begin{equation}
C_{s,i}^{\mathrm{req}}
  \defeq \sum_{t=1}^{T_s}\mathbf{1}\{i\in S_{s,t}\},
\qquad
\sum_{i=1}^{N}C_{s,i}^{\mathrm{req}}=KT_s.
\label{eq-routing-requested-expert-counts}
\end{equation}
Aggregate over sources to obtain the global requested count
$C_i^{\mathrm{req}}\defeq\sum_s C_{s,i}^{\mathrm{req}}$.  If $\rho(i)$ is the
expert-parallel rank that owns expert $i$, the total demand placed on
destination rank $q$ is
\begin{equation}
R_q^{\mathrm{req}}
  \defeq
  \sum_{i:\rho(i)=q}C_i^{\mathrm{req}}.
\label{eq-routing-requested-rank-counts}
\end{equation}
Thus the same top-$K$ decisions define two workload distributions: counts over
experts and sums over the experts colocated on each destination rank.

Capacity acts between requested routes and execution.  Let
$C_{s,i}^{\mathrm{keep}}$ be the routes admitted by the capacity policy and
define
\begin{equation}
\begin{aligned}
D_{s,i}
  &\defeq C_{s,i}^{\mathrm{req}}-C_{s,i}^{\mathrm{keep}},
&
C_i^{\mathrm{keep}}
  &\defeq\sum_s C_{s,i}^{\mathrm{keep}},\\
D_i
  &\defeq C_i^{\mathrm{req}}-C_i^{\mathrm{keep}},
&
M_i
  &\defeq C_i^{\mathrm{keep}},
&
R_q^{\mathrm{keep}}
  &\defeq\sum_{i:\rho(i)=q}C_i^{\mathrm{keep}}.
\end{aligned}
\label{eq-routing-kept-and-dropped-counts}
\end{equation}
Here $D_{s,i}$ and $D_i$ count source-specific and aggregate dropped routes,
$M_i$ is the row count actually presented to expert $i$ (the per-expert row
count of Section~\ref{sec-grouped-gemm}; its subscript is an expert index,
not the rank count $M_E$), and
$R_q^{\mathrm{keep}}$ is the actual receive and compute load on rank $q$.
Write $\boldsymbol{C}^{x}=(C_1^x,\ldots,C_N^x)$ for
$x\in\{\mathrm{req},\mathrm{keep}\}$.  In a no-drop step,
$C_{s,i}^{\mathrm{keep}}=C_{s,i}^{\mathrm{req}}$; otherwise the distinction is
essential.  A kept histogram can appear benign only because excess demand was
dropped, while a requested histogram alone does not describe the work that
ran.  Table~\ref{tab-routing-three-quantities} records which part of the
training system produces and consumes each routing quantity.

\begin{table}[!htb]
\centering
\footnotesize
\renewcommand{\arraystretch}{1.14}
\begin{tabularx}{\linewidth}{@{}
  >{\raggedright\arraybackslash}p{0.20\linewidth}
  >{\raggedright\arraybackslash}p{0.25\linewidth}
  >{\raggedright\arraybackslash}X@{}}
\toprule
\textbf{Quantity} & \textbf{Created by} & \textbf{Consumed by} \\
\midrule
Router score or logit & Router projection and gating function & The
configured score or logit drives top-$K$; selected unbiased affinities or
logits determine mixture weights; normalized all-expert scores form LBL's
differentiable term \\
\arrayrulecolor{black!22}\hline\arrayrulecolor{black}
Requested count $C^{\mathrm{req}}$ & Top-$K$ expert identifiers & LBL's
detached hard term, post-batch bias feedback, and capacity admission; on the
rowwise path, all-gathered requested splits form the keep matrix that
determines dispatch destination ranks and row offsets \\
\arrayrulecolor{black!22}\hline\arrayrulecolor{black}
Kept count $C^{\mathrm{keep}}$ & Capacity keep/drop policy & Dispatch and
combine payload; expert grouped GEMM row counts on real-count paths; and
executed-work diagnostics \\
\bottomrule
\end{tabularx}
\caption{\textbf{Three routing quantities participate in different parts of
the training system.}  The current LBL and post-batch bias controller observe
requested selections before expert-parallel capacity is enforced.}
\label{tab-routing-three-quantities}
\end{table}

Olmo uses \textbf{token-choice routing}: every token selects experts
\citep{shazeer2017outrageously}.
BASE Layers instead solves a balanced assignment, and Expert Choice lets each
expert select tokens; these alternatives change which side of the
token--expert relation imposes the constraint
\citep{lewis2021base,zhou2022expertchoice}.  The
remainder of this section concerns Olmo's token-choice path.

\needspace{7\baselineskip}
\subsection{The Costs of Imbalance Depend on Where It Appears}
\label{sec-routing-imbalance-costs}

For any nonzero load vector
$\boldsymbol{x}\in\mathbb{R}_{\geq0}^{n}$ with $n>1$,
define its \textbf{max/mean imbalance} as
\begin{equation}
I_{\max}(\boldsymbol{x})
  \defeq
  \frac{\max_j x_j}{\frac{1}{n}\sum_jx_j}
  = n\frac{\lVert\boldsymbol{x}\rVert_\infty}
           {\lVert\boldsymbol{x}\rVert_1}.
\label{eq-routing-max-mean-imbalance}
\end{equation}
Applying this definition to requested expert counts, kept expert counts,
requested destination-rank totals, and kept destination-rank totals answers
four different questions (Table~\ref{tab-routing-imbalance-questions}).  The
expert-level pair describes selection and executed expert shapes; the
rank-level pair describes demand and admitted work at the rank that can
become a straggler.

\begin{table}[!htb]
\centering
\footnotesize
\renewcommand{\arraystretch}{1.14}
\begin{tabularx}{\linewidth}{@{}
  >{\raggedright\arraybackslash}p{0.17\linewidth}
  >{\raggedright\arraybackslash}X
  >{\raggedright\arraybackslash}X@{}}
\toprule
\textbf{Counts} & \textbf{Over experts} & \textbf{Over destination ranks} \\
\midrule
Requested & Selection concentration: the training signal the router asks for
& Demand on the hottest rank: capacity pressure and drop risk \\
\arrayrulecolor{black!22}\hline\arrayrulecolor{black}
Kept & Executed expert shapes presented to grouped GEMM & Admitted receive,
compute, and combine load on the rank that can straggle \\
\bottomrule
\end{tabularx}
\caption{\textbf{Max/mean imbalance over requested and kept routes.}
Requested and kept counts, over experts or over destination ranks, separate
selection concentration, executed expert shape, demand pressure, and
admitted straggler load.}
\label{tab-routing-imbalance-questions}
\end{table}

Max/mean identifies the hottest coordinate but does not describe how the
remaining load is distributed. Let
$\pi_j=x_j/\sum_\ell x_\ell$ be the normalized load distribution and
$H(\boldsymbol{\pi})=-\sum_j\pi_j\log\pi_j$ its entropy.  The number of
never-selected experts, coefficient of variation, and normalized entropy
deficit $I_H(\boldsymbol{x})=1-H(\boldsymbol{\pi})/\log n$ capture
complementary aspects of concentration.  The source--destination request
matrix contains more communication information than its row or column sums:
it resolves which source rank sends how many rows to which destination rank,
the per-pair traffic that row and column totals average away.  Drop rate and
receiving-buffer utilization then show how capacity transformed demand into
executed work.

The \textbf{model-side} selection pattern starts with requested expert counts, but
routed-expert activations and gradients follow kept counts.  An expert can be requested yet receive
no signal from a dropped route.  At the other extreme, requiring every small
or domain-homogeneous microbatch to be uniform can oppose useful
specialization.  The desirable distribution is therefore scope-dependent:
avoiding starvation over the training population does not require every local
microbatch to look uniform
\citep{qiu2025demons}.

The \textbf{systems-side} effect starts with kept counts.  Even at a fixed rank total,
concentrating rows into fewer experts changes the shapes presented to grouped
GEMM.  Section~\ref{sec-grouped-gemm} measured this directly: as the expert-load
histogram concentrated, normalized grouped-GEMM throughput declined with
entropy deficit, and median throughput in the most-concentrated entropy range
ran about seven points below the least concentrated
(Figure~\ref{fig-grouped-gemm-entropy-sweep}).  A matched
appendix comparison further shows that a single-spike and a three-hot-expert
histogram with the same maximum load can have different grouped-GEMM times
(Figure~\ref{fig-grouped-gemm-histogram-shapes}).  These fixed-work
microbenchmarks isolate local expert computation; they do not include
dispatch, cross-rank waiting, or a learned-router trajectory.

\textbf{Destination-rank imbalance} adds those omitted costs.  A rank with high
$R_q^{\mathrm{req}}$ consumes capacity faster and is more likely to drop
routes.  A rank with high $R_q^{\mathrm{keep}}$ receives more rows, performs
more expert work, returns more outputs, and can delay peers or a pipeline
stage.  Expert-level and rank-level imbalance can also move independently: a
rank may receive a balanced total that is concentrated in one of its local
experts, or several moderately hot local experts may make the whole rank a
straggler, so diagnostics need both levels.

\subsection{Rank-Wide Capacity Bounds Work Without Balancing It}
\label{sec-routing-rank-capacity}

Let $T_{\mathrm{route}}=KT_s$ be the number of routes produced by one source
expert-parallel rank.  When source ranks hold equal token counts and each
destination owns the same number of experts, balanced expert routing places
an average of $T_{\mathrm{route}}$ requested rows on each destination rank:
the factor $M_E$ from all sources cancels the probability $1/M_E$ of choosing
any one destination.  Olmo's rowwise sync-free path reserves
the destination-rank budget
\begin{equation}
C_{\mathrm{rank}}
  \defeq
  \max\!\left(1,\left\lceil c\,T_{\mathrm{route}}\right\rceil\right),
\label{eq-routing-rank-capacity}
\end{equation}
where $c$ is the capacity factor.  This is \textbf{one shared limit for all experts
owned by the destination rank}, not one independent limit per local expert.
If the rank owns $N_{\mathrm{local}}$ experts, its balanced per-expert mean is
$T_{\mathrm{route}}/N_{\mathrm{local}}$.  The rank budget alone can therefore
fit one expert at approximately
\begin{equation}
\frac{C_{\mathrm{rank}}}
     {T_{\mathrm{route}}/N_{\mathrm{local}}}
  \approx cN_{\mathrm{local}},
\label{eq-routing-rank-capacity-hot-expert}
\end{equation}
times that mean if its neighbors are sufficiently cool.  This is a bound on
what can fit, not evidence that a trained router reaches it.

The shared budget lets a hot local expert use capacity left unused by cooler
local experts while the aggregate admitted demand remains within
$C_{\mathrm{rank}}$.  A conventional independent expert capacity instead
assigns each of the $N_{\mathrm{local}}$ experts a limit near
\begin{equation}
C_{\mathrm{expert}}
  \approx
  \left\lceil
    \frac{c\,T_{\mathrm{route}}}{N_{\mathrm{local}}}
  \right\rceil.
\label{eq-routing-per-expert-capacity}
\end{equation}
At approximately the same aggregate reserved rows, that policy can drop routes to
one hot expert while neighboring experts leave their allocations unused
\citep{lepikhin2021gshard,fedus2022switch}.  The rank-wide rule trades an
individual expert bound for flexible sharing within a destination-rank bound.

Reserved capacity and executed rows are also distinct.  Olmo's rowwise
NVSHMEM path allocates capacity-shaped receive and combine buffers but packs
valid rows densely at the front.  If the kept local-expert counts are
$(M_1,\ldots,M_{N_{\mathrm{local}}})$, the grouped-GEMM offsets are
\begin{equation}
o_j\defeq\sum_{i=1}^{j}M_i,
\qquad j=1,\ldots,N_{\mathrm{local}}.
\label{eq-routing-packed-expert-offsets}
\end{equation}
Only $[0,o_{N_{\mathrm{local}}})$ belongs to expert compute.  Unused capacity
rows lie outside every expert interval and may contain stale values.
On this path, communication uses fixed capacity but grouped GEMM processes no
padding rows: the device-side real-count offsets described in
Sections~\ref{sec-sync-free-training} and~\ref{sec-grouped-gemm} separate
physical allocation from logical work.  Other dispatch paths may instead pass
capacity-padded splits to expert compute.

On every capacity-enabled sync-free forward, all expert-parallel ranks
all-gather their requested split counts and independently derive the same
deterministic \textbf{keep matrix}.  For each destination, the current policy flattens
counts in local-expert-major, then source-rank order, cumulatively admits routes
until it reaches $C_{\mathrm{rank}}$, and keeps a prefix within each
source/expert split.  Restoration and combine mask the tail-dropped routes.
This makes overflow bounded and reproducible, but the ordering can affect
which routes survive; a capacity evaluation should therefore report the
distribution of drops, not only their total.

The aggregate \textbf{route-drop fraction} is
\begin{equation}
\delta_{\mathrm{route}}
  \defeq
  \frac{\sum_{s,i}D_{s,i}}
       {\sum_{s,i}C_{s,i}^{\mathrm{req}}}
  = \frac{\sum_{s,i}D_{s,i}}{\sum_sKT_s}.
\label{eq-routing-route-drop-rate}
\end{equation}
For $K>1$, this is not the fraction of tokens that lose every routed
contribution.  A model-quality analysis should separately report tokens that
lose at least one route, tokens that lose all $K$, and the selected mixture
weight removed by dropping.

\begin{keyidea}[Capacity Bounds Admitted Work]
Capacity bounds the receive-buffer storage and the worst-case admitted rank
work after an outlying routing event. It filters requests after selection;
the bound itself provides no guarantee of balanced requested load.
The intended regime is \emph{nearly dropless under the observed routing
distribution}: balancing keeps normal demand within the fixed budget, while
capacity handles rare excursions.  This is not an unconditional dropless
guarantee.
\end{keyidea}

\subsection{A Shared Rank Budget Drops No More Routes than Matched Expert Budgets}
\label{sec-routing-capacity-policy-comparison}

\begin{figure}[!htb]
\centering
\includegraphics[width=\linewidth]{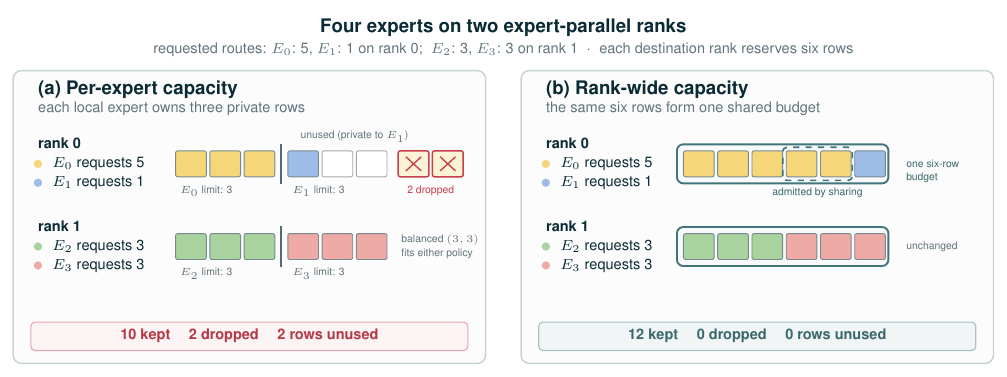}
\caption{\textbf{Rank-wide capacity shares a destination rank's unused rows
among its local experts.}  Both policies reserve six rows per destination
rank for the same requested routes.  (a)~Independent three-row expert limits
drop two $E_0$ routes while $E_1$'s two private rows sit unused.  (b)~One
shared six-row budget admits every request; the dashed outline marks the two
routes that only pooling admits.  Rank 1's balanced demand is identical under
both policies.  Each colored block is one requested route; a crossed block is
dropped, and an empty block is a reserved, unused row.  Conceptual
illustration with no measured quantities.}
\label{fig-routing-capacity-unused-sharing}
\end{figure}

Two capacity policies can reserve the same aggregate rows on a destination
rank yet admit different subsets of a skewed request histogram.
Figure~\ref{fig-routing-capacity-unused-sharing} shows the mechanism on one
four-expert example, and Table~\ref{tab-routing-capacity-policies} compares
the two allocation policies.

\begin{table}[!htb]
\centering
\footnotesize
\renewcommand{\arraystretch}{1.18}
\renewcommand{\tabularxcolumn}[1]{m{#1}}
\begin{tabularx}{\linewidth}{@{}
  >{\raggedright\arraybackslash}m{0.15\linewidth}
  >{\raggedright\arraybackslash}m{0.18\linewidth}
  >{\raggedright\arraybackslash}m{0.27\linewidth}
  >{\raggedright\arraybackslash}X@{}}
\toprule
\textbf{Policy} & \textbf{Allocation unit} & \textbf{Unused capacity and overflow} &
\textbf{Expert compute} \\
\midrule
Independent per expert & One limit for each local expert & Rows reserved by a
cool expert cannot admit routes for a hot neighbor; overflow is expert-local &
Padded BMM may execute capacity rows; grouped GEMM can still use real kept
counts \\
\arrayrulecolor{black!22}\hline\arrayrulecolor{black}
Shared rank-wide & One aggregate limit for the destination rank & Local
experts draw from one budget; overflow begins only when their aggregate fills the rank
budget & Olmo's rowwise path packs kept rows and uses real cumulative offsets;
another sync-free path can execute capacity-padded splits \\
\bottomrule
\end{tabularx}
\caption{\textbf{Rank-wide and per-expert policies allocate a matched
aggregate row budget differently.} Pooling lets a hot expert use rows that
an independent expert budget would leave unused.}
\label{tab-routing-capacity-policies}
\end{table}

At exactly matched aggregate rows, the drop advantage from pooling unused capacity is
algebraic.  Given a destination-rank total budget $B$ and independent
expert budgets $b_j$ satisfying $\sum_jb_j=B$, the total drops for requested
local-expert counts $C_j^{\mathrm{req}}$ are
\begin{equation}
\begin{aligned}
D^{\mathrm{rank}}
  &= \left[\sum_jC_j^{\mathrm{req}}-B\right]_+,\\
D^{\mathrm{expert}}
  &= \sum_j\left[C_j^{\mathrm{req}}-b_j\right]_+,
\end{aligned}
\qquad [z]_+\defeq\max(z,0).
\label{eq-routing-capacity-policy-drops}
\end{equation}
Because $[\sum_jz_j]_+\leq\sum_j[z_j]_+$,
\begin{equation}
D^{\mathrm{rank}}\leq D^{\mathrm{expert}}.
\label{eq-routing-rank-capacity-dominance}
\end{equation}
A shared rank budget therefore \textbf{cannot drop more routes in total}
than independent expert budgets with the same aggregate rows. The budgets
must match after integer rounding; a shared nominal capacity factor alone need
not give equal aggregate rows.  The inequality is
strict when a hot expert exceeds its private budget while a neighbor leaves
rows unused.  It does not prove a universally better training policy: rank-wide
tail-drop may preserve a different set of tokens and deliberately gives up the
stronger per-expert workload bound.

Per-expert capacity is the standard drop-and-pad formulation in GShard and
Switch Transformer
\citep{lepikhin2021gshard,fedus2022switch}.  Megatron-Core likewise documents
both a per-expert capacity factor and, for the HybridEP and NCCL-EP backends
of its flex token dispatcher, a per-expert-rank capacity factor
\citep{nvidia2026megatronconfig}. Comparisons therefore need to specify
the capacity policy for the backend and revision being compared.

Capacity reacts to overload after selection.  Balancing instead changes
future requested demand, beginning with the token scope over which that demand
is measured.

\subsection{Load-Balancing Scope Is Part of the Training Objective}
\label{sec-lbl-scope-versus-surrogate}

The \emph{balance batch} is the set of tokens used to construct the requested
hard frequency vector.  It need not coincide with the optimizer batch, and
changing it changes the training objective even when the language-modeling
batch, optimizer, and model are unchanged.  A narrow scope asks every small
token group to use the experts uniformly; a broad scope allows imbalance
within one sequence, rank, domain, or microbatch as long as it is compensated
elsewhere in the balance batch.  The latter can reduce conflict between load
balancing and specialization when microbatches are small or domain-homogeneous
\citep{qiu2025demons}.

For a balance batch of $T$ tokens, $N$ routed experts, and $K$ selections per
token, let $f_i=C_i^{\mathrm{req}}/(KT)$ be expert $i$'s fraction of requested
routes before capacity, and let $P_i=T^{-1}\sum_{t=1}^{T}p_{t,i}$ be its mean
router score, with $p_{t,i}$ normalized across experts for each token.  The
Switch-style load-balancing loss couples these two statistics
\citep{lepikhin2021gshard,fedus2022switch}:
\begin{equation}
\mathcal{L}_{\mathrm{LBL}}
  \defeq N\sum_{i=1}^{N}f_iP_i
  = N\boldsymbol{f}^{\mathsf T}\boldsymbol{P}.
\label{eq-lbl-switch}
\end{equation}
The hard fractions $f_i$ are treated as constants; gradients flow through
$P_i$.  The factor $N$ keeps the balanced value equal to one as the expert
count changes.  The external coefficient multiplying this loss in the
training objective is omitted here.

Table~\ref{tab-lbl-scope-taxonomy} separates five relevant scopes, including
three distributed or accumulation-aware variants that are often conflated
under ``global load balancing.''  Only the first two are implemented as Olmo
LBL granularities today.

\begin{table}[!htb]
\centering
\footnotesize
\renewcommand{\arraystretch}{1.12}
\begin{tabularx}{\linewidth}{@{}
  >{\raggedright\arraybackslash}p{0.16\linewidth}
  >{\raggedright\arraybackslash}X
  >{\raggedright\arraybackslash}p{0.23\linewidth}
  >{\raggedright\arraybackslash}p{0.15\linewidth}@{}}
\toprule
\textbf{Scope} & \textbf{Statistics and semantics} &
\textbf{Coordination and cost} & \textbf{Olmo status} \\
\midrule
Instance & One per-sequence hard histogram, reduced across tensor- and
context-parallel shards as applicable, paired with each shard's local
differentiable soft-score contribution & Reductions induced by sequence
sharding & Implemented as
\texttt{instance} \\
\arrayrulecolor{black!22}\hline\arrayrulecolor{black}
Rank-local microbatch & Current rank's current microbatch; unrelated local
sequences can compensate, while CP shards remain local & No additional count
communication & Implemented as \texttt{local\_batch} \\
\arrayrulecolor{black!22}\hline\arrayrulecolor{black}
Synchronized current microbatch & Current hard counts reduced over the
stage-local token balance group; each rank contributes its local soft term &
One $N$-element count all-reduce per routed-layer forward & Design only \\
\arrayrulecolor{black!22}\hline\arrayrulecolor{black}
GA-buffered online global & Synchronized counts accumulated through gradient
accumulation; microbatch $m$ uses the prefix through $m$ & Same all-reduce plus
an $N$-element per-router buffer & Design only; analogous to Megatron
\texttt{global\_aux\_loss} \\
\arrayrulecolor{black!22}\hline\arrayrulecolor{black}
Exact optimizer-batch global & Final hard counts and soft-score sums over all
ranks and all accumulation microbatches & Delayed backward, retained graphs,
or a routing prepass; stale statistics instead define an approximation & Not
implemented \\
\bottomrule
\end{tabularx}
\caption{\textbf{Load-balancing scopes and their current Olmo status.}
``Online global'' is accumulation-aware but is not the exact optimizer-batch
objective.}
\label{tab-lbl-scope-taxonomy}
\end{table}

The last two rows are not interchangeable.  Let $m=1,\ldots,M$ index the
microbatches in one gradient-accumulation window.  Let
$C_{m,i}^{\mathrm{req}}$ be expert $i$'s requested count after synchronization
over balance group $\mathcal{G}$, let $T_m$ be the corresponding valid-token
count, and let
\begin{equation}
Q_{m,i}
  \defeq
  \sum_{r\in\mathcal{G}}\sum_{t\in(r,m)}p_{t,i}
\label{eq-lbl-global-soft-sum}
\end{equation}
be the conceptual global soft-score sum, where $p_{t,i}$ is token $t$'s
normalized router score for expert $i$.  A token-weighted online variant uses
the frequency and microbatch loss
\begin{equation}
\widetilde f_{m,i}
  \defeq
  \frac{\sum_{j=1}^{m}C_{j,i}^{\mathrm{req}}}
       {K\sum_{j=1}^{m}T_j},
\qquad
P_{m,i}\defeq\frac{Q_{m,i}}{T_m},
\qquad
\mathcal{L}^{\mathrm{online}}_m
  \defeq N\sum_{i=1}^{N}\widetilde f_{m,i}P_{m,i}.
\label{eq-lbl-online-global}
\end{equation}
This applies Equation~\ref{eq-lbl-switch} with accumulating hard counts and
current-microbatch soft scores.  Section~\ref{sec-token-gerrymandering}
analyzes why this hard/soft coupling need not track requested-load balance.
Microbatch one sees only its own counts; microbatch two sees the first two;
and so forth.  Once an early microbatch has backpropagated, its gradient is not
recomputed when later counts arrive.  The method expands the effective scope
online but is not the exact optimizer-batch objective unless $M=1$.  The
buffered approximation analyzed by %
\citet{qiu2025demons} accumulates routing counts across microbatches.
The Megatron implementation maintains the microbatch-number mean
$\overline C_{m,i}=m^{-1}\sum_{j\leq m}C_{j,i}^{\mathrm{req}}$ and uses
$N\sum_i\operatorname{sg}(\overline C_{m,i})Q_{m,i}/(KT_m^2)$, where
$\operatorname{sg}$ denotes stop-gradient
\citep{nvidia2026megatronrouterbuffer}.  It equals
Equation~\ref{eq-lbl-online-global} when all microbatches contain the same
number of valid tokens.  With padding, masking, or variable-size packed
microbatches, Equation~\ref{eq-lbl-online-global} is a proposed token-weighted
generalization rather than a description of the literal Megatron formula.

Equation~\ref{eq-lbl-online-global} defines a conceptual global loss.  A
per-rank implementation can use local soft-score sums and rely on distributed
gradient reduction, but the scale of that local contribution follows from the
gradient-averaging convention of the DDP implementation---here
\texttt{MultiGroupDistributedDataParallel}
(Section~\ref{sec-ddp-vs-fsdp})---so the correct group-size factor differs
across training stacks rather than transferring between them.

In contrast, define final optimizer-batch statistics
\begin{equation}
C_i^{\mathrm{step}}
  \defeq\sum_{m=1}^{M}C_{m,i}^{\mathrm{req}},
\qquad
Q_i^{\mathrm{step}}\defeq\sum_{m=1}^{M}Q_{m,i},
\qquad
T^{\mathrm{step}}\defeq\sum_{m=1}^{M}T_m.
\label{eq-lbl-exact-global-statistics}
\end{equation}
The exact optimizer-batch objective is
\begin{equation}
\mathcal{L}_{\mathrm{LBL}}^{\mathrm{step}}
  =
  \frac{N}{K(T^{\mathrm{step}})^2}
  \sum_{i=1}^{N}
  \operatorname{sg}\!\left(C_i^{\mathrm{step}}\right)
  Q_i^{\mathrm{step}}.
\label{eq-lbl-exact-global}
\end{equation}
Every microbatch's soft-score
sum is now weighted by the same final count vector.  Olmo backpropagates each
gradient-accumulation microbatch as it is processed, so implementing this
objective literally would require retaining every graph until the final count
is known, performing an additional routing prepass, or substituting stale
statistics.  These choices add memory, compute, or approximation error.

The synchronization group must follow router-token sharding rather than
expert-parameter sharding.  The intended Olmo design uses the stage-local
DP$\times$CP group: it includes distinct data- and context-parallel token
shards for the same router layer, excludes other pipeline stages, and operates
on counts over global expert identifiers before expert-local slicing.  Using
\texttt{WORLD} under pipeline parallelism would mix different layers and can
violate collective ordering; using an expert replica or placement group would
answer a different question.  The count payload is only $N$ elements, but its
collective occurs once per routed layer and microbatch, so its cost is set by
launch latency and collective ordering rather than by bytes.

The scope choice changes both specialization pressure and systems behavior.
Smaller scopes suppress immediate local skew but can discourage specialization.
Broader scopes permit locally skewed batches and require communication or
delayed information; a globally balanced count can still leave one
expert-parallel rank transiently hot. Changing the scope changes the
training recipe, including both its balancing pressure and coordination cost.

\subsection{Auxiliary Gradients and Hard-Count Feedback Are Different Controllers}
\label{sec-lbl-aux-versus-loss-free}

Switch-style LBL adds a differentiable term whose detached requested counts
become coefficients on the router's soft-score gradient.  DeepSeek-style
auxiliary-loss-free balancing instead treats requested load as feedback to a
non-trainable selection controller
\citep{wang2024lossfreebalancing,deepseekai2024deepseekv3}.  Let $a_{t,i}$ be
the unbiased quantity used for selection, let $\omega_{t,i}$ be the unbiased
affinity used for weighting, and let $b_i$ be a persistent expert bias.
Neither $a_{t,i}$ nor $\omega_{t,i}$ includes $b_i$.  Selection and mixture
weighting use
\begin{equation}
S_t
  = \operatorname{TopK}_i\!\left(a_{t,i}+b_i,K\right),
\qquad
w_{t,i}
  = \operatorname{Normalize}_{j\in S_t}
      \!\left(\omega_{t,j}\right)_i.
\label{eq-lbl-loss-free-selection}
\end{equation}
The bias changes which experts are selected, while their mixture weights
come from the original affinities. For top-$K$ softmax routing, $a_{t,i}=z_{t,i}$ and
$\omega_{t,i}=\exp z_{t,i}$; score-based routing can use
$a_{t,i}=\omega_{t,i}=p_{t,i}$.  DeepSeek uses sigmoid affinities
\citep{deepseekai2024deepseekv3}; Olmo applies the same biased-selection
pattern to the configured router scores.

Over one feedback window, define the requested count and its mean as
\begin{equation}
C_i^{\mathrm{fb}}
  \defeq \sum_t\mathbf{1}\{i\in S_t\},
\qquad
\overline C^{\mathrm{fb}}
  \defeq \frac{1}{N}\sum_i C_i^{\mathrm{fb}}.
\label{eq-lbl-loss-free-feedback-counts}
\end{equation}
The controller updates outside autograd,
\begin{equation}
b_i
  \leftarrow
  b_i+\gamma\,
  \operatorname{sign}\!\left(
    \overline C^{\mathrm{fb}}-C_i^{\mathrm{fb}}
  \right).
\label{eq-lbl-loss-free-update}
\end{equation}
Underloaded experts receive a positive selection offset and overloaded experts
a negative one.  The update acts on future requested assignments; it does not
observe which routes a downstream capacity policy kept.

Olmo enables this controller whenever the bias-update rate is positive.
Requested counts accumulate across the microbatches of one training batch.  A
post-batch hook applies Equation~\ref{eq-lbl-loss-free-update} after the final
backward and before the optimizer step, then resets the accumulator.  The
router's count reduction defaults to the \texttt{WORLD} process group unless
a balance group is explicitly configured. The next batch selects from
unbiased scores plus the persistent bias, while mixture weights still come
from unbiased affinities.

Under PP, the controller requires a stage-local balance group: the default
\texttt{WORLD} group mixes counts from different router layers. This report makes no claim about
the controller's model quality, stability, feedback-scope behavior, or
systems cost.

When used without LBL, the controller removes the balancing-loss gradient
and its potential interference with cross-entropy. Its behavior still
depends on the update rate and feedback scope.
A large $\gamma$ can make loads oscillate around the target; a small value can
lag behind a changing router; and a broad feedback scope can hide per-sequence
or rank-local hotspots %
\citep{wang2024lossfreebalancing}.  DeepSeek-V3 reports
retaining a very small sequence-level
auxiliary safeguard alongside its bias controller
\citep{deepseekai2024deepseekv3}.  Its update rate and safeguard coefficient
are recipe choices rather than universal constants.

When bias feedback and LBL are enabled together, the bias affects the
requested counts used by LBL: hard selections depend on
$\boldsymbol{a}+\boldsymbol{b}$, while LBL's soft mean still uses the unbiased
normalized scores $\boldsymbol{p}$. The selected experts can therefore differ
from those favored by the scores entering the soft term. The coefficient
$\lambda_{\mathrm{LBL}}$ and bias-update rate $\gamma$ control these different
feedback paths.

We next analyze the auxiliary-loss controller, separating what its scalar
value measures from the effect of its gradient.

\subsection{Token Gerrymandering: The Model Can Learn to Hack the LBL Objective}
\label{sec-token-gerrymandering}

\begin{figure}[!htb]
\centering
\includegraphics[width=0.92\linewidth]{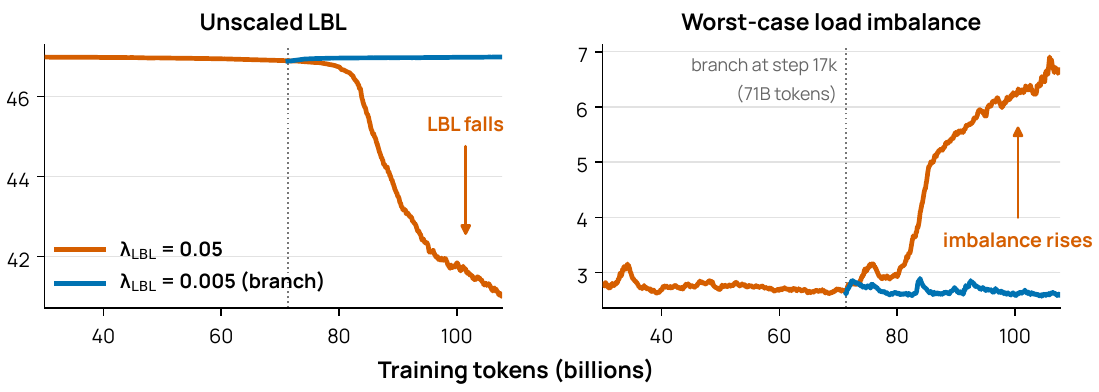}
\caption{\textbf{The load-balancing loss falls while load imbalance rises.}
Orange is the original $\lambda_{\mathrm{LBL}}=0.05$ trajectory; blue reloads
the step-17,000 checkpoint with $\lambda_{\mathrm{LBL}}=0.005$.  Unscaled LBL
is the sum over the 47 MoE blocks; imbalance is the worst max/mean expert
count over blocks and ranks.  Smoothed traces only;
Figure~\ref{fig-token-gerrymandering-training} adds the raw measurements and
the cross-entropy panel, and Section~\ref{sec-lbl-empirical-collapse} gives
the protocol.}
\label{fig-token-gerrymandering-intro}
\end{figure}

Figure~\ref{fig-token-gerrymandering-intro} previews the empirical result of
Section~\ref{sec-lbl-empirical-collapse}: after a checkpoint branch, the trajectory that keeps the
larger LBL coefficient drives its load-balancing loss down while its load
imbalance climbs, and the branch with the smaller coefficient keeps its
imbalance from rising.
This subsection and Section~\ref{sec-lbl-gradient-and-joint-objective}
explain why the objective rewards that trade.

\begin{keyidea}[The LBL Scalar Is a Signed Cross-Term, Not a Balance Certificate]
The standard objective couples requested hard-route frequency to the mean soft
router-score vector.  After centering both vectors at uniform load, the
coupling is a signed cross-term rather than a distance from uniform requested
load.  Frequently selected experts can win many tokens by small margins while
rarely or never selected experts absorb soft mass without winning---the
failure mode we call \textbf{Token Gerrymandering}.
\end{keyidea}

\subsubsection{Requested Hard Assignments and Soft Mass}
\label{sec-lbl-two-distributions}

Consider one routed layer and one balance batch containing $T$ tokens.  Let
$N$ be the number of routed experts, let each token select $K$ experts, and let
$S_t$ be token $t$'s selected set.  For this theoretical block, write
$C_i\equiv C_i^{\mathrm{req}}$ for the requested top-$K$ count aggregated over
the chosen balance batch.  Capacity is downstream of this statistic and does
not enter the current LBL implementation.

\begin{definitionbox}[Requested Hard Load and Soft Mass]
The requested route count, normalized requested fraction, and mean normalized
router score for expert $i$ are
\begin{equation}
\begin{aligned}
C_i\equiv C_i^{\mathrm{req}}
  &\defeq \sum_{t=1}^{T}\mathbf{1}\{i\in S_t\},
&
f_i &\defeq \frac{C_i}{KT},
&
P_i &\defeq \frac{1}{T}\sum_{t=1}^{T}p_{t,i}.
\end{aligned}
\label{eq-lbl-hard-soft-definitions}
\end{equation}
Both $\boldsymbol{f}$ and $\boldsymbol{P}$ are probability vectors:
$\sum_i f_i=\sum_iP_i=1$.  The hard vector $\boldsymbol{f}$ describes which
experts top-$K$ selected before capacity.  The soft vector $\boldsymbol{P}$ is
a differentiable summary of router scores; it is neither a selected-route
histogram nor an executed workload.  If no route is dropped,
$C_i^{\mathrm{req}}=C_i^{\mathrm{keep}}$; the LBL algebra itself does not
assume that equality.
\end{definitionbox}

A direct requested-load metric used by Olmo is the maximum expert count
divided by the mean expert count.  It is the expert-level instance of
Equation~\ref{eq-routing-max-mean-imbalance}:
\begin{equation}
I_{\max}^{\mathrm{req}}
  \defeq \frac{\max_i C_i}{\frac{1}{N}\sum_i C_i}
  = N\max_i f_i.
\label{eq-lbl-hard-imbalance}
\end{equation}
Perfect requested balance gives $I_{\max}^{\mathrm{req}}=1$.  The number of
never-selected experts, the coefficient of variation of the $C_i$, and the
sum of counts placed on each
expert-parallel rank provide complementary views.  Requested counts determine
demand; capacity converts demand into the kept counts that determine expert
compute and actual communication.

The Switch-style loss in Equation~\ref{eq-lbl-switch} instead couples the
hard and soft vectors.  Olmo constructs $C_i^{\mathrm{req}}$ from
top-$K$ indices before capacity under no-gradient bookkeeping and differentiates
Equation~\ref{eq-lbl-switch} only through $P_i$.  For sigmoid routing, the
scores used by this loss are normalized across experts before the dot product,
so the analysis below applies to both normalized sigmoid and softmax scores.

The standard presentation associates the minimum of this objective with
uniform requested and soft routing
\citep[Section~2.2 and Appendix~F]{fedus2022switch}.  That conclusion would
follow immediately if one could identify the soft proxy with requested load.
If $\boldsymbol{P}=\boldsymbol{f}$, then
\begin{equation}
\mathcal{L}_{\mathrm{LBL}}
  = N\lVert\boldsymbol{f}\rVert_2^2
  = 1 + N\lVert\boldsymbol{f}-\boldsymbol{u}\rVert_2^2,
\qquad
u_i\defeq\frac{1}{N},
\label{eq-lbl-if-hard-equals-soft}
\end{equation}
and uniform requested load is the unique minimizer.  Top-$K$ routing does not impose
$\boldsymbol{P}=\boldsymbol{f}$.  It imposes only an ordering constraint within
each token: selected experts must outrank unselected experts.  Aggregating
those token-wise inequalities does not make the two expert-wise marginals
equal.

\subsubsection{The Centered Form Exposes the Surrogate}
\label{sec-lbl-centered-form}

Let $\boldsymbol{u}=(1/N,\ldots,1/N)$ be uniform load.  Because both
$\boldsymbol{f}-\boldsymbol{u}$ and
$\boldsymbol{P}-\boldsymbol{u}$ sum to zero, expanding
Equation~\ref{eq-lbl-switch} gives
\begin{equation}
\boxed{
\mathcal{L}_{\mathrm{LBL}}
  = 1
  + N(\boldsymbol{f}-\boldsymbol{u})^{\mathsf T}
      (\boldsymbol{P}-\boldsymbol{u})
  = 1 + N^2
      \operatorname{Cov}_{i\sim\operatorname{Unif}([N])}(f_i,P_i).
}
\label{eq-lbl-centered}
\end{equation}
Here the covariance is taken across expert coordinates within one balance
scope.  It is not a stochastic claim about correlation over training time.

In Equation~\ref{eq-lbl-centered}, if the hard and soft deviations point in
the same direction, the centered
inner product is positive and the loss exceeds one.  If either vector is
uniform, or if their deviations are orthogonal, the loss equals one.  If the
deviations are anti-aligned, the loss is below one.  In particular,
\begin{equation}
\boldsymbol{P}=\boldsymbol{u}
\quad\Longrightarrow\quad
\mathcal{L}_{\mathrm{LBL}}=1
\quad\text{for every probability vector }\boldsymbol{f}.
\label{eq-lbl-uniform-soft-no-certificate}
\end{equation}
With deterministic tie breaking, uniform scores can pair exactly with a
collapsed requested-route histogram; with strict routing margins, that pair can be
approached arbitrarily closely by giving the winning expert an infinitesimal
advantage.  Thus collapse can have the same limiting scalar value as perfect
balance.  More strongly, the compatible anti-aligned constructions below
score \emph{better} than perfect balance.

The scalar thus fails as a balance metric even though its stop-gradient
construction can still push score mass away from frequently selected
experts. %
\citet{liu2024mixtureofexperts}
gave a compatible one-dead-expert counterexample in which the scalar falls
below one, while %
\citet{su2025moeloadbalancing} emphasized that the scalar value
and its balancing gradient are different objects.  In peer-reviewed work,
\citet{lange2026shinkaevolve} identified a related blind spot in which
global-batch LBL can appear balanced even when some experts remain barely used.
The centered form makes that blind spot exact:
Equation~\ref{eq-lbl-switch} controls one signed cross-term, but it controls
neither
$\lVert\boldsymbol{f}-\boldsymbol{u}\rVert_2$ nor
$\lVert\boldsymbol{P}-\boldsymbol{u}\rVert_2$ separately. The following
constructions quantify how far the scalar can fall below one under
imbalanced routing.

\subsubsection{A Six-Token Counterexample}
\label{sec-lbl-six-token-counterexample}

The smallest useful example has two experts, top-1 routing, and six tokens.
No capacity filter is present, so requested and kept routes coincide.  Ignore
the batch dimension and use the normalized score matrix
\begin{equation}
\begin{bmatrix}
1.0 & 0.0 \\
1.0 & 0.0 \\
0.4 & 0.6 \\
0.4 & 0.6 \\
0.4 & 0.6 \\
0.4 & 0.6
\end{bmatrix}.
\label{eq-lbl-six-token-scores}
\end{equation}
Expert 1 wins the first two tokens, while expert 2 wins the other four.  The
requested hard and soft marginals are
\begin{equation}
\boldsymbol{C}^{\mathrm{req}}=(2,4),
\qquad
\boldsymbol{f}=\left(\frac{1}{3},\frac{2}{3}\right),
\qquad
\boldsymbol{P}=\left(\frac{3}{5},\frac{2}{5}\right).
\label{eq-lbl-six-token-marginals}
\end{equation}
The requested imbalance is $I_{\max}^{\mathrm{req}}=4/3$, yet
\begin{equation}
\mathcal{L}_{\mathrm{LBL}}
 = 2\left(
   \frac{1}{3}\frac{3}{5}
   + \frac{2}{3}\frac{2}{5}
   \right)
 = \frac{14}{15}
 \approx 0.9333
 < 1.
\label{eq-lbl-six-token-value}
\end{equation}

\begin{figure}[!htb]
\centering
\includegraphics[width=\linewidth]{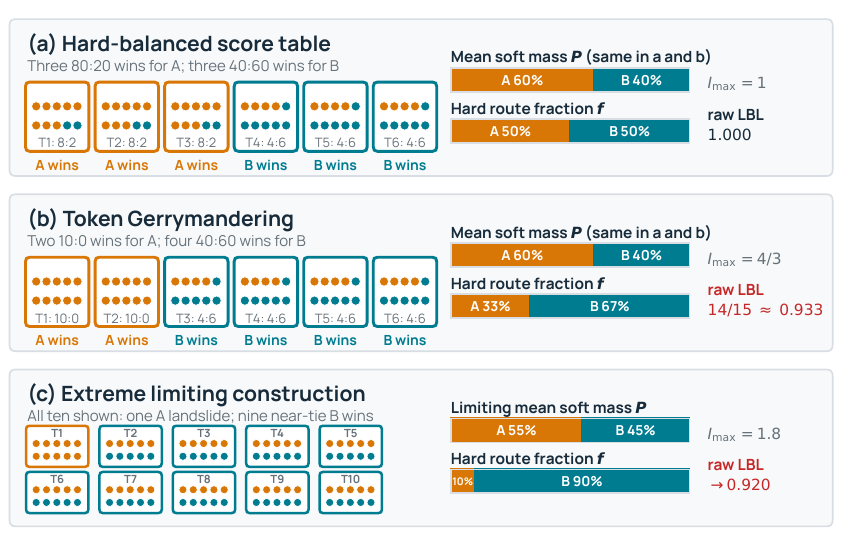}
\caption{\textbf{Soft router mass and requested routes can diverge sharply.}
Orange denotes expert 1 (A), and teal denotes expert 2 (B).  Each outlined
tile is one token in a two-expert, top-1 router.  Its ten interior dots
discretize the normalized scores used to form $\boldsymbol{P}$; the tile border
marks the sole selected route, which is also executed in this capacity-free
construction.  Panels (a) and
(b) hold $\boldsymbol{P}=(3/5,2/5)$ fixed.  Panel (a) is hard-balanced, with
$\boldsymbol{f}=(1/2,1/2)$ and unscaled LBL equal to one.  Panel (b)
concentrates expert 1's soft mass into two decisive wins while expert 2 wins
four tokens $0.6$--$0.4$, giving $\boldsymbol{f}=(1/3,2/3)$,
$I_{\max}^{\mathrm{req}}=4/3$, and LBL $=14/15$.  Panel (c) draws all ten tokens in an
extreme limit: one $(1,0)$ expert-1 win and nine
$(1/2-\epsilon,1/2+\epsilon)$ expert-2 wins yield
$\boldsymbol{f}=(0.1,0.9)$, $\boldsymbol{P}\to(0.55,0.45)$,
$I_{\max}^{\mathrm{req}}=1.8$, and LBL $\to0.92$.  Panel (c)'s five-orange/five-teal dot
patterns show the $\epsilon\to0^+$ limiting soft mass; their teal borders mark
the strict expert-2 wins for every $\epsilon>0$.  Dots belonging to the losing
expert are soft scores, not selected routes.  Exact zeros, ones, and
$\epsilon\to0^+$ are boundary idealizations approachable with finite logits;
the panels are feasible score tables, not a training trajectory, and the
displayed values are counterexamples rather than minima.}
\label{fig-token-gerrymandering-election}
\end{figure}

Panels (a) and (b) of Figure~\ref{fig-token-gerrymandering-election} recast the
six-token example as equal winner-take-all ``elections.''  Both panels hold $\boldsymbol{P}=(3/5,2/5)$ fixed; moving from the
hard-balanced panel (a) to panel (b) changes the requested route fraction from
$(1/2,1/2)$ to $(1/3,2/3)$.  The same soft mass is packed into two
decisive expert-1 wins while expert 2 wins four by smaller margins; the
requested expert load becomes less balanced as the auxiliary scalar falls from one to
$14/15$.

After aggregation, the lightly loaded expert holds more soft mass and the
heavily loaded expert less, so their centered marginals are exactly
anti-aligned, the property quantified by Equation~\ref{eq-lbl-centered}. The
election analogy describes only this concentration of score mass; the tokens
themselves stay fixed.

This example disproves any monotonic interpretation of
$\mathcal{L}_{\mathrm{LBL}}$ as requested-route imbalance: a lower value
coexists with a higher value of the direct requested-load metric.  It is not
itself an optimum.  The
next result shows that the discrepancy is a global score-level property rather
than an isolated numerical choice.

\subsubsection{Perfect Balance Is Not the Global Score-Level Optimum}
\label{sec-lbl-two-expert-infimum}

For two experts and top-1 routing, relabel the underloaded expert as expert 1
and write
\begin{equation}
\boldsymbol{f}=(x,1-x),
\qquad 0\leq x\leq\frac{1}{2}.
\label{eq-lbl-two-expert-load}
\end{equation}
For one token, let $q$ be its soft probability for expert 1.  The token's
contribution to $\boldsymbol{f}^{\mathsf T}\boldsymbol{p}_t$ is
\begin{equation}
xq+(1-x)(1-q)
  = (1-x)+(2x-1)q.
\label{eq-lbl-two-expert-token-contribution}
\end{equation}
Since $2x-1\leq0$, this quantity decreases as $q$ increases.  On the fraction
$x$ of tokens routed to expert 1, the smallest contribution is approached by
$q\to1$, giving contribution $x$.  On tokens routed to expert 2, top-1 routing
requires $q<1/2$; the smallest contribution is approached by
$q\to1/2$ from below, giving contribution $1/2$.  Therefore, for a fixed
requested load $x$,
\begin{equation}
\inf \mathcal{L}_{\mathrm{LBL}}(x)
  = 2\left[x\cdot x+(1-x)\frac{1}{2}\right]
  = 1-x+2x^2.
\label{eq-lbl-two-expert-infimum}
\end{equation}
Differentiating the final quadratic gives its minimum at $x=1/4$:
\begin{equation}
\inf_{x,\{\boldsymbol{p}_t\}}
\mathcal{L}_{\mathrm{LBL}}
  = \frac{7}{8},
\qquad
\boldsymbol{f}=\left(\frac{1}{4},\frac{3}{4}\right).
\label{eq-lbl-two-expert-global-infimum}
\end{equation}
Thus the global score-level infimum has a 1:3 requested-load split, whereas perfect
balance has value one.  For the fixed 2:4 split in
Equation~\ref{eq-lbl-six-token-marginals}, the infimum is $8/9$; replacing the
four $(0.4,0.6)$ rows by
$(1/2-\epsilon,1/2+\epsilon)$ approaches that value.

The result is an \emph{infimum}: a finite softmax cannot produce an exact
zero or one, and strict top-1 assignment approaches the overloaded expert's
tie from one side. Finite $T$ also discretizes $x$. The result concerns free
per-token normalized score tables with compatible requested counts; it does
not claim that every shared linear router attains a smooth finite-logit minimum. A
balanced positive-margin assignment is a flat, non-strict local plateau with
value one while its requested assignments remain unchanged; it is simply not the
global score-level optimum.

\subsubsection{Token Gerrymandering Scales to Many Experts}
\label{sec-lbl-many-expert-gerrymandering}

The two-expert gap is bounded, but the same mechanism becomes much stronger
with many experts.  Choose $A$ \emph{active} experts and
$D=N-A$ \emph{dead}, or never-selected, experts.  Route tokens uniformly among
the active experts, so an active expert has $f_j=1/A$ and a dead expert has
$f_d=0$.  For $0<\epsilon<1$, on a token routed to active expert $j$, choose
the limiting score pattern
\begin{equation}
p_{t,j}=\frac{1+D\epsilon}{D+1},
\qquad
p_{t,d}=\frac{1-\epsilon}{D+1}
\ \text{for every dead expert }d,
\qquad
p_{t,r}\to0
\ \text{for other active experts }r,
\label{eq-lbl-many-expert-scores}
\end{equation}
then take $\epsilon\to0^+$.  These scores sum to one in the zero-tail limit,
and expert $j$ beats every dead expert by $\epsilon$.  The dead experts
collectively absorb almost all soft mass, but none is selected.  An active
expert wins only $1/A$ of the tokens, so
\begin{equation}
P_j\longrightarrow\frac{1}{A(D+1)}
\quad\Longrightarrow\quad
\mathcal{L}_{\mathrm{LBL}}
\longrightarrow
\frac{N}{A(N-A+1)}.
\label{eq-lbl-many-expert}
\end{equation}

\begin{figure}[!htb]
\centering
\includegraphics[width=\linewidth]{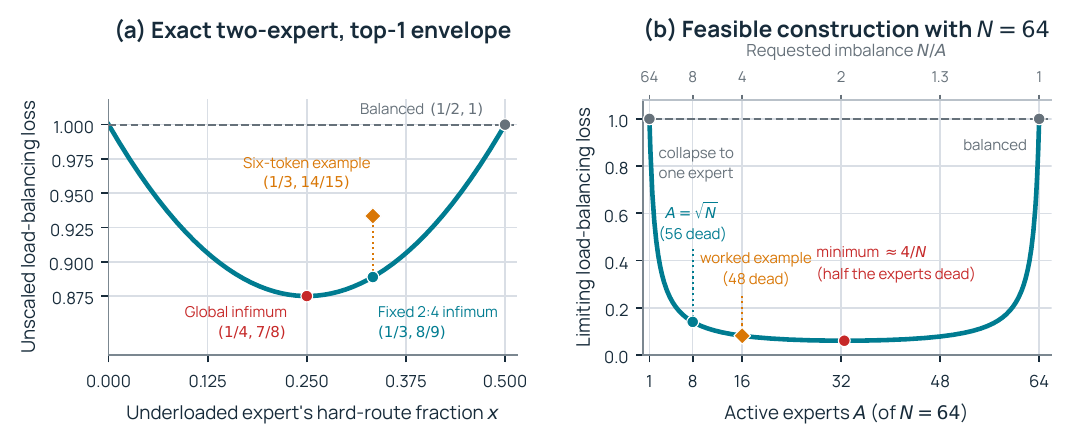}
\caption{\textbf{The Switch-style scalar can prefer imbalanced requested routes.}
Panel~(a): for two experts and top-1 routing, the fixed-load score-level
infimum is $1-x+2x^2$.  Its minimum is $7/8$ at a 1:3 route split; the supplied
six-token score matrix lies above that envelope at $14/15$, while perfect
balance has value one.  Panel~(b): the limiting value $N/(A(N-A+1))$ of
the explicit $N=64$ construction in Equation~\ref{eq-lbl-many-expert} as the
active-expert count $A$ varies. The interior of the curve lies far below the
balanced value of one while the requested imbalance $\tfrac{N}{A}$ on the top axis
grows; the marks show the $A=\sqrt{N}$ comparison, the worked $A=16$
example, and the minimizing $A\approx(N+1)/2$. At both endpoints the scalar
returns to exactly one, so even total collapse onto a single active expert
matches the balanced score.  Both panels are analytical;
they contain no training measurements.}
\label{fig-token-gerrymandering-theory}
\end{figure}

For $N=64$ and $A=16$, the limiting loss is
$64/(16\cdot49)\approx0.0816$, even though 48 experts are dead and
$I_{\max}^{\mathrm{req}}=\tfrac{N}{A}=4$.  Choosing $A\approx(N+1)/2$ makes the constructed value
approximately $4/N$ while roughly half the experts are unused.  An even
sharper asymptotic comparison takes $A\approx\sqrt{N}$: then
$\mathcal{L}_{\mathrm{LBL}}=O(N^{-1/2})$ approaches zero while
$I_{\max}^{\mathrm{req}}=\Theta(N^{1/2})$ diverges and almost every expert is dead.  These are
feasible limiting constructions, not claims about the typical route
distribution of a trained model.
Figure~\ref{fig-token-gerrymandering-theory}b plots this family over the full
range of $A$.

The construction also extends to top-$K$.  Select each token's $K$ winners
uniformly from $A\geq K$ active experts.  For $0<\epsilon<1/K$, give each
winner score $(1+D\epsilon)/(D+K)$, each dead expert score
$(1-K\epsilon)/(D+K)$, and the other active experts scores approaching zero,
then take $\epsilon\to0^+$.  Every active expert has $f_j=1/A$ and
$P_j\to K/[A(D+K)]$, yielding
\begin{equation}
\mathcal{L}_{\mathrm{LBL}}
\longrightarrow
\frac{NK}{A(N-A+K)}.
\label{eq-lbl-many-expert-topk}
\end{equation}
The construction upper-bounds the general score-level infimum and is enough to
refute uniform global optimality.  We do not claim that
Equation~\ref{eq-lbl-many-expert-topk} is the exact optimum for arbitrary
$(N,K,T)$, tie breaking, grouped selection, or score-correction bias.

The many-expert construction preserves the same requested-load/soft-mass
anti-alignment as the two-expert example.

\subsubsection{A Larger Scope Does Not Repair the Surrogate}
\label{sec-lbl-scope-invariance}

Changing the balance scope leaves the scalar's limitation intact. For any
scope in which $\boldsymbol{f}$ and $\boldsymbol{P}$ are normalized over the same
tokens, Equation~\ref{eq-lbl-centered} still holds.  A global-batch LBL gives
the router more tokens over which to satisfy the constraint; it does not turn
$N\boldsymbol{f}^{\mathsf T}\boldsymbol{P}$ into a norm of global requested
load.  Instance-level balancing prevents cancellation between sequences, but
it does not remove the anti-alignment construction within one sequence.
Likewise, synchronized-microbatch, online-global, and exact optimizer-batch
statistics change which tokens may compensate for one another without
changing the signed cross-term itself.

\subsection{Gradient Dynamics and the Joint Objective}
\label{sec-lbl-gradient-and-joint-objective}

The requested counts are detached, so within a region where top-$K$ assignments do
not change, $\boldsymbol{f}$ is constant.  For top-1 softmax logits $z_{t,i}$,
differentiating Equation~\ref{eq-lbl-switch} gives
\begin{equation}
\frac{\partial\mathcal{L}_{\mathrm{LBL}}}{\partial z_{t,i}}
  = \frac{N}{T}p_{t,i}
    \left(f_i-\boldsymbol{f}^{\mathsf T}\boldsymbol{p}_t\right).
\label{eq-lbl-logit-gradient}
\end{equation}
For every token, the auxiliary gradient transfers soft mass toward experts
whose requested load $f_i$ is below the score-weighted average.  This is the
intended balancing pressure. A task-preferred, frequently selected expert
can nevertheless remain the winner by an arbitrarily small
margin while cold or dead experts absorb much of the differentiable mass.  The
gradient can therefore improve the surrogate faster than it improves the
requested-selection histogram.

Routing boundaries make the dynamics discontinuous.  When two scores cross,
the hard vector $\boldsymbol{f}$ changes abruptly and the gradient field is
recomputed.  A shared router may cross, cycle around, or avoid the limiting
score tables above.  The theory establishes feasible objective geometry; by
itself it does not prove that stochastic gradient descent reaches a stable
gerrymandered routing pattern.

The connection to training appears only after restoring the auxiliary
coefficient.  Write the optimized objective as
\begin{equation}
\mathcal{L}_{\mathrm{train}}
  = \mathcal{L}_{\mathrm{CE}}
  + \lambda_{\mathrm{LBL}}\mathcal{L}_{\mathrm{LBL}}
  + \mathcal{L}_{\mathrm{other}},
\label{eq-lbl-joint-objective}
\end{equation}
where $\mathcal{L}_{\mathrm{CE}}$ is next-token cross-entropy (CE) and
$\mathcal{L}_{\mathrm{other}}$ contains every other enabled term, such as a
router z-loss %
\citep{zoph2022stmoe} or language-model z-loss
\citep{chowdhery2023palm}.  Compare the model before and after a routing
transition on the same evaluation batch with fixed loss normalization.  If
the transition hurts language modeling but reduces LBL, then
\begin{equation}
\Delta\mathcal{L}_{\mathrm{CE}}>0,
\qquad
\Delta\mathcal{L}_{\mathrm{LBL}}<0.
\label{eq-lbl-collapse-direction}
\end{equation}
The post-transition model has a lower scalar objective exactly when
\begin{equation}
\Delta\mathcal{L}_{\mathrm{train}}<0
\quad\Longleftrightarrow\quad
\lambda_{\mathrm{LBL}}
\left(-\Delta\mathcal{L}_{\mathrm{LBL}}\right)
  > \Delta\mathcal{L}_{\mathrm{CE}}
    + \Delta\mathcal{L}_{\mathrm{other}}.
\label{eq-lbl-collapse-threshold}
\end{equation}
Whether an LBL coefficient is \emph{high} depends on the net CE and
other-term cost of a reachable routing transition, the number of routed layers
contributing auxiliary loss, and the implementation's layer and microbatch
normalization; there is no universal numerical threshold. A larger
coefficient does not change the minimizers of LBL
alone; it changes which compromises lower the joint objective.  When such a
compromise wins, the observable pattern is the \textbf{surrogate-hacking
signature}: the optimized objective falls while both language-modeling loss
and requested imbalance worsen.  This
before-and-after comparison is not an optimizer-dynamics theorem: momentum and
stochastic gradients need not lower the instantaneous loss of every training
batch, and losses from different batches do not define the deltas above.

\subsubsection{Empirical Signature of Reachable Collapse}
\label{sec-lbl-empirical-collapse}

\begin{figure}[!htb]
\centering
\includegraphics[width=\linewidth]{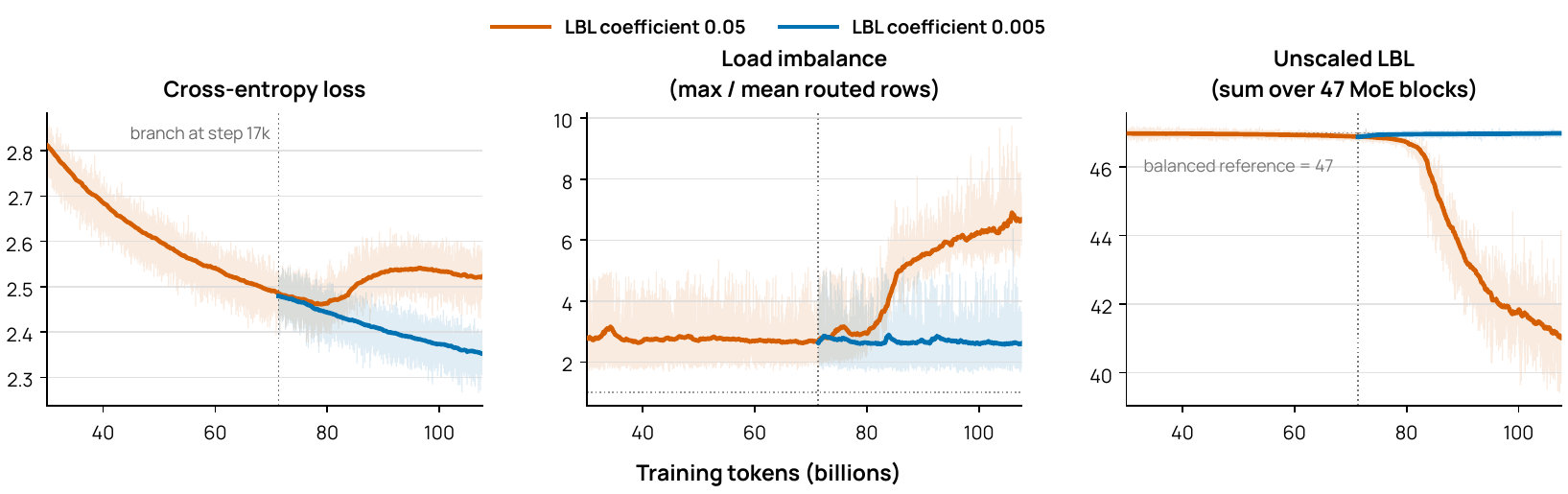}
\caption{\textbf{Checkpoint-branched training with different LBL coefficients.}
With the higher coefficient, the surrogate falls while CE and imbalance
rise. Orange shows the
original $\lambda_{\mathrm{LBL}}=0.05$ trajectory; blue shows the branch that
reloads the step-17,000 checkpoint with $\lambda_{\mathrm{LBL}}=0.005$.
After the checkpoint branch, the high-coefficient trajectory eventually
develops worsening CE and sharply rising worst-block/rank max/mean
expert-count imbalance even as its unscaled LBL falls.  The low-coefficient
branch instead continues improving CE while its LBL and imbalance remain
stable.  Light traces show every raw training measurement; thick traces use a
time-weighted exponential moving average (TWEMA) with a
0.75-billion-token half-life.  The dotted
vertical line marks the step-17,000 branch point.  This is a checkpoint-branched result for one configuration,
not a universal coefficient threshold.}
\label{fig-token-gerrymandering-training}
\end{figure}

The score-level constructions do not show that stochastic training reaches a
gerrymandered routing pattern.  Figure~\ref{fig-token-gerrymandering-training} provides
a checkpoint-branched test in a 48-layer, 64-expert, top-4 MoE.  The original
trajectory uses instance-level LBL with
$\lambda_{\mathrm{LBL}}=0.05$.  At step 17,000 (71.3 billion training
tokens), before the observed divergence, a second trajectory
reloads the same model and optimizer checkpoint and changes the coefficient to
$0.005$.  This intervention controls the starting point more
tightly than an independent-from-scratch coefficient sweep.  Dispatch is dropless in these runs, so requested counts are also the executed
expert loads.  Block 0 is dense,
so the logged LBL is the sum of the instance-level loss over the 47 MoE
blocks; the logged load
imbalance is the worst max/mean expert count over those blocks and distributed
ranks.  These runs use the production HSDP pretraining stack rather than the
expert-parallel paths of this report.
Appendix~\ref{app-token-gerrymandering-run} records the full run
recipe and the archived-history provenance.

The comparison establishes the surrogate--workload mismatch
(Table~\ref{tab-token-gerrymandering-branch}).  Between the early and late
windows, the high-coefficient trajectory lowers its unscaled LBL sum by about
twelve percent while its worst-case load imbalance more than doubles and its
CE worsens; the low-coefficient branch improves CE while its LBL stays near
the balanced reference and its imbalance does not rise.  The effect is not
isolated to one layer: median imbalance increases in all 47 MoE blocks on the
high-coefficient trajectory, and over the branched interval its unscaled LBL
and worst-case load imbalance have Spearman correlation $-0.87$.

\begin{table}[!htb]
\centering
\footnotesize
\renewcommand{\arraystretch}{1.14}
\begin{tabular}{@{}lrrrrrr@{}}
\toprule
& \multicolumn{3}{c}{\textbf{High coefficient} ($\lambda_{\mathrm{LBL}}=0.05$)}
& \multicolumn{3}{c}{\textbf{Low-coefficient branch} ($0.005$)} \\
\cmidrule(lr){2-4}\cmidrule(l){5-7}
\textbf{Median over window} & Early & Late & Change & Early & Late & Change \\
\midrule
CE loss & 2.473 & 2.523 & $+2.0\%$ & 2.469 & 2.358 & $-4.5\%$ \\
Unscaled LBL sum & 46.864 & 41.278 & $\mathbf{-11.9\%}$
& $\approx47$ & $\approx47$ & stable \\
Worst-case load imbalance & 2.748 & 6.317 & $\mathbf{+129.9\%}$
& 2.655 & 2.544 & $-4.2\%$ \\
\bottomrule
\end{tabular}
\caption{\textbf{Loss and imbalance before and after the observed collapse.}
Medians over steps 17,001--17,999 (early) and
24,000--25,669 (late), both after the step-17,000 branch.  Unscaled LBL is the
sum over the 47 MoE blocks, whose balanced reference is 47; imbalance is the
worst max/mean expert count over blocks and ranks.  The low-coefficient
branch's LBL is reported only as its proximity to that reference.}
\label{tab-token-gerrymandering-branch}
\end{table}

\begin{figure}[!htb]
\centering
\includegraphics[width=0.55\linewidth]{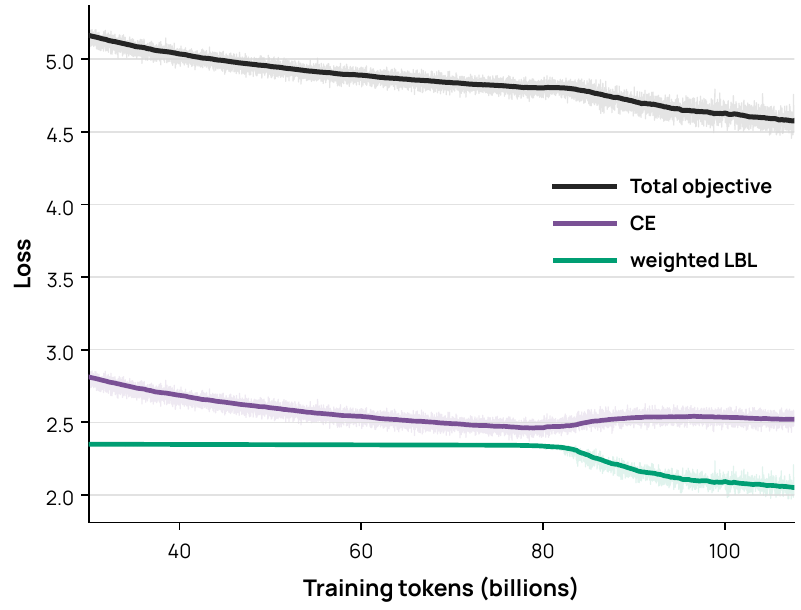}
\caption{\textbf{The reconstructed objective decreases despite worsening CE.}
This panel shows only the original $\lambda_{\mathrm{LBL}}=0.05$
trajectory; black, purple, and green now denote the total and its distinct
components rather than different coefficient settings.  Across the collapse
interval, the decrease in coefficient-weighted LBL more than offsets the
increase in CE, so the reconstructed optimized objective---CE plus weighted
LBL plus the enabled LM-head z-loss---continues downward.  Light traces show
every raw training measurement; thick traces use a TWEMA
with a 0.75-billion-token half-life.}
\label{fig-token-gerrymandering-objective}
\end{figure}

The unscaled LBL and CE curves in
Figure~\ref{fig-token-gerrymandering-training} have different weights in
the optimized objective. Figure~\ref{fig-token-gerrymandering-objective}
combines them with those weights for the high-coefficient trajectory.

Over the same windows, the median coefficient-weighted LBL contribution falls
from $2.343$ to $2.064$, enough for the median reconstructed optimized
objective to decrease from $4.820$ to $4.591$ ($-4.8\%$) despite the CE
regression.  Olmo injects LBL
through a custom autograd term rather than adding it to one monolithic
forward-loss scalar.  The plotted objective therefore reconstructs the
optimized scalar from the separately logged CE, coefficient-weighted LBL, and
LM-head z-loss.  Router z-loss is disabled in these runs.

This establishes reachability of the predicted surrogate-hacking signature
for this configuration: training can lower the effective objective while
making both language modeling and routed workload worse.  It does not
locate a universal coefficient threshold or establish how frequently the
failure occurs.  These are also stochastic training-batch curves rather
than losses reevaluated on one fixed batch, and the archived metrics do not
contain same-scope $\boldsymbol{f}$ and $\boldsymbol{P}$ vectors.  Attributing
the trajectory to the exact hard/soft anti-alignment construction would
require the centered term in Equation~\ref{eq-lbl-centered}, or
equivalent evidence from soft mass on \mbox{never-selected} experts and
top-$K$--top-$(K+1)$ margins. A lower LBL scalar did not certify a more
balanced expert workload.

\subsection{Measuring Routing Transients and Outcomes}
\label{sec-lbl-reporting-implications}

Routing is nonstationary, especially early in training. A router can move
from near-random assignments through sharp transients before reaching a more
stable distribution, and both capacity utilization and step time can move
with it.  Uniform or random expert assignment is useful for isolating an
optimistic balanced systems path, but it is not a substitute for trained
routing. Interpreting throughput therefore requires knowing whether the
routes are artificially balanced, early-transient, or drawn from a declared
stable training window. Three diagnostic questions connect those routing
conditions to the model and systems outcomes.

\paragraph{What did the router request, and what did capacity admit?}
$\boldsymbol{C}^{\mathrm{req}}$ records selected top-$K$ routes before
capacity, and $\boldsymbol{C}^{\mathrm{keep}}$ records the rows admitted for
expert execution.  When no route is dropped the two hard histograms coincide.
Otherwise both histograms are needed. Route-drop rate, the fraction of tokens
losing at least one or
all selected routes, selected weight removed by dropping, and receiving-buffer
utilization describe how capacity changed the workload and removed model
contributions.

\paragraph{Which experts or ranks became bottlenecks?}
Max/mean and a full-distribution statistic for requested expert counts,
the number of never-selected experts, and the maximum requested load on one
expert-parallel rank distinguish expert concentration from rank pressure.
The corresponding kept-count statistics describe the work that actually ran.
Grouped-GEMM and step time connect kept counts to computation; the
source--destination request matrix and rank waiting connect demand and admitted
work to communication and straggling.

\paragraph{Did balance improve, or only the surrogate?}
The soft mean $\boldsymbol{P}$ summarizes differentiable router mass rather
than selected or executed routes.
Diagnosing Token Gerrymandering requires the requested hard fraction
$\boldsymbol{f}$ and soft mean $\boldsymbol{P}$ from the same routed layer and
load-balancing scope. Their centered inner product, together with
$\lVert\boldsymbol{f}-\boldsymbol{u}\rVert_2$ and
$\lVert\boldsymbol{P}-\boldsymbol{u}\rVert_2$, separates anti-alignment from
the magnitude of each deviation; top-$K$--top-$(K+1)$ margins and
soft mass on never-selected experts provide complementary evidence.  Pearson
correlation alone is insufficient because it is undefined when either vector
is uniform and discards the magnitudes that determine LBL.

The effect on the optimized objective depends on the effective LBL
contribution after layer and microbatch normalization, together with every
enabled auxiliary term. An unscaled LBL curve beside
CE can disprove the scalar's use as a balance certificate, but it cannot
establish why the optimizer preferred a transition.

The evidence in this section establishes the objective geometry, the
fixed-work grouped-GEMM sensitivity,
and the reachability of the surrogate-hacking signature in one
checkpoint-branched training comparison. The requested and kept workloads,
rather than the auxiliary scalar alone, connect routing to the expert GEMMs
and communication that the training system executes.

\section{MXFP8 Training}
\label{sec-mxfp8-training}

Routing determines which expert GEMMs run and how their work is distributed.
MXFP8 changes the precision and volume of the operands those GEMMs compute
with and communicate.

\textbf{Microscaling (MX)} formats pair a low-precision payload with one
shared power-of-two scale per small block of values; MXFP8 is the member
with an FP8 payload and 32-value blocks, standardized by the Open Compute
Project (OCP) together with the FP8 payload formats
\citep{micikevicius2022fp8,rouhani2023microscaling,ocpMXFP8spec}. In
training, MXFP8 requires changes to operand quantization, scale layouts,
saved activations, expert transport, and weight refresh after each optimizer
update. Those changes can accelerate sufficiently large GEMMs, reduce
storage for selected saved activations, and shrink expert-communication
payloads. They also add
conversion, layout, transport-management, caching, and backward work.
Section~\ref{sec-mxfp8-benefits} quantifies the potential savings and
explains why persistent training-state memory stays approximately unchanged.

Blackwell is the main hardware target for the implementation described here.
Hopper supports FP8 Tensor Core operations; Blackwell adds native support for
the 32-value, E8M0-scaled block representation used by MXFP8
\citep{ocpMXFP8spec,nvidiaCublasBlockScaling}. The Tensor Core operation can
consume quantized operands and their block scales together. Software must
still create both components in the layouts expected by the instruction.

Table~\ref{tab-mxfp8-hardware} separates ordinary FP8 support from native
MXFP8 support. Hopper accelerates E4M3 and E5M2 FP8, but it does not provide
the block-scaled matrix instruction that consumes an E8M0 scale per 32 payload
values. The Blackwell variants in the table do. Their instruction families
differ: data-center Blackwell uses \texttt{tcgen05.mma}, while SM120 RTX
Blackwell exposes block scaling through \texttt{mma.sync}.

\begin{table}[!htb]
\centering
\scriptsize
\setlength{\tabcolsep}{2pt}
\renewcommand{\arraystretch}{1.15}
\begin{tabularx}{\linewidth}{@{}
  >{\RaggedRight\arraybackslash}X
  >{\centering\arraybackslash}m{0.05\linewidth}
  *{3}{>{\raggedleft\arraybackslash}m{0.072\linewidth}}
  >{\centering\arraybackslash}m{0.085\linewidth}
  >{\RaggedRight\arraybackslash}m{0.21\linewidth}@{}}
\toprule
& & \multicolumn{3}{c}{\textbf{Dense peak (TFLOP/s)}} & & \\
\cmidrule(lr){3-5}
\textbf{GPU} & \textbf{CC} & \textbf{BF16} & \textbf{FP8} & \textbf{FP4} &
\makecell[c]{\textbf{Native}\\\textbf{MXFP8}} &
\makecell[c]{\textbf{Matrix-instruction}\\\textbf{family}} \\
\midrule
H100 SXM & 9.0 & 989.5 & 1979 & \textemdash & No &
\texttt{wgmma.mma\_async} \\
HGX B200, per GPU & 10.0 & 2250 & 4500 & 9000 & Yes &
\texttt{tcgen05.mma} \texttt{block\_scale} \\
HGX B300, per GPU & 10.3 & 2250 & 4500 & 13500 & Yes &
\texttt{tcgen05.mma} \texttt{block\_scale} \\
RTX PRO 6000 Blackwell, 600 W & 12.0 & 503.8 & 1007.6 & 2015.2 & Yes &
\texttt{mma.sync} \texttt{block\_scale} \\
\bottomrule
\end{tabularx}
\caption{\textbf{FP8 support alone does not imply native MXFP8 support.}
Peak columns report dense theoretical Tensor Core TFLOP/s. H100 dense values
are half of NVIDIA's published 2:4-sparse figures. HGX B200/B300 system
figures are normalized to one of eight GPUs and to dense execution; the HGX
page lists FP8 and BF16 as sparse peaks with dense as half, and lists FP4 as
both sparse and dense values for B200 and B300. The RTX row uses the
Workstation Edition 600-W specifications. Compute capabilities and precision
support follow NVIDIA's TensorRT matrix; instruction names follow the CUTLASS
and PTX documentation
\citep{nvidiaH100spec,nvidiaHGXB200spec,nvidiaRTXBlackwellArchitecture,nvidiaTensorRT10141SupportMatrix,nvidiaBlackwellSM100Gemm,nvidiaPTX88}.}
\label{tab-mxfp8-hardware}
\end{table}

\subsection{Benefits of MXFP8 Training}
\label{sec-mxfp8-benefits}

\begin{keyidea}[Where MXFP8 Saves Time and Memory]
MXFP8 can make large GEMMs faster, selected saved activations smaller, and
expert communication lighter. Persistent memory grows slightly
instead: the recipe retains two quantized weight orientations alongside
the FP32 main weights and optimizer states.
\end{keyidea}

\subsubsection{Faster GEMMs}

A BF16 GEMM uses two bytes per input value and the BF16 Tensor Core path. The
Olmo MXFP8 recipe uses one E4M3 payload byte per value plus block scales, and on
Blackwell GPUs it uses the higher-throughput FP8 Tensor Core path. Peak
Tensor Core specifications list an exact $2\times$ FP8-to-BF16 ratio on both
Hopper and Blackwell (Table~\ref{tab-mxfp8-hardware})
\citep{nvidiaH100spec,nvidiaHGXB200spec}. That ratio is an arithmetic
ceiling, not an end-to-end speedup: quantization (Q), dequantization (DQ),
scale construction, layout conversion, kernel launches, small GEMMs,
communication, and all non-GEMM work remain on the step timeline.

\begin{figure}[!htb]
  \centering
  \includegraphics[width=\linewidth]{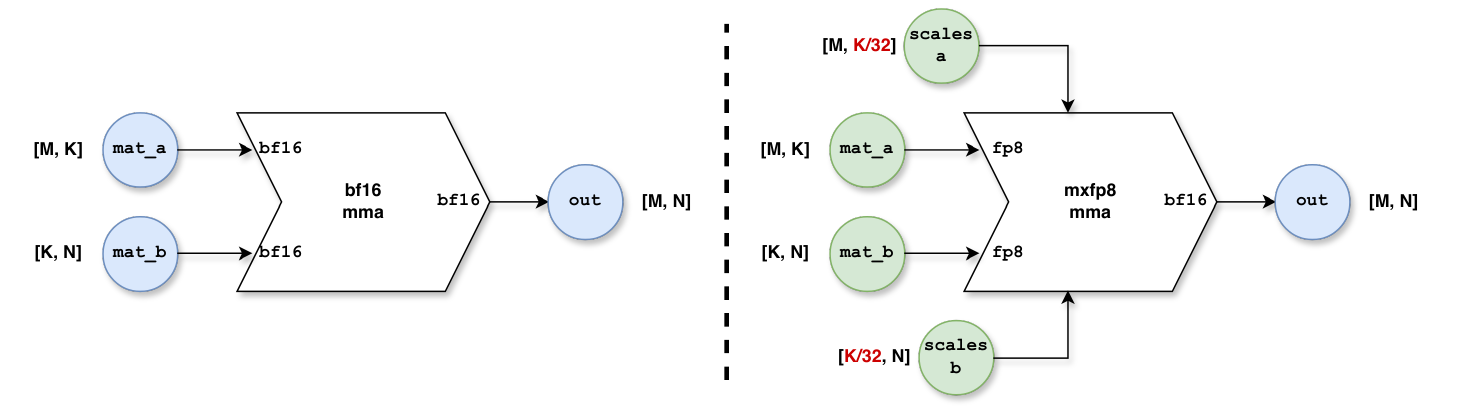}
  \caption{\textbf{BF16 and native MXFP8 Tensor Core paths.}
  For $A\in\mathbb{R}^{M\times K}$ and
  $B\in\mathbb{R}^{K\times N}$, the MXFP8 operation consumes E4M3 qdata and
  logical scale grids of shape $M\times(K/32)$ and $(K/32)\times N$ before
  scale swizzling. ``MMA'' names the native Tensor Core operation accurately:
  Blackwell provides \texttt{tcgen05.mma} instructions with the
  \texttt{kind::mxf8f6f4} and \texttt{block\_scale} modifiers for this
  computation. Olmo invokes the higher-level \texttt{scaled\_mm} operator,
  whose kernel includes the MMA main loop and an epilogue that converts the
  accumulator to the requested BF16 output. Thus, BF16 labels the completed
  operator output, not the raw MMA accumulator.}
  \label{fig-mxfp8-bf16-native-gemm}
\end{figure}

Figure~\ref{fig-mxfp8-bf16-native-gemm} contrasts the operands supplied to
the BF16 and native MXFP8 Tensor Core paths. The BF16 path consumes two BF16
matrices. The MXFP8 path consumes E4M3 qdata together with one E8M0 scale per
32 values along the reduction dimension. On Blackwell, the instruction-level
primitive is a fifth-generation Tensor Core matrix multiply--accumulate (MMA):
the MXFP8 path uses the block-scaled \texttt{tcgen05.mma} family. The scale
shapes in the figure are the logical pre-swizzle grids; the physical blocked
layouts consumed by the instruction are more specialized. Appendix~\ref{app-blackwell-tcgen05} separates the PyTorch operator, GEMM kernel, and
Tensor Core instruction interfaces in more detail.

The attainable speedup depends on shape. Large dense GEMMs can amortize
operand quantization and scale-layout work over substantial matrix
multiplication. A routed layer may instead present many smaller expert GEMMs,
where the same fixed work occupies a larger fraction of kernel time. Weight
quantization is easier to amortize because its caches are rebuilt after an
optimizer update and reused across the following microbatches. Activation
quantization is paid on each invocation.

\needspace{6\baselineskip}
\subsubsection{Smaller Saved Activations}

\begin{figure}[!htb]
\centering
\includegraphics[width=0.30\linewidth]{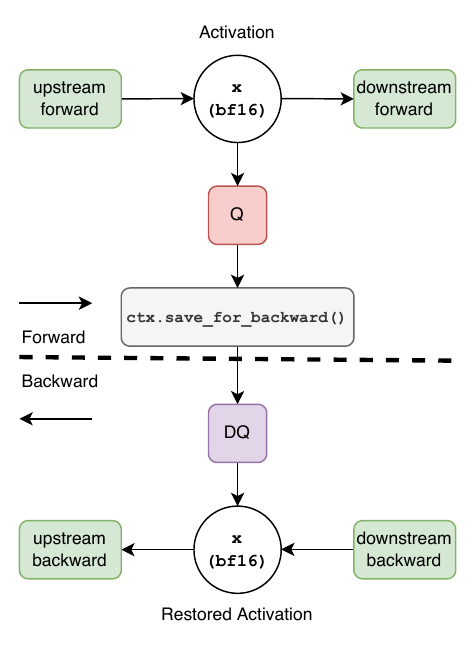}
\caption{\textbf{Saving a quantized activation for backward.}
Forward quantizes the BF16 activation before \texttt{save\_for\_backward};
backward dequantizes it only when a BF16 consumer needs the value. This saves
memory only when the quantized representation replaces, rather than
supplements, another retained BF16 copy.}
\label{fig-mxfp8-quantized-activation-save}
\end{figure}

For a block of 32 values, the nominal MXFP8 representation contains 32 bytes
of E4M3 qdata and one byte of E8M0 scale. Ignoring blocked-layout padding, its
storage per value is

\begin{equation}
B_{\mathrm{MXFP8}} = 1 + \frac{1}{32} = \frac{33}{32}
\quad\text{bytes/value},
\label{eq-mxfp8-bytes-per-value}
\end{equation}

compared with two bytes per BF16 value. Replacing a \emph{unique} BF16 saved tensor
with this representation saves $31/32$ bytes per value, a nominal reduction
of $1-33/64\approx48.4\%$. If another operator already keeps the same BF16
activation alive, adding an MXFP8 copy instead costs another $33/32$ bytes per
value. The implementation must therefore determine whether another operator
already retains each BF16 tensor before choosing its saved precision. In our
attention path, for example, the QKV projection and output projection use
different save policies because their BF16 inputs are retained by different
parts of the backward graph. Figure~\ref{fig-mxfp8-quantized-activation-save} (Section~\ref{sec-mxfp8-recipe})
shows the resulting boundary between forward-time packing and backward-time
unpacking.

\subsubsection{Smaller Dispatch and Combine Payloads}

Rowwise EP normally moves one hidden row for each routed token--expert
assignment. A BF16 row of width $d$ carries $2d$ payload bytes. When $d$ is a
multiple of 32, its nominal MXFP8 representation carries
$d+d/32=33d/32$ bytes, again a $48.4\%$ reduction in the row payload.

This does not halve the whole communication operation. Route indices, route
weights, counters, and synchronization are unchanged. Qdata and scales also
need destinations and transfers of their own. MXFP8 reduces the numerical
payload for dispatch and combine, and that
reduction is useful when the saved transport time exceeds the added
quantization, scale, and transfer-management work.

\subsubsection{Persistent Training-State Memory}

Each linear weight is consumed in two orientations. Forward computes with
$W^\top$, while dgrad computes with $W$. Each orientation needs its own qdata
and scale layout. We call these qdata-and-scale representations
\textbf{MXFP8 compute-weight caches}, or simply \textbf{weight caches} in this
section. At nominal storage, the two MXFP8 views require
$2\times33/32=2.0625$ bytes per original weight value. This is $3.125\%$ more
than one two-byte BF16 compute weight even before blocked-scale padding.

Our optimizer also retains an FP32 main weight and FP32 optimizer states.
After the MXFP8 caches have been initialized, the full BF16 initialization
anchor can be released, but the two caches take its place. MXFP8 therefore
does not remove the persistent training-state memory wall by itself. EP, PP, and distributed
optimizer sharding remain responsible for dividing parameters, gradients, and
optimizer tensors, as
described in Part~\ref{part-parallelization}.

\begin{figure}[!htb]
\centering
\includegraphics[width=0.9\linewidth]{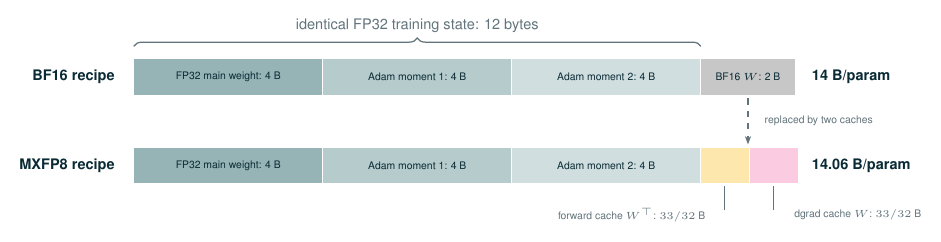}
\caption{\textbf{Persistent memory with BF16 weights and two MXFP8 weight caches.}
Nominal bytes per parameter: both recipes keep the FP32 main
weight and two FP32 Adam moments; the BF16 recipe adds one two-byte compute
weight, while the MXFP8 recipe replaces it with the forward and dgrad weight
caches at $33/32$ bytes each.  Gradient buffers are identical in both recipes
and excluded, and blocked-scale padding is ignored.  Conceptual illustration
of the nominal accounting with no measured quantities.}
\label{fig-mxfp8-persistent-state}
\end{figure}

Figure~\ref{fig-mxfp8-persistent-state} compares the two recipes. MXFP8
replaces a two-byte BF16 compute weight with two cache orientations of
nearly the same combined size, so the composition of persistent training state changes
while its total stays approximately constant.

\subsection{Understanding the MXFP8 Format}
\label{sec-mxfp8-format}

\begin{figure}[!htb]
\centering
\includegraphics[width=0.86\linewidth]{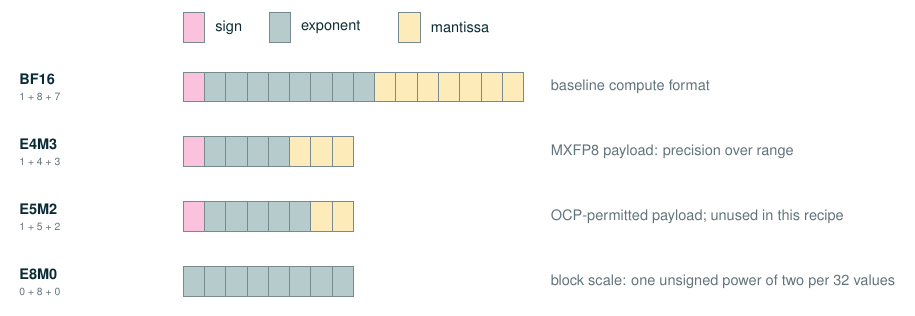}
\caption{\textbf{Bit layouts of the baseline, payload, and scale formats.}
BF16 keeps eight exponent bits and seven mantissa bits. The E4M3
payload trades exponent range for mantissa precision; E5M2 makes the opposite
trade and is permitted by OCP MX but unused in the Olmo recipe.  The E8M0
scale is an exponent-only format: one unsigned power of two per 32-value
block, with no sign and no mantissa.}
\label{fig-mxfp8-formats}
\end{figure}

An MXFP8 tensor is defined by more than its one-byte E4M3 payload. Its E8M0
scales, the number of payload values that share each scale, and the physical
scale layout consumed by a particular operand role are equally important.
These details explain why conversion has separate quantization and swizzle
costs, why qdata and scales cross operator boundaries together, and why native
MXFP8 GEMM requires Blackwell-specific hardware support.
Figure~\ref{fig-mxfp8-formats} compares the bit layouts involved: the BF16
baseline, the two payload formats the OCP MX specification permits, and the
E8M0 scale.

\subsubsection{Quantization, Dequantization, and Swizzle}

For a BF16 block $x$, quantization selects a power-of-two scale $s$ and stores

\begin{equation}
q = \operatorname{E4M3}(x/s),
\qquad
\widehat{x} = s q.
\label{eq-mxfp8-quantization}
\end{equation}

\begin{figure}[!htb]
  \centering
  \includegraphics[width=0.64\linewidth]{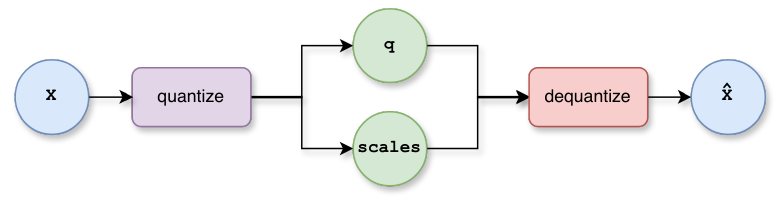}
  \caption{\textbf{Quantization produces a qdata\textendash{}scale pair.}
  Dequantization consumes both components and reconstructs
  $\widehat{x}=s q$, an approximation to the original $x$; it does not invert
  quantization exactly.}
  \label{fig-mxfp8-quantize-dequantize}
\end{figure}

The qdata $q$ and the scale $s$ together encode one MXFP8 tensor; neither
part is usable alone. Dequantization applies the second relation
when an operator requires BF16 or FP32 values
(Figure~\ref{fig-mxfp8-quantize-dequantize}).

\begin{figure}[!htb]
\centering
\includegraphics[width=\linewidth]{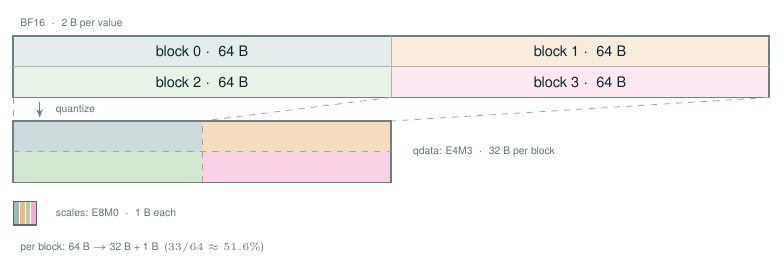}
\caption{\textbf{Quantization produces one contiguous qdata array and one
contiguous scale array.}  Sizes are drawn to byte scale for a two-row tensor
with two 32-value blocks per row: each 64-byte BF16 block becomes a 32-byte
E4M3 qdata block, half as wide, and the flattened scale array adds one E8M0
byte per block, each scale $1/32$ the width of a qdata block.  The four
scales are the narrow strip at the lower left, drawn at true width; color
marks which block each scale covers, and the dashed spans are logical:
neither array is fragmented.  Conceptual illustration with no measured
quantities.}
\label{fig-mxfp8-block-quantization}
\end{figure}

\begin{figure}[!htb]
\centering
\includegraphics[width=\linewidth]{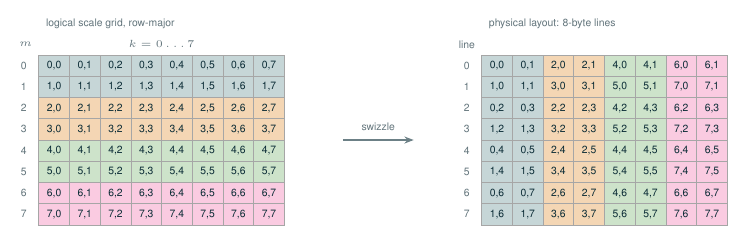}
\caption{\textbf{The scale order a GEMM consumes is a tiled, interleaved
layout.}  Cells are labeled $m,k$ for the scale of row $m$ and block column
$k$, and colors mark row groups.  Before the swizzle the grid is row-major;
after it, each physical line interleaves the four row groups while a row's
scales within a tile stay adjacent.  The drawing is a structure-preserving
miniature with two-row groups and two-column tiles; the hardware layout
applies the same rule with 32-row groups and four-column tiles, rearranging
$128\times4$ tiles of scales into 32 sixteen-byte lines so that rows $m$,
$m{+}32$, $m{+}64$, and $m{+}96$ become physically adjacent
\citep{nvidiaCublasBlockScaling}.  The result resembles a transpose because
whole row-group sub-blocks are relocated from a vertical stack to a
horizontal line, but it is not one: inside every sub-block, $m$ still
advances downward and $k$ rightward, and each scale keeps its $m,k$
identity; only its address changes.  Conceptual illustration with no
measured quantities.}
\label{fig-mxfp8-scale-swizzle}
\end{figure}

The scale order associated with adjacent 32-value input blocks is not
necessarily the blocked order required by a GEMM. Scaled matrix-multiply
kernels consume swizzled scale layouts
chosen for Tensor Core access. Quantization therefore has two distinct tasks:
compute the scale and qdata, then arrange the scales for the operand role and
matrix layout used by the kernel. Rowwise communication may use a different
addressable scale layout from scaled GEMM
\citep{nvidiaTransformerEngineMXFP8}.
Figure~\ref{fig-mxfp8-block-quantization} draws the block decomposition, and
Figure~\ref{fig-mxfp8-scale-swizzle} draws the layout distinction.

\begin{definitionbox}[A Logical MXFP8 Tensor Has Two Physical Components]
One logical MXFP8 tensor is represented by two physical components:
\textbf{qdata} and \textbf{E8M0 scales}, together with the \textbf{layout that
relates them}. Operators consume qdata and scales together using their
associated layout.
\end{definitionbox}

\subsubsection{Scaling Granularity}

\begin{figure}[!htb]
\centering
\includegraphics[width=\linewidth]{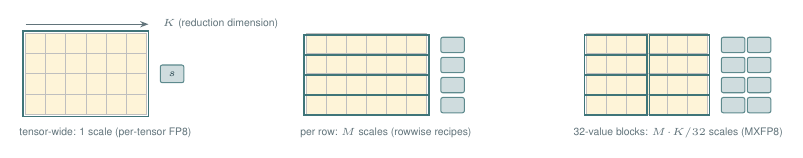}
\caption{\textbf{Scale granularity is a spectrum along the reduction
dimension.}  One tensor-wide scale must cover the largest value in the
tensor, per-row scales track variation more closely, and 32-value block
scales track it most tightly at the highest metadata cost.  Conceptual
illustration with no measured quantities.}
\label{fig-mxfp8-scale-granularity}
\end{figure}

The number of values sharing a scale affects both numerical precision and
storage cost. A tensor-wide scale is cheap to store but must cover the
largest value in the tensor. Per-row or per-channel scales track variation
more closely at greater metadata and conversion cost. Block scaling lies
between those choices. The OCP MX specification permits E4M3 and E5M2 payload
formats. Olmo uses the E4M3 MXFP8 variant: each adjacent block of 32 E4M3
values shares one power-of-two E8M0 scale %
\citep{ocpMXFP8spec}.
Figure~\ref{fig-mxfp8-scale-granularity} contrasts the three granularities.

Smaller scale groups reduce the range that any one scale must cover, but they
also increase the number of scales that must be produced, moved, and arranged.
The 32-value group is therefore part of both the numerical recipe and the
systems interface. Shape divisibility and scale layout are not optional kernel
details.

\subsubsection{Native Blackwell MXFP8 GEMM}

Native block scaling applies the E8M0 scales inside the matrix operation
\citep{nvidiaCublasBlockScaling}, avoiding a separate materialization of
dequantized operands before GEMM. It accelerates the scaled matrix
multiplication itself; software still determines how often values are
converted into and out of MXFP8. Those conversions are the first challenge
examined in Section~\ref{sec-mxfp8-challenges}.

\subsection{Challenges in MXFP8 Training}
\label{sec-mxfp8-challenges}

Around the native block-scaled GEMM, the training system must produce
correctly oriented operands for forward,
dgrad, and Wgrad; pass qdata and scales through autograd; refresh derived
weight caches after each optimizer update; and avoid paying Q, DQ, or scale
swizzles more often than necessary. The following challenges determine
whether the hardware throughput advantage survives in end-to-end training.

\needspace{7\baselineskip}
\subsubsection{Quantization and Layout Overhead}

\begin{figure}[!htb]
  \centering
  \includegraphics[width=\linewidth]{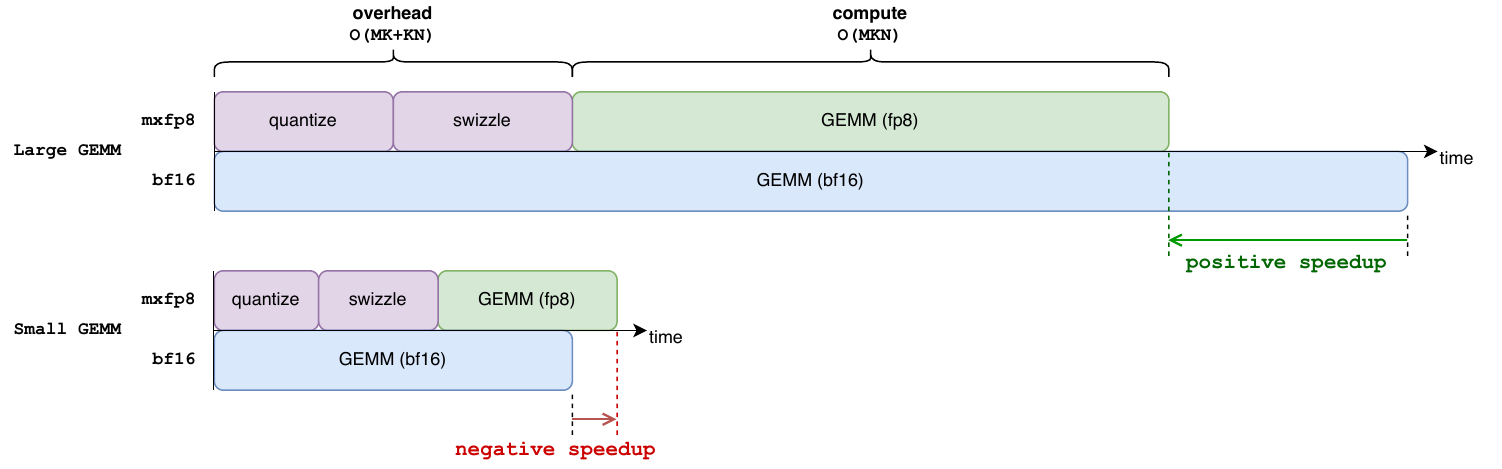}
  \caption{\textbf{Conversion and layout costs in MXFP8 execution.}
  MXFP8 is faster only when its GEMM savings exceed these costs.
  This is a schematic cost model, not measured timing. The
  displayed $O(MK+KN)$ conversion term assumes both operands are quantized on
  the critical path. In Olmo training, the $K\times N$ weight conversion is
  normally paid during cache refresh and reused across microbatches, leaving
  activation conversion as the principal per-invocation cost. For small
  GEMMs, that remaining overhead can still turn the expected speedup into a
  slowdown.}
  \label{fig-mxfp8-speedup-overhead}
\end{figure}

For one candidate GEMM (Figure~\ref{fig-mxfp8-speedup-overhead}), MXFP8 is
faster only when

\begin{equation}
T_{\mathrm{Q}} + T_{\mathrm{layout}} + T_{\mathrm{MXFP8\ GEMM}}
  + T_{\mathrm{DQ}}
< T_{\mathrm{BF16\ GEMM}}.
\label{eq-mxfp8-crossover}
\end{equation}

Not every term is present on every path. A cached weight avoids Q on each
microbatch, and a scaled GEMM that returns BF16 may not need a separate output
DQ kernel. A saved activation or communicated row requires Q when it is
produced and DQ if its consumer requires BF16. These costs are absent from
the Tensor Core peak ratio.

The crossover depends on the matrix dimensions and on expert load. Dense
linears usually produce a few large GEMMs; MoE grouped GEMMs divide tokens
across experts and can become too small or uneven to amortize conversion and
launch overhead. Our single-GPU microbenchmark sweeps these dimensions to
locate the crossover.

\conceptlead{Prequantized MXFP8 overtakes BF16 at smaller sizes than
conversion-inclusive execution.} Small square projections do not recover the conversion
cost within the single-B300 sweep. Table~\ref{tab-mxfp8-crossover-thresholds}
separates the measured crossover points for prequantized GEMM and cached-weight
execution with per-call activation quantization. The sweep also records Q,
scale swizzle, Q-plus-swizzle, DQ, and fully dynamic execution alongside
the BF16 and MXFP8 GEMMs.

Table~\ref{tab-mxfp8-crossover-sweep} answers a different question: it compares
operator boundaries at the larger assignment size $M=32{,}768$, rather than
locating crossovers. Charging LHS quantization at the routed boundary narrows
every advantage and leaves the $1024$ and $2048$ squares and the OLMoE-family down
projection slower than BF16.
\begin{table}[!htb]
\centering
\small
\renewcommand{\arraystretch}{1.12}
\begin{tabularx}{\linewidth}{@{}
  >{\raggedright\arraybackslash}X
  >{\centering\arraybackslash}p{0.27\linewidth}
  >{\centering\arraybackslash}p{0.29\linewidth}@{}}
\toprule
\textbf{Projection family} & \textbf{Prequantized GEMM faster from}
& \textbf{Cached RHS, LHS Q included} \\
\midrule
Rectangular projections & $M=2{,}048$ & $M=8{,}192$ \\
$4096\rightarrow4096$ & $M=4{,}096$ & $M=16{,}384$ \\
$1024\rightarrow1024$, $2048\rightarrow2048$ & --- & No crossover in sweep \\
\bottomrule
\end{tabularx}
\caption{\textbf{Conversion-inclusive execution crosses later than
prequantized GEMM.} Reported crossover points relative to BF16 from the
single-B300 $M=128$--$16{,}384$ sweep, with three warm-ups and five repeats.
The dash means no prequantized threshold is reported here; it is not a claim
that those raw GEMMs never cross. The rectangular row gives the size from which
every rectangular shape is faster; prequantized $16384\rightarrow2048$ already
reaches $1.08\times$ at $M=1{,}024$.}
\label{tab-mxfp8-crossover-thresholds}
\end{table}

Submission overhead also separates measured kernel work from eager wall
time. CUDA-Graph timing shows that Python submission gaps
can dominate short eager sequences; removing those gaps measures device-service
time, not training wall time. A reported speedup therefore depends on the timing boundary as well as on
which conversions it includes.

\begin{table}[!htb]
\centering
\small
\renewcommand{\arraystretch}{1.12}
\begin{tabularx}{\linewidth}{@{}
  >{\raggedright\arraybackslash}X
  *{4}{>{\centering\arraybackslash}p{0.135\linewidth}}@{}}
\toprule
\textbf{Projection} & \textbf{Raw MXFP8 GEMM} & \textbf{Shared, cached RHS}
& \textbf{Routed, prequantized LHS} & \textbf{Routed, LHS Q charged} \\
\midrule
$1024\rightarrow1024$ & 1.00$\times$ & 0.25$\times$ & 0.52$\times$ & 0.31$\times$ \\
$2048\rightarrow2048$ & 1.76$\times$ & 0.84$\times$ & 1.31$\times$ & 0.91$\times$ \\
$4096\rightarrow4096$ & 1.84$\times$ & 1.45$\times$ & 1.70$\times$ & 1.43$\times$ \\
$4096\rightarrow6144$ & 1.84$\times$ & 1.53$\times$ & 1.71$\times$ & 1.49$\times$ \\
$4096\rightarrow8192$ & 1.83$\times$ & 1.62$\times$ & 1.76$\times$ & 1.46$\times$ \\
$8192\rightarrow4096$ & 1.92$\times$ & 1.52$\times$ & 1.76$\times$ & 1.46$\times$ \\
$2048\rightarrow16384$ & 1.76$\times$ & 1.66$\times$ & 1.49$\times$ & 1.33$\times$ \\
$16384\rightarrow2048$ & 1.79$\times$ & 1.18$\times$ & 1.77$\times$ & 1.12$\times$ \\
OLMoE $1536\rightarrow3072$ & 1.72$\times$ & 0.99$\times$ & 1.34$\times$ & 1.02$\times$ \\
OLMoE $1536\rightarrow1536$ & 1.60$\times$ & 0.48$\times$ & 0.95$\times$ & 0.55$\times$ \\
\bottomrule
\end{tabularx}
\caption{\textbf{At $M=32{,}768$, conversion and layout work separate the raw
GEMM speedup from what an operator boundary delivers.}  Speedups are relative
to the matching BF16 primitive with three warm-ups and five repeats.  Shared
columns use dense matrix multiplication with the production two-copy scale
swizzle; routed columns use grouped GEMM over 16 balanced local experts, so
the two families are not directly comparable.  The last column additionally
charges left-hand-side quantization to the routed boundary.  The OLMoE rows
are OLMoE-family shapes of this single-B300 benchmark, not the published
OLMoE configuration.  Table~\ref{tab-sync-graph-seed} (Appendix~\ref{app-sync-verification})
times the same routed $1536\rightarrow3072$ configuration with longer eager
windows (25-ms warm-up, 100-ms timed windows) and measures $1.43\times$
prequantized and $1.21\times$ with Q charged; the ratios therefore depend on
the timing protocol.}
\label{tab-mxfp8-crossover-sweep}
\end{table}

\needspace{7\baselineskip}
\subsubsection{Forward, Dgrad, and Wgrad Need Different Operands}

For a linear layer,

\begin{equation}
Y=XW^\top,
\qquad
\nabla_X=\nabla_Y W,
\qquad
\nabla_W=\nabla_Y^\top X.
\label{eq-mxfp8-linear-three-gemms}
\end{equation}

Forward and dgrad use the same weight parameter but opposite matrix
orientations. Olmo consequently stores one prequantized RHS cache for
$W^\top$ and another for $W$. They are derived from one FP32 main weight and
must be refreshed together; they are not independent parameters.

The activation has a similar layout problem. In forward, $X$ is the left-hand
operand. In Wgrad, the saved $X$ is the right-hand operand of
$\nabla_Y^\top X$. The LHS and RHS scale layouts are not interchangeable, so
an activation quantized for forward cannot automatically be reused for Wgrad
\citep{nvidiaTransformerEngineMXFP8}.
When Olmo saves $X$ in MXFP8 for Wgrad, it produces the RHS-oriented
representation explicitly.

Dense \texttt{MXFP8Linear} uses scaled MXFP8 operations for forward, dgrad,
and Wgrad. The current routed grouped-Wgrad path is different: it dequantizes
the needed operands and uses BF16 grouped matrix multiplication. The tested
natural grouped-Wgrad contraction
and scale layouts were slower on the measured development shapes.

\begin{figure}[!htb]
\centering
\includegraphics[width=\linewidth]{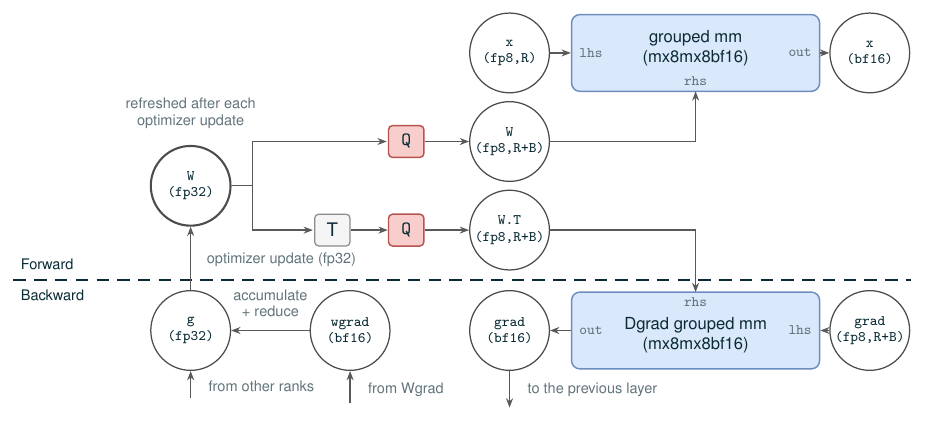}
\caption{\textbf{One FP32 main weight becomes two oriented MXFP8 caches.}
After every optimizer update, the FP32 main weight is quantized into the
forward cache and, through an explicit transpose, the dgrad cache, each
consumed as the RHS of its scaled grouped GEMM and reused across
microbatches until the next update.  The figure treats $W$ as the
right-hand operand ($xW$), so its $W$ and $W$.T labels are $W^\top$ and $W$ in
the notation of the text.  The gather that assembles the sharded
main weight stages through a transient BF16 copy freed immediately after
quantization; under the FP8-only expert-parameter mode evaluated here,
forward and backward hold no weight other than the two caches.  The loop closes through the weight-gradient path: the BF16 Wgrad
output accumulates into the FP32 gradient buffer and is reduced across ranks
before the optimizer update writes the FP32 main weight and triggers the next
refresh; the dgrad output flows to the previous layer.
Notation: Q quantizes to MXFP8 and T transposes; (fp8,\,R) is qdata with
rowwise scales; (fp8,\,R+B) keeps rowwise plus blocked (swizzled) scales;
mx8mx8bf16 names the two MXFP8 inputs and BF16 output of the grouped GEMM.
Conceptual illustration with no measured quantities.}
\label{fig-mxfp8-weight-cache-path}
\end{figure}

Figure~\ref{fig-mxfp8-weight-cache-path} draws the resulting weight path:
one FP32 main weight quantized after each update into the two oriented
caches that forward and dgrad consume.  It is a zoom of the full routed
pipeline in Figure~\ref{fig-mxfp8-routed-autograd} (Section~\ref{sec-mxfp8-recipe})
and uses that figure's
notation.

\subsubsection{Autograd Must Carry qdata and Scales as One Value}
PyTorch autograd normally treats one tensor as one differentiable value.
MXFP8 instead encodes one BF16 tensor as qdata and scales. A custom
autograd interface must carry or save the two components together, remember
their layout, and produce the gradient of the original BF16 tensor. It
must not expose either component alone as if it were the activation.

This interface also removes avoidable conversions. In the default fused routed
path, adjacent backward stages thread explicit qdata and scales inside one
custom-autograd region. An MXFP8 tensor wrapper can carry the same
qdata-and-scale pair
between split operators. When adjacent layouts agree, the consumer uses the
pair directly and avoids
MXFP8$\rightarrow$BF16 dequantization followed immediately by
BF16$\rightarrow$MXFP8 requantization.

\subsubsection{DDP Must Be MXFP8-Aware}

Olmo's DDP implementation follows the synchronization and precision policy
of each parameter group (Section~\ref{sec-ddp-vs-fsdp}). It must also
manage MXFP8 weight-cache refresh: qdata and scale caches are not ordinary
\texttt{nn.Parameter} objects, and they must never be updated directly.

Olmo represents each MXFP8 weight parameter with an \texttt{FP8WeightStore}.
The store records the parameter name and shape, the required cache orientations,
and a high-precision gradient sink. Under the precision policy of Section~\ref{sec-ddp-vs-fsdp},
BF16 Wgrad outputs accumulate into the FP32 gradient buffer and are reduced
over the parameter's replica group. The distributed optimizer then updates
the sharded FP32 main weight.

After initialization and every optimizer copy-back, Olmo reconstructs the
local weight values and rebuilds all required MXFP8 caches. Training-resume
and model-only training loads also perform this refresh before first use. The
BF16 initialization anchor may be released after the
FP32 main weight and caches exist. It is not a second source of truth. If the
main weight advances to step $t+1$ while a cache remains at step $t$, the next
forward pass uses the previous cache.

\begin{keyidea}[Derived Representations of the FP32 Main Weight]
The FP32 main weight is the source of truth. Forward caches, dgrad caches, BF16
views, and checkpoint views are derived representations and must be rebuilt
when the main weight changes.
\end{keyidea}

\subsubsection{Scale Selection Is a Training Recipe}
\label{sec-mxfp8-scale-rounding}

Cache refresh determines when weights are quantized; scale selection determines
the range and spacing of the values they can represent.
The OCP MX v1.0 specification %
\citep{ocpMXFP8spec} and
the NVIDIA training recipe %
\citep{mishra2025mxfp8recipes} choose the E8M0
scale differently. OCP's baseline conversion first
rounds the block maximum down to a power of two and divides by $2^8$, the
largest power of two representable by E4M3. The round-ceiling
(\texttt{rceil}) recipe instead divides by the actual E4M3 maximum, 448, and
rounds the resulting scale upward to a power of two.
Table~\ref{tab-mxfp8-floor-rceil-pseudocode} shows the
two conversions in the form implemented by Olmo
\citep{ocpMXFP8spec,mishra2025mxfp8recipes}.

\begin{table}[!htb]
\centering
\small
\setlength{\tabcolsep}{8pt}
\begin{tabularx}{\linewidth}{>{\raggedright\arraybackslash}X>{\raggedright\arraybackslash}X}
\toprule
\textbf{OCP v1.0: floor} & \textbf{NVIDIA: round-ceiling (\texttt{rceil})} \\
\midrule
\begin{minipage}[t]{\linewidth}
\ttfamily\small
a = max(abs(float(v[0:32])))\par
\textcolor{red}{e = floor(log2(a)) - 8}\par
e = clamp(e, -127, 127)\par
s = 2**e\par
for i in 0..31:\par
\hspace*{1em}q[i] = sat\_E4M3(\par
\hspace*{2em}RNE(float(v[i]) / s))\par
return q, E8M0(e + 127)
\end{minipage}
&
\begin{minipage}[t]{\linewidth}
\ttfamily\small
a = max(abs(float(v[0:32])))\par
\textcolor{red}{e = ceil(log2(a / 448))}\par
e = clamp(e, -127, 127)\par
s = 2**e\par
for i in 0..31:\par
\hspace*{1em}q[i] = sat\_E4M3(\par
\hspace*{2em}RNE(float(v[i]) / s))\par
return q, E8M0(e + 127)
\end{minipage}
\\
\bottomrule
\end{tabularx}
\caption{\textbf{Floor and round-ceiling differ only in scale selection.}
The pseudocode assumes a finite BF16 input block. Both paths store 32 E4M3
values and one E8M0 scale, clamp the unbiased scale exponent to E8M0's
representable range, and use saturating round-to-nearest-even (RNE) conversion
for qdata. Zero and non-finite special cases are omitted for clarity.}
\label{tab-mxfp8-floor-rceil-pseudocode}
\end{table}

The \texttt{rceil} rule chooses the smallest power-of-two scale that prevents
the block maximum from exceeding 448. The \texttt{floor} rule can choose a
scale that is twice as fine, but then the largest values saturate when the
normalized block maximum lies above 448. For $a=500$, for example,
\texttt{floor} selects $s=1$ and clips at 448, whereas \texttt{rceil} selects
$s=2$ and avoids clipping.

OCP v1.0 says that its floor-style mechanism should be supported, while also
allowing other conversion algorithms. NVIDIA's experiments identify scale
rounding as a training-recipe choice and report that rounding the ratio upward
avoids the degradation observed with the floor recipe
\citep{mishra2025mxfp8recipes}. Olmo implements both modes and uses
\texttt{rceil} by default, matching Transformer Engine's documented scaling rule
\citep{nvidiaTransformerEngineMXFP8};
\texttt{floor} remains available for controlled comparisons.

\begin{figure}[!htb]
\centering
\includegraphics[width=\linewidth]{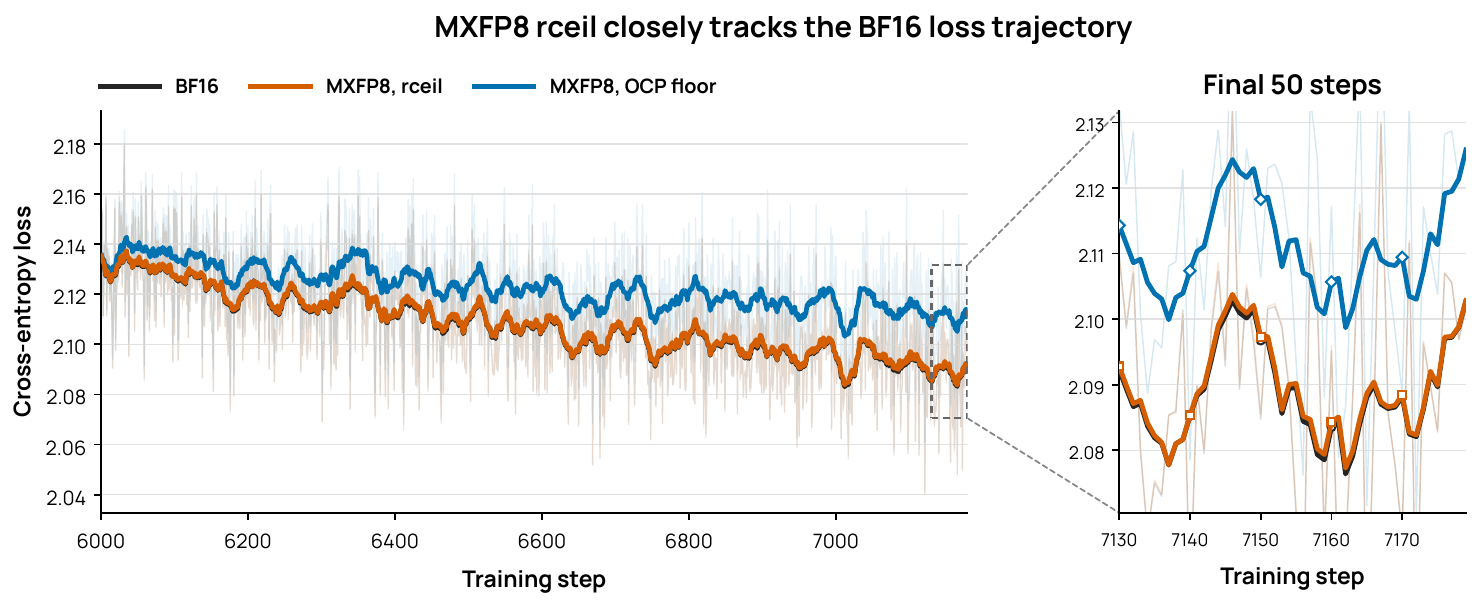}
\par\vspace{1.5em}
\includegraphics[width=\linewidth]{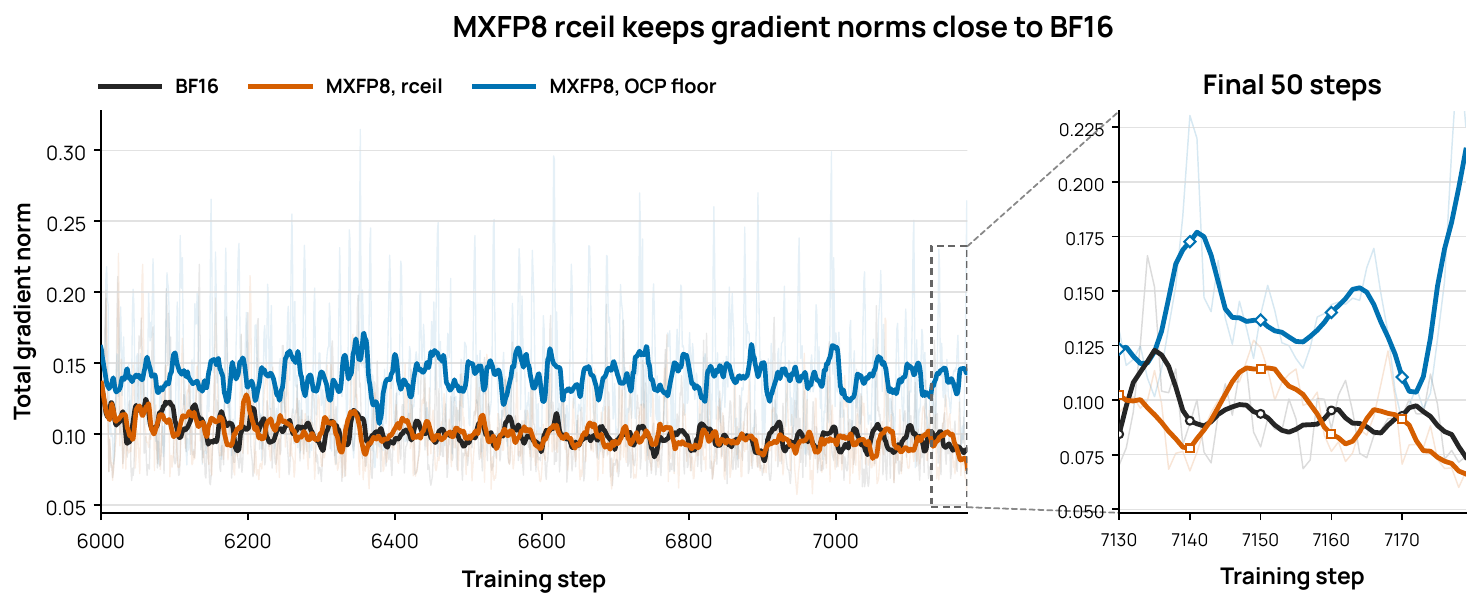}
\caption{\textbf{Round-ceiling MXFP8 follows the BF16 training trajectory.}
The top plot reports cross-entropy loss; the bottom reports total gradient
norm. The model has 8.36B active and 166.72B total parameters across 31 layers,
with top-4 routing over 128 routed experts and one shared expert. The common
step-6000 checkpoint was trained from scratch with MXFP8 OCP-floor scaling on
approximately 56B tokens. Step 7179 corresponds to approximately 67B tokens,
so the plotted continuation spans approximately 11B tokens. The main panels use a 21-step
centered moving average, while the final-50-step panels use a lighter 7-step
centered moving average; faint lines show the raw measurements. Over the final
50 steps, \texttt{rceil} differs from BF16 by +0.00048 mean loss and $-1.9\%$
mean gradient norm, whereas \texttt{floor} differs by +0.02200 mean loss and
+51.1\% mean gradient norm.}
\label{fig-mxfp8-rceil-training}
\end{figure}

To test whether this local conversion difference affects training, we fork a
31-layer, 8.36B-active/166.72B-total MoE with top-4 routing over 128 routed
experts and one shared expert. The common step-6000 checkpoint was trained from
scratch with MXFP8 using OCP floor scaling. From that checkpoint onward, the
three branches use BF16, MXFP8 \texttt{rceil}, and MXFP8 \texttt{floor},
respectively. Figure~\ref{fig-mxfp8-rceil-training} compares the resulting loss
and gradient-norm trajectories.

The direction of this result matches the NVIDIA recipe study: round-ceiling
scaling follows the higher-precision baseline, whereas floor scaling produces
a visible loss degradation %
\citep{mishra2025mxfp8recipes}.

\subsection{The Olmo MXFP8 Recipe for MoE}
\label{sec-mxfp8-recipe}

The optimizer keeps the main weights, the source of truth, in FP32.
Forward and dgrad use MXFP8 weight caches. Custom-autograd regions accept and
return BF16 at ordinary model boundaries, while selected internal activations
and backward gradients remain in MXFP8. Wgrad results accumulate into FP32
gradient buffers; gradient reduction, clipping, and the optimizer update are
also performed in high precision. The next forward pass uses caches rebuilt
from the updated FP32 main weights, the loop that
Figure~\ref{fig-mxfp8-weight-cache-path} isolated.

\subsubsection{Routed Experts}

The routed path combines MXFP8 compute with rowwise EP. In forward, the router
produces token--expert assignments. Dispatch quantizes each selected BF16
token row into qdata and scales and places those components in the destination
expert's buffers. The up/gate grouped GEMM consumes the dispatched activation
and the cached MXFP8 expert weights. A fused SwiGLU-plus-quantization step
produces the hidden representation needed by the down projection without
requiring a separate BF16-to-MXFP8 pass. The down grouped GEMM then produces a
BF16 expert output. That row is quantized for combine transport, and combine
reconstructs the route-weighted BF16 token result on the originating rank.

Backward follows the data dependencies in reverse. Combine backward applies
the route weights while quantizing the output gradient for transfer to the
ranks that hold the selected experts. Dgrad uses the orientation-specific
MXFP8 weight caches. Where two
adjacent backward stages inside the fused autograd region agree on layout,
their qdata and scales pass directly between those stages, avoiding a DQ--Q
cycle. Routed Wgrad currently dequantizes the saved operands and uses BF16
grouped matrix
multiplication before accumulating the resulting weight gradient into the
high-precision sink. Dispatch backward quantizes the input gradient to MXFP8
for transfer and returns it in BF16 to the originating token ranks.

The routed-expert recipe is therefore mixed precision:
MXFP8 where scaled GEMM or transport benefits, BF16 at ordinary model
interfaces and the current grouped-Wgrad contraction, and FP32 for accumulated
gradients, FP32 main weights, and optimizer states.

\begin{figure}[!htb]
\centering
\includegraphics[width=\linewidth]{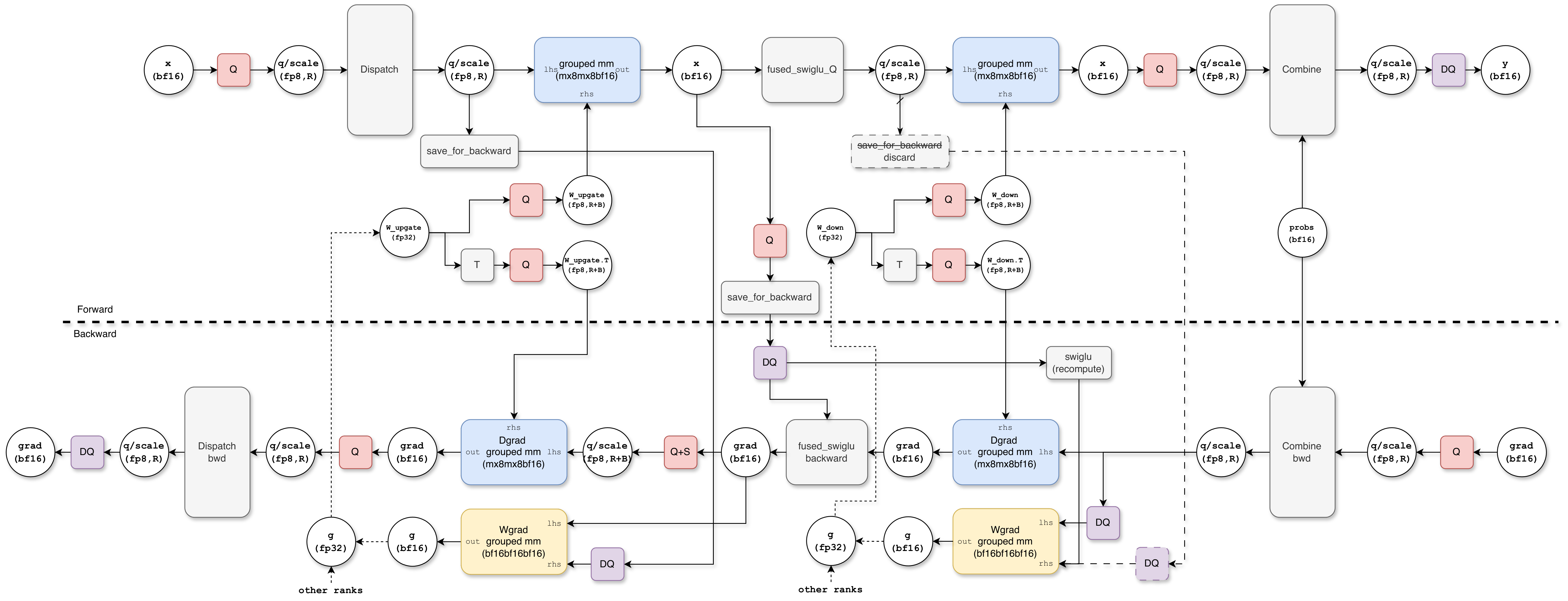}
\caption{\textbf{Routed-expert MXFP8 forward and backward.}
Forward quantizes routed rows for dispatch, applies MXFP8 grouped GEMMs for the
up/gate and down projections, and quantizes the combine payload. Backward
reverses the communication path, uses MXFP8 weight caches for dgrad, and uses
the current BF16 grouped-GEMM path for Wgrad before accumulation into FP32
gradient buffers that are then reduced with the other ranks' contributions
(dashed arrows). The drawing also identifies saved, discarded, recomputed,
quantized, and dequantized values across the custom-autograd boundary.
Notation: (fp8,\,R) is qdata with rowwise scales; (fp8,\,R+B) keeps rowwise
plus blocked (swizzled) scales; Q, DQ, and T are quantize, dequantize, and
transpose; Q+S is a fused quantize-plus-swizzle; mx8mx8bf16 and bf16bf16bf16
name each grouped GEMM's two input dtypes and output dtype. Weight labels
treat $W$ as the right-hand operand ($xW$), so $W$ and $W$.T correspond to
$W^\top$ and $W$ in the notation of the text.}
\label{fig-mxfp8-routed-autograd}
\end{figure}

Figure~\ref{fig-mxfp8-routed-autograd} puts these boundaries into one autograd
view. The upper path follows forward execution from dispatch through combine;
the lower path follows the corresponding dgrad, Wgrad, and communication work
in backward.

\subsubsection{Dense and Shared-Expert Linears}

The same cache-and-update scheme applies outside the routed experts. Olmo replaces
selected attention projections and shared-expert linears with
\texttt{MXFP8Linear}. Each weight has forward and dgrad caches derived
from its FP32 main weight, and Wgrad is routed to the same high-precision
accumulation path used by the DDP optimizer.

The saved-input policy is operator-specific. The packed QKV projection saves
its Wgrad input in the MXFP8 RHS layout. The attention output projection keeps
its Wgrad input in BF16 because the surrounding attention graph already keeps
the relevant BF16 activation alive; adding an MXFP8 copy would not provide the
same memory benefit. Attention projections thus use operator-specific
compute and saved-activation precision.

Shared experts use the same forward/dgrad cache principle, but their shapes
and replica groups follow the shared-expert parameter group rather than the
routed-expert group. The recipe evaluated here uses one shared expert with
MXFP8-only compute-weight storage, and both shared-expert linears save
their Wgrad inputs in MXFP8.

\subsubsection{MXFP8 Activation Saving Without MXFP8 Compute}

Storage precision and compute precision are independent choices. Olmo's
saved-tensor hooks can pack selected BF16 activations, including the query, key, and value tensors saved for attention backward, into MXFP8 and dequantize them only when
backward consumes them. The surrounding forward GEMM does not need to use
MXFP8 for this mechanism to reduce saved-activation memory.

This option is profitable only when the packed tensor replaces a BF16 tensor
that would otherwise remain live. It also trades Q during forward and DQ
during backward for lower peak memory. Storage-only MXFP8 is therefore a configuration choice
separate from MXFP8 compute.

\subsubsection{Kernel Problems and Fusions}

MXFP8 creates useful fusion opportunities because an intermediate often needs
to be transformed before its next consumer. The integrated routed path fuses
SwiGLU with row quantization, and combine backward fuses route weighting with
gradient quantization. Both avoid writing a large BF16 intermediate solely to
read it again for Q. Weight caches similarly move weight quantization from the
microbatch path to the post-update boundary.

The current recipe transfers qdata and scales separately. Paired or packed
qdata-plus-scale transport and direct weighted-quantize-plus-transfer kernels
are prototypes outside the evaluated recipe. For routed Wgrad, global DQ followed by BF16 grouped GEMM
was faster than the natural MXFP8 contraction on the measured development
shapes, whose reduction dimension and scale layout do not align well with the
native block-scaled operation.  Table~\ref{tab-mxfp8-kernel-status}
summarizes these choices.

\begin{table}[!htb]
\centering
\small
\begin{tabularx}{\linewidth}{lXX}
\toprule
\textbf{Operation} & \textbf{Current path} & \textbf{Reason or alternative} \\
\midrule
SwiGLU + activation Q & Integrated fusion & Avoid a separate BF16
intermediate and Q pass \\
Route weighting + gradient Q & Integrated fusion & Avoid a route-expanded
BF16 gradient before dispatch \\
Qdata + scale transport & Separate transfers & Paired and packed transport
alternatives are outside the evaluated recipe \\
Routed grouped Wgrad & DQ, then BF16 grouped GEMM & The natural MXFP8 path was
slower on the measured development shapes \\
\bottomrule
\end{tabularx}
\caption{\textbf{Current implementation choices and nearby alternatives.}
``Integrated'' identifies the active code path; it does not isolate that
fusion's performance contribution.}
\label{tab-mxfp8-kernel-status}
\end{table}

\subsection{Throughput Benchmark}
\label{sec-mxfp8-benchmark}

We compare four modes to separate the attention-projection contribution
from the combined MLP-and-EP contribution: a BF16 reference; MXFP8 for the
QKV and output projections in
attention; MXFP8 for the dense, shared, and routed MLPs together with MXFP8
rowwise EP transport; and the combined recipe.

\begin{figure}[!htb]
  \centering
  \includegraphics[width=0.98\linewidth]{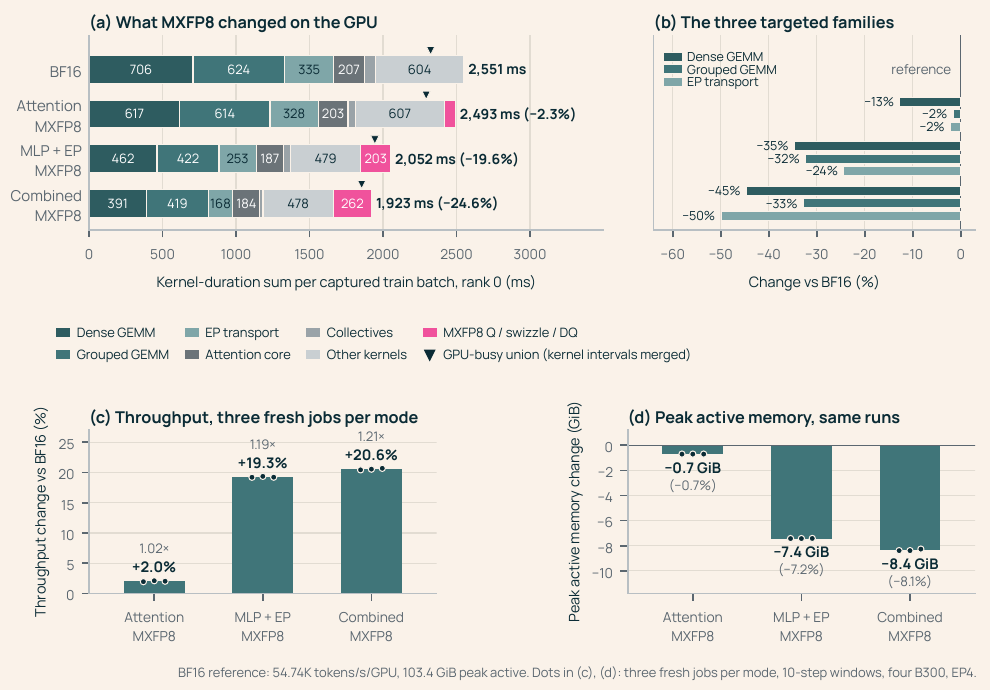}
  \caption{\textbf{Kernel-time breakdown, throughput, and memory across the
  four MXFP8 modes.} The combined recipe reduces the rank-0 kernel-duration
  sum by a quarter. Panels (a,b) use profiled kernel durations to identify
  changed work; panel (c) uses unprofiled runs for throughput because tracing
  perturbs the launch-heavy combined path.
  (a)~Kernel-duration sums per captured train batch on rank~0 by kernel
  family, means of three captured batches per mode.  Teal families are the
  ones MXFP8 targets, gray families are not targeted, and pink is the added
  quantization, scale-swizzle, and dequantization work (75.5, 203.1, and
  262.1~ms per batch; 3.0\%, 8.0\%, and 10.3\% of the BF16 sum).  Family
  sums overlap across streams and are not an additive wall-time model; the
  triangle marks the union of kernel-busy intervals, which shrinks by
  1.3\%, 16.3\%, and 20.2\%.  (b)~Change of the three targeted families
  relative to BF16; the bars in each row are, top to bottom, dense GEMM,
  grouped GEMM, and EP transport, in the colors of~(a).  The EP-transport family includes barrier kernels; the
  payload-moving kernels alone fall by 41\% in both the MLP+EP and combined
  modes, so part of the combined $-50\%$ is reduced waiting.
  (c)~Throughput change over three fresh distributed jobs per
  mode; bars are means of the three ten-step windows and dots are the
  repetitions.  The BF16 reference is 54.74K tokens/s/GPU and the largest
  sample standard deviation is 0.14K, smaller than the dot markers.
  (d)~Peak active memory change in the same runs from the 103.40~GiB BF16
  reference.  Four B300 GPUs, EP${}=4$, uniform routing, and identical
  initialized weights, inputs, and routing in every mode
  (Table~\ref{tab-mxfp8-benchmark-setup}); these are systems measurements at
  one operating point, not training-quality evidence
  (Section~\ref{sec-mxfp8-scale-rounding}).}
  \label{fig-mxfp8-recipe-throughput-memory}
\end{figure}

Figure~\ref{fig-mxfp8-recipe-throughput-memory} reports a controlled
four-B300 experiment whose configuration is summarized in
Table~\ref{tab-mxfp8-benchmark-setup}.  Throughput is total work divided by
wall time over the timed window, which avoids both first-use compilation and
an average of already-rounded per-step rates.

\begin{table}[!htb]
\centering
\footnotesize
\renewcommand{\arraystretch}{1.14}
\begin{tabularx}{\linewidth}{@{}
  >{\raggedright\arraybackslash}p{0.24\linewidth}
  >{\raggedright\arraybackslash}X@{}}
\toprule
\textbf{Setting} & \textbf{Value} \\
\midrule
Model & 8 layers; 3.172B active and 13.037B total parameters \\
\arrayrulecolor{black!22}\hline\arrayrulecolor{black}
Experts and routing & 32 routed experts, top-4, one shared expert \\
\arrayrulecolor{black!22}\hline\arrayrulecolor{black}
Hardware and parallelism & Four B300 GPUs, EP${}=4$ \\
\arrayrulecolor{black!22}\hline\arrayrulecolor{black}
Global step & 524,288 tokens in eight gradient-accumulation microbatches \\
\arrayrulecolor{black!22}\hline\arrayrulecolor{black}
Matched across modes & Fixed-seed initialization and initialized weights,
generated input batches, routing decisions, fused attention implementation,
optimizer precision, compilation, and recompute policy \\
\arrayrulecolor{black!22}\hline\arrayrulecolor{black}
Timing window & Steps 1--14 untimed; throughput from total work and wall
time over steps 15--24 \\
\arrayrulecolor{black!22}\hline\arrayrulecolor{black}
Repetitions & Three fresh distributed jobs per mode \\
\bottomrule
\end{tabularx}
\caption{\textbf{Configuration of the four-mode MXFP8 recipe benchmark.}
The four modes differ only in where MXFP8 is applied: nowhere (BF16
reference), attention projections, MLPs plus rowwise EP transport, or both.}
\label{tab-mxfp8-benchmark-setup}
\end{table}

Quantizing attention projections alone gives a modest benefit at this shape:
quantization and layout work offset part of the GEMM savings. The MLP and EP
path supplies nearly all
of the gain. It accelerates the large dense and expert GEMMs and cuts the fixed
rowwise EP buffer allocation from 5.00 GiB in BF16 to 2.836 GiB for MXFP8 qdata
and scales. Combining both paths gives the highest throughput, about 21\%
above the BF16 reference, and the lowest peak active memory.

The rank-0 profiles in Figure~\ref{fig-mxfp8-recipe-throughput-memory}(a,b)
support this attribution. The three families MXFP8 targets account for 65\%
of BF16 kernel time and fall by about a quarter to a half once the MLP path is
quantized, while the added conversion work costs a tenth of BF16 kernel time.
The untargeted families contribute a much smaller share of the reduction.

These short performance runs verify the systems comparison, not long-horizon
training quality. Their first logged loss is identical and their first gradient
norms are close under the matched initialization, while the longer continuation
experiment in Section~\ref{sec-mxfp8-scale-rounding} provides the
training-trajectory evidence for the selected scale recipe.

\begin{sectiontakeaway}
MXFP8's compute, activation-memory, and communication benefits depend on
the cost of Q/DQ, scale layout, cache refresh, and the autograd interface
that carries qdata and scales together. Olmo keeps one FP32 main weight
as the source of truth and treats every MXFP8 representation as a derived cache or
payload, so persistent training-state memory does not shrink. In the measured
four-mode ablation, the MLP and expert-transport path supplies nearly all of
the end-to-end throughput gain, and the combined recipe also lowers peak
active memory.
\end{sectiontakeaway}

\FloatBarrier

\section{Topology-Agnostic Checkpoint Saving and Loading}
\label{sec-checkpointing}

\conceptlead{Checkpoint tensors are independent of rank placement.} A checkpoint should identify the model and optimizer
tensors required to resume a particular training step, without requiring the
next run to use the same rank layout. Topology-independent checkpoint tensors and load-time resharding are
also explicit goals of Universal Checkpointing and PyTorch Distributed
Checkpoint %
\citep{lian2025universalcheckpointing,pytorchDistributedCheckpoint}.
Olmo saves FP32 main weights, FP32 optimizer states, and the small
amount of optimizer and trainer metadata needed to continue the run. In MoE,
dense tensors are sharded over DP, expert tensors over EP-MP and EP-DP, and
layers over PP. Changing those degrees changes which slice each rank holds,
not the underlying tensor values.

\begin{table}[!htb]
\centering
\small
\begingroup
\renewcommand{\arraystretch}{1.10}
\renewcommand{\tabularxcolumn}[1]{m{#1}}
\setlength{\tabcolsep}{4pt}
\begin{tabularx}{\linewidth}{
  >{\raggedright\arraybackslash}m{0.19\linewidth}
  >{\centering\arraybackslash}m{0.09\linewidth}
  >{\centering\arraybackslash}m{0.11\linewidth}
  >{\centering\arraybackslash}m{0.18\linewidth}
  >{\raggedright\arraybackslash}X}
\toprule
\textbf{Saved item} & \textbf{Format} & \textbf{Size} &
\makecell[c]{\textbf{166.7B example}\\\textbf{(GiB)}} & \textbf{What it restores} \\
\midrule
FP32 main weights & FP32 & $4P$ bytes & 621.10 & The source-of-truth parameter values updated by the optimizer. \\
\arrayrulecolor{black!22}\hline\arrayrulecolor{black}
FP32 first moment & FP32 & $4P$ bytes & 621.10 & Adam's exponential moving average of gradients. \\
\arrayrulecolor{black!22}\hline\arrayrulecolor{black}
FP32 second moment & FP32 & $4P$ bytes & 621.10 & Adam's exponential moving average of squared gradients. \\
\arrayrulecolor{black!22}\hline\arrayrulecolor{black}
Optimizer bookkeeping & FP32 scalars & Negligible & $<0.04$ & Per-parameter step counters and rolling windows of recent losses and gradient norms. \\
\arrayrulecolor{black!22}\hline\arrayrulecolor{black}
Trainer and run metadata & Mixed & Negligible & $<0.04$ & Step, token and epoch counters; data-loader position; random-number-generator state; callbacks; configuration; and data manifest. \\
\midrule
\textbf{Total} & & $\boldsymbol{\approx 12P}$ \textbf{bytes} &
\makecell[c]{\textbf{1,863.33}\\{\footnotesize predicted 1,863.29}} & BF16 compute weights and MXFP8 compute-weight caches are reconstructed after loading. \\
\bottomrule
\end{tabularx}
\endgroup
\caption{\textbf{Contents of a training checkpoint.} Here $P$ is the total
parameter count. The $12P$ term dominates because the checkpoint contains one
FP32 main-weight array and two FP32 Adam-moment arrays. The example column is
the step-10,000 checkpoint of a run with the 31-layer, 8.36B-active/166.72B-total MoE
configuration also used in Section~\ref{sec-mxfp8-scale-rounding}; the
predicted value is three times the unrounded FP32 array size (621.096~GiB), and the two negligible
categories together with distributed-checkpoint metadata fill the remainder.}
\label{tab-checkpoint-contents}
\end{table}

In Table~\ref{tab-checkpoint-contents}, the complete 166.7B-parameter checkpoint occupies
\textbf{1,863.33~GiB} by file size, within 0.04~GiB of three times its FP32
array size. All remaining checkpoint contents fit in that difference.
Topology changes alter the files and chunks written by each rank, but not the
dominant $12P$ payload.

\subsection{Checkpoint Tensors Are Topology-Independent}

Olmo identifies each global checkpoint tensor by a \textbf{stable key},
global shape, dtype, tensor role (for example, main weight or optimizer
moment), canonical element order, and offsets within the global tensor. The physical checkpoint files
may still contain chunks shaped by the save topology; topology independence
comes from describing where those chunks belong in the global tensor. For
example, changing from EP8 to EP4 changes the number of experts held by each
EP-MP rank, not the expert identities or their order. FP32 main weights and
optimizer states are persisted, while BF16 compute weights, MXFP8 compute-weight caches,
communication buffers, and process groups are derived for the active run.

\subsection{Low-Memory Saving and Load-Time Resharding}

During saving, dense tensors are already exposed in their checkpoint DTensor
layouts %
\citep{pytorch2026dtensor}. Expert tensors use a \textbf{checkpoint-only flat view} that orders the local
EP-MP expert shards and records EP-DP sharding or replication. In the normal
path this view aliases the live optimizer storage. The distributed checkpoint
writer stages each rank's local chunks on CPU and serializes them with their
global coordinates, writing each replicated chunk once from its lowest rank,
without first gathering the complete expert tensor on GPU. PP contributes globally
named layer keys from each stage rather than encoding PP rank IDs in those
keys. In one controlled MXFP8 development run, save-time peak
allocated and reserved memory remained at their pre-save values of 93.1 and
121.4 GiB; this is a save-side measurement only.

During loading, the destination run first constructs the shards required by
its requested topology %
\citep{pytorchDistributedCheckpoint}. The reader stages saved chunks on CPU, copies the
intersections needed by those targets, and restores the FP32 main weights
maintained by the distributed optimizer. A direct training load then reconstructs
the BF16 compute weights and MXFP8 compute-weight caches. Unlike saving, this
step all-gathers the DP and EP-DP shards of the main weights on GPU, because
each rank needs complete compute weights; this report does not measure
load-time memory.

\subsection{Topology Changes and Model Conversion}

Topology-agnostic loading changes placement, not the model schema. The source
and destination must agree on parameter names, shapes, tensor roles, and
canonical ordering. Changing DP, EP-MP, EP-DP, or PP is a \textbf{topology
change} that redistributes those same global tensors; changing the number of
experts, the model or expert width, or the layer structure is a \textbf{model conversion}.

As a result, one checkpoint can resume or resize an experiment under a
different DP, EP-MP, EP-DP, or PP layout, and it can be inspected or exported to
another format without a separate full-checkpoint GPU job. The evidence in this
section covers the mechanism and the save path; this report includes no
measured restore under a changed topology.

\FloatBarrier

\section{Activation Recompute}
\label{sec-recompute}

\begin{figure}[!htb]
  \centering
  \includegraphics[width=\linewidth]{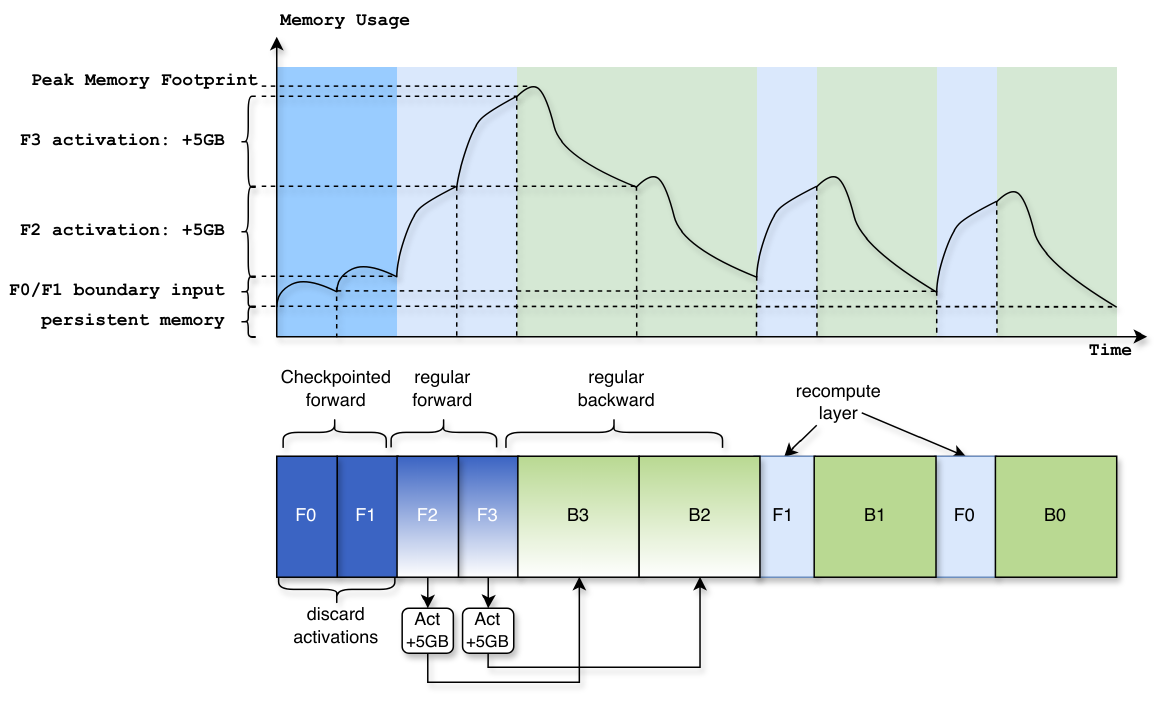}
  \caption{\textbf{Layer recompute trades additional forward work for lower
  activation memory.} In this illustrative schedule, F0 and F1 are
  checkpointed, whereas F2 and F3 retain their backward-needed activations.
  Checkpointing discards F0/F1's internal activations but retains their boundary
  inputs; the two forwards are therefore recomputed immediately before B1 and
  B0. The 5~GB increments and smooth memory curve illustrate the accounting and
  are not measurements from a training run.}
  \label{fig-layer-recompute}
\end{figure}

\conceptlead{Recompute converts retained-activation memory into repeated
forward computation.}  Persistent training state---parameters, gradient
buffers, and optimizer states---is fixed by the model and its sharding
(Sections~\ref{sec-parallelism-overview} and~\ref{sec-ddp-vs-fsdp}).  Saved
activations behave differently: they grow with the rank microbatch, the
sequence length, and the number of blocks whose backward has not yet run, and
they dominate the transient part of peak memory.  Activation checkpointing
reduces this component by retaining a checkpoint region's boundary input
instead of all of its internal forward activations %
\citep{chen2016sublinear}.
Before the region's backward computation, the forward operations are
recomputed to regenerate the values that backward needs.
Figure~\ref{fig-layer-recompute} shows this trade for a mixed four-layer
example.

\begin{figure}[!htb]
  \centering
  \includegraphics[width=0.82\linewidth]{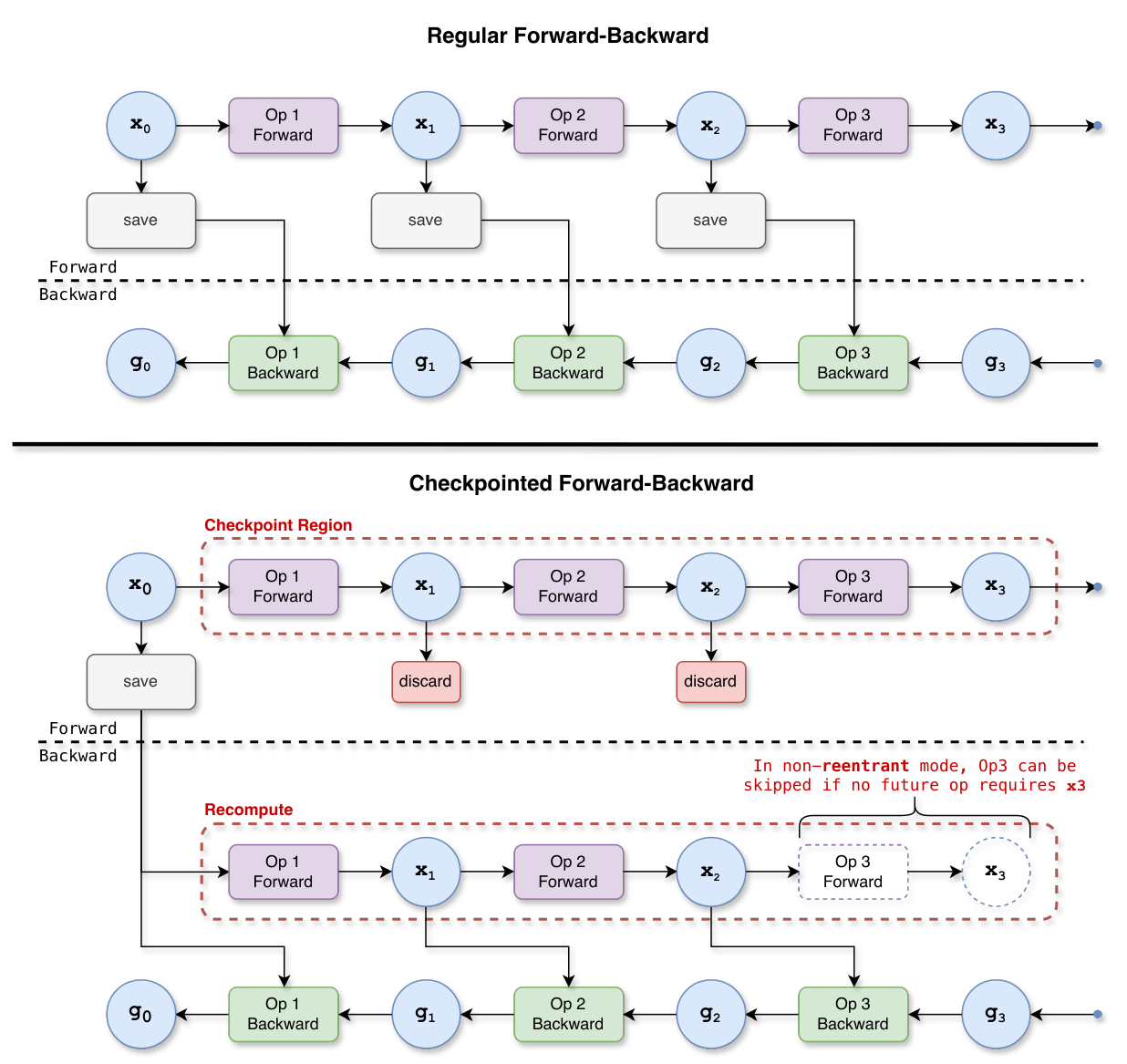}
  \caption{\textbf{Non-reentrant checkpointing can stop recomputation early.}
  A regular forward saves the intermediate values required by each backward
  operator. A checkpointed forward instead retains the region input $x_0$ and
  discards $x_1$ and $x_2$. During backward, the region is recomputed until those
  required values have been regenerated. In this simplified example, Op~3's
  forward need not be recomputed because Op~3 backward requires $x_2$, while the
  original checkpoint output $x_3$ remains available to downstream
  computation. The general stopping condition is determined by backward's
  tensor requirements, not merely by the position of an operator in the
  region.}
  \label{fig-selective-recompute}
\end{figure}

The amount of recomputation also depends on the checkpoint implementation.
PyTorch's non-reentrant variant stops recomputing once it has regenerated every tensor needed
by the pending backward computation, rather than necessarily executing the
entire checkpointed region again %
\citep{pytorchActivationCheckpoint}.
Figure~\ref{fig-selective-recompute} illustrates this early-stop behavior.

\subsection{Choosing Recompute Granularity}
\label{subsec-recompute-granularity}

The choice of checkpoint region determines which activations are retained and
which operations repeat %
\citep{korthikanti2023recompute,jain2020checkmate}.
The DDP training stack selects among block-level and sub-block regions
through its model configuration, summarized in
Table~\ref{tab-recompute-granularities}.  Per-block recompute, which the production recipes
call per-layer recompute, wraps every transformer block, so backward holds only one block's regenerated
activations at a time.  Chunk recompute instead wraps the entire block stack
in a single region: only the embedding output is retained, and the stack
recomputes from that boundary during backward.  Selected-block recompute wraps
only blocks named by key, and sub-block options wrap the attention
region, the routed-expert compute and unpermute region, or, on the block
all-to-all EP path, only the post-combine unpermute, inside an otherwise
unwrapped block.  The stack rejects chunk and per-block recompute together, and
TBO (Section~\ref{subsec-tbo-lesson}) rejects configurations that enable per-block recompute.

\begin{table}[!htb]
\centering
\small
\renewcommand{\arraystretch}{1.15}
\renewcommand{\tabularxcolumn}[1]{m{#1}}
\begin{tabularx}{\linewidth}{@{}
  >{\raggedright\arraybackslash}m{0.21\linewidth}
  >{\raggedright\arraybackslash}m{0.30\linewidth}
  >{\raggedright\arraybackslash}X@{}}
\toprule
\textbf{Policy} & \textbf{Checkpointed region} & \textbf{Retained at the
boundary} \\
\midrule
Per-block & Every transformer block & Each block's input; used by the
measured endpoint below and, per layer, by the Ultra production recipe of
Section~\ref{sec-production-throughput} \\
\arrayrulecolor{black!22}\hline\arrayrulecolor{black}
Chunk & The whole block stack as one region & Only the embedding output
between forward and backward; backward regenerates the whole stack's
activations at once \\
\arrayrulecolor{black!22}\hline\arrayrulecolor{black}
Selected blocks & Blocks named by key & The named blocks' inputs; other
blocks save normally \\
\arrayrulecolor{black!22}\hline\arrayrulecolor{black}
Attention region & The attention sub-region of a block & The sub-region
input; expert compute and EP transport save normally \\
\arrayrulecolor{black!22}\hline\arrayrulecolor{black}
MoE expert region & The routed-expert compute and the unpermute that follow
dispatch, on the block all-to-all EP path & The permuted expert input; dispatch
and combine transport save normally \\
\bottomrule
\end{tabularx}
\caption{\textbf{Block-level and sub-block recompute policies of the DDP
stack.}  All regions except a routed block on the DeepEP v2 path use
PyTorch's non-reentrant checkpoint, so the early-stop behavior of
Figure~\ref{fig-selective-recompute} applies inside them; that routed block
uses the reentrant variant and recomputes the whole block. The table omits the narrow
post-combine-unpermute option. Appendix~\ref{app-base-transformer-recompute} separately
describes the base transformer's alternatives outside the DDP stack.}
\label{tab-recompute-granularities}
\end{table}

Recomputation must be deterministic to be correct.  The DDP block rejects
dropout by construction, router jitter is disabled in every recipe in this
report, and the checkpoint wrappers restore the random-number-generator
state, so random expert assignment, used by the headline runs of
Section~\ref{sec-production-throughput}, replays the same routes.
Routed blocks on the DeepEP v2 path use the reentrant checkpoint variant,
which always recomputes the complete region; that path was validated under
compilation.

\subsection{Recomputing a Routed Block Repeats Its Communication}
\label{subsec-recompute-moe}

For a dense block, recomputation repeats local compute.  For a routed block, the
checkpointed region contains the router, dispatch, expert compute, and
combine, so recomputation can also \textbf{re-execute the expert-parallel
communication} during backward, as it does in the measured configuration below. The recomputed forward reproduces the
same scores and the same top-$K$ routes from the same boundary input, so the
re-executed dispatch moves the same rows to the same destination ranks.  The
memory--time trade for a routed block therefore counts repeated transport,
not only repeated FLOPs, and the measured cost below includes both.

Checkpointing also changes the rowwise path's buffer policy.  The lease
system of Section~\ref{subsec-rowwise-ep-buffer-reuse} exists to keep a
symmetric-memory payload alive from forward to its backward consumer.  Under
a checkpointed forward, the first pass discards its
intermediates, and the recompute pass regenerates payloads immediately
before they are consumed.  The rowwise path therefore detects that it is
running inside a checkpoint region and switches from leased symmetric
buffers to per-call scratch buffers for combine outputs, rather than holding
pool blocks whose saved contents no backward will read.

\textbf{Metrics are accumulated only once.} Router statistics, load-balancing
metrics, and auxiliary-loss accumulators are forward side effects, and a
recomputed forward would count every routed microbatch twice.  The
checkpoint context therefore distinguishes the original forward from
recomputation, and the router and rowwise paths accumulate metrics and
auxiliary-loss statistics \textbf{only in the original pass}.  This keeps the
balance-scope accounting of Section~\ref{sec-routing-load-balancing} intact
under every recompute policy in eager execution; under compilation, the
recompute detection falls back to a heuristic check.

One recompute mechanism in the router is independent of block policy.  The
router computes its logits from an FP32 cast of the block's BF16 activation.
Saving that cast would retain a four-byte copy of every routed token solely
for the router linear's backward, so the router materializes the FP32 input
through an \textbf{output-discard checkpoint}: the storage is released after the
logits are produced and regenerated immediately before the router linear's
backward consumes it.  The BF16 activation remains the only persistent copy.

\subsection{Composition with Pipeline Schedules}
\label{subsec-recompute-composition}

In the pipeline schedules of Section~\ref{sec-pipeline-parallelism}, a stage retains
activations for every in-flight microbatch, so its activation footprint is
\textbf{the product of the in-flight count and the per-microbatch
footprint}.
Recompute shrinks the second factor: a stage that checkpoints its blocks
holds boundary inputs for each in-flight microbatch instead of full
activation sets, at the cost of recomputing those blocks inside the backward
phase of the schedule.  This is the regime in which the production Ultra
operating points run: at 58B active parameters with PP8, the recipe of
Section~\ref{sec-production-throughput} combines MXFP8 with per-layer
recompute to stay within the memory budget. Recomputation can also repeat
communication: regenerating the needed activations may require rerunning
collectives, such as context-parallel attention or a routed block's dispatch
and combine. Sub-block policies can keep dispatch and combine outside the
recomputed region, but the attention region still contains any
context-parallel attention collectives.

\subsection{Measured Memory\textendash{}Throughput Trade-Off}
\label{subsec-recompute-measured}

We measured full per-block recompute in a matched four-GPU training run
whose configuration is summarized in
Table~\ref{tab-recompute-benchmark-setup}.  Throughput is total useful
tokens divided by total elapsed time over the timed window, repeated in three
fresh processes with the execution order balanced across variants.

\begin{table}[!htb]
\centering
\footnotesize
\renewcommand{\arraystretch}{1.14}
\begin{tabularx}{\linewidth}{@{}
  >{\raggedright\arraybackslash}p{0.24\linewidth}
  >{\raggedright\arraybackslash}X@{}}
\toprule
\textbf{Setting} & \textbf{Value} \\
\midrule
Model & One dense block followed by seven MoE blocks; $d_{\text{model}}=4096$;
expert intermediate width 4096 \\
\arrayrulecolor{black!22}\hline\arrayrulecolor{black}
Experts and routing & 32 routed experts, top-4, one shared expert; uniform
routes and zero token drops \\
\arrayrulecolor{black!22}\hline\arrayrulecolor{black}
Hardware and parallelism & Four B300 GPUs, EP${}=4$ \\
\arrayrulecolor{black!22}\hline\arrayrulecolor{black}
Batch & Two 8192-token sequences per rank microbatch; 524,288 tokens per
optimizer step \\
\arrayrulecolor{black!22}\hline\arrayrulecolor{black}
Matched across variants & BF16 execution, identical normally initialized
weights, identical input batches \\
\arrayrulecolor{black!22}\hline\arrayrulecolor{black}
Timing window & Steps 1--14 untimed (compilation and caches); throughput
over steps 15--24 \\
\arrayrulecolor{black!22}\hline\arrayrulecolor{black}
Repetitions & Three fresh processes per variant \\
\bottomrule
\end{tabularx}
\caption{\textbf{Configuration of the per-block recompute benchmark.}  The
two variants differ only in whether every transformer block is wrapped in a
non-reentrant checkpoint region.}
\label{tab-recompute-benchmark-setup}
\end{table}

\begin{figure}[!htb]
  \centering
  \includegraphics[width=0.9\linewidth]{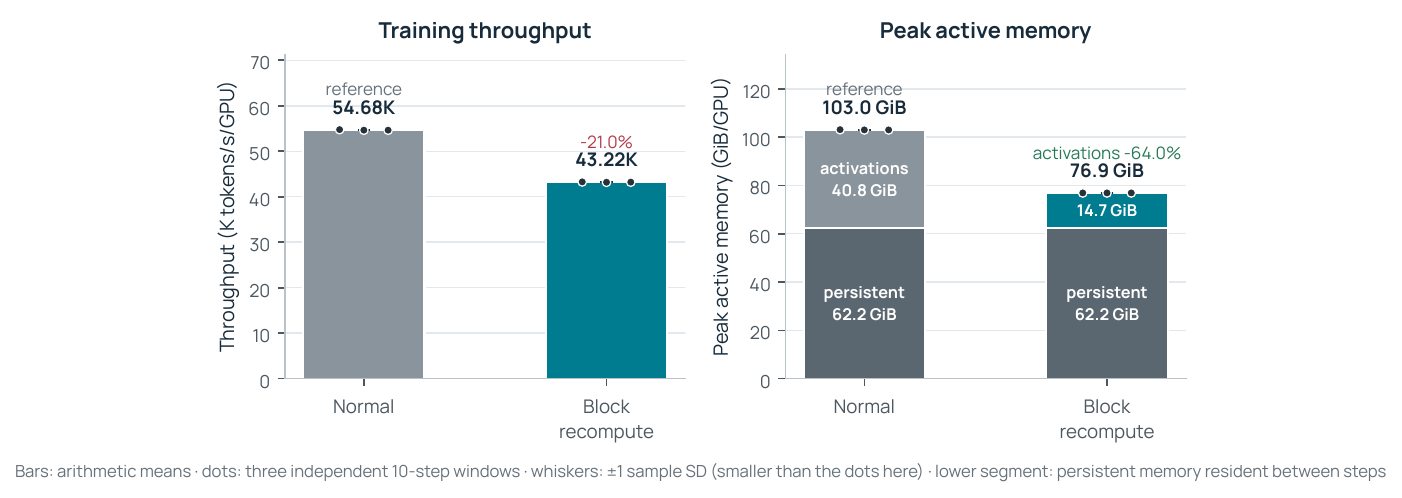}
  \caption{\textbf{Block recompute removes nearly two thirds of activation
  memory at a one-fifth throughput cost in this matched configuration.} Bars
  show the arithmetic mean of three fresh-process repetitions; dots show the
  individual ten-step windows, and the sample standard deviations are smaller
  than the dot markers.  Normal training in this campaign sustains 54.68K tokens/s/GPU at
  103.03~GiB peak active memory; recomputing every transformer block sustains
  43.22K tokens/s/GPU at 76.90~GiB, 20.95\% lower throughput for 26.13~GiB
  (25.36\%) less peak memory.  The lower segment of each memory bar, labeled
  ``persistent'', is the memory resident between steps (parameters, gradient
  buffers, optimizer states, and persistent buffers), 62.21~GiB in both modes; the
  reduction therefore falls entirely on the transient activation part, which
  shrinks from 40.82 to 14.69~GiB.  Measurements use four NVIDIA B300 GPUs and
  the fixed model, batch, routing, and timing protocol described in the
  text.}
  \label{fig-recompute-throughput-memory}
\end{figure}

The saving comes entirely out of activations.  Persistent state---the
parameters, gradient buffers, optimizer states, and persistent buffers
resident between steps---is identical in both variants, so the 26.13~GiB
reduction removes nearly two thirds of the transient part of the peak.

The profile shows which work repeats. A normal
\texttt{train\_batch} contains 64 projected block-forward ranges: eight blocks
across eight rank-local microbatches. Per-block recompute raises that count to
128 because each block forward is recomputed during backward, and for the seven
routed blocks each recomputation includes dispatch and combine. The summed GEMM
and attention kernel duration rises by 24.8\%, while the projected rank-0 GPU
span rises by 26.7\%, closely matching the independently measured 26.5\%
increase in time per unit of useful work. These duration sums are work
indicators rather than an additive wall-time decomposition because kernels on
different streams can overlap.

Full per-block recompute is useful when the 26.13~GiB reduction enables a larger microbatch, a deeper
pipeline stage, or a model that would otherwise not fit, and this is how the
Ultra production recipe uses it. When memory already fits, the measured
20.95\% throughput cost favors retaining activations or selecting a narrower
checkpoint region from Table~\ref{tab-recompute-granularities}. The report makes no
timing claim for the chunk, selected-block, or sub-block policies; their
checkpoint mechanism is the same, but their region boundaries and retained
inputs differ.

\subsection{Choosing a Recompute Policy}
\label{subsec-recompute-choosing}

Transient memory can also be reduced through microbatch size and activation
precision. A
smaller rank microbatch reduces every saved activation at once but pushes
grouped GEMMs toward the throughput crossover of
Section~\ref{sec-grouped-gemm}, since fewer rows reach each expert.  MXFP8
activation saving (Section~\ref{sec-mxfp8-training}) stores selected saved
activations as qdata and scales instead of BF16, shrinking them without
repeating any forward work.  Recompute removes saved activations entirely,
at the price of the recomputed region's forward time.  The Ultra recipe of
Section~\ref{sec-production-throughput} combines MXFP8 compute with per-layer
recompute
to keep the model within device memory.

Choosing a region starts with how much memory must be freed and which
operations would repeat. A useful starting point is the smallest region
that meets the memory budget, preferably excluding expert-parallel transport.
The sub-block regions of
Table~\ref{tab-recompute-granularities} repeat attention or
the routed-expert compute and unpermute, and neither repeats expert-parallel
transport.  Selected blocks cover a moderate shortfall at a cost proportional
to the share of blocks named. Per-block recompute can accommodate a deep
pipeline stage with many in-flight microbatches, at about a quarter more time per unit of useful work in
this shape.  Chunk
recompute, which retains only the embedding output between forward and
backward, helps a pipeline stage with very many in-flight microbatches; its
backward regenerates the whole stack's activations at once, so for a single
microbatch its peak approaches that of no recompute.

\begin{sectiontakeaway}
Checkpoint-region selection determines both memory savings and repeated work.
For a complete routed block, that work can include dispatch and combine;
metrics, however, are accumulated only on the original forward. At the
measured operating point, per-block recompute removes nearly two thirds of
activation memory, or one quarter of the peak, at a one-fifth throughput cost.
The Ultra recipes accept this cost because the model otherwise exceeds the
memory budget.
\end{sectiontakeaway}

\FloatBarrier

\section{Production Training Throughput}
\label{sec-production-throughput}

The preceding sections measured individual mechanisms and controlled
combinations. Here we report \textbf{complete training configurations}, from
Tiny to Ultra, with their model scale, batch, topology, precision, recompute
policy, and useful-model TFLOP/s/GPU. These runs establish feasible operating
points; their different configurations limit comparisons between them.

\subsection{Production Run Matrix}

\conceptlead{The headline runs use random routing.} Random routing removes the changing expert-load distribution of a
learned router and gives the training system a reference that stabilizes
quickly once a run is long enough.
It does not measure model quality or the throughput ultimately reached by a
learned router, so every row in Table~\ref{tab-production-throughput-headline}
is labeled as a systems reference rather than a time-to-quality result. All
runs use sequence length 8192, top-4 routing, one shared expert, NVIDIA B300
GPUs, and eight GPUs per node. Tiny uses 16 GPUs, Small and Medium use 128,
and Large and Ultra use 512.

Each run contributes a representative rate read from its throughput curve
(for the nine-step Ultra-256E run, its best step), and we select the \textbf{highest observed rate} across each configuration's
runs. Some configurations were launched more than once;
environmental problems were suspected in the lower-rate repeats, but their
causes were not established. These readings are neither repetition averages
nor estimates of the hardware ceiling, and they do not come from one
synchronized timing window.

We report \textbf{BF16-reference MFU} (Equation~\ref{eq-intro-mfu}), using
the B300's dense BF16 peak of 2,250 TFLOP/s/GPU
\citep{nvidiaHGXB200spec}. These percentages are derived from the throughput
readings, not separately measured. We keep the same denominator for MXFP8,
whose execution mixes precisions; its percentage is therefore a BF16 reference,
not utilization of the FP8 peak. The table also gives the parameter activation
ratio $\alpha_P = P_{\mathrm{active}}/P_{\mathrm{total}}$ defined in
Equation~\ref{eq-intro-activation-ratios}.

\conceptlead{These measurements use an eight-GPU NVLink domain.} Each node provides one
eight-GPU NVLink domain; communication beyond those eight GPUs crosses the
inter-node fabric (Section~\ref{subsec-tested-nvl8}). The Megatron-Core MoE
report evaluates GB200 on NVL72, where a much larger EP group can remain
inside one NVLink domain. It also reports GB300 NVL results, but does not
specify a 72-GPU domain for those measurements
\citep{nvidia2026megatroncoremoe}. Some of
its reported per-GPU throughput values are higher than ours, but the different
GPU platform, NVLink domain, model, batch, topology, and software recipe prevent an
apples-to-apples framework comparison.

\begin{table}[!htb]
\centering
\small
\begingroup
\setlength{\tabcolsep}{3.5pt}
\renewcommand{\arraystretch}{1.18}
\renewcommand{\tabularxcolumn}[1]{m{#1}}
\begin{tabularx}{\linewidth}{
  @{}>{\RaggedRight\arraybackslash\bfseries}m{0.142\linewidth}
  >{\centering\arraybackslash}m{0.185\linewidth}
  >{\centering\arraybackslash}m{0.075\linewidth}
  >{\centering\arraybackslash}m{0.085\linewidth}
  >{\centering\arraybackslash}m{0.105\linewidth}
  >{\centering\arraybackslash}m{0.108\linewidth}
  >{\centering\arraybackslash}X@{}}
\toprule
\rowcolor{black!5}
\textbf{Operating point} &
\makecell[c]{\textbf{Active @ total}\\\footnotesize Parameter activation} & \textbf{GBS} &
\textbf{EP / PP} & \textbf{Precision} & \textbf{Recompute} &
\makecell[c]{\textbf{TFLOP/s/GPU}\\\footnotesize MFU (BF16 ref.)} \\
\midrule
Tiny-64E       & \makecell[c]{1.59B @ 12.91B\\\footnotesize $\alpha_P=12.3\%$}  & 8 Mi  & 1 / 1  & BF16   & No  & \makecell[c]{\textbf{903}\\\footnotesize 40.1\%} \\
Small-64E      & \makecell[c]{4.29B @ 40.86B\\\footnotesize $\alpha_P=10.5\%$}  & 24 Mi & 8 / 1  & BF16   & No  & \makecell[c]{\textbf{841}\\\footnotesize 37.4\%} \\
Medium-64E     & \makecell[c]{7.37B @ 70.22B\\\footnotesize $\alpha_P=10.5\%$}  & 16 Mi & 8 / 2  & BF16   & No  & \makecell[c]{\textbf{853}\\\footnotesize 37.9\%} \\
Medium-96E     & \makecell[c]{7.38B @ 103.75B\\\footnotesize $\alpha_P=7.11\%$} & 24 Mi & 8 / 2  & BF16   & No  & \makecell[c]{\textbf{855}\\\footnotesize 38.0\%} \\
Medium-128E    & \makecell[c]{7.38B @ 137.27B\\\footnotesize $\alpha_P=5.38\%$} & 24 Mi & 8 / 4  & BF16   & No  & \makecell[c]{\textbf{843}\\\footnotesize 37.5\%} \\
Large-128E     & \makecell[c]{15.14B @ 295.99B\\\footnotesize $\alpha_P=5.12\%$} & 32 Mi & 8 / 4 & BF16   & No  & \makecell[c]{\textbf{768}\\\footnotesize 34.1\%} \\
Ultra-128E     & \makecell[c]{58.36B @ 1.200T\\\footnotesize $\alpha_P=4.86\%$} & 64 Mi & 8 / 8  & MXFP8 & Yes & \makecell[c]{\textbf{858}\\\footnotesize 38.1\%} \\
Ultra-256E     & \makecell[c]{58.41B @ 2.380T\\\footnotesize $\alpha_P=2.45\%$} & 64 Mi & 32 / 8 & MXFP8 & Yes & \makecell[c]{\textbf{689}\textsuperscript{$\ddagger$}\\\footnotesize 30.6\%} \\
\bottomrule
\end{tabularx}
\endgroup
\caption{\textbf{Headline random-routing throughput.} GBS is the global
batch size in binary millions of tokens ($1\ \mathrm{Mi}=2^{20}$). EP and PP
are expert- and pipeline-parallel degrees. Each parameter cell gives active
and total counts, then their ratio $\alpha_P$; each throughput cell gives useful
active-model TFLOP/s/GPU, then MFU normalized to the dense BF16 peak of
2,250 TFLOP/s/GPU. This common reference also applies to MXFP8 and does not
measure FP8-peak utilization. These unmatched operating points
differ in model shape, batch, topology, and precision and therefore do not
form a controlled scaling curve. \textsuperscript{$\ddagger$}The 2.380T
Ultra-256E run used EP32 with the DeepEP v2 path; its value is the best
observed in a nine-step run whose curve had not stabilized. Full configurations
and run provenance appear in
Appendix~\ref{app-production-throughput-runs}.}
\label{tab-production-throughput-headline}
\end{table}

\subsection{Throughput Across Model Sizes}

\conceptlead{Stored expert parameter size grows faster than active compute.} The
three Medium configurations hold active parameters nearly constant while
increasing total parameter size from 70.22B to 137.27B. Their selected observed
throughputs remain in a narrow 843--855 TFLOP/s/GPU band, but their global
batches and PP degrees are not all matched. The result shows feasible
operating points at each total parameter size; it does not isolate the cost of adding
experts.

\conceptlead{Learned routing runs below the random reference.} The
learned-routing companions in Appendix~\ref{app-production-throughput-runs}
show a gap of about one percent at Tiny without
EP, and about nine percent at Small with EP8, where a learned router's
changing expert loads can reduce throughput. These companions show how optimistic
the random-routing reference can be; Section~\ref{sec-routing-load-balancing}
examines the associated load imbalance.

\conceptlead{Memory capacity limits the largest operating points.} Large-128E
runs at the smallest rank microbatch of any BF16 operating point in the
matrix, a choice made under memory pressure. At Ultra scale, MXFP8 also saves activation and
EP-buffer memory, and the recipe still
requires per-layer recomputation (Section~\ref{sec-recompute}). Recomputing
those activations adds forward work, so despite MXFP8, Ultra-128E's 858 TFLOP/s/GPU stays close to the smaller
BF16 points. Its 64-Mi-token
global batch and PP8 topology also differ from Large-128E's 32-Mi-token batch
and PP4 topology, preventing a precision-only comparison. Ultra-256E further
changes EP from 8 to 32 and uses the DeepEP v2 path; its nine measured steps
establish only that the configuration ran at the reported rate during a short,
still-changing window.

Across these configurations, the stack supports 12.9 billion to
1.2 trillion total parameters, with the 2.38-trillion-parameter extension
demonstrated in the short run described above.

\section{Optimization Attempts and Lessons}
\label{sec-optimization-attempts}

We also tested approaches that we did not adopt in the training recipes.
They targeted saved activations, exposed communication, conversion kernels,
host launch overhead, and the number of separate kernels in the MoE path.
In several cases, reducing one cost increased another: offload exceeded the
host-link bandwidth budget, overlap slowed concurrent kernels, and preserving
the natural MXFP8 layout made Wgrad slower. Other attempts required more
integration work than their prototypes justified. CUDA Graph replay needed
pipeline-aware activation storage, while the megakernel needed a complete
distributed training runtime.

In its supported BF16 rowwise configuration, two-batch overlap gave a small
gain that did not justify the added coordination.
Table~\ref{tab-optimization-attempts-summary} summarizes the decisions.

\begin{table}[!htb]
\centering
\small
\renewcommand{\arraystretch}{1.18}
\setlength{\tabcolsep}{4pt}
\begin{tabularx}{\linewidth}{@{}>{\bfseries\raggedright\arraybackslash}p{0.18\linewidth}
                              >{\raggedright\arraybackslash}p{0.21\linewidth}
                              >{\raggedright\arraybackslash}p{0.31\linewidth}
                              X@{}}
\toprule
\textbf{Direction} & \textbf{Intended benefit} & \textbf{Why it was not adopted} & \textbf{Limiting requirement} \\
\midrule
CPU activation offload & Lower saved-activation memory & The GPU produces
activation bytes about six times faster than PCIe can move them to and from
host memory;
pinned host memory and prefetch scheduling add substantial machinery. & Host
bandwidth or compression sufficient to match activation production. \\
\arrayrulecolor{black!22}\hline\arrayrulecolor{black}
Wave overlap & Hide dispatch and combine under expert compute & Waves reduce
expert-GEMM size, while concurrent GPU communication stretches both the
communication and GEMM regions on many tested shapes. & Enough hidden
communication to offset smaller GEMMs and concurrent-kernel slowdown. \\
\arrayrulecolor{black!22}\hline\arrayrulecolor{black}
Two-batch overlap & Pipeline two half-batches through adjacent MoE blocks &
The rowwise NVSHMEM kernels and the other lane's compute compete for the same
GPU resources; the extra lane also increases host scheduling. & Lower
communication interference or less host work through replay. \\
\arrayrulecolor{black!22}\hline\arrayrulecolor{black}
MegaMoE megakernel & Coordinate communication and expert compute inside
persistent CUDA kernels & The kernel grew into a distributed training-runtime project
before it had a production backward and inter-node path. & Production-shaped
forward scheduling, full backward, and inter-node transport. \\
\arrayrulecolor{black!22}\hline\arrayrulecolor{black}
Natural-layout MXFP8 Wgrad & Eliminate global dequantization before Wgrad &
The naturally produced scales do not match Wgrad's reduction layout; the
prototype lost to global dequantization plus tuned BF16 grouped GEMM. &
Wgrad-oriented requantization plus native block-scaled GEMM faster than the
converted path. \\
\arrayrulecolor{black!22}\hline\arrayrulecolor{black}
Static CUDA Graphs & Remove repeated host launch work & Stable graph addresses
conflict with several live pipeline microbatches; safe staging adds copies and
memory. & Schedule-indexed saved-context storage with replay savings that
exceed staging and retention costs. \\
\bottomrule
\end{tabularx}
\caption{\textbf{Optimization attempts and reasons for not adopting them.}
The last column identifies the bandwidth, execution, or storage requirement
that limited each tested approach.}
\label{tab-optimization-attempts-summary}
\end{table}

\subsection{CPU Activation Offload Was Limited by PCIe}
\label{subsec-cpu-activation-offload}

CPU activation offload exchanges GPU memory for host-link traffic. After a
forward layer produces tensors that autograd will need later, the runtime copies
selected tensors from device to host (D2H), releases their GPU storage, and
copies them back from host to device (H2D) before the corresponding backward
operation. Every offloaded byte therefore crosses the host link twice over one
training step. Overlap can hide these copies only if the link drains tensors at
least as quickly as the forward pass produces them and returns them within the
backward prefetch window. Activation offload follows a line of work exemplified
by vDNN %
\citep{rhu2016vdnn}; the decision here concerns its service-rate
balance on the measured Olmo workload.

Let $A$ be the selected activation bytes produced during one layer interval,
$t_f$ the corresponding forward production interval, and $t_b$ the usable
backward prefetch interval. If $B_{\mathrm{D2H}}$ and $B_{\mathrm{H2D}}$ are
the sustained pinned-memory bandwidths, necessary service-rate conditions are

\begin{equation}
  B_{\mathrm{D2H}} \geq \frac{A}{t_f},
  \qquad
  B_{\mathrm{H2D}} \geq \frac{A}{t_b}.
  \label{eq-cpu-offload-bandwidth}
\end{equation}

When $A/t_f$ exceeds $B_{\mathrm{D2H}}$, each layer produces a larger copy
backlog than PCIe can retire. Moving the copy earlier cannot remove that
backlog, even on a separate CUDA stream; it only changes where the eventual
wait appears. We write the required-to-available D2H bandwidth ratio as
$\rho=(A/t_f)/B_{\mathrm{D2H}}$.

Table~\ref{tab-cpu-offload-budget} works the inequality for the routed
block of the four-B300 benchmark of Section~\ref{sec-recompute}, using that
profile's GPU-projected layer timings and an activation inventory implied by
the block's autograd graph at its shape ($T=16{,}384$ tokens per rank
microbatch, $d=h=4096$, top-4 routing, one shared expert, BF16).  The
inventory counts valid routed rows only. As an aggregate consistency check,
the inventory summed over the seven routed blocks, the dense block, and the
output logits is within a few percent of the measured 40.8~GiB transient
peak. The logits are not offload candidates, so the table's
whole-microbatch traffic excludes them.

\begin{table}[!htb]
\centering
\footnotesize
\begingroup
\definecolor{offband}{HTML}{EDE6DA}
\definecolor{offrequired}{HTML}{B8236B}
\definecolor{offavailable}{HTML}{2E6B70}
\definecolor{offverdict}{HTML}{DCE9EA}
\renewcommand{\arraystretch}{1.22}
\setlength{\tabcolsep}{5pt}
\begin{tabularx}{\linewidth}{
  >{\raggedright\arraybackslash}X
  >{\raggedleft\arraybackslash}p{0.12\linewidth}
  >{\raggedleft\arraybackslash}p{0.13\linewidth}}
\toprule
\textbf{One routed block, one rank microbatch} & \textbf{Size} & \textbf{Value} \\
\midrule
\rowcolor{offband}
\multicolumn{3}{l}{\textbf{Saved activations} \quad multiples of $T\!\times\!d$ BF16 elements} \\
Attention: block input, normed input, grouped-query attention $q$/$k$/$v$, attention output & $4.5\,Td$ & 0.60 GB \\
MoE norm inputs: residual and normed & $2\,Td$ & 0.27 GB \\
Routed experts: received rows, up/gate output, SwiGLU output, expert outputs ($K=4$) & $20\,Td$ & 2.68 GB \\
Shared expert: up/gate output, SwiGLU output & $3\,Td$ & 0.40 GB \\
\arrayrulecolor{black!30}\hline\arrayrulecolor{black}
\textbf{Candidate bytes $A$} & $\mathbf{29.5\,Td}$ & \textbf{3.96 GB} \\
\rowcolor{offband}
\multicolumn{3}{l}{\textbf{Timing} \quad GPU-projected ranges of the Section~\ref{sec-recompute} profile} \\
Forward of one routed block, $t_f$ & & 10.3 ms \\
Backward of one routed block, $t_b$, its share of the 198~ms microbatch backward & & 22 ms \\
One microbatch's share of the profiled step (forward, backward, gradient reduction) & & 295 ms \\
\rowcolor{offband}
\multicolumn{3}{l}{\textbf{Required versus available} \quad GB/s per direction} \\
\textcolor{offrequired}{Required:} D2H during forward, $A/t_f$ & & \textcolor{offrequired}{\textbf{384}} \\
\textcolor{offrequired}{Required:} H2D before backward, $A/t_b$ & & \textcolor{offrequired}{\textbf{180}} \\
\textcolor{offrequired}{Required:} averaged over the whole microbatch, 30.6 GB each way & & \textcolor{offrequired}{\textbf{104}} \\
\textcolor{offavailable}{Available:} PCIe Gen5 x16, specification & & \textcolor{offavailable}{\textbf{63}} \\
\textcolor{offavailable}{Available:} pinned copy, one GPU alone, measured on a Gen5 host & & \textcolor{offavailable}{\textbf{53}} \\
\midrule
\rowcolor{offverdict}
\textbf{Required over available, $\rho=(A/t_f)/B_{\mathrm{D2H}}$, specification link} & & \textbf{6.1} \\
\rowcolor{offverdict}
\textbf{Selective offload bound, $f\leq1/\rho$} & & $\boldsymbol{\approx 1/6}$ \\
\bottomrule
\end{tabularx}
\endgroup
\caption{\textbf{The estimated activation-offload rate exceeds PCIe
bandwidth by six to seven times for one GPU.} GB is $10^9$ bytes. The
inventory is analytic (saved tensors implied by the block's autograd graph,
uniform routing, valid rows only); the timings are the Section~\ref{sec-recompute}
profile on B300 GPUs; the link rates are the PCIe Gen5 x16 specification and
pinned-memory copies measured on an H100 host with the same link generation.  Against
the measured 53~GB/s copy rate rather than the specification, $\rho$ rises
to 7.2.  The B300 host link itself was not measured.}
\label{tab-cpu-offload-budget}
\end{table}

A routed block produces its 3.96~GB of
candidate activations in 10.3~ms, a 384~GB/s stream, against a 63~GB/s
link, so $\rho\approx6$ with the link entirely to itself and above~7 at the
copy rate actually measured on a host of the same link generation. Even
spread over the entire 295~ms microbatch, the traffic is
still 104~GB/s in each direction, and the measured copies lose a further
15--25\% of their rate when both directions run at once.  A node's
eight GPUs offloading together would push about 3~TB/s into host memory,
more than a two-socket DDR5 host can absorb regardless of link generation;
a PCIe Gen6 host link would halve $\rho$ but leave both the per-GPU
forward-rate mismatch and the host-memory limit.

Faster compute further narrows the offload window unless host bandwidth
increases with it. The activation-production rate in
Table~\ref{tab-cpu-offload-budget} is Blackwell's, since
the block timings come from B300 GPUs, while the link is the same PCIe Gen5
x16 that served Hopper hosts.  At Hopper's roughly $2.3\times$ lower BF16
peak (Table~\ref{tab-mxfp8-hardware}), scaling the block time inversely with
peak arithmetic throughput would put $\rho$ near~3, still above the available
bandwidth. This is an estimate, not a Hopper measurement.

A selective policy can offload at most approximately a fraction
$f\leq 1/\rho$ of the candidate bytes without building a sustained D2H
backlog.  At $\rho\approx6$ that is about one sixth, roughly 0.65~GB per
block or about 5~GB over a microbatch: less than a fifth of what per-block
recompute removes at the same operating point
(Figure~\ref{fig-recompute-throughput-memory}), for an additional subsystem with
its own failure modes.  That reduction is too small to justify the
subsystem in the current recipe.

\subsection{Wave Overlap Was Not a Reliable Win}
\label{subsec-wave-overlap-lesson}

Wave scheduling divides one MoE invocation into
several route groups. While experts compute one wave, another CUDA stream can
dispatch a later wave or combine an earlier one. This creates an opportunity
to hide communication, but it also replaces one larger grouped GEMM with
several smaller grouped GEMMs and runs communication beside compute on the
same GPU. Fine-grained overlap systems such as COMET and Lancet show why this
direction is attractive %
\citep{zhang2025comet,jiang2024lancet}. Its benefit
depends on how much both communication and compute slow down when run together.

The controlled measurements in Section~\ref{subsec-ep-overlap} show why stream
concurrency alone did not guarantee a win. In an isolated two-node EP-MP${}=16$ experiment,
rowwise NVSHMEM dispatch or combine slowed the concurrent expert GEMM by about
23--27\% across the tested communication-block sweep. The communication region
also stretched. A wide 128-block launch still shortened the pair's makespan by
22--25\% relative to sequential execution, but even a one-block communication kernel did not protect the
GEMM; it mainly made communication take longer. A DeepEP v2 sweep found the
wave path slower than the no-wave path at the tested communication settings of
4, 6, 16, 32, and 64 SMs on its own two-node EP-MP${}=16$ shape; without a
wave-sequential control, that loss cannot be split between smaller GEMMs and
concurrent-kernel interference.

The measurements do not identify the interference mechanism. The rowwise path executes
GPU warps that issue one-sided transfers and stream barriers, so candidate
causes include SM issue pressure, NVSHMEM progress work, L2 or memory-fabric
contention, and barrier overhead. Wave partitioning independently reduces GEMM
efficiency. Other measured shapes include both wins and losses, so the result
does not show that communication--compute overlap is never useful. Concurrent
kernels can still take longer to finish %
\citep{lee2025characterizing}. For the tested setup,
phase-sequential execution is the appropriate reference; the current evidence
does not establish that overlap improves the complete forward and backward
path.

\subsection{Two-Batch Overlap}
\label{subsec-tbo-lesson}

\begin{figure}[!htb]
  \centering
  \includegraphics[width=\linewidth]{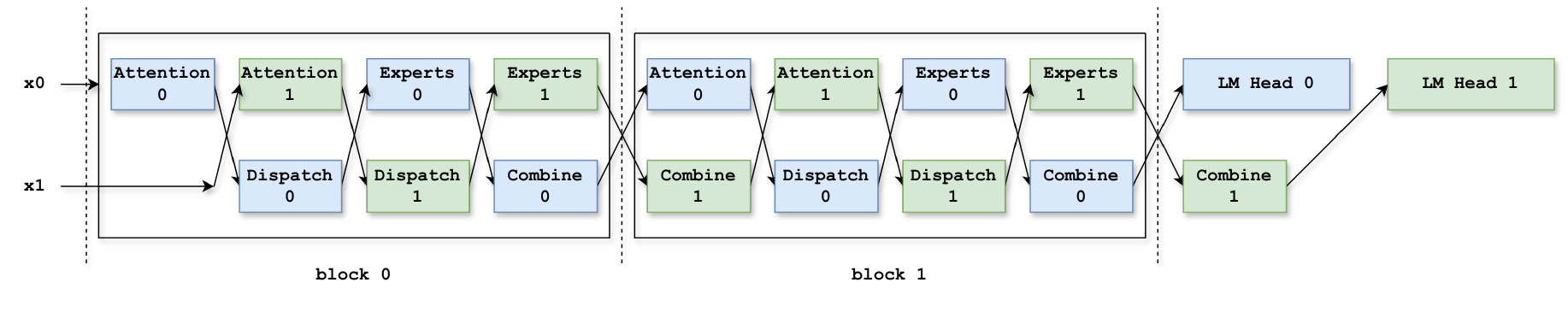}
  \caption{\textbf{Two-batch overlap interleaves two half-microbatch lanes
  across adjacent MoE blocks.} Blue and green denote lanes 0 and 1. The
  schedule interleaves attention, expert computation, and expert-parallel
  dispatch and combine so that communication from one lane can run alongside
  computation from the other. The arrows show data dependencies; the visual
  overlap alone does not imply a wall-time speedup.}
\label{fig-two-batch-overlap}
\end{figure}

\textbf{Two-batch overlap (TBO)} applies the overlap idea at a larger
granularity. It
splits the local microbatch into two half-batch lanes and interleaves their work
across adjacent MoE blocks (Figure~\ref{fig-two-batch-overlap}). While one
lane runs attention or expert GEMMs, rowwise
dispatch or combine for the other lane can remain pending on the EP
communication stream. We implemented this BF16 rowwise schedule and tried it
during development, but decided not to use it in our current training recipes.
The implementation requires BF16 rowwise NVSHMEM EP and rejects rowwise MXFP8,
per-block recompute, context parallelism, and the DeepEP v2 path, so the
MXFP8, per-layer-recompute Ultra recipes of
Section~\ref{sec-production-throughput} could not use it; the adoption
decision concerns the BF16 rowwise recipes.
DeepEP documents a contrasting legacy TBO path designed around
low-interference communication
\citep{deepseekDeepEPLegacy}.

\begin{figure}[!htb]
  \centering
  \includegraphics[width=0.72\linewidth]{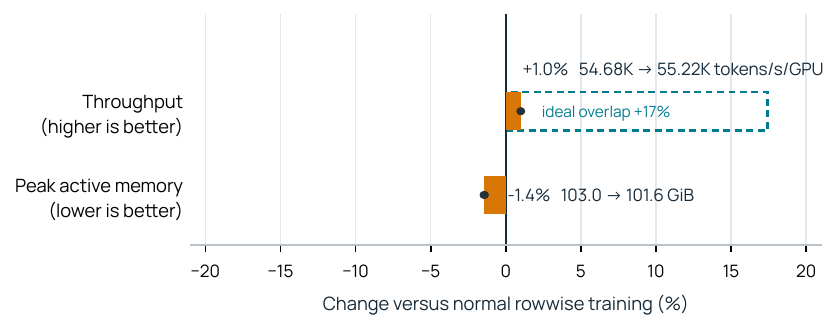}
  \caption{\textbf{Two-batch overlap improves throughput by 1\%, compared
  with a 17\% ideal-overlap estimate.} Filled bars show TBO's change of
  the three-repetition mean relative to normal rowwise training, with the
  individual ten-step windows as dots.  The dashed outline is an ideal-overlap
  upper bound: EP transport occupies
  14.8\% of the normal run's GPU span on rank~0, so hiding all of it with no
  slowdown of the concurrent compute would raise throughput by at most 17\%;
  about 40~ms of that transport already overlaps compute in normal training,
  so a tighter bound is about 15\%.  Normal training sustains 54.68K tokens/s/GPU at 103.03~GiB peak
  active memory; TBO sustains 55.22K at 101.57~GiB.  Same four-GPU B300
  configuration, matched model, batch, routing, and timing protocol as
  Figure~\ref{fig-recompute-throughput-memory}.}
  \label{fig-tbo-throughput-memory}
\end{figure}

Figure~\ref{fig-tbo-throughput-memory} reports a matched end-to-end
comparison using the four-GPU configuration and timing protocol of
Section~\ref{sec-recompute}: TBO improves throughput by one percent and
trims peak active memory by about 1.5~GiB.  The figure also draws what
ideal overlap would gain.  EP transport occupies a seventh of the normal
run's GPU span, so even a scheme that hid every millisecond of it without
slowing the compute beside it would gain at most seventeen percent, or about
fifteen percent after excluding the transport that already overlaps compute
in normal training; TBO achieves a one-percent gain. This is a small positive
result for one operating point, not evidence that TBO is generally faster.

\conceptlead{Concurrent kernels take longer under TBO.} Olmo's rowwise NVSHMEM communication is not an SM-free direct memory
access (DMA) transfer. Its route-level PUT and GET kernels use SM scheduling
and memory-system resources also needed by expert GEMMs. In the profile, TBO
overlaps 503.6~ms of communication and compute, equal to 72.2\% of TBO's own EP
transport interval. At the same time, the summed EP-transport duration grows
by 98.9\%, the GEMM duration by 20.4\%, and attention duration by 16.5\%.
These sums combine contention with the lower efficiency of two smaller lane
kernels and can overlap with one another; they are not an additive wall-time
decomposition. Taking the union of the intervals, the projected GPU span
falls by 0.9\%, consistent with the independent 1.00\% throughput gain.
Thus contention is real, but it does not quite erase the overlap benefit in
this configuration. NCCL All-to-All can have a different contention profile,
but it also retains the packing and unpacking permutations that rowwise EP
removes.

\conceptlead{Splitting the microbatch roughly doubles the host scheduling
work.} Each half-batch lane has its own attention, dispatch, expert, combine,
and synchronization stages. The profile records 89.8\% more kernel launches
and 126.4\% more CUDA API calls than normal execution. Kernel-idle gaps of at
least 10~$\mu$s also become 2.09$\times$ as frequent, although total idle time
within the projected range rises by only 9.7~ms. The CPU remains far enough
ahead at this rank-local microbatch size, so launch work does not erase the
small win. At smaller microbatches or on a faster kernel path, the same extra
submission work moves the system closer to a launch-limited regime. CUDA Graph
replay could lower this cost, but the current stack has no stable graph path
that captures this schedule end to end.

\conceptlead{TBO adds coordination to the pipeline schedule.} The implementation carries
two live lanes, separate communication buffers, and stream-ordering dependencies
across neighboring blocks. Pipeline parallelism (PP) already imposes its own
microbatch order and point-to-point dependencies, so combining the two
schedules requires additional coordination rather than a local MoE-layer
change. The code accepts TBO with PP, but we measured only a no-PP
configuration, and its one-percent gain did not justify adopting the
additional schedule-dependent coordination.
A different microbatch size, transport, interconnect, or pipeline schedule
can change the balance between hidden communication and the costs of
contention, smaller kernels, and extra launches.

\subsection{The MegaMoE Kernel Became a Runtime Project}
\label{subsec-megamoe-lesson}
To control this interference directly, a persistent MoE kernel can assign
explicit roles to communication,
scheduling, and tensor-core work. \textbf{MegaMoE}, our BF16 prototype of such a
persistent MoE megakernel in the style of DeepGEMM's Mega MoE
(Section~\ref{subsec-ep-overlap}), explored this direction by bringing route
dispatch, persistent expert scheduling, the two expert linear layers, the
activation, and weighted combine under one Olmo-controlled CUDA
implementation.

\conceptlead{The prototype needed a complete distributed execution path.}
A usable training backend still needed a production tensor-core and
data-movement scheduler, cross-rank signaling, reusable peer workspaces,
returned expert outputs, backward for activations and both expert weights, and
inter-node transport.
These pieces are coupled: a fast local forward body cannot replace the rowwise
path if backward or remote movement returns to a separate, incompatible
schedule. The prototype grew across Python, C++, CUDA, and distributed-runtime
code before it had a production training backward.

We removed that path and kept rowwise EP as the end-to-end baseline. The
prototype's tensor-core body ran as a serialized debug version, so its timing
does not estimate the performance of the intended persistent implementation.

\subsection{Wgrad Needed a Different Scale Layout}
\label{subsec-natural-mxfp8-wgrad}

The rowwise MXFP8 forward path naturally represents activations and
output gradients with scales
$s_A[m,k_b]$ and $s_{dY}[m,n_b]$, where $m$ is a routed-token row and $k_b$ and
$n_b$ are feature blocks. Expert weight-gradient computation is

\begin{equation}
  dW_e = A_e^{\mathsf T} dY_e,
\end{equation}

so its reduction dimension is the token index $m$. The natural scale factors
therefore change along the Wgrad reduction. Blackwell's native block-scaled
matrix instructions expect scale factors arranged according to their supported
GEMM reduction layout; the already-available rowwise representation is not
that layout.

The exact-natural prototype handled the mismatch by dequantizing tiles into
BF16 shared memory and using BF16 tensor-core arithmetic. Table~\ref{tab-natural-mxfp8-wgrad}
shows that it was substantially slower than two global dequantization kernels
followed by the tuned BF16 grouped Wgrad. Despite doing a separate global
conversion, that path benefits from the faster GEMM implementation.

\begin{table}[!htb]
\centering
\small
\renewcommand{\arraystretch}{1.14}
\setlength{\tabcolsep}{8pt}
\begin{tabular}{@{}rcc@{}}
\toprule
$K=N$ & \textbf{Exact-natural prototype} & \textbf{Global DQ + BF16 grouped Wgrad} \\
\midrule
2,048 & 0.380 ms & 0.118 ms \\
4,096 & 1.449 ms & 0.334 ms \\
8,192 & 5.522 ms & 1.232 ms \\
\bottomrule
\end{tabular}
\caption{\textbf{Preserving the natural MXFP8 scales was slower than converting
both operands globally.} The B200 prototype used four experts,
256 rows per expert, 1,024 total rows, and BF16 output. This report makes no claim about
current-stack Wgrad performance.}
\label{tab-natural-mxfp8-wgrad}
\end{table}

The current path therefore dequantizes the saved MXFP8 activation and output
gradient, then calls BF16 grouped GEMM for Wgrad. This does not imply that
native MXFP8 Wgrad is impossible. Such a path would deliberately requantize or
repack the operands into a Wgrad-oriented scale layout before calling a native
block-scaled grouped GEMM. Its cost includes that conversion as part of the
complete backward region. The prototype measurements favor the converted
layout and tuned BF16 GEMM over directly consuming the forward layout.

\subsection{Static CUDA Graphs Need Pipeline-Aware Storage}
\label{subsec-static-cudagraph-lesson}

CUDA Graph replay targets a different cost: repeated host submission. As
Section~\ref{sec-sync-free-training} explains, replay can be attractive when
the GPU executes small regions faster than Python and the CUDA runtime can
submit them. The static PyTorch capture/replay path considered here records
a fixed launch sequence with stable tensor addresses, then replays that
sequence with much less host work. CUDA also provides graph-update APIs that
can change node parameters, including memory addresses; those mechanisms are
outside this static prototype
\citep{nvidiaCUDAGraphs,pytorchCUDASemantics}.

Host submission can dominate a short operation. In a standalone single-GPU
grouped-linear microbenchmark, the captured device sequence for a short
MXFP8 expert operation runs more than twice as fast as its BF16 baseline
while under ordinary eager submission the same operation is slower than
BF16
(Table~\ref{tab-sync-graph-seed} in Appendix~\ref{app-sync-verification}).

\begin{figure}[!htb]
\centering
\includegraphics[width=\linewidth]{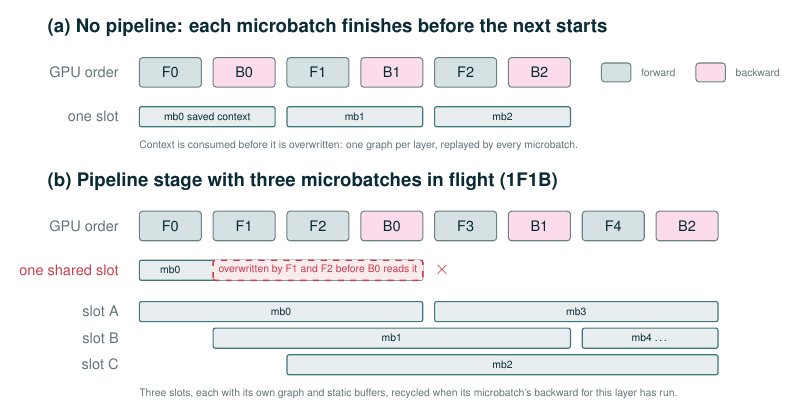}
\caption{\textbf{Under a pipeline schedule, static per-layer replay needs one
saved-context slot per in-flight microbatch.}  (a)~Without a pipeline the
backward of each microbatch consumes the layer's saved context before the
next forward overwrites it, so one graph per layer serves every microbatch.
(b)~With three microbatches in flight, a shared slot is overwritten by F1 and
F2 before B0 reads it; three slots, each with its own graph and static
buffers, are recycled as each microbatch's backward for the layer completes.
Conceptual illustration after the design described for \mbox{Megatron-Core}
\citep{nvidia2026megatroncoremoe}; no measured quantities.}
\label{fig-cudagraph-pp-slots}
\end{figure}

\conceptlead{Stable graph addresses conflict with live pipeline
activations.} A pipeline stage can execute several microbatch forward passes
before their corresponding backward passes. If every replay of one layer uses
the same activation slot, a later forward overwrites data that an earlier
backward will read. A pipeline-compatible design can assign a graph runner
and static activation slot to every simultaneously live
$(\text{layer},\text{microbatch})$ interval, or capture a larger fixed schedule
containing the forward--backward reuse order. Separate slots increase graph
pools and retained activation memory. Copying live tensors into graph-owned
staging storage stabilizes their addresses but adds device-to-device copies.
Staging, graph-pool memory, activation retention, and capture cost all count
against replay savings. Figure~\ref{fig-cudagraph-pp-slots} illustrates the
separate-slot design with three in-flight microbatches.

The prototype could capture a narrow rowwise forward core in non-pipeline and
smoke-test configurations, but it did not establish a beneficial complete PP
training path. Marking more tensors as static did not solve the scheduling
problem, because the number and reuse order of safe slots depends on the
pipeline schedule. We therefore retired the graph-specific backend rather than
maintain a second rowwise implementation with an incomplete memory trade-off.

Megatron-Core implements the separate-slot design: one
graph per layer and microbatch under pipeline parallelism, graphs captured in
execution order so that they share one memory pool, and static buffers
reused by the pipeline order, for which it reports about a ten percent
end-to-end speedup at roughly 7~GB of extra memory on a DeepSeek-V3
configuration %
\citep{nvidia2026megatroncoremoe}. This result includes the
schedule-dependent storage that our isolated forward capture lacked.

\subsection{Local Savings Must Survive the Training Schedule}
\label{subsec-negative-result-record}

These decisions favor the complete training path over savings in an isolated
region. The offload budget exceeds host-link bandwidth; wave overlap must pay
for smaller GEMMs and interference; and TBO's small gain did not justify its
additional schedule coordination. For Wgrad, global conversion followed by a
tuned GEMM beat direct consumption of the forward layout. The graph and
megakernel prototypes lacked the storage or distributed execution needed to
carry their local operations through training.

\FloatBarrier

\clearpage
\part{Findings and Outlook}
\label{part-outlook}
The remaining sections examine training-recipe choices and a separate
benchmarking lesson. We begin with learning-rate scaling and upcycling,
turn to operand-value effects in GEMM measurements, then return to
shared-expert structure and activation merging. We close with post-training
and a comparison with related systems.

\section{Learning Rate Scaling with Global Batch Size}
\label{sec-lr-batch-scaling}

Batch-size warmup increases the amount of data processed by each optimizer
step as training progresses.  It can reduce the number of synchronization and
optimizer steps required to process a fixed token budget, but the learning
rate must change with the batch. In our batch-size-warmup run, square-root
scaling left visible changes in the loss curve at later batch increases.

A common recipe uses the square-root
rule %
\citep{hoffer2017train}, which %
\citet{malladi2022sdes} derive for Adam:
if the batch grows from $B$ to $NB$, then

\begin{equation}
  \eta(NB)=\eta(B)N^{\alpha}, \qquad \alpha=\tfrac{1}{2}.
  \label{eq-lr-batch-scaling}
\end{equation}

\noindent This is the rule used by the batch-size-warmup experiments of
\citet{merrill2025criticalbatch}, where the batch grows only after the measured
critical batch size has increased. The derivation of %
\citet{malladi2022sdes} rests on a
stochastic-differential-equation analysis whose same-trajectory condition
also scales Adam's moment coefficients and $\epsilon$ together with the
learning rate. Olmo changes only
the learning rate. These results motivate square-root scaling, but do not
imply that a single exponent preserves the training trajectory at every batch
transition.

\subsection{Critical Batch Size Defines the Useful Batch Regime}

Let $g_i$ be the gradient from one training example, $G=\mathbb{E}[g_i]$ the
full gradient, and $\Sigma=\operatorname{Cov}(g_i)$ the per-example gradient
covariance.  For a batch of $B$ independently sampled examples,

\begin{equation}
  \widehat G_B=\frac{1}{B}\sum_{i=1}^{B}g_i,
  \qquad
  \mathbb{E}[\widehat G_B]=G,
  \qquad
  \operatorname{Cov}(\widehat G_B)=\frac{\Sigma}{B}.
  \label{eq-lr-batch-gradient-noise}
\end{equation}

Larger batches therefore give less noisy gradient estimates.  Under a local
quadratic approximation with Hessian $H$, the expected loss after one
stochastic-gradient update is

\begin{equation}
  \mathbb{E}\!\left[L(\theta-\eta\widehat G_B)\right]
  \approx L(\theta)-\eta\lVert G\rVert^2
  +\frac{\eta^2}{2}
   \left(G^\top H G+\frac{\operatorname{tr}(H\Sigma)}{B}\right).
  \label{eq-lr-batch-quadratic-loss}
\end{equation}

Minimizing Equation~\ref{eq-lr-batch-quadratic-loss} gives

\begin{align}
  \eta_{\mathrm{opt}}(B)
    &=\frac{\eta_{\max}}{1+\mathcal{B}_{\mathrm{noise}}/B},
  &
  \eta_{\max}
    &=\frac{\lVert G\rVert^2}{G^\top H G},
  \nonumber\\
  \mathcal{B}_{\mathrm{noise}}
    &=\frac{\operatorname{tr}(H\Sigma)}{G^\top H G}.
  \label{eq-lr-batch-noise-optimum}
\end{align}

This model predicts two regimes %
\citep{mccandlish2018largebatch}.  When
$B\ll\mathcal{B}_{\mathrm{noise}}$, the optimal stochastic gradient descent
(SGD) step grows approximately in proportion to $B$.  When
$B\gg\mathcal{B}_{\mathrm{noise}}$, it approaches $\eta_{\max}$, so enlarging
the batch yields diminishing optimization benefit.  The often-used
\textbf{simplified noise scale},

\begin{equation}
  \mathcal{B}_{\mathrm{simple}}
    =\frac{\operatorname{tr}(\Sigma)}{\lVert G\rVert^2},
  \label{eq-lr-batch-simple-noise}
\end{equation}

\noindent follows only after removing the Hessian weighting.  It is therefore
a useful intuition for the turnover, not an exact critical-batch estimator.
Large empirical studies likewise find that the useful degree of data
parallelism depends on the workload, optimizer, tuning protocol, and target
quality %
\citep{shallue2019dataparallelism}.

For language-model training, rather than inferring critical batch size from
gradient noise, %
\citet{merrill2025criticalbatch}
branch from a checkpoint and directly find the largest batch whose loss still
matches a smaller-batch reference after the same number of tokens.  For an
Adam branch with batch multiplier $k$, they set $B_k=kB$ and
$\eta_k=\sqrt{k}\eta$, measure its smoothed loss $L_k$ after a fixed token
budget, and, with a small loss tolerance $\varepsilon$, define

\begin{equation}
  k^*=\max\!\left\{k:
    L_k\leq L_j+\varepsilon\ \text{for every }j<k\right\},
  \qquad B^*=k^*B.
  \label{eq-lr-batch-branched-cbs}
\end{equation}

They find that this \textbf{empirical critical batch size} starts small,
grows rapidly, and then plateaus.  They also find that the simplified noise scale can
underestimate the directly measured value by orders of magnitude and can
follow a different trend.  Equation~\ref{eq-lr-batch-branched-cbs}
tests the useful batch range while taking square-root learning-rate scaling as
an input.  Critical batch size therefore answers \emph{when} a larger batch
remains token-efficient; it does not determine the exact Adam learning-rate
multiplier needed at that transition.

\subsection{Adam Does Not Imply One Universal Exponent}

\begin{keyidea}[Square-Root Scaling Is a Local Approximation]
Two independent effects keep the batch exponent from being universal: the
optimal step saturates as the batch approaches the turnover scale, and Adam's
slowly adapting second moment lags an abrupt change in gradient noise.  The
first makes the exponent depend on where the batch sits relative to the
turnover; the second can push it above one half at a batch transition.  Neither
fixes its value.
\end{keyidea}

For learning-rate searches, %
\citet{mccandlish2018largebatch} use a more
general saturating curve, which we write as

\begin{equation}
  \eta(B)=\frac{\eta_\infty}{(1+B_{\mathrm{turn}}/B)^{\alpha}},
  \label{eq-lr-batch-saturating-rule}
\end{equation}

\noindent where $B_{\mathrm{turn}}$ is the fitted turnover scale and
$\eta_\infty$ is the large-batch limit.  These names distinguish this fitted
curve from the directly measured $B^*$ above.  Its local log--log slope is

\begin{equation}
  \alpha_{\mathrm{eff}}(B)
  =\frac{\mathrm{d}\log\eta}{\mathrm{d}\log B}
  =\alpha\frac{B_{\mathrm{turn}}}{B+B_{\mathrm{turn}}}.
  \label{eq-lr-batch-effective-exponent}
\end{equation}

The curve behaves like a power law well below the turnover, but its
\textbf{effective exponent} decreases toward zero as the batch grows.  A fixed power law is thus
at best a local approximation, even before accounting for optimizer state.

Adam adds another reason for the exponent to vary.  For coordinate $i$, its
update has the form

\begin{equation}
  \Delta\theta_i=-\eta
  \frac{\mathbb{E}_{\beta_1}[G_i]}
       {\sqrt{\mathbb{E}_{\beta_2}[G_i^2]}+\epsilon_{\mathrm{Adam}}}.
  \label{eq-lr-batch-adam-update}
\end{equation}

Ignoring the exponential moving averages and $\epsilon_{\mathrm{Adam}}$ for
intuition gives

\begin{equation}
  \Delta\theta_i\approx -\eta\,
  \frac{\operatorname{sign}(\mathbb{E}[G_i])}
       {\sqrt{1+s_i/\mathbb{E}[G_i]^2}},
  \label{eq-lr-batch-adam-noise}
\end{equation}

\noindent where $s_i$ is the step-to-step variance of that gradient
coordinate.  If batch sampling dominates this variance, then $s_i\propto
1/B$.  This motivates square-root scaling for Adam.  In practice, however,
$\beta_2$ is commonly close to one, so the second-moment accumulator reacts
slowly to an abrupt change in gradient noise.  %
\citet{mccandlish2018largebatch}
argue that
this lag moves the useful exponent from $0.5$ toward the SGD limit of $1.0$;
their measured Adam and RMSProp exponents lie between these values before the
learning rate eventually flattens.  This theory supports our suspicion that
$\alpha=0.5$ can slightly under-correct a batch transition, but it does not
predict a specific replacement exponent.  Their lag argument also assumes
$\beta_2$ near $0.999$, for which the second moment needs on the order of a
thousand steps to react.  The run in Figure~\ref{fig-lr-batch-warmup} uses
$\beta_2=0.95$, whose second moment averages over roughly twenty steps, so this
lag alone cannot explain a curvature change that stays visible for thousands
of steps.

\subsection{An Empirical Exponent of 0.53 Smoothed Batch Transitions in Internal Tuning}

\begin{figure}[!htb]
\centering
\includegraphics[width=0.94\linewidth]{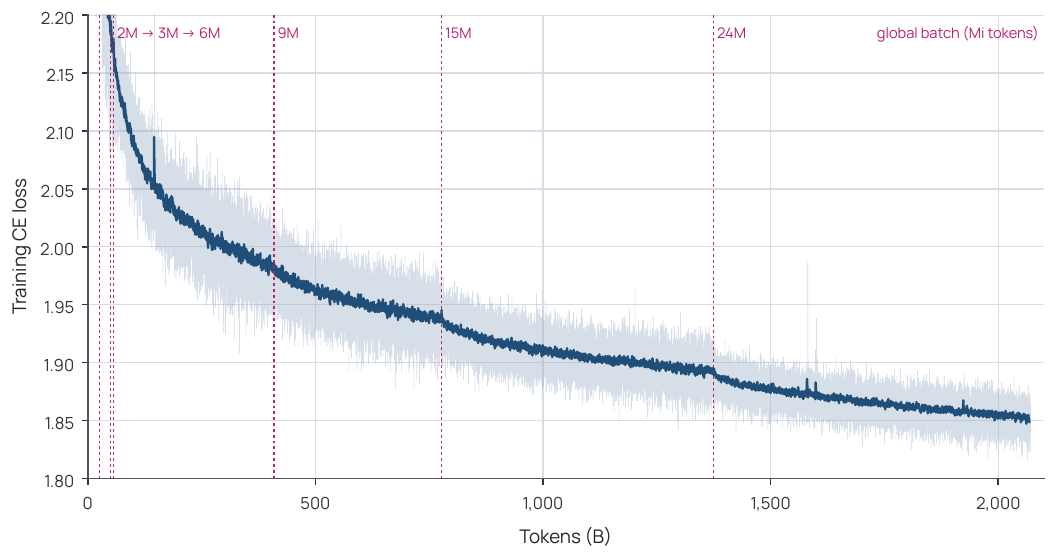}
\caption{\textbf{Square-root learning-rate scaling leaves visible curvature
changes at large batch-size transitions.} Training cross-entropy loss is
shown against cumulative tokens for a $3.21\text{B}@23.10\text{B}$ MoE trained
for approximately 2.07 trillion tokens.  The light trace is the raw loss and
the dark trace is a time-weighted exponential moving average (TWEMA).  Dotted
lines mark each global-batch increase, labeled with the new batch in Mi tokens (written as M in the figure); the batch progresses through 1, 2, 3, 6, 9, 15, and 24 Mi tokens,
and the later transitions occur at approximately 409B, 777B, and 1.375T
cumulative tokens.
Appendix~\ref{app-lr-batch-warmup-run} records the run's recipe and
reproduction artifacts.}
\label{fig-lr-batch-warmup}
\end{figure}

Figure~\ref{fig-lr-batch-warmup} follows one
$3.21\text{B}@23.10\text{B}$ (active@total parameters) MoE with 24 layers,
48 routed experts, top-$4$ routing, and one shared expert, trained with AdamW
($\beta_1=0.9$, $\beta_2=0.95$).  Its global batch increases through 1, 2, 3,
6, 9, 15, and 24 Mi tokens while the learning rate follows
Equation~\ref{eq-lr-batch-scaling}.  At the later, larger-batch transitions,
the smoothed loss shows a visible downward change in curvature.  The effect is
consistent with the effective learning rate becoming slightly smaller at the
transition: multiplying by $N^{1/2}$ did not fully preserve the local training
trajectory.

Across repeated internal tuning runs, we found $\alpha=0.53$ to be a useful
empirical setting: it removed the visible change in curvature at the batch
increase more consistently than $\alpha=0.5$.  This is a small correction; for
a batch doubling, it raises the learning rate only
$2^{0.03}-1\approx2.1\%$ above square-root scaling.  Its direction is
consistent with the slow-adapting Adam second moment discussed above, but its
value was selected from experiments rather than derived from the theory or
fitted to the single curve in Figure~\ref{fig-lr-batch-warmup}.  We therefore
present $0.53$ as an Olmo training recipe, not as a universal scaling exponent.

\begin{sectiontakeaway}
Critical batch size identifies when a larger batch remains useful; the Adam
learning-rate multiplier still needs tuning. In our internal tuning runs (not
plotted here), $\alpha=0.53$ preserved the local loss trajectory at batch-size
transitions more closely than square-root scaling.
\end{sectiontakeaway}

\section{Scaling Expert Learning Rate with Sparsity Did Not Help}
\label{sec-lr-sparsity-scaling}

Each routed expert sees only a fraction of the global tokens.  This motivates
a simple heuristic: if $K$ of $N$ experts are selected for each token, reduce
the expert learning rate by the square root of the expert activation ratio
$\alpha_E=K/N$,

\begin{equation}
  \eta_{\mathrm{expert}}
    = \eta_{\mathrm{main}}\sqrt{K/N}.
  \label{eq-expert-lr-sparsity-scaling}
\end{equation}

\noindent We are not aware of an established expert-learning-rate rule of this
form; we tested it as a direct analogy to the square-root batch-size rule of
Section~\ref{sec-lr-batch-scaling}.  For top-$4$ routing over 64 experts,
this sets the expert learning rate to one quarter of the main learning rate.
This analogy assumes that expert-gradient noise follows the same scaling as an ordinary
independently sampled batch.  Routed tokens, router specialization, Adam
preconditioning, and correlations between experts need not satisfy that
assumption. OLMoE's routing analyses, for example, find domain and vocabulary
specialization rather than identical random subsamples across experts
\citep{muennighoff2024olmoe}.

We tested the heuristic with $0.43\text{B}@2.51\text{B}$ (active@total
parameters) MoEs using 64 routed experts and top-$4$ routing.  Each point
trains one model to 150B tokens with AdamW ($\beta_1=0.9$, $\beta_2=0.95$).  One
sweep uses the same learning rate for all parameters; the other applies
Equation~\ref{eq-expert-lr-sparsity-scaling} to the routed-expert parameters
while sweeping the main learning rate.  The unscaled sweep reached a lower best validation loss than the
sparsity-scaled sweep, even after the main learning rate was retuned for
each: the smaller expert learning rate did not improve this controlled model
family.

\begin{figure}[!htb]
\centering
\includegraphics[width=0.85\linewidth]{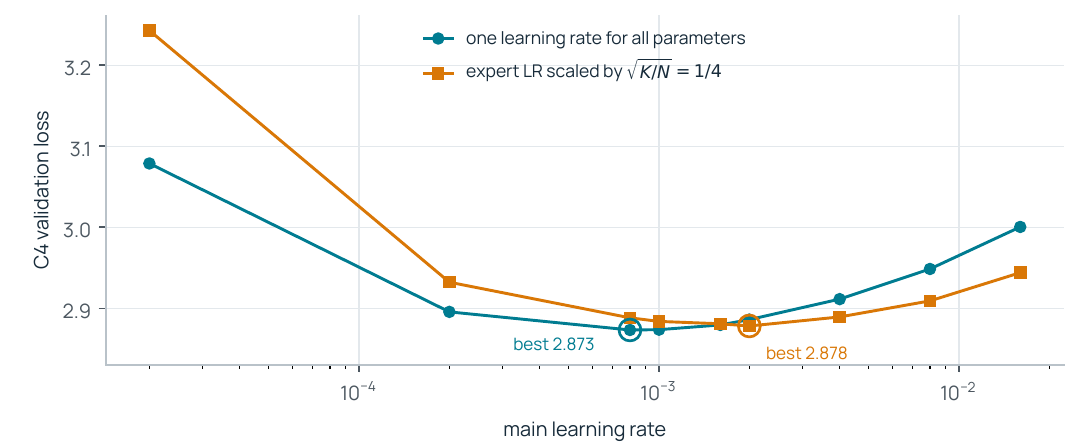}
\caption{\textbf{Sparsity-scaled expert learning rates did not improve
validation loss in the tested sweeps.} Each point is one
$0.43\text{B}@2.51\text{B}$, top-$4$-of-$64$ MoE trained for 150B tokens. The
teal sweep uses one learning rate for all parameters; the orange sweep sets
$\eta_{\mathrm{expert}}=\eta_{\mathrm{main}}\sqrt{K/N}
=\eta_{\mathrm{main}}/4$.  Both sweeps cover the same nine main learning
rates; markers are observed runs and connecting lines are visual guides.
After independently tuning the main learning rate, the best unscaled run
reaches C4 %
\citep{raffel2020exploring} validation loss 2.873 at
$8\times10^{-4}$, while the best
sparsity-scaled run reaches 2.878 at $2\times10^{-3}$.  Each point is a
single run without seed replication, so the comparison is directional rather
than a precise estimate of the gap.}
\label{fig-lr-sparsity-sweep}
\end{figure}

In these sweeps, the expert activation ratio alone did not justify square-root
down-scaling. Our default is therefore to use the same
learning rate for routed experts and the rest of the model unless a matched
sweep for the target model supports a different choice.

\section{Why We Do Not Upcycle}
\label{sec-upcycling}

\conceptlead{We train the target MoE architecture from the beginning.}
This is a recipe choice for the long training horizon considered in this
report, not a general conclusion that upcycling is ineffective. In
\textbf{sparse upcycling}, a pretrained dense checkpoint is converted into an
MoE, typically by copying or mapping its feed-forward weights into multiple
experts, initializing a router, and then continuing training
\citep{komatsuzaki2023sparseupcycling}. This can provide a useful head start
when a suitable dense checkpoint already exists. When dense pretraining and
sparse continuation share one fixed compute budget, however, the dense prefix
also consumes tokens that could have trained the target sparse model directly.

The published OLMoE study tested this trade-off. It converted an OLMo-1B
checkpoint trained for 2T tokens into a top-2-of-8 MoE, then compared continued
training with a sparse model trained from scratch. The upcycled model was
better early in the continuation, but the scratch model caught up after roughly
500B tokens of sparse training and began to outperform it near 600B tokens
\citep{muennighoff2024olmoe}. That crossover matters for our setting: like the
5T-token plan of the published OLMoE recipe, our training horizon is far
longer than the roughly 500B tokens over which the upcycled model led.
A dense checkpoint was
therefore not a compelling prerequisite for the training run.

Upcycling also couples the target model to the source architecture and its
initialization. In the OLMoE study, for example, the source checkpoint lacked
query-key normalization and used normal initialization, and the authors
report that both hurt stability. More generally, copying one dense multilayer perceptron into several routed
experts starts those experts from the same function; the router and subsequent
gradients must then create specialization. Olmo-core includes prototype
conversion paths for copying the dense MLP into routed or shared branches and
for the virtual-group initialization that %
\citet{he2024upcycling} introduce
for upcycling into fine-grained experts, but changing the mapping does not
remove the underlying budget decision. For a new model whose sparse
architecture and
training recipe are known in advance, direct sparse training lets the router
and independently initialized experts adapt together over the entire run.

This decision is conditional. Drop-Upcycling shows that partially
reinitializing copied experts can improve specialization and long-run learning
relative to direct copying %
\citep{nakamura2025dropupcycling}. Scaling-law
results likewise show that the preferred path depends on how much compute has
already been spent on the dense checkpoint and how much sparse continuation
remains %
\citep{liew2025scalingupcycling}. Upcycling may therefore be attractive
when a compatible dense checkpoint already exists and the continuation is
short. For our long training horizon and target architecture, the available
evidence did not justify making it the default.

\section{GEMM Timing Depends on Operand Values}
\label{sec-gemm-operand-values}

While comparing rowwise expert parallelism (EP) with DeepEP v2, we
initially observed different expert-GEMM times after matching tensor shapes
and routing.  The two paths did not, however, contain comparable weights: one
used normally initialized BF16 values, while the other consumed storage
created with \texttt{torch.empty()}, whose documented contents are
uninitialized %
\citep{pytorchEmptyDocs}. Once both paths used the same normal
initialization and routing, the apparent rowwise advantage disappeared.

The effect reproduced without MoE, communication, routing, or Olmo-core.  A
minimal Blackwell benchmark measured one PyTorch grouped GEMM and one dense
matrix multiplication while changing only the BF16 weight values.  Table~\ref{tab-gemm-value-timing}
shows the resulting median per-call GPU times, measured with CUDA events
around each call.

\begin{table}[!htb]
\centering
\small
\renewcommand{\arraystretch}{1.14}
\setlength{\tabcolsep}{7pt}
\begin{tabular}{@{}lrr@{}}
\toprule
\textbf{Weight values} & \textbf{Grouped GEMM} & \textbf{Dense GEMM} \\
\midrule
Normal random & 31.059 ms & 24.724 ms \\
Uniform random & 31.682 ms & 25.829 ms \\
Random sign, fixed magnitude & \textbf{25.979 ms} & 24.160 ms \\
All zeros & \textbf{18.896 ms} & \textbf{19.424 ms} \\
Constant fill & \textbf{20.606 ms} & 24.350 ms \\
Uninitialized storage & 30.213 ms & 24.321 ms \\
\bottomrule
\end{tabular}
\caption{\textbf{Operand values changed GEMM timing at fixed shape, layout,
and dtype.} The grouped operation used eight groups with 16,384 rows per
group and $8192\times16384$ BF16 weights; the dense operation used the
corresponding $M=131{,}072$, $K=8192$, $N=16{,}384$ shape.  Measurements use
one NVIDIA B300 GPU, two warmup iterations, and five timed iterations; the
table reports their median. Bold marks the times that depart clearly from
the normal-random baseline.
Uninitialized storage is shown only to expose the original benchmark error;
it is not a meaningful model-like weight-value distribution. Its contents
depend on prior memory use, so this row need not reproduce the gap seen in
the original comparison. The reproducer, recovered
summary data, and provenance notes are bundled with the report's experiment
artifacts.}
\label{tab-gemm-value-timing}
\end{table}

Relative to normally distributed weights, zeros reduced the grouped-GEMM time
by 39\%, a constant fill reduced it by 34\%, and random signs at a fixed
magnitude reduced it by 16\%.  The dense GEMM also took 21\% less time with zero
weights, but its constant-fill time remained close to the normal case.  The
effect is therefore neither unique to grouped GEMM nor a simple monotonic law
of value entropy.  This is a single run with a fixed case order that
preserved medians but not
individual samples, clocks, power, temperature, or the complete software
environment. The percentages characterize this run rather than a general
Blackwell performance law.

\subsection{Operand-Dependent Hardware Behavior Has Prior Evidence}

On an NVIDIA A100,
\citet{bhalachandra2022inputentropy} measured the same $16{,}384^3$ DGEMM with
fixed and random matrix values.  The fixed-valued case reached 19.4 TFLOP/s,
whereas the random case reached 18.6 TFLOP/s, about 4\% lower, while drawing
67\% more average power, close to the GPU's thermal design power. Their measurements show that operand values affect
the physical workload even when matrix dimensions match.

Other studies also find value-dependent power, without always observing a
runtime difference. ALUPower measured substantial value-dependent energy variation
in GPU arithmetic %
\citep{lucas2016alupower}.  More recently,
\citet{gregersen2024inputdependent} varied GEMM value distributions, bit
similarity, placement, and sparsity on an A100 GPU and repeated selected
experiments on H100, V100, and RTX 6000 GPUs.
They found power changes approaching 40\%.  In their main A100 experiments,
run at a matrix size chosen to avoid consistent throttling, however, the average CUTLASS GEMM runtime for a fixed datatype remained
consistent to the microsecond level.
Together, these studies establish that operand bits can change physical GPU
activity.  They do not establish that every GPU, shape, or power regime will
turn that change into a runtime difference.

\subsection{Possible Mechanisms and Their Evidence}

Our benchmark did not isolate a hardware mechanism. A plausible explanation
is data-dependent switching activity. Similar bit patterns and low Hamming
weight can reduce bit transitions through arithmetic and memory datapaths,
which changes dynamic power.  If a power-limited GPU responds by changing its
operating frequency, the same instruction stream can then take a different
amount of time.  This explanation is consistent with the prior power studies,
but our run did not record power or SM clocks and therefore does not test the
causal chain.

Separate B300 stress measurements show that power limiting can reduce clocks
and throughput. The grouped-GEMM diagnostic of Section~\ref{sec-grouped-gemm}
(Appendix~\ref{app-grouped-gemm-large-m-power}) drove B300 GPUs against
their board power limit with very large expert shapes: the software
power-cap limiter was active in effectively every telemetry sample, the median SM clock fell from 1174 to 960~MHz, and median grouped-GEMM
throughput fell from 1366 to 1163~TFLOP/s, while per-trial throughput per MHz
(each trial's throughput divided by its mean SM clock) had per-shape medians
within 1.16--1.19 TFLOP/s/MHz on the same kernel template.  That run varied shape rather than
operand values, so it speaks to how this hardware responds rather than to
what triggered the response in Table~\ref{tab-gemm-value-timing}. In the stress diagnostic, throughput fell with the lower clock.
These measurements do not establish why operand values changed timing in
the original run. Table~\ref{tab-gemm-value-evidence-chain} collects the
observations relevant to the power-and-clock hypothesis. Matching operands, as described in
Section~\ref{subsec-gemm-operand-rule}, controls the observed timing effect
without requiring a choice among mechanisms.

\begin{table}[!htb]
\centering
\footnotesize
\begingroup
\definecolor{chainband}{HTML}{EDE6DA}
\renewcommand{\arraystretch}{1.25}
\setlength{\tabcolsep}{5pt}
\begin{tabularx}{\linewidth}{
  >{\raggedright\arraybackslash}p{0.30\linewidth}
  >{\raggedright\arraybackslash}X
  >{\raggedright\arraybackslash}p{0.17\linewidth}}
\toprule
\rowcolor{chainband}
\textbf{Observation} & \textbf{What was measured} & \textbf{Source} \\
\midrule
\textbf{Operand values change dynamic power} & Random values drew 67\% more
average power than fixed values in a $16{,}384^3$ DGEMM; value distributions
moved GEMM power by up to about 40\% & A100, prior work \\
\arrayrulecolor{black!22}\hline\arrayrulecolor{black}
\textbf{A power-limited GPU lowers its clock} & Power-cap limiter active in
99.3\% of telemetry samples; SM clock 1174 to 960~MHz & B300, Section~\ref{sec-grouped-gemm}, Appendix~\ref{app-grouped-gemm-large-m-power} \\
\arrayrulecolor{black!22}\hline\arrayrulecolor{black}
\textbf{Lower throughput tracks the lower clock} & Throughput 1366 to
1163~TFLOP/s at a nearly constant 1.16--1.19~TFLOP/s per MHz (per-shape medians), same kernel template
& B300, Appendix~\ref{app-grouped-gemm-large-m-power} \\
\arrayrulecolor{black!22}\hline\arrayrulecolor{black}
\textbf{Operand values change time} & Grouped-GEMM time fell by 16--39\% and dense-GEMM
time by up to 21\% at fixed shape, layout, and dtype & B300, Table~\ref{tab-gemm-value-timing} \\
\bottomrule
\end{tabularx}
\endgroup
\caption{\textbf{Measurements relevant to a possible power-and-clock
explanation.} The prior-work row summarizes %
\citet{bhalachandra2022inputentropy}
and %
\citet{gregersen2024inputdependent}. The two clock-related rows come from the B300
grouped-GEMM stress diagnostic, which varied shape rather than operand
values: they show how this hardware trades clock for power at the board
limit.  That diagnostic held each shape under sustained load (an eight-second
conditioning region, then a 1.5-second measured region), whereas the
operand-value run timed five calls of 19--32~ms after two warmup calls, so
the report does not establish whether that run reached the power limit.
The report draws its benchmarking rule from the last row alone and
attributes the effect to no single mechanism.}
\label{tab-gemm-value-evidence-chain}
\end{table}

Inline compression is another possible mechanism, but we have no evidence
that it was active in this experiment. NVIDIA documents that Hopper can transfer compressible global
memory using fewer physical bytes and thereby exceed nominal uncompressed
bandwidth %
\citep{nvidiaHopperInlineCompression}.  This is an opt-in property of
an individual allocation, requested through the CUDA Driver API; it is not a
general promise that ordinary tensors are compressed
\citep{nvidiaCompressibleMemory}. Our benchmark used
ordinary PyTorch allocations on a Blackwell B300 and never requested the
compressible-memory attribute.  A later check on the same GPU class with the
current PyTorch allocator found that an ordinary tensor, when it could be
queried at all, reported no compression attribute; that check describes the
current allocator rather than the environment of the original run.
The high arithmetic intensity of
these GEMMs and the different constant-fill behavior of grouped and dense
kernels also argue against assuming a single bandwidth-only explanation.

The public GEMM interfaces do not declare that zero or constant operands may
skip mathematical work, so we do not interpret the result as semantic zero
skipping.  We also did not preserve a kernel-identity or instruction trace
that would rule out every library- or kernel-specific effect.  The claim this
report makes is limited to the observation: operand values changed the time
of an otherwise identical GEMM. The measurements do not identify which
combination of switching power, clock response, and memory-system behavior
produced it.

\subsection{Use the Same Initialization in Every Comparison}
\label{subsec-gemm-operand-rule}

\begin{keyidea}[Operands Are Part of the Benchmark Configuration]
For a backend comparison, use identical loaded weights whenever possible.
Otherwise, use the same initializer, seed, scale, dtype, and layout on every
path.  Matching only the GEMM dimensions is insufficient.
\end{keyidea}

This control should be applied after sharding, conversion, or expert placement
so that no path retains uninitialized local storage.  End-to-end comparisons
should load the same checkpoint.  Synthetic comparisons should use the same
initialization method and random seed.  Zeros, constant fills, and
\texttt{torch.empty()} are useful diagnostic cases, but they are not substitutes
for model-like values.  If a communication change appears to alter an
otherwise unchanged GEMM, operand statistics, clocks, and power should be
checked before attributing the difference to the communication backend.

\section{One Shared Expert Is All You Need}
\label{sec-ungated-shared-experts}
\begin{figure}[!htb]
  \centering
  \includegraphics[width=0.75\linewidth]{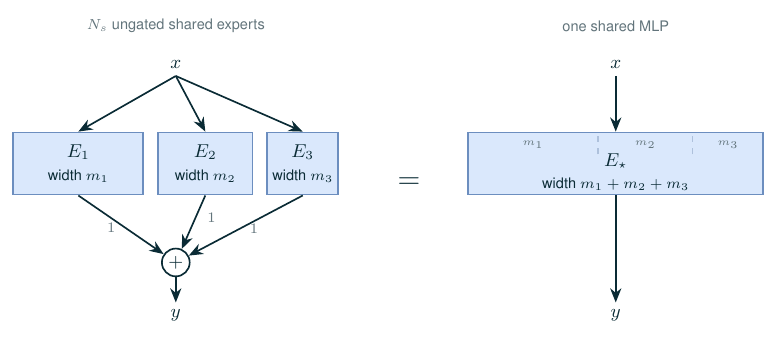}
  \caption{\textbf{Combining ungated shared experts into one wider MLP.}
  Every token reaches all $N_s$ shared experts, and their
  outputs are added with fixed unit coefficients, so the sum equals one MLP
  whose intermediate width is the sum of the branch widths
  (Equation~\ref{eq-ungated-shared-expert-collapse}). The dashed marks show
  where each original partition sits after concatenating the up and gate
  projections and stacking the down projections; box widths are drawn
  proportional to intermediate widths.}
  \label{fig-ungated-shared-collapse}
\end{figure}

\conceptlead{Multiple ungated shared experts can be merged.}  Several recent
MoE designs expose the number of always-active shared experts as a choice;
Olmo-core supports at most one \emph{ungated} shared expert (several shared
experts require a learned shared-expert gate), because any larger ungated
count expresses the same model class as one wider shared MLP.
Suppose every token is sent to several shared experts, every shared expert
receives the same input, and their outputs are added without
expert-specific, input- or task-dependent gates. The same function can be
computed by one shared multilayer perceptron (MLP) whose intermediate width is
the sum of their widths (Figure~\ref{fig-ungated-shared-collapse}). The number
of named shared experts is therefore only
a partitioning choice; the architectural quantity is the \textbf{total shared
width}.

MoEfication uses the same intermediate-dimension partition of a Transformer
feed-forward network, and DSMoE writes the equivalence explicitly for a
SwiGLU MLP %
\citep{zhang2022moefication,lv2025dsmoe}. At
fixed total shared width, changing only the number of ungated shared modules
does not change the function class.

\subsection{The Collapse Is Exact}

Let shared expert $j$ have intermediate width $m_j$. Using row-vector
notation, write its SwiGLU computation as %
\citep{shazeer2020glu}
\begin{equation}
  E_j(x)
  = \left[\operatorname{SiLU}\!\left(xW_g^{(j)}\right)
      \odot \left(xW_u^{(j)}\right)\right]W_d^{(j)}.
  \label{eq-shared-expert-swiglu}
\end{equation}
The gate projection $W_g^{(j)}$ (part of SwiGLU, not an output gate) and the
up projection each have $m_j$ columns, and the down projection has $m_j$
rows. Form one wider MLP by concatenating the gate and up projections
and stacking the down projections:
\begin{equation}
  W_g^\star=\left[W_g^{(1)}\ \cdots\ W_g^{(N_s)}\right],
  \qquad
  W_u^\star=\left[W_u^{(1)}\ \cdots\ W_u^{(N_s)}\right],
  \qquad
  W_d^\star=
  \begin{bmatrix}
    W_d^{(1)}\\[-2pt]
    \vdots\\[-2pt]
    W_d^{(N_s)}
  \end{bmatrix}.
  \label{eq-shared-expert-concatenation}
\end{equation}
Because SiLU and the elementwise product act independently along the
intermediate dimension, multiplication by the stacked down projection adds
the partitions:
\begin{equation}
  E_\star(x)
  = \left[\operatorname{SiLU}(xW_g^\star)
      \odot (xW_u^\star)\right]W_d^\star
  = \sum_{j=1}^{N_s}E_j(x),
  \qquad
  \operatorname{width}(E_\star)=\sum_{j=1}^{N_s}m_j.
  \label{eq-ungated-shared-expert-collapse}
\end{equation}

The same construction applies to an ordinary two-layer MLP. Fixed
expert-specific coefficients do not change the represented function, although
they change training dynamics (Section~\ref{sec-merge-expert-activations}): for
$\sum_j c_jE_j(x)$, absorb each constant $c_j$ into the corresponding rows of
$W_d^{(j)}$. A single input- or task-dependent gate shared by the entire
branch also does not require several experts, since
\begin{equation}
  a(x)\sum_j E_j(x)=a(x)E_\star(x).
\end{equation}
Among coefficient choices, only distinct input- or task-dependent coefficients
$a_j(x,t)$ generally prevent this fixed reparameterization; the next
subsection lists other mechanisms that also preserve expert identity.

The wide and partitioned forms have the same parameters and ideal model
FLOPs, and their parameter gradients map slice by slice. When initialization,
optimizer settings, and optimizer variables are also mapped in the same way,
an element-wise optimizer such as AdamW follows the same training trajectory
in both forms; a matrix-wise optimizer such as Muon
(Section~\ref{sec-ddp-vs-fsdp}) generally does not, because it updates the
wide matrix differently from its slices.
Separate kernels may accumulate floating-point terms in a different order,
and the two layouts may have different systems performance, but neither effect
creates a new represented function.

\subsection{Dynamic Gates Preserve Expert Identity}

Multiple shared experts become a real architectural choice when the branches
are no longer fixed partitions of one MLP.  Let $h_j(x)$ denote expert $j$'s
intermediate activation of width $m_j$, and let $g_j(x,t)$ be its coefficient
for input $x$ and, when applicable, task $t$.  Concatenate the activations as
$H(x)=[h_1(x)\ \cdots\ h_{N_s}(x)]$.  The gated merge can then be written as

\begin{align}
  y(x,t)
    &= \sum_{j=1}^{N_s}g_j(x,t)h_j(x)W_d^{(j)}
     = H(x)D(g(x,t))W_d^\star,
  \nonumber\\
  D(g)
    &= \operatorname{blockdiag}\!\left(
      g_1I_{m_1},\ldots,g_{N_s}I_{m_{N_s}}\right).
  \label{eq-dynamic-shared-expert-gate}
\end{align}

When the coefficients differ by expert and vary with the input or task,
$D(g(x,t))$ cannot generally be absorbed into the fixed down projection
$W_d^\star$.  This is the precise boundary of the collapse.  A common scalar
$a(x,t)$ applied to the whole sum does not change this: then $D=aI$, the experts still
collapse into $E_\star$, and only the outer branch gate remains.  Conditional
selection, per-expert normalization, different inputs, and branch-specific
operations also preserve expert identity.

Earlier MoE architectures illustrate this boundary.  One terminology
distinction is important: in multi-task learning, a ``shared expert'' usually
means an expert whose parameters are available to several tasks.  It does not
necessarily mean the always-active shared branch used in recent language-model
MoEs.  Table~\ref{tab-gated-shared-expert-precedents} therefore lists these
models as precedents for \emph{separately gated} shared experts, not as
counterexamples to the ungated equivalence.

\begin{table}[!htb]
\centering
\footnotesize
\renewcommand{\arraystretch}{1.18}
\setlength{\tabcolsep}{4pt}
\begin{tabular}{@{}
  >{\raggedright\arraybackslash}m{0.23\linewidth}
  >{\raggedright\arraybackslash}m{0.20\linewidth}
  >{\raggedright\arraybackslash}m{0.25\linewidth}
  >{\raggedright\arraybackslash}m{0.20\linewidth}@{}}
\toprule
\textbf{Architecture} &
\textbf{Expert structure} &
\textbf{Gate and evaluation} &
\textbf{Why experts stay distinct} \\
\midrule
Adaptive local experts\newline%
\citep{jacobs1991adaptive} &
Common expert bank; every expert receives the input &
Normalized, input-dependent responsibilities for individual experts &
Each expert receives a different dynamic coefficient \\
\arrayrulecolor{black!22}\hline\arrayrulecolor{black}
Hierarchical MoE\newline%
\citep{jordan1994hierarchical} &
Experts organized beneath a tree of gates &
Input-dependent coefficients composed along each path &
The effective coefficient varies by expert and input \\
\arrayrulecolor{black!22}\hline\arrayrulecolor{black}
Sparsely-gated MoE\newline%
\citep{shazeer2017outrageously} &
Globally available feed-forward expert bank &
Noisy top-$K$ routing evaluates and weights a selected subset &
Both the coefficients and executed experts vary \\
\arrayrulecolor{black!22}\hline\arrayrulecolor{black}
One-gate Mixture-of-Experts (OMoE)\newline%
\citep{ma2018mmoe} &
One expert pool supplies all task towers &
One input-dependent \emph{vector} gate forms a common dense mixture &
One gating network still emits one coefficient per expert \\
\arrayrulecolor{black!22}\hline\arrayrulecolor{black}
Multi-gate Mixture-of-Experts (MMoE)\newline%
\citep{ma2018mmoe} &
Expert parameters are shared across tasks &
Each task has its own input-dependent softmax vector &
The same input can expose a different mixture to each task \\
\arrayrulecolor{black!22}\hline\arrayrulecolor{black}
CGC / PLE\newline%
\citep{tang2020ple} &
Multiple shared experts coexist with task-specific experts &
Task gates separately weight both groups; PLE stacks these units &
Every shared expert remains separately addressable \\
\bottomrule
\end{tabular}
\caption{\textbf{Earlier architectures retain multiple experts through
expert-specific dynamic gates.} Customized Gate Control (CGC) and Progressive
Layered Extraction (PLE) are the clearest older examples with an explicitly
named bank of multiple shared experts.  Their meaning of ``shared'' is
parameter sharing across tasks, whereas recent language-model shared experts
are typically always-active branches.  None of the listed models satisfies
the fixed-coefficient assumptions of Equation~\ref{eq-ungated-shared-expert-collapse}.}
\label{tab-gated-shared-expert-precedents}
\end{table}

In the one-gate baseline, ``one gate'' means one
network that emits a vector of expert coefficients, not one scalar multiplying the
sum.  Multi-gate Mixture-of-Experts (MMoE) makes the task dependence explicit,
while Customized Gate Control and PLE add separately gated task-specific and
shared banks.  Dense vector gating and sparse top-$K$ routing both prevent the
collapse for the same algebraic reason; sparsity additionally changes which
expert computations execute.

\subsection{Expert Count Is Not Shared Width}

The equivalence changes how shared-expert designs and ablations should be
reported. For an ungated shared branch, the primary parameter-size variable is
\begin{equation}
  m_s=\sum_{j=1}^{N_s}m_j,
  \label{eq-total-shared-width}
\end{equation}
not $N_s$ alone. Increasing the number of fixed-width shared experts increases
$m_s$, so it tests additional unconditional parameters and compute rather than
an independent benefit from having more expert identities. A count ablation
isolates expert partitioning only when total shared width remains fixed; under
the assumptions above, such an ablation should differ only through numerical
and systems effects.

DeepSeekMoE provides a prominent always-active shared-expert construction
\citep{dai2024deepseekmoe}. The collapse is compatible with useful shared
parameters, since an unconditional branch may still capture common knowledge;
it shows only that several unweighted shared partitions can be represented as one
wider branch with the same total parameter size.

\subsection{Shared and Routed Widths Are Independent}

This result directly motivates Olmo's representation. Olmo uses one shared
MLP and chooses its intermediate width independently from the width of one
routed expert. There is no mathematical reason for the two widths to match.
The shared width controls the unconditional parameters applied to every token;
the routed width controls the parameters contributed by each selected route.

For example, the study configuration analyzed in
Section~\ref{sec-merge-expert-activations} uses routed-expert width 2,560 and
shared width 1,280. Calling the latter ``half of one routed expert'' is only a
comparison of widths. It is already the complete shared branch. Splitting it
into two width-640 modules and adding their outputs would produce no new model
function; combining two width-1,280 modules would simply create one
width-2,560 shared MLP in partitioned form.

One wider MLP can also expose a larger GEMM and fewer module launches, but the
mathematical identity alone does not predict throughput. Tensor placement,
quantization scales, and available kernels can make one physical layout faster
than another. Olmo's use of one shared MLP follows from the functional
equivalence, not a measured performance advantage. Its total width remains a
model-recipe choice independent of routed-expert width.

\FloatBarrier

\section{Merging Routed and Shared Expert Activations}
\label{sec-merge-expert-activations}

Section~\ref{sec-ungated-shared-experts} established that several ungated
shared-expert partitions collapse into one wider shared MLP. After choosing
that shared branch, the model recipe must specify how to weight it against
the token-dependent routed branches.

An MoE layer may contain both \emph{routed experts}, selected separately for
each token, and a \emph{shared expert}, evaluated for every token. The routed
and shared branches are projected back to the same model width before they
are added, but that does not determine how strongly each branch should enter
the sum. A normalization that appears innocuous can shrink the MoE output,
change its Jacobian, and rescale the gradients of routed and shared parameters
by different amounts.

\begin{figure}[!htb]
  \centering
  \includegraphics[width=0.55\linewidth]{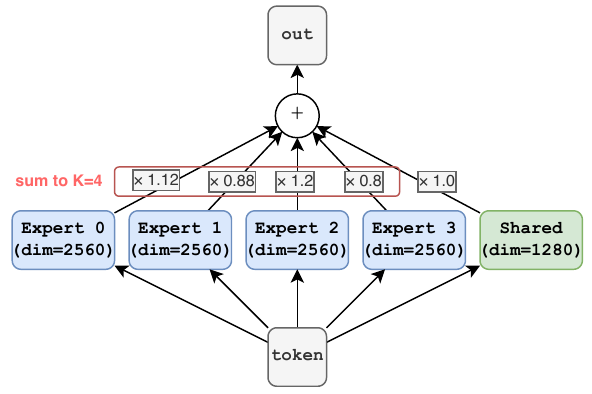}
  \caption{\textbf{Merging four routed-expert outputs with one shared-expert
  output.}
  In the Olmo rule, selected routed coefficients vary by token but sum to
  $K=4$, so their mean is one; the shared-expert coefficient is also one.
  Routed and shared experts may have different intermediate widths because
  every branch returns a model-width tensor before the addition. Conceptual
  illustration with no measured quantities.}
  \label{fig-merge-expert-activations}
\end{figure}

For token $x$, let $\mathcal{S}(x)$ contain the $K$ selected routed experts,
let $f_i(x)\in\mathbb{R}^{d_{\mathrm{model}}}$ be routed expert $i$'s output,
and let $g(x)\in\mathbb{R}^{d_{\mathrm{model}}}$ be the shared-expert output.
We write the merge as
\begin{equation}
  y_{\mathrm{MoE}}(x)
  = \sum_{i\in\mathcal{S}(x)} r_i(x) f_i(x) + s(x)g(x),
  \qquad |\mathcal{S}(x)|=K.
  \label{eq-merge-routed-shared}
\end{equation}
The coefficients $r_i(x)$ determine the routed contribution, while $s(x)$
sets the gain of the shared branch
(Figure~\ref{fig-merge-expert-activations}). As
Appendix~\ref{app-expert-merge-practices}
shows, public MoE implementations make substantially different choices for
both quantities %
\citep{deepseekai2024deepseekv3,muennighoff2024olmoe,nvidia2026megatroncoremoe}.
We therefore treat the merge rule as part of the model and
initialization recipe.

\subsection{A Dense MLP Provides the Scale Reference}

A dense SwiGLU MLP provides a useful reference because
its intermediate dimension can be partitioned exactly. Split its up, gate,
and down-projection weights into $K$ blocks of width $d_r$, the routed-expert
intermediate width, and one block of width $d_s$, the total shared width
($m_s$ in Equation~\ref{eq-total-shared-width}). Matrix multiplication
distributes across this partition, giving
\begin{equation}
  f_{\mathrm{dense}}(x)
  = \sum_{i=1}^{K} f_i(x) + g(x),
  \qquad d_{\mathrm{act}}=K d_r+d_s.
  \label{eq-dense-mlp-block-decomposition}
\end{equation}
Unit coefficients exactly recover a dense MLP formed by concatenating the
active intermediate blocks.
For the initialization-scale experiment below, $K=4$, $d_r=2560$, and
$d_s=1280$, so the matched dense width is
\begin{equation}
  d_{\mathrm{act}}=4\times2560+1280=11{,}520.
\end{equation}

Routing replaces the four unit routed coefficients with data-dependent route
weights. Let $p_i(x)$ denote the selected router probabilities after L1
normalization, so $\sum_i p_i(x)=1$. Olmo's production
recipes configure the Olmo-core router to use
\begin{equation}
  r_i(x)=Kp_i(x), \qquad s(x)=1.
  \label{eq-olmo-expert-merge-rule}
\end{equation}
The combination weight $p_{t,e}$ of Equation~\ref{eq-routed-moe} corresponds
to $r_i$ here, not to $p_i$. If the selected probabilities are uniform, $p_i=1/K$ and every active branch
again has coefficient one. The routed mixture therefore approaches the exact
dense decomposition continuously as the selected probabilities approach a
uniform distribution.

This choice also accounts for different expert widths without an additional
width multiplier. Define the \textbf{width-weighted mean coefficient}
\begin{equation}
  \bar a_w
  = \frac{d_r\sum_i r_i+d_s s}{K d_r+d_s}.
  \label{eq-width-weighted-merge-coefficient}
\end{equation}
Equation~\ref{eq-olmo-expert-merge-rule} gives $\bar a_w=1$ for any $d_r$ and
$d_s$. Since the width-weighted mean is already one, a further common factor
would rescale the whole MoE output rather than compensate for unequal widths.
The narrower shared
expert already contributes fewer independently initialized intermediate
units, so multiplying its output by a second width fraction would count the
same width difference twice under this initialization.

\begin{keyidea}[Matching the Coefficients of a Partitioned Dense MLP]
Selected router probabilities have mean $1/K$. Multiplying them by $K$ makes
the mean routed coefficient one, matching the coefficient of each block in an
additively partitioned dense MLP. The shared branch then enters the same sum
with unit gain.
\end{keyidea}

\subsection{Predicting Variance at Initialization}

The dense decomposition also yields a simple scale prediction. Assume the
randomly initialized branch outputs are centered and approximately
independent, and that output variance is proportional to intermediate width
under the common initialization. Conditional on the router coefficients, the
MoE-to-dense output-variance ratio is
\begin{equation}
  R_{\mathrm{var}}
  \approx
  \frac{d_r\sum_i r_i^2+d_s s^2}{K d_r+d_s}.
  \label{eq-expert-merge-variance-ratio}
\end{equation}
For the Olmo rule,
\begin{equation}
  R_{\mathrm{var}}^{\mathrm{Olmo}}
  \approx
  \frac{d_r K^2\sum_i p_i^2+d_s}{K d_r+d_s}.
  \label{eq-olmo-expert-merge-variance-ratio}
\end{equation}
Uniform routing gives $\sum_i p_i^2=1/K$ and hence
$R_{\mathrm{var}}^{\mathrm{Olmo}}=1$. Router concentration increases the
ratio according to the concentration term $\sum_i p_i^2$.

Table~\ref{tab-expert-merge-normalizations} contrasts this rule with the three
alternatives used in our controlled experiment. All four define valid model
parameterizations, but they do not begin at the same forward or backward
scale.

\begin{table}[!htb]
  \centering
  \small
  \renewcommand{\arraystretch}{1.22}
  \renewcommand{\tabularxcolumn}[1]{m{#1}}
  \setlength{\tabcolsep}{5pt}
  \begin{tabularx}{\linewidth}{>{\raggedright\arraybackslash}m{0.25\linewidth}
      >{\centering\arraybackslash}m{0.32\linewidth} X}
    \toprule
    \textbf{Convention} & \textbf{Coefficients} & \textbf{Immediate effect} \\
    \midrule
    Olmo
      & $r_i=Kp_i,\quad s=1$
      & Unit-average routed coefficients and a unit-gain shared branch. \\
\arrayrulecolor{black!22}\hline\arrayrulecolor{black}
    Routed unit sum
      & $r_i=p_i,\quad s=1$
      & The routed mixture receives one unit in total, while the shared branch
        independently receives unit gain. \\
\arrayrulecolor{black!22}\hline\arrayrulecolor{black}
    Equal branch mass
      & $r_i=\tfrac{1}{2}p_i,\quad s=\tfrac{1}{2}$
      & Routed and shared contributions split one unit of coefficient mass. \\
\arrayrulecolor{black!22}\hline\arrayrulecolor{black}
    Width-proportional unit sum
      & \rule{0pt}{18pt}$r_i=\dfrac{K d_r}{d_{\mathrm{act}}}p_i,\quad
         s=\dfrac{d_s}{d_{\mathrm{act}}}$\rule[-9pt]{0pt}{0pt}
      & A single unit of coefficient mass is divided in proportion to active
        intermediate width. \\
    \bottomrule
  \end{tabularx}
  \caption{\textbf{Merge conventions evaluated in the initialization-scale
  study.} The selected probabilities satisfy $\sum_i p_i=1$, and
  $d_{\mathrm{act}}=K d_r+d_s$.}
  \label{tab-expert-merge-normalizations}
\end{table}

\subsection{The Measured Forward and Backward Scales Agree}

\begin{figure}[!htb]
  \centering
  \includegraphics[width=\linewidth]{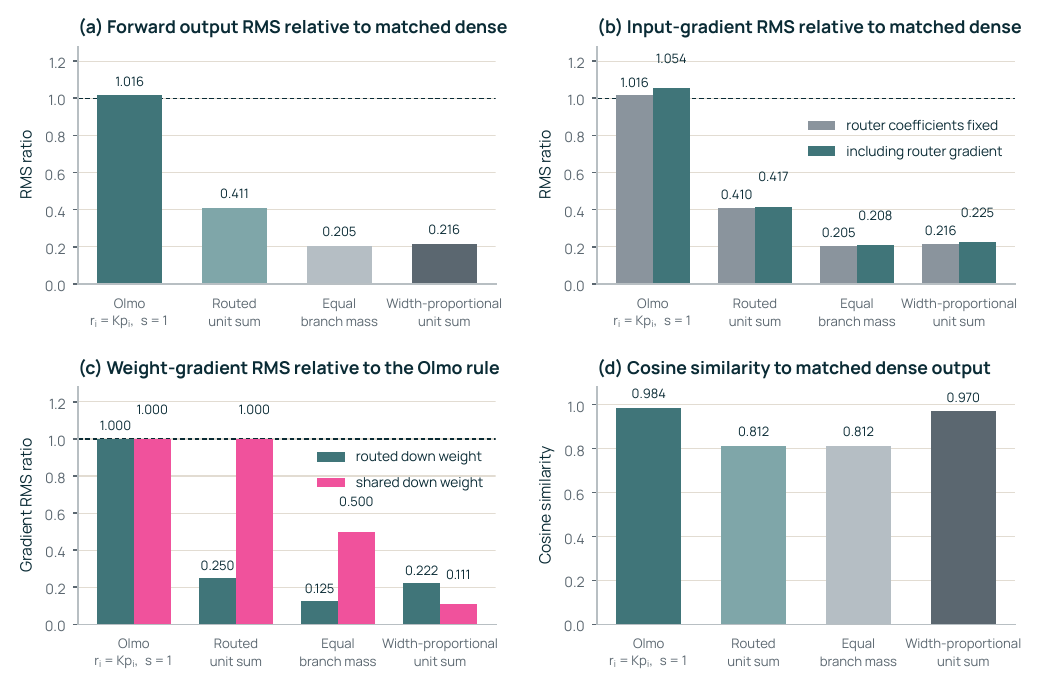}
  \caption{\textbf{Olmo's merge rule preserves dense-like scale at random
  initialization.} The four conventions are those of
  Table~\ref{tab-expert-merge-normalizations}.  (a)~The Olmo output RMS is
  $1.016\times$ the matched dense
  reference, corresponding to a $1.032\times$ variance ratio. (b)~Its
  fixed-router input-gradient RMS is also $1.016\times$ dense; differentiating
  through the router raises the ratio to $1.054\times$. (c)~Alternatives that
  constrain the routed or total coefficient mass to one reduce the routed
  down-projection gradient to a quarter or less of the Olmo value while the
  shared gradient stays at one or falls with it. (d)~Olmo's output also stays
  closest in direction to the dense output.  Forward measurements use the
  full-size initialization configuration; backward measurements use the
  dimensionally scaled surrogate described in the text.}
  \label{fig-expert-merge-forward-backward-scale}
\end{figure}

We test the prediction in a controlled random-initialization experiment
summarized in Figure~\ref{fig-expert-merge-forward-backward-scale}. The
forward study uses the layer shape of an earlier Olmo development recipe, not
the production recipe: $d_{\mathrm{model}}=2560$, $d_r=2560$, $d_s=1280$, and
$K=4$ experts selected from $N=48$ router candidates. It instantiates only the
four selected experts and evaluates 512
random tokens for each of eight seeds. The dense reference concatenates the
same selected and shared weights into an exact width-11,520 SwiGLU MLP, so
differences arise
from merge coefficients rather than different sampled parameters.

The backward study instantiates all 48 routed experts and gathers each
token's four. To make repeated gradient measurements practical, it scales
every dimension down by a factor of ten and raises the initialization
standard deviation by $\sqrt{10}$, which preserves the leading-order
projection variances. It uses 1024 tokens for
each of twelve seeds and applies the same random upstream gradient to every
normalization. We report input gradients with router coefficients both fixed
and differentiable, separating the expert-path Jacobian from the additional
gradient through the router.

The selected probabilities have
$\mathbb{E}[\sum_i p_i^2]=0.2589$, only slightly above the uniform top-4 value
of 0.25. Substituting this measurement into
Equation~\ref{eq-olmo-expert-merge-variance-ratio} predicts a variance ratio
of 1.032, matching the measured value to within 0.03 percentage points. The
shared-to-routed branch RMS ratio is $0.7082\pm0.0028$, close to
$\sqrt{1280/2560}=0.7071$. This confirms that intermediate width already
controls branch scale under the tested initialization.

The backward results expose a second advantage of the Olmo parameterization.
With router coefficients fixed, its input-Jacobian scale remains close to the
dense reference. Its routed and shared down-projection gradients also retain
the unit-coefficient scale: the dense reference's routed and shared
down-projection gradient RMS values are $0.98\times$ and $1.00\times$ the Olmo
values (Figure~\ref{fig-expert-merge-forward-backward-scale}(c) is normalized
to Olmo, so it does not plot these ratios). By contrast, routed-unit-sum normalization reduces
the routed down-projection gradient RMS to $0.25\times$ the Olmo value for
$K=4$, while leaving the shared down-projection gradient unchanged. Changing
the merge rule is therefore also a parameter-family-specific learning-rate
change unless the initialization and optimizer are adjusted with it.

\subsection{Scope of the Initialization Result}

Olmo's $r_i=Kp_i,\ s=1$ rule preserves dense-like forward and
expert-path backward scale under the stated architecture, initialization, and
approximately uniform selected routing. It has a direct additive
interpretation, handles unequal expert widths without a second width penalty,
and makes its forward departure from dense scale predictable through
$\sum_i p_i^2$.

The result does not establish a universal MoE normalization. Public models
use retained routed mass below one, unit routed mass, learned shared gates,
and model-specific routed scales; Appendix~\ref{app-expert-merge-practices}
provides a source-pinned comparison. These recipes also differ in
initialization, upcycling, router form, and expert width, so their coefficients
cannot be ranked in isolation.

Fixed coefficient changes can often be absorbed into expert down projections,
but that reparameterization does not make the training recipes identical. It
changes initial parameter scale, optimizer states, weight decay, gradient
clipping, and effective learning rates. The measurements here characterize
random initialization and local backward scale; they do not measure trained
language-model quality.

\section{Beyond Pretraining}
\label{sec-beyond-pretraining}

When the same distributed MoE learner is used for post-training, parameter
placement and gradient synchronization are still needed for a model whose total parameter size exceeds one device's memory, but supervision,
additional model copies, and rollout generation change the workload. We map
those requirements onto the sparse learner and identify the systems needed
around it. \textbf{This section is a design mapping:}
we report no integrated MoE post-training run, throughput, or quality result.

Ai2's open-instruct codebase provides implementations of several
post-training workloads, while the T\"ulu~3 report documents an earlier
actor-critic reinforcement-learning recipe
\citep{allenai2026openinstruct,lambert2024tulu3}. This section considers
four workloads: supervised fine-tuning (SFT), Direct Preference
Optimization (DPO), Group Relative Policy Optimization (GRPO)
\citep{shao2024deepseekmath}, and Proximal Policy Optimization (PPO) with a
critic %
\citep{schulman2017ppo}. The open-instruct implementations use different
backends: cached-reference DPO uses Accelerate and one GRPO path uses
Olmo-core HSDP, while open-instruct's SFT support
includes OLMoE models. Reusing a post-training objective does not require
copying its original distributed backend. Table~\ref{tab-post-training-components} decomposes each
workload into its components and marks what this report's mechanisms
could contribute to each. Every workload contains a learner. Preference and
reinforcement-learning workloads add reference policies or other model
copies, and reinforcement learning also needs a rollout inference system.

\begin{table}[!htb]
\centering
\footnotesize
\begingroup
\definecolor{ptband}{HTML}{EDE6DA}
\renewcommand{\arraystretch}{1.22}
\setlength{\tabcolsep}{4pt}
\newcommand{\yes}{\textcolor{ai2accent}{\textbf{\checkmark}}}
\newcommand{\ext}{\textcolor{black!45}{$\circ$}}
\newcommand{\no}{\textcolor{black!35}{--}}
\begin{tabularx}{\linewidth}{
  >{\raggedright\arraybackslash}p{0.17\linewidth}
  >{\centering\arraybackslash}p{0.04\linewidth}
  >{\centering\arraybackslash}p{0.045\linewidth}
  >{\centering\arraybackslash}p{0.05\linewidth}
  >{\centering\arraybackslash}p{0.04\linewidth}
  >{\raggedright\arraybackslash}p{0.11\linewidth}
  >{\raggedright\arraybackslash}X}
\toprule
\rowcolor{ptband}
& \multicolumn{4}{c}{\textbf{Component in workload}} & & \\
\rowcolor{ptband}
\textbf{Component} & \textbf{SFT} & \textbf{DPO} & \textbf{GRPO} & \textbf{PPO} & \textbf{Transfer} & \textbf{What transfers} \\
\midrule
\textbf{Policy learner}: the trained MoE & \yes & \yes & \yes & \yes & Learner mechanisms &
The mechanisms in Parts~\ref{part-parallelization} and
\ref{part-efficient-training} apply to the learner: distributed data parallel
(DDP) synchronization and the sharded optimizer (Section~\ref{sec-ddp-vs-fsdp}),
expert parallel (EP) placement and sync-free rowwise movement
(Section~\ref{sec-expert-parallelism}), pipeline parallel (PP) layer placement
(Section~\ref{sec-pipeline-parallelism}), grouped GEMM
(Section~\ref{sec-grouped-gemm}), MXFP8 (Section~\ref{sec-mxfp8-training}),
recompute (Section~\ref{sec-recompute}), and topology-agnostic checkpoints
(Section~\ref{sec-checkpointing}). Response-only SFT changes the learner's
label tensor and loss denominator while retaining these mechanisms. \\
\arrayrulecolor{black!22}\hline\arrayrulecolor{black}
\textbf{Reference policy}: frozen, forward only & \no & \yes & \yes & \yes & Forward path only &
The forward path only: sync-free dispatch, grouped GEMM, and the MXFP8
forward caches. DPO can
cache reference log-probabilities before updates, as open-instruct does,
avoiding a separate resident reference model. A resident reference adds
parameter storage and forward activations, but no gradient buffers or
optimizer states. \\
\arrayrulecolor{black!22}\hline\arrayrulecolor{black}
\textbf{Critic}: trained value model & \no & \no & \no & \yes & Second learner &
A trained value model. If it uses the same MoE backbone as the policy,
the learner mechanisms apply to both; its actual architecture determines
the added memory. GRPO removes this model by using group-relative
advantages. \\
\arrayrulecolor{black!22}\hline\arrayrulecolor{black}
\textbf{Reward}: verifier or reward model & \no & \no & \ext & \ext & None needed, or forward path only &
A CPU-based checker needs no learner GPU, but verification cost depends on
the task and execution environment. A learned reward model adds a separate
forward-only workload. \\
\arrayrulecolor{black!22}\hline\arrayrulecolor{black}
\textbf{Rollout generation}: inference engine & \no & \no & \ext & \ext & Insight only, not code &
An inference system outside this report's measurements. At small decode
batches, rows per expert can fall below the grouped-GEMM crossover
(Section~\ref{sec-grouped-gemm}); batching and routing determine that count.
Host-visible route metadata (Section~\ref{sec-sync-free-training}), load
imbalance (Section~\ref{sec-routing-load-balancing}), and operand-dependent
GEMM time (Section~\ref{sec-gemm-operand-values}) are relevant costs, but
the training measurements do not quantify them in a serving engine. \\
\arrayrulecolor{black!22}\hline\arrayrulecolor{black}
\textbf{Weight transfer}: learner to inference & \no & \no & \ext & \ext & Substrate only &
The topology-agnostic checkpoints of Section~\ref{sec-checkpointing}
are the substrate, since they already describe every saved chunk by its
place in the global tensor, independent of the EP and PP layout that
produced it; their latency inside a rollout loop is not
measured here. \\
\bottomrule
\end{tabularx}
\endgroup
\caption{\textbf{Components of post-training workloads and their mapping to
the sparse learner.} Transfer categories describe reuse of learner mechanisms,
not demonstrated end-to-end integration with these trainers.
A check mark (\textcolor{ai2accent}{\checkmark}) marks a
component that the workload contains and that this report's mechanisms
serve; an open circle ($\circ$) marks a component the workload contains
outside the learner, which this report does not measure; a
dash marks a component the workload does not have. The reference row assumes
reference-regularized GRPO and PPO. SFT, cached-reference DPO, and GRPO
orchestration follow open-instruct
\citep{allenai2026openinstruct}; the PPO configuration follows the
T\"ulu~3 RLVR recipe %
\citep{lambert2024tulu3}. Both RL examples use vLLM
rollout engines
\citep{kwon2023vllm} and verifiable rewards.  Reinforcement learning from a
learned reward model adds that model to the reward row.}
\label{tab-post-training-components}
\end{table}

\subsection{Response-Only Supervision Uses Label Masking}

Supervised fine-tuning
\citep{wei2022finetuned,ouyang2022training} trains on supervised
input--target examples, such as instructions paired with demonstrations.

Here we consider a response-only objective: the prompt conditions the
prediction, while only response tokens contribute direct loss terms.
Olmo's label-mask path can express this choice by marking prompt labels
as ignored. The DDP train module excludes those labels from the loss
denominator and identifies them through
\texttt{label\_ignore\_index} (default $-100$); it derives per-batch labels
when the data loader does not supply them. A supervised batch
therefore differs from a pretraining batch in the contents of the label tensor
and in the resulting denominator, not in the forward path, the routed expert
computation, or the gradient reduction.

\conceptlead{Masked positions still execute the MoE layer.}
Packed, padded, and interleaved fixed-sequence-length datasets are available in
the data layer, and the choice among them determines how many real tokens each
rank microbatch carries. This matters more under sparse routing than under
dense training. Every position present in the microbatch is routed, dispatched
to its selected experts, computed, and combined, whether or not its label
contributes to the loss. Padding and masked prompt positions consume dispatch
bandwidth, expert GEMM rows, and activation memory even when they have no
direct loss term. Under EP, masked rows still cross the
interconnect during dispatch and combine. Sequence packing reduces this
overhead by reducing padding. Post-training throughput should therefore count
supervised tokens (the useful tokens of
Table~\ref{tab-appendix-metric-definitions}) separately from routed rows
before it is compared with the useful-model rates of
Section~\ref{sec-production-throughput}.

\conceptlead{Instance-level filtering keeps removed tokens in the denominator.}
A separate, optional instance filter, which Olmo's pretraining recipes enable
and its SFT recipes do not, removes whole sequences from the objective but
adds every label position of each removed sequence back into the loss
denominator. Those positions therefore contribute zero loss over a nonzero
count, lowering the reported loss relative to an exact denominator over the
retained instances; the bias depends on how many instances the filter
removes. Loss curves from a filtered and an unfiltered stage are not
directly comparable.

\subsection{Preference Training Scores Two Sequences Against a Reference}

\conceptlead{DPO scores paired sequences under a policy and a reference model.} DPO
computes its loss from a chosen and a rejected
continuation of the same prompt, scored under both the trained policy and a
frozen reference policy %
\citep{rafailov2023dpo}. One logical example becomes
two sequences. Their combined lengths determine the routed token count,
so the effective rank microbatch must be re-derived; the memory headroom
measured for pretraining does not transfer
unchanged. The paired sequences also share a prompt prefix whose routed
computation is performed twice unless the implementation deliberately shares
it. Reference log-probabilities require either a second resident model, which
uses memory alongside the learner, or a precomputation pass whose outputs
are cached and read back. Precomputation removes the resident model's memory
cost but fixes the reference policy for the run.

The cached-reference DPO implementation in open-instruct scores the dataset
with the initial policy before updating it, then reads those scores during
training %
\citep{allenai2026openinstruct,lambert2024tulu3}. This avoids
allocating a second resident reference copy. Its existing trainer uses
Accelerate; a sparse DPO adaptation could instead supply the objective to
the learner described here. The paired-sequence and reference accounting
would still apply, but this report makes no throughput or quality claim for
that adaptation.

\subsection{RLVR Adds Systems Outside the Learner}

\conceptlead{RLVR surrounds the learner with a rollout and verification
system.} Reinforcement Learning with Verifiable Rewards (RLVR)
\citep{lambert2024tulu3} generates rollouts from the current policy, verifies
them against a programmatic checker, and updates the policy from the
resulting rewards. GRPO and the actor-critic PPO configuration considered
here differ inside the learner: PPO trains a value model alongside the
policy, whereas GRPO estimates advantages from a group of rollouts
\citep{schulman2017ppo,shao2024deepseekmath}. A sparse GRPO policy therefore
needs one learner; PPO adds value-model training, with a cost set by the
chosen critic architecture. DeepSeek-R1 is a public example applied to an
MoE policy %
\citep{guo2025deepseekr1}, and HybridFlow makes the boundary
between the rollout system and the learner explicit
\citep{sheng2025hybridflow}. The learner described in this report is one
component of that arrangement. The
additional components---rollout generation at inference precision and batch
shape, reward verification, actor orchestration, and repeated transfer of
updated weights from the learner to the inference engine---are a separate
systems problem whose costs are not addressed by any measurement in this
report.

\conceptlead{Online weight transfer requires a separate latency budget.}
Olmo-core can already export a trained MoE checkpoint into a Hugging
Face model definition, and it contains a generation module and sampling path.
Those provide an offline route from a training checkpoint to an inference
engine for evaluation. Online reinforcement learning
requires repeated, low-latency propagation of updated
expert weights from the training topology into a serving topology whose expert
placement is chosen for inference rather than for training. The
topology-agnostic checkpoint representation of Section~\ref{sec-checkpointing}
is the natural substrate for that transfer, since it already describes every
saved chunk by its place in the global tensor, independent of the EP and PP
layout that produced it. Whether it is fast enough
to sit inside a rollout loop is untested here.

\subsection{Supervised and Routed Token Rates Measure Different Work}

Comparisons across stages should state the base checkpoint, learner
configuration, batch construction, and token accounting. The same number
of supervised tokens can require different numbers of routed tokens because
prompts, padding, and preference pairs still execute the model. Reporting
supervised tokens per second alongside routed tokens per second makes that
difference visible.

\section{Related MoE Systems Make Different Trade-Offs}
\label{sec-related-work}

MoE systems differ in which tensors they move, which weights remain in memory
between uses, how long buffers remain valid, and which numerical
representations determine the optimizer update. We compare those choices
here, including systems that use the same terms---expert parallelism,
distributed optimization, FP8, or overlap---for different implementations.
Table~\ref{tab-related-work-design-axes} summarizes four design axes: weight
materialization, route representation, update precision, and overlap resource
control. Buffer lifetime and replica groups are discussed in
Sections~\ref{subsec-related-route-representations}
and~\ref{subsec-related-replica-groups}.
The comparison covers the mechanism families used in this report rather than
an exhaustive survey of MoE implementations.
Relative performance requires matched experiments; another system's reported
results are not performance measurements for Olmo-core. The quantitative
results discussed here apply to their stated models, software, and hardware;
they do not establish portability or scaling beyond those settings.

\begin{table}[!htb]
\centering
\footnotesize
\begingroup
\definecolor{rwband}{HTML}{EDE6DA}
\renewcommand{\arraystretch}{1.22}
\setlength{\tabcolsep}{4pt}
\begin{tabularx}{\linewidth}{
  >{\raggedright\arraybackslash}p{0.15\linewidth}
  >{\raggedright\arraybackslash}X
  >{\raggedright\arraybackslash}X
  >{\raggedright\arraybackslash}X
  >{\raggedright\arraybackslash}X}
\toprule
\rowcolor{rwband}
\textbf{System} & \textbf{Weight materialization} & \textbf{Route representation} & \textbf{Update tensor precision} & \textbf{Overlap resource control} \\
\midrule
\mbox{Olmo-core} DDP (this report) & Rank-local compute weights stay
resident across microbatches; experts partitioned by EP & Fixed-capacity,
route-indexed rowwise movement with device-side agreement inside a node;
optional DeepEP v2 across nodes; no host split
lists & FP32 main weights, optimizer states, and gradients are the source of truth; BF16 and
MXFP8 are derived caches and payloads & Phase-sequential by default; wave
and two-batch overlap measured and not adopted \\
\arrayrulecolor{black!22}\hline\arrayrulecolor{black}
\mbox{Megatron-Core} MoE & Distributed optimizer: resident local compute
weights; FSDP is a separate sharded-weight option & Dispatch as a backend choice,
collective and optimized, drop-and-pad or dropless, including a sync-free
dropless path & Configurable; example:
FP32 main weights and gradients, BF16 moments; FP32 update arithmetic & Hiding
conditioned on sufficient independent work; SM carve-out with a reported GEMM
cost \\
\arrayrulecolor{black!22}\hline\arrayrulecolor{black}
PyTorch FSDP & Full-reshard policy: gather weights before forward and backward;
release gathered copies after each & Not MoE-specific & FSDP2: original-dtype
parameter shards for optimizer; compute and reduction dtypes configurable
& Parameter prefetch; not an MoE dispatch concern \\
\arrayrulecolor{black!22}\hline\arrayrulecolor{black}
\mbox{DeepSpeed-MoE} & Gather/retention policy not specified in cited MoE paper
& MoE-specific collectives at scale & Main-weight, gradient, and moment dtypes
not specified in cited MoE paper & Not its emphasis \\
\arrayrulecolor{black!22}\hline\arrayrulecolor{black}
MegaBlocks & Gather/retention policy not specified in cited paper
& Block-sparse expert kernels keep
every assignment without capacity padding & Megatron-LM mixed precision;
update-tensor dtypes not specified in cited paper & Not its
emphasis \\
\arrayrulecolor{black!22}\hline\arrayrulecolor{black}
\mbox{DeepSeek-V3} and DeepEP & V3 training: ZeRO-1 retains full local compute
weights; experts partitioned by EP & V3 training: optimized
dispatch and combine kernels; DeepEP v2 lowers communication's SM footprint &
V3 training: FP32 master weights and accumulation gradients; BF16 AdamW moments
& V3 training: communication SMs assigned
explicitly; cache and compute interference acknowledged \\
\bottomrule
\end{tabularx}
\endgroup
\caption{\textbf{Design choices in related MoE systems.} Cells are taken from each system's
own account in the source versions cited in this section; none is a measured comparison.  ``Not its
emphasis'' marks an axis that the cited description does not treat as a
design choice. Weight materialization describes compute-weight gather/retention,
and update precision refers to optimization tensors under the stated policy;
in the combined DeepSeek row, ``V3 training'' marks the DeepSeek-V3 training
report and ``DeepEP v2'' the DeepEP library.}
\label{tab-related-work-design-axes}
\end{table}

\subsection{Conditional Compute Shifts Work to Systems}

Sparse activation separates stored expert parameter size from per-token computation.
Sparsely gated MoE layers established this construction
\citep{shazeer2017outrageously}; Switch Transformers simplified routing to
top-1 %
\citep{fedus2022switch}; GShard demonstrated automatic sharding of
conditional computation %
\citep{lepikhin2021gshard}; and GLaM applied sparse
activation to large-scale language models %
\citep{du2022glam}.

Inactive experts still contribute parameter, optimizer, and checkpoint
storage, and selected tokens must reach the ranks holding their experts.
Conditional computation reduces active compute while adding expert
placement, token movement, and distinct gradient-reduction and optimizer
requirements.

\subsection{Token and Weight Movement Costs Differ}

PyTorch FSDP is a general answer to model-memory pressure. It shards
parameters, gradients, and optimizer state and materializes parameters for
computation %
\citep{zhao2023pytorchfsdp,pytorch2026fsdp2docs}. Large dense matrix multiplications can
amortize those gather and reshard operations. For routed experts, however, the stored pool may
be much larger than the weights used by any token. Olmo-core's DDP path keeps
assigned weights in device memory across microbatches to avoid repeatedly
materializing those expert weights.

ZeRO established the broader optimizer-state and gradient-partitioning design
space %
\citep{rajbhandari2020zero}. DeepSpeed-MoE then combined these ideas with
expert parallelism and MoE-specific communication at scale
\citep{rajbhandari2022deepspeedmoe}. Olmo's distributed optimizer belongs to
this family, while its step ordering and multi-group reducer are specific to
the implementation described here.

Keeping local compute weights materialized increases the persistent
training-state footprint, so FSDP can remain useful for memory-constrained
MoEs. EP can reduce that footprint
by partitioning experts, but then selected token rows must travel to the ranks
that hold them. How long full weights remain materialized and where experts are placed are
separate decisions, which can in principle be composed in several ways. The
current measurements do not fully isolate the cost of keeping assigned weights
materialized from expert placement, so they establish neither a universal
wall-time advantage nor a universal memory advantage for resident weights over
full-reshard FSDP.

Megatron-Core's parallel folding allows dense and sparse portions of a layer
to use different parallel mappings. Its MoE technical report
\citep{nvidia2026megatroncoremoe} describes this design in Megatron-LM
\citep{nvidia2026megatronlm}, together with its interactions with dispatch,
expert compute, load balance, memory, distributed optimization,
checkpointing, reduced precision, and overlap.

Olmo-core represents the same dense\textendash{}sparse distinction with a
dense mesh and an MoE mesh with EP-DP and EP-MP axes. Comparing these mappings
requires following their effects on gradient reduction, optimizer state,
runtime caches, and checkpoint views.

\subsection{Route Representations Trade Compactness for Regularity}
\label{subsec-related-route-representations}

A routed layer must represent a data-dependent number of token assignments.
One design family packs tokens and launches variable-size collectives described
by per-rank counts. This represents the accepted payload
compactly, but a host-facing collective or grouped-GEMM interface can make GPU
route counts part of the CPU launch dependency. Another family reserves bounded
capacity and keeps a GPU-side map from each accepted route to its
destination. That removes the split-list dependency, but requires capacity,
route metadata, agreement, and a precisely defined overflow policy.

Tutel studies adaptive MoE dispatch and parallel placement across different
cluster and model configurations %
\citep{hwang2023tutel}. MegaBlocks instead
uses block-sparse kernels to preserve token assignments without padding every
expert to the same capacity %
\citep{gale2023megablocks}. These systems illustrate
that token layout, communication, and expert-kernel shape must be considered
together; the collective's name alone determines none of their
costs.

Megatron-Core presents token dispatch as a first-class backend choice, including
collective and optimized alternatives
\citep{nvidia2026megatroncoremoe,nvidia2026megatronlm}. The Megatron-Core MoE
report also describes a sync-free dropless path, in which device-initiated grouped GEMM and
dispatch into preallocated upper-bound buffers remove the host synchronization
on token counts. Olmo-core's rowwise
backend uses fixed-capacity route-indexed movement
and GPU-side agreement, with communication payloads in preallocated symmetric
buffers managed by the rowwise runtime. DeepEP, which Olmo-core can also select
as its inter-node EP backend (Section~\ref{subsec-ep-deepep-v2}), supplies
another optimized dispatch design and, in its v2 path,
explicitly reduces the SM footprint devoted to communication
\citep{zhao2025deepep}.

A compact variable payload may
expose a host dependency, while a bounded payload may reserve unused rows or
drop overflow; an optimized GPU transport can also compete with expert
compute for device resources. Comparing these representations requires holding accepted
work and expert shape fixed, then accounting for host gaps, bytes, drops,
capacity, and maximum-rank time.

Dynamic shape also raises a buffer-reuse question that transport APIs alone do
not answer. A dispatched activation saved for Wgrad may outlive the call that
moved it, especially under recomputation or a pipeline schedule. Olmo makes
that dependency explicit with autograd-visible leases on symmetric payloads
(Section~\ref{subsec-rowwise-ep-buffer-reuse}).
Keeping more dispatched activations for backward requires more simultaneous
symm buffers. Reusing a buffer sooner saves memory, but is safe only after every consumer
operator of its previous contents has finished.

\subsection{Placement Determines Replica Groups}
\label{subsec-related-replica-groups}

Once experts are partitioned, a local model part contains different kinds of
replication. Dense parameters may agree over the folded dense group, while a
local expert shard agrees only with copies of that shard. A system can encode
that distinction in module decomposition, parallel mappings, reducer metadata,
optimizer state, or some combination. The choice changes both the stored
tensors and the gradient traffic for each rank.

Megatron-Core includes distributed optimization and checkpointing inside its
MoE system boundary %
\citep{nvidia2026megatroncoremoe}. Olmo-core uses a
per-parameter process-group selector in its DDP reducer, carries the same
parameter families into high-precision optimizer state, and exposes
topology-independent checkpoint views. This path saves without assembling any
full expert tensor on one GPU; in one controlled MXFP8 development run of
Section~\ref{sec-checkpointing}, save-time peak memory stayed at its pre-save
values (a save-side measurement only).

GShard and DeepSpeed-MoE likewise couple expert placement to the process groups
that synchronize or shard model state
\citep{lepikhin2021gshard,rajbhandari2022deepspeedmoe}. Like Olmo-core, these systems derive their synchronization and sharding
groups from the chosen parallel placement; Olmo-core differs in the
particular reducer, optimizer-state, and checkpoint interfaces described
here.

\subsection{Conversion Costs and Communication\textendash{}Compute Interference}

Reduced precision lowers arithmetic, communication, or storage cost only by
introducing another encoding of each model value. Megatron-Core treats
reduced precision and recomputation as parts of its integrated MoE performance
problem %
\citep{nvidia2026megatroncoremoe}. In Olmo-core, FP32 main weights,
optimizer states, and accumulated gradients determine the training update.
BF16 compute weights and oriented MXFP8 weight caches are derived from those
main weights; MXFP8 activation payloads are derived runtime representations. Conversion,
cache refresh, transport, memory, and loss behavior are therefore all part of
the BF16-versus-MXFP8 comparison; raw GEMM time alone does not characterize the
training trade-off.

FP8 training and the later OCP microscaling formats provide the numerical and
hardware context for this design
\citep{micikevicius2022fp8,rouhani2023microscaling,ocpMXFP8spec}, and
DeepSeek-V3 is the prominent production example of FP8 training with
fine-grained block scaling %
\citep{deepseekai2024deepseekv3}. The
Olmo-core-specific questions are where quantization occurs, which
representations cross operator or communication boundaries, and whether the
subsequent GEMM or transfer saves more time than the conversion takes.

DeepSeek-V3 and Megatron-Core issue token
movement beside independent expert computation, but both expose the resource
budget behind the overlap: DeepSeek-V3 assigns communication SMs and discusses
cache/compute interference, while Megatron-Core conditions hiding on sufficient
independent work and reports a GEMM cost from its SM carve-out
\citep{deepseekai2024deepseekv3,nvidia2026megatroncoremoe}. DeepEP v2 lowers
communication's SM footprint %
\citep{zhao2025deepep}; fewer communication SMs
do not guarantee that concurrent kernels retain standalone
speed.

COMET makes the same concern explicit: coarse waves and separate-stream
scheduling can reduce expert-compute efficiency, motivating finer-grained
fused resource control %
\citep{zhang2025comet}. A broader distributed-training
study observes compute stretch under communication while still finding overlap
beneficial on average in its own suite %
\citep{lee2025characterizing}. These
results depend on how resources are allocated for the measured shape,
transport, and platform.

Lancet schedules MoE communication and computation jointly, while
MegaScale-MoE combines several parallel and communication optimizations for
large training runs %
\citep{jiang2024lancet,jin2025megascale_moe}. Both treat
overlap as a scheduling and resource-allocation problem, with explicit
coordination between communication and compute.

Olmo evaluates overlap in two separate studies. A model-level two-batch overlap (TBO) study
evaluates two lanes in flight. A separate same-batch EP-wave study compares
wave-overlapped execution with an unsplit no-wave baseline; it lacks the
wave-sequential control that would separate smaller GEMMs from concurrency
(Section~\ref{subsec-ep-overlap}).
The unsplit baseline is needed because an overlapped timeline can still lose to a
larger, more efficient unsplit GEMM.

\subsection{Expert Placement Connects Token Movement and Updates}

Olmo-core DDP combines resident rank-local compute weights with route-indexed
token movement and replica-specific updates. EP determines which experts a
rank stores and which token rows it receives; the reducer synchronizes each
expert's gradients only with replicas of that expert. Keeping those choices
consistent is what lets the resident-weight design extend from forward
execution to optimizer updates and checkpoint export.

\section{Conclusion}
\label{sec-conclusion}

Olmo-core's DDP stack reduces the parameter movement that slowed our early MoE
baselines and removes the per-block host synchronization of the original MoE
path.  In the preliminary capacity sweep (one eight-GPU node, an eight-layer
model, random routing, one run per point), the composed
DDP and EP stack keeps throughput within a modest margin of the dense reference as
the expert pool grows.  Separate production-scale runs cover a 12.9-billion-parameter
model on 16 GPUs through a 1.2-trillion-parameter model on 512 GPUs, with a
short 2.38-trillion-parameter extension.  Those
headline rates use random routing and different operating configurations;
they demonstrate the stack's supported model sizes and throughput, not a controlled
scaling curve or a time-to-quality result.

\subsection{Reducing Costs That Repeat Across the Step}

Sparse routing limits the compute per token, but it does not by itself
limit weight movement, gradient traffic, or optimizer memory.  Keeping
rank-local model partitions resident removes repeated weight materialization;
EP reduces the expert parameters held by each rank and routes token rows to
them instead.  The trillion-parameter recipes also use sharded optimizer state, MXFP8
activations and EP payloads, and per-layer recompute to fit in device memory.  These changes
affect both how fast a configuration runs and whether it fits at all.

An operation's total cost scales with how often it runs. Routing runs for
every token in every MoE layer, buffer allocation and reuse must accommodate
each microbatch, and kernel launches repeat on every rank.
In the original MoE path, all-to-all split sizes and grouped-GEMM group sizes
both required routing counts on the CPU.  Copying those counts back to the
host repeatedly interrupted submission of GPU work.  Device-resident
metadata removes this recurring wait without changing the routing
calculation.

A bandwidth budget can also rule out an optimization without an end-to-end
benchmark. For the
routed block analyzed in the offload study, the estimated activation
production rate is about six times the theoretical host-link bandwidth.
Scheduling can shift when copies occur, but cannot sustain full offload at
that production rate.  This is a bandwidth budget for the stated workload
and link, not a claim that all forms of activation offload are ineffective.

\subsection{What the Measurements Do and Do Not Establish}

Several optimizations looked promising in isolation but offered little or no
end-to-end benefit.  Communication and expert compute can overlap on a
timeline while still slowing each other.  Quantizing attention projections
alone gave only a modest gain, because conversion and layout work offset much
of the GEMM savings; it remains in the adopted recipe, where the MLP and EP
path supplies nearly all of the benefit.  A weight-gradient kernel that
preserves the natural MXFP8 scale layout avoids a conversion, yet in a B200
kernel-level prototype it was slower than two global dequantization kernels
followed by tuned BF16 grouped GEMM.  Component timings explain these outcomes; they cannot replace a
measurement of the full execution path.

Even a kernel benchmark needs more than shapes and dtypes to be
reproducible.  In one B300 run, we measured different GEMM times with identical shapes,
layouts, and precision when the operand values changed.  The effect was measured only at the kernel level, but it changes
how we construct and interpret microbenchmarks.

A related measurement problem arises in load balancing.  The Switch-style
scalar is a cross-term between soft router scores and hard expert counts.
It can decrease as soft score mass shifts even when the requested routes
remain imbalanced.  A lower scalar therefore does not establish better load
balance.  Requested and kept route counts must be inspected directly; the
scalar's value, its gradient, and the resulting routing behavior are
distinct parts of the analysis.

\subsection{Precision, Memory, and Scheduling Interact}

MXFP8 affects more than GEMM time: it reduces EP payloads and selected saved
activations, slightly increases persistent memory through its two weight
caches, and changes the operand layouts required by the GEMMs.
Recomputing a routed block repeats dispatch and combine as well as expert
compute.  Concurrent dispatch and expert kernels share SMs, caches, and
memory bandwidth; the measurements show the resulting slowdown but do not
identify which shared resource causes it.  These interactions explain why the combined recipe must
be measured even when each component has already been benchmarked.

Pipeline scheduling adds a correctness constraint.  Activations from several
microbatches can remain live for backward, along with the communication
payloads they reference. Those buffers can be reused only after the last
consumer has finished, even if their addresses are fixed for graph replay
(Sections~\ref{subsec-rowwise-ep-buffer-reuse} and
\ref{sec-optimization-attempts}).

These dependencies need to be explicit at component interfaces.  A quantizer
must produce a layout the transport and GEMM can consume without avoidable
conversion.  A buffer pool must be sized for the pipeline schedule's maximum
concurrency, not just one forward pass.  Device memory, SM capacity, network
bandwidth, and host submission time all constrain the same training step.
We report unsuccessful optimizations as well as adopted ones because a
small timing gain may not justify the additional scheduling or memory
complexity.

\subsection{What Carries Forward}

The measured crossovers are specific to the hardware and configurations in
this report.  A different fabric, expert shape, or GEMM implementation can
change whether overlap helps.  Faster activation production would make full
offload harder if host-link bandwidth stayed fixed; a different link or a
more selective offload policy changes that calculation.  Reusing a result
requires checking those conditions, including the frequency of each cost
and the maximum number of live microbatches.

The training mechanisms also offer a starting point for the learner in
supervised fine-tuning, preference training, and reinforcement learning.
Those workloads add different combinations of frozen models, generation,
and data movement; the pretraining measurements do not establish their
end-to-end performance.  Routing, expert shapes, and operand values also matter for the surrounding
inference systems; this report makes no claim about inference performance.

The result of this work is a training stack with substantially less repeated
parameter movement, GPU-resident routing metadata, and memory policies that
support trillion-parameter configurations.  Its performance depends on how
these mechanisms are combined.  The configurations, profiles, and negative
results in this report document those dependencies so that subsequent
changes can be evaluated against the complete training step.

\phantomsection
\section*{Acknowledgments}
\addcontentsline{toc}{section}{Acknowledgments}
This material is based upon work supported by the National Science Foundation
under Award No. 2413244.

\clearpage
\appendix
\part*{Appendices}
\addcontentsline{toc}{part}{Appendices}
\section{Kernel Interfaces and Profiling Reference}
\label{app-kernel-interface-reference}

This appendix documents the interfaces and profiling methods used by the
expert-parallel, grouped-GEMM, and MXFP8 implementations
(Sections~\ref{sec-expert-parallelism}, \ref{sec-grouped-gemm},
and~\ref{sec-mxfp8-training}). It records the operation boundaries, payload
requirements, and measurements needed to reproduce and maintain these paths.

\subsection{Rowwise Dispatch and Combine Use PUT/GET}

The rowwise operation stack has four conceptual levels: model-level MoE phase,
Python autograd function, Python C++/CUDA extension wrapper, and CUDA kernel
using NVSHMEM PUT/GET primitives %
\citep{nvidiaNVSHMEMUsing,nvidia2026nvshmemmemory}.

The phase name does not determine the transport primitive; dispatch and
combine each use PUT in one direction and GET in the other:

\begin{center}
\small
\begin{tabularx}{\linewidth}{lcX}
\toprule
\textbf{MoE phase} & \textbf{Transport} & \textbf{Why} \\
\midrule
Dispatch forward & PUT & Source-token rank holds the input row and pushes it to the rank that stores the selected expert. \\
Combine forward & GET & Source-token rank produces the combined output and pulls expert rows back. \\
Combine backward & PUT & Source-token rank pushes weighted output gradients to expert-output gradients. \\
Dispatch backward & GET & Source-token rank pulls route gradients and sums them into token gradients. \\
\bottomrule
\end{tabularx}
\end{center}

The autograd node names below are private implementation classes and change
with the code; the wrapper names are the stable interface. The BF16 separate
autograd nodes are:

\begin{center}
\small
\begin{tabularx}{\linewidth}{lXX}
\toprule
\textbf{Autograd node} & \textbf{Forward wrapper} & \textbf{Backward wrapper} \\
\midrule
\texttt{\_DispatchRowwiseAutograd} & \texttt{rowwise\_dispatch\_put} & \texttt{rowwise\_combine\_get} \\
\texttt{\_RowwiseCombineWeightedAutograd} & \texttt{rowwise\_combine\_get} & \texttt{rowwise\_dispatch\_put} \\
\bottomrule
\end{tabularx}
\end{center}

The FP8 separate autograd nodes are:

\begin{center}
\small
\begin{tabularx}{\linewidth}{lXX}
\toprule
\textbf{Autograd node} & \textbf{Forward wrapper} & \textbf{Backward wrapper} \\
\midrule
\texttt{\_DispatchRowwiseFP8Autograd} & \texttt{rowwise\_dispatch\_put\_scaled} & \texttt{rowwise\_combine\_get\_scaled} \\
\texttt{\_RowwiseCombineWeightedFP8Autograd} & \texttt{rowwise\_combine\_get\_scaled} & \texttt{rowwise\_dispatch\_put\_scaled} \\
\bottomrule
\end{tabularx}
\end{center}

The fused FP8 node, \texttt{\_RowwiseFP8DispatchExpertsCombineAutograd}, exposes
the whole routed expert region as one autograd node. Its forward half covers
scaled rowwise dispatch, up/gate scaled grouped GEMM, SwiGLU, down scaled
grouped GEMM, and scaled rowwise combine. Its backward half runs the adjoint
sequence: combine backward through scaled dispatch, down dgrad/wgrad, SwiGLU
backward, up/gate dgrad/wgrad, and dispatch backward through scaled combine.

\needspace{10\baselineskip}
\subsection{Rowwise Wrappers Enforce Runtime Requirements}

The main Python wrappers are:

\begin{center}
\small
\begin{tabularx}{\linewidth}{lX}
\toprule
\textbf{Wrapper} & \textbf{Requirement} \\
\midrule
\texttt{rowwise\_dispatch\_put} & Push rows from local \texttt{input} to peer \texttt{out} according to \texttt{dst\_ranks} and \texttt{dst\_rows}; optional \texttt{probs} applies route weighting. \\
\arrayrulecolor{black!22}\hline\arrayrulecolor{black}
\texttt{rowwise\_dispatch\_put\_scaled} & Quantize high-precision rows to MXFP8 qdata/scales, then PUT qdata and scales into separate peer buffers. \\
\arrayrulecolor{black!22}\hline\arrayrulecolor{black}
\texttt{rowwise\_combine\_get} & GET remote expert rows into local route scratch and reduce \texttt{[N, K, D]} into \texttt{[N, D]}, optionally weighted by route probabilities. \\
\arrayrulecolor{black!22}\hline\arrayrulecolor{black}
\texttt{rowwise\_combine\_get\_scaled} & GET qdata and scales, dequantize/reduce locally, and optionally save gathered routes for router-probability gradients. \\
\arrayrulecolor{black!22}\hline\arrayrulecolor{black}
\texttt{rowwise\_gather\_get} & One-to-one GET helper used by scaled combine and other route-gather cases. \\
\arrayrulecolor{black!22}\hline\arrayrulecolor{black}
\texttt{rowwise\_combine\_get\_fused} & BF16/half extension path that combines GET and reduction in one CUDA kernel. \\
\bottomrule
\end{tabularx}
\end{center}

The regular path materializes two physical route tensors,
\texttt{dst\_ranks[N, K]} and \texttt{dst\_rows[N, K]}, where \texttt{N} is
the number of local token rows (the $T$ of
Section~\ref{subsec-rowwise-ep-motivation}) and \texttt{K} is top-$K$. Combine wrappers may
pass those same tensors under source-oriented argument names; there are not
four independent route maps.

Invalid routes use negative rank or row values. PUT kernels skip invalid routes. BF16 combine can zero-fill invalid gathered rows. FP8 scaled combine masks invalid routes before reduction.

The usual barrier defaults are:

\begin{center}
\small
\begin{tabularx}{\linewidth}{lXX}
\toprule
\textbf{Operation} & \textbf{Pre-barrier} & \textbf{Post-barrier} \\
\midrule
Dispatch PUT & false & true \\
Combine GET & true & false \\
\bottomrule
\end{tabularx}
\end{center}

Call sites may override these defaults when TBO, buffer reuse, or lease release order requires a different synchronization point.

\subsection{CUDA/NVSHMEM Kernels Operate per Route}

The rowwise CUDA implementation is route-oriented. A route id is
\texttt{token\_row * top\_k + k}.

The important CUDA kernels are:

\begin{center}
\small
\begin{tabularx}{\linewidth}{lXX}
\toprule
\textbf{CUDA kernel} & \textbf{Transport primitive} & \textbf{Role} \\
\midrule
\texttt{dispatchRowsPut} & \texttt{nvshmemx\_putmem\_warp} & Unweighted row dispatch PUT. \\
\texttt{dispatchRowsPutWeighted} & \texttt{nvshmemx\_putmem\_warp} & Weighted row PUT when route probabilities are supplied. \\
\texttt{gatherRowsGet} & \texttt{nvshmemx\_getmem\_warp} & Gather remote rows into local route scratch. \\
\texttt{combineRowsReduceKernel} & local CUDA & Reduce gathered \texttt{[N, K, D]} into \texttt{[N, D]}. \\
\texttt{combineRowsGetKernel} & \texttt{nvshmemx\_getmem\_warp} & Optional fused BF16/half GET plus reduce path. \\
\bottomrule
\end{tabularx}
\end{center}

For dispatch, each route derives the source row, peer rank, and destination row,
then PUTs \texttt{input[src\_row]} into \texttt{out[dst\_row]} on the peer. For
gather/combine, each route derives the peer and source row, GETs
\texttt{expert\_out[src\_row]} into local route scratch, and reduces gathered
\texttt{[token\_row, :, :]} routes into \texttt{output[token\_row, :]}.

\subsection{FP8 Transport Requires Compatible Layouts}

Rowwise FP8 communication carries two payload families: MXFP8 qdata and MXFP8
row scales, usually with block size 32. BF16 dispatch writes
\texttt{dispatch\_out[capacity\_rows, hidden]}, while FP8 dispatch writes
\texttt{dispatch\_out\_q[capacity\_rows, hidden]} and
\texttt{dispatch\_out\_scales[capacity\_rows, hidden / block\_size]}.

The exact scale layout depends on the consuming grouped GEMM path, not just
the numerical format. Transformer Engine, for example, uses different scale
layouts for communication and GEMM and swizzles between them
\citep{nvidiaTransformerEngineMXFP8}.
In Olmo's rowwise path, the received qdata and scales are passed to grouped
GEMM through its prequantized-LHS interface. This reuses the quantized payload;
the consumer still converts unblocked row scales to its blocked GEMM layout.

\subsection{Blackwell MXFP8 Uses Block-Scaled Tensor Core MMA}
\label{app-blackwell-tcgen05}

Three related terms describe different levels of the MXFP8 compute path. Olmo
calls PyTorch's \texttt{scaled\_mm} operator and requests a BF16 result. The
selected device kernel implements a tiled general matrix multiplication
(GEMM), including operand movement, a Tensor Core main loop, and an output
epilogue. Within that main loop, the Blackwell instruction-level primitive is
the fifth-generation Tensor Core matrix multiply--accumulate (MMA), exposed by
the PTX \texttt{tcgen05.mma} family
\citep{nvidiaBlackwellSM100Gemm,nvidiaTCGen05MMA}.

\begin{table}[!htb]
\centering
\small
\begin{tabularx}{\linewidth}{lX}
\toprule
\textbf{Level} & \textbf{Term and responsibility} \\
\midrule
PyTorch operator & \texttt{scaled\_mm}: accepts qdata and scale tensors, selects a backend, and returns the requested output dtype. \\
\arrayrulecolor{black!22}\hline\arrayrulecolor{black}
Device kernel & Block-scaled GEMM: tiles the matrices, moves operands and scales, runs the reduction over $K$, and applies the output epilogue. \\
\arrayrulecolor{black!22}\hline\arrayrulecolor{black}
Tensor Core instruction & \texttt{tcgen05.mma}: computes one matrix tile update into an accumulator. \\
\bottomrule
\end{tabularx}
\caption{\textbf{Operator, kernel, and instruction levels of MXFP8 GEMM.} The operator,
kernel, and instruction names refer to different levels of the same MXFP8
matrix multiplication.}
\label{tab-blackwell-mxfp8-levels}
\end{table}

For BF16 operands, the corresponding Blackwell instruction family uses the
\texttt{kind::f16} modifier. Native MXFP8 uses
\texttt{kind::mxf8f6f4} together with \texttt{block\_scale}; its full PTX
family includes the choice of one- or two-CTA execution:

\begin{center}
\small
\texttt{tcgen05.mma.cta\_group::\{1|2\}.kind::mxf8f6f4.block\_scale}.
\end{center}

For the MXFP8 case, the instruction applies the two block-scale tensors during
the reduction. Its numerical operation is

\begin{equation}
D_{ij} = C_{ij} + \sum_{k=0}^{K-1}
\left(S^{(A)}_{i,\lfloor k/32\rfloor} A_{ik}\right)
\left(S^{(B)}_{\lfloor k/32\rfloor,j} B_{kj}\right),
\end{equation}

where each E8M0 entry in $S^{(A)}$ or $S^{(B)}$ scales a block of 32 consecutive E4M3
values along the GEMM-$K$ dimension. The logical scale grids shown in
Figure~\ref{fig-mxfp8-bf16-native-gemm} are the operator-facing view. The
Blackwell main loop stages qdata in shared memory, copies scale factors into
Tensor Memory, and keeps the FP32 accumulator in Tensor Memory; the scale
grids must therefore be swizzled into the blocked physical layout expected by
the kernel
\citep{nvidiaTCGen05MMA}.

The raw block-scaled MMA does not directly define a BF16 result. It updates an
FP32 accumulator in Tensor Memory. The GEMM epilogue reads that accumulator
and converts it to the output dtype requested by \texttt{scaled\_mm}. In the
current Olmo path, that requested dtype is BF16.

\subsection{Availability of Device-Side Grouped GEMM}
\label{app-device-grouped-gemm-availability}

The PyTorch source snapshot audited for this report distinguishes the public
grouped-matrix-multiplication operation from the backend selected at runtime.
For CUDA, the architecture gate used by the device-offset fast path admits
only compute-capability major versions 9 (SM90) and 10 (SM100 family)
\citep{pytorchGroupedBlasSource}.
Table~\ref{tab-device-grouped-gemm-availability} summarizes the resulting
behavior.

\begin{table}[!htb]
\centering
\small
\begin{tabularx}{\linewidth}{lXX}
\toprule
\textbf{PyTorch entry point} & \textbf{CC major 9 or 10} &
\textbf{Other CUDA architectures} \\
\midrule
BF16 \texttt{torch.\_grouped\_mm} & Device-offset fast path; no offset
copy to the host & Host-loop fallback; offsets are copied to the CPU before
the per-group GEMMs \\
\arrayrulecolor{black!22}\hline\arrayrulecolor{black}
Scaled \texttt{torch.\_scaled\_grouped\_mm} & Architecture gate passes &
Rejected by the architecture gate rather than routed to the host-loop
fallback \\
\bottomrule
\end{tabularx}
\caption{\textbf{Architecture availability in the audited PyTorch snapshot.}
The gate checks the compute-capability major version, so B300 at 10.3 follows
the major-10 path, while SM120 at 12.0 does not.}
\label{tab-device-grouped-gemm-availability}
\end{table}

The BF16 fallback makes the behavioral difference explicit: it calls
\texttt{offs.cpu()} and then launches one matrix multiplication per group
\citep{pytorchGroupedMMFallbackSource}.  The fast CUDA implementation, by
contrast, instantiates CUTLASS SM90 and SM100-family grouped kernels.  This is
a boundary of the current PyTorch integration, not evidence that SM120 lacks
Tensor Core matrix multiplication.  The CUTLASS source bundled with the same
PyTorch snapshot contains an SM120 grouped-GEMM example
\citep{nvidiaCUTLASSSM120GroupedGEMM}, and Table~\ref{tab-mxfp8-hardware}
records the SM120 \texttt{mma.sync} instruction family.  In particular, the
restriction is not a lack of ``WMMA'' support: WMMA is a
programming interface, while the underlying matrix-instruction family differs
across GPU architectures.

\subsection{Common Grouped-GEMM Shapes in Training}
\label{app-common-grouped-gemm-shapes}

The matrix row count seen by grouped GEMM follows the rank-local microbatch
and the routing multiplicity.  For $B_\mu$ sequences of length $L$ and
top-$K$ routing, balanced expert parallelism gives

\begin{equation}
M_{\mathrm{rank}} \simeq B_\mu L K,
\qquad
\overline{M}_e = \frac{M_{\mathrm{rank}}}{N_{\mathrm{local}}},
\qquad
N_{\mathrm{local}} = \frac{N}{M_E}.
\label{eq-common-grouped-gemm-shapes}
\end{equation}

The EP degree does not divide $M_{\mathrm{rank}}$: an EP-MP group of size
$M_E$ contributes that many source microbatches
before distributing the routed rows over the same number of destination
ranks.  Table~\ref{tab-common-grouped-gemm-shapes} shows the resulting shapes
for representative training configurations.  Shared-expert computation is
separate, and load imbalance can move an individual rank away from the
balanced expectation.

\begin{table}[!htb]
\centering
\begingroup
\scriptsize
\setlength{\tabcolsep}{2.7pt}
\renewcommand{\arraystretch}{1.18}
\begin{tabular}{@{}lrrrrrrrrrr@{}}
\toprule
& \multicolumn{3}{c}{\textbf{Rank-local microbatch}}
& \multicolumn{3}{c}{\textbf{Expert layout}}
& \multicolumn{2}{c}{\textbf{Projection}}
& \multicolumn{2}{c}{\textbf{Routed rows}} \\
\cmidrule(lr){2-4}\cmidrule(lr){5-7}\cmidrule(lr){8-9}\cmidrule(l){10-11}
\textbf{Scale} & $B_\mu$ & $L$ & $K$
& $M_E$ & $N$ & $N_{\mathrm{local}}$
& $d$ & $h$ & $M_{\mathrm{rank}}$ & $\overline{M}_e$ \\
\midrule
Small-2  & 3 & 8K & 4 & 8  & 64  & 8  & 2048 & 2560  & 96K  & 12K \\
Small-3  & 4 & 4K & 8 & 8  & 64  & 8  & 2048 & 2560  & 128K & 16K \\
Small-4  & 3 & 8K & 8 & 8  & 128 & 16 & 1536 & 1536  & 192K & 12K \\
Medium-2 & 3 & 8K & 4 & 8  & 128 & 16 & 3584 & 4096  & 96K  & 6K \\
Large-2  & 1 & 8K & 4 & 8  & 128 & 16 & 6144 & 8192  & 32K  & 2K \\
Ultra-2  & 2 & 8K & 4 & 32 & 128 & 4  & 8192 & 10240 & 64K  & 16K \\
\bottomrule
\end{tabular}
\endgroup
\caption{\textbf{Common grouped-GEMM shapes in training.}  $d$ is the model
width and $h$ is the routed-expert intermediate width.  $M_{\mathrm{rank}}$ is the
expected valid routed-row count received by one EP rank, and
$\overline{M}_e$ is the corresponding mean per local expert.  A
capacity-padded BMM at factor 1.25 executes approximately
$1.25\overline{M}_e$ rows per expert.  Total routed rows alone do not define
the complete GEMM shape because local-expert count and projection dimensions
also vary. Row names indicate approximate model scale only; these representative shapes are not the measured configurations of
Section~\ref{sec-production-throughput}.}
\label{tab-common-grouped-gemm-shapes}
\end{table}

The smaller configurations in the table use expert widths between 1536 and
2560 and route 96K--192K rows per rank.  To complement the $d=h=4096$ result
in the main text, we repeat the same isolated BF16 projection benchmark with
$d=h=2048$, 16 local experts, and $M=4\mathrm{K}$ through 256K.  The BMM path
again executes exactly 1.25$\times$ per-expert capacity.  Its regular kernel
is faster at 4K and 8K total rows; grouped GEMM becomes faster between 8K and
16K and remains faster through the rest of the sweep.  At 128K total rows,
BMM takes 15.4\% longer.

\begin{figure}[!htb]
  \centering
  \includegraphics[width=\linewidth]{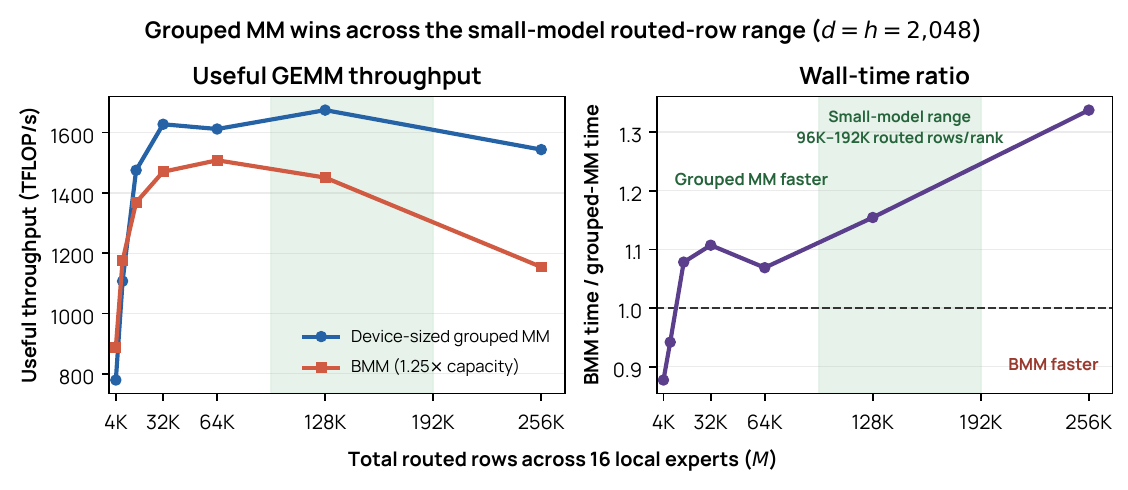}
  \caption{\textbf{Grouped GEMM also wins across the small-model training
  range.}  This one-B300 experiment fixes 16 local experts and a square BF16
  projection with $d=h=2048$.  The green band marks the 96K--192K routed-row
  range represented by the small configurations in
  Table~\ref{tab-common-grouped-gemm-shapes}; their exact widths and local
  expert counts vary, so the band is an operating-range guide rather than an
  exact reconstruction of every model.  BMM reserves 1.25$\times$ capacity
  per expert, while grouped GEMM executes only valid rows.  Each point averages
  20 queued CUDA-event samples after 10 warm-ups.}
  \label{fig-grouped-gemm-capacity-sweep-dh2048}
\end{figure}

\FloatBarrier
\subsection{Equal Compute Can Require More HBM Traffic}
\label{app-expert-roofline}

Small expert batches perform less arithmetic per weight load. A roofline
model makes this limit explicit before routing, synchronization, or kernel
scheduling overhead is counted. For one GPU operation, let $F_g$ be its
arithmetic work, $Q_g$ its high-bandwidth memory (HBM) traffic, $T_g$ its
elapsed time, and $I=F_g/Q_g$ its arithmetic intensity. With compute ceiling
$C_{\mathrm{peak}}$ and HBM bandwidth ceiling $B_{\mathrm{HBM}}$, the idealized
roofline gives %
\citep{austin2025rooflines}
\begin{equation}
T_g \geq \max\!\left(\frac{F_g}{C_{\mathrm{peak}}},
                     \frac{Q_g}{B_{\mathrm{HBM}}}\right),
\qquad
\frac{F_g}{T_g}\leq\min(C_{\mathrm{peak}},\,I B_{\mathrm{HBM}}).
\label{eq-intro-roofline}
\end{equation}
For a BF16 expert projection
$[M_e,d_{\mathrm{in}}][d_{\mathrm{in}},d_{\mathrm{out}}]
\to[M_e,d_{\mathrm{out}}]$, $M_e$ is the token-row count and
$d_{\mathrm{in}},d_{\mathrm{out}}$ are the input and output widths.
For this forward projection, reading both operands once from HBM and
writing a BF16 result gives
\begin{equation}
I=\frac{2M_e d_{\mathrm{in}}d_{\mathrm{out}}}
        {2(M_e d_{\mathrm{in}}+d_{\mathrm{in}}d_{\mathrm{out}}
           +M_e d_{\mathrm{out}})}.
\label{eq-expert-projection-intensity}
\end{equation}
When $M_e\ll d_{\mathrm{in}},d_{\mathrm{out}}$, this is approximately
$M_e$ FLOP/byte: fewer rows per expert reduce weight reuse. For a stated
BF16 compute reference and HBM bandwidth, the balance point is
\begin{equation}
I_\star=\frac{C_{\mathrm{BF16}}}{B_{\mathrm{HBM}}}
\quad\text{FLOP/byte}.
\label{eq-intro-roofline-ridge}
\end{equation}
Under this HBM traffic model, even a zero-overhead projection below that intensity
cannot reach the compute peak. Cache reuse, tiling, operand precision, and
actual HBM traffic change the bound. This compute--bandwidth balance point is
distinct from the measured grouped-GEMM-versus-BMM crossover in
Section~\ref{subsec-grouped-gemm-crossover}.

\paragraph{Dense and MoE forward GEMMs with equal FLOPs.}
Figure~\ref{fig-dense-moe-roofline} fixes the original token count $T$, input
width $d$, and useful GEMM FLOPs. The dense FFN has intermediate width $Kh$;
the MoE has $N$ experts of intermediate width $h$ and selects $K$ per token.
Assume balanced routes with every expert used, so each expert receives
$M_e=TK/N$ rows. Dense executes the fused up/gate product
$[T,d][d,2Kh]$ followed, after the pointwise activation, by the down product
$[T,Kh][Kh,d]$. Grouped GEMM executes $N$ corresponding pairs with
shapes $[TK/N,d][d,2h]$ and $[TK/N,h][h,d]$. Both perform
\begin{equation}
F_{\mathrm{up}}=4TKdh,\qquad F_{\mathrm{down}}=2TKdh,
\qquad F_{\mathrm{pair}}=6TKdh.
\end{equation}
For BF16 arrays, Table~\ref{tab-dense-moe-roofline-traffic} counts the bytes
read and written by these GEMMs. The MoE inputs are separate,
already-materialized route rows; the writes that created them are outside
this accounting. The down GEMM writes $TK$ output rows before combine,
whereas dense writes $T$; combine is not folded into these GEMM writes.

\begin{table}[!htb]
\centering
\small
\begin{tabular}{lccc}
\toprule
\textbf{HBM traffic category} & \textbf{Dense bytes} & \textbf{MoE bytes} &
\textbf{MoE / dense} \\
\midrule
Weight reads & $6Kdh$ & $6Ndh$ & $N/K$ \\
FFN input reads & $2Td$ & $2TKd$ & $K$ \\
Output writes, before combine & $2Td$ & $2TKd$ & $K$ \\
Up/gate writes + down input reads & $6TKh$ & $6TKh$ & $1$ \\
\bottomrule
\end{tabular}
\caption{\textbf{More HBM traffic for the same forward GEMM FLOPs.}
One-read HBM accounting for BF16 inputs, weights, and outputs. Each ratio
compares a single category, not total HBM traffic. The last row counts only
GEMM-side accesses to hidden activations, not the pointwise activation's
own reads and writes.}
\label{tab-dense-moe-roofline-traffic}
\end{table}

The equal last-row count follows from the matched-compute widths, not from
eliminating MoE intermediates. Dense produces $[T,2Kh]$ up/gate values and
consumes $[T,Kh]$ activated values; the corresponding routed tensors are
$[TK,2h]$ and $[TK,h]$. Each pair has equal element counts. In a grad-enabled
BF16 forward, both paths execute a separate SwiGLU expression between the GEMMs. Its own accesses,
any temporary tensors, and cache effects are outside the GEMM-only model;
tensor shapes alone do not establish measured HBM traffic.

For each layout, separate the HBM bytes of the two GEMMs:
\begin{equation}
\begin{aligned}
Q_{\mathrm{dense,up}} &= 2(Td+2Kdh+2TKh), &
Q_{\mathrm{dense,down}} &= 2(TKh+Khd+Td), \\
Q_{\mathrm{MoE,up}} &= 2(TKd+2Ndh+2TKh), &
Q_{\mathrm{MoE,down}} &= 2(TKh+Nhd+TKd).
\end{aligned}
\end{equation}
The two GEMM passes are sequential, so their modeled times add:
\begin{equation}
\widehat t=\sum_{j\in\{\mathrm{up,down}\}}
\max\!\left(\frac{F_j}{C_{\mathrm{BF16}}},
\frac{Q_j}{B_{\mathrm{HBM}}}\right),\qquad
\widehat R=\frac{F_{\mathrm{pair}}}{\widehat t}.
\label{eq-dense-moe-gemm-pair-roofline}
\end{equation}
The figure plots $\widehat R/C_{\mathrm{BF16}}$ as $T$ varies, with $N=64$,
$K=4$, and $d=h=4096$. It uses the nominal B300 BF16 compute-to-HBM-bandwidth
ratio %
\citep{nvidiaHGXB200spec,nvidia2026hgxb300}; the vertical axis labels the
common reference as peak FLOPs, not an achievable kernel rate. This is a
roofline expressed against token count rather than arithmetic intensity.
The hardware reference is identical; the different HBM byte counts produce
different curves. At small batches, weight reads dominate and MoE reuses each weight
across only $TK/N$ tokens instead of $T$. Larger batches amortize those reads,
and both modeled curves eventually reach the common compute ceiling for
these widths.

\paragraph{The HBM traffic model defines the scope.}
This is a single-GPU forward GEMM-pair comparison, not the complete FFN or
training step. Balanced, all-used routing is an analytical assumption, not
a guarantee for random or learned routes. Equal FLOPs do not imply equal
total parameter size or model quality. The model assumes no cache reuse
between operand arrays and omits tiling-induced rereads; caching or fused
indexed input access can reduce the counted HBM traffic, while rereads can
increase it. Packing, route merging, activation functions, backward,
inter-GPU communication, gradient synchronization, and optimizer work are
excluded. Fusion or pipelining across the two GEMM passes can change their
HBM traffic and timing. The curves are ceilings within these assumptions, not measured
rates or an end-to-end MFU bound. Measured projection efficiency appears in
Section~\ref{subsec-grouped-gemm-crossover}.

\subsection{The Maximum Load Does Not Specify the Expert Histogram}
\label{app-grouped-gemm-histogram-shapes}

\begin{figure}[!htb]
  \centering
  \includegraphics[width=\linewidth]{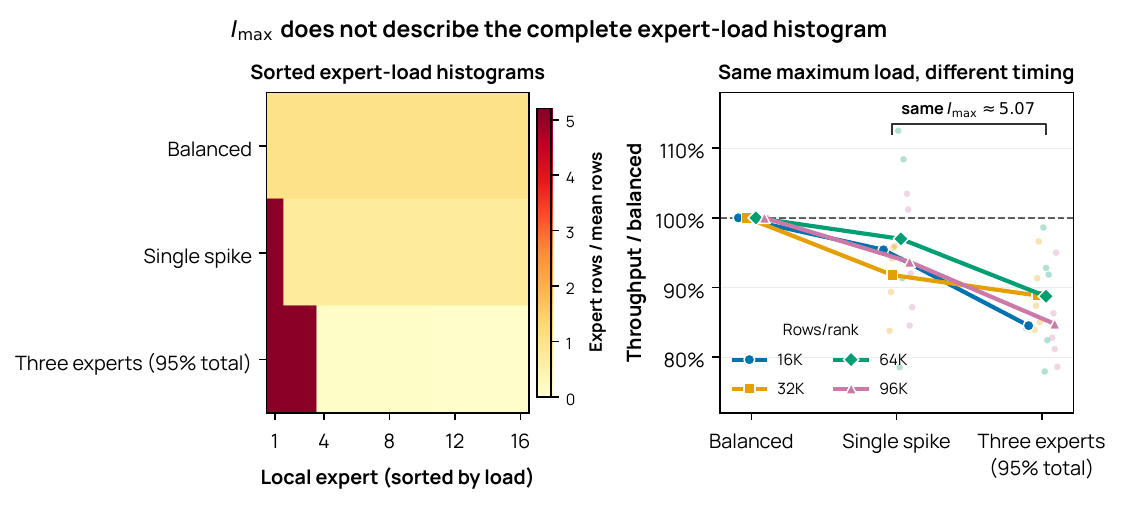}
  \caption{\textbf{The same $I_{\max}$ can describe different grouped GEMM
  workloads.}  \textbf{Left:} sorted expert loads for a representative 64K-row
  case.  \textbf{Right:} rank-local grouped GEMM throughput normalized to the
  balanced assignment in the same process.  Large markers report arithmetic
  means, while small translucent points show five independent runs on the same
  B300.  In each run, let $a$ be the base expert index.  The matched single
  spike overloads expert $a$, while the three-hot assignment overloads experts
  $(a,a+5,a+10)\bmod16$; the five values of $a$ are 0, 3, 6, 9, and 12.  Every run
  averages 20 queued CUDA-event samples after 10 warm-ups.  The benchmark
  isolates one BF16 $d=h=4096$ projection and excludes dispatch,
  communication, and cross-rank waiting.}
  \label{fig-grouped-gemm-histogram-shapes}
\end{figure}

The maximum-to-mean ratio $I_{\max}$ records only the largest expert load.
It does not record how much work is concentrated among the other experts.
Figure~\ref{fig-grouped-gemm-histogram-shapes} constructs a direct
counterexample with 16 local experts.  The matched single-spike assignment
gives one expert 31.67\% of all rows and divides the remainder across 15
experts.  The three-hot assignment gives the same number of rows to each
of three experts, so those three receive approximately 95\% in total and the
other 13 divide the remaining 5\%.  After eight-row alignment, both have
$I_{\max}\approx5.07$, the same total rows, and the same useful FLOPs.

The three-hot assignment has lower mean grouped GEMM throughput at all four
row counts.  Its additional loss relative to the matched single spike ranges
from 3 to 11 percentage points.  The condition ranges overlap at some shapes,
so the experiment does not make $I_{\max}$ or any other single summary a
complete timing model.  It establishes the narrower point: two assignments
with the same maximum load can present different collections of matrix
problems to grouped GEMM and take different amounts of time.

\FloatBarrier
\subsection{Grouped-GEMM Capacity Benchmark Details}
\label{app-grouped-gemm-crossover-details}

The benchmark in Section~\ref{sec-grouped-gemm} uses 16 local experts, $d=h=4096$, and
BF16 operands.  It isolates one square forward projection and compares exact
device-sized grouped GEMM with BMM at
$C=\lceil\,1.25\,\overline{M}_e\,\rceil$ rows per expert. We fix capacity at
1.25 times the mean rather than the observed maximum, so balanced routes
still incur the configured padding. Both paths receive the same
seeded valid activations and expert weights and write into preallocated
outputs.  Routing, communication, activation functions, and backward are
outside the timed region, so the result identifies a kernel crossover rather
than predicting complete expert-MLP time.

The raw per-call timings, validation output, and manifests are retained in
the experiment package bundled with this report.

\subsection{Matched Grouped\textendash{}Dense Projection Benchmark}
\label{app-grouped-dense-matched-projection}

This benchmark compares three mathematically equivalent BF16 projections.  We
initialize one canonical
$W_{\mathrm{dense}}\in\mathbb{R}^{8192\times81920}$ and partition its output
columns in two ways: four expert matrices of shape $8192\times20480$, or eight
expert matrices of shape $8192\times10240$.  Both grouped layouts contain the
exact dense BF16 values.  One seeded input $X$ is physically repeated for each
grouped problem, and sampled output rows match bitwise at every measured $T$.

All three paths execute $2T\times8192\times81920$ useful FLOPs and access the
same number of weight and output elements.  They do not have identical input
traffic: the grouped paths supply $X$ to four or eight GEMMs.  The experiment
isolates this projection and excludes the activation function, down projection,
routing, communication, shared experts, and backward.

We sweep powers of two from $T=1$ through $65536$ in eight paired trials, one
on each of eight B300 GPUs.  Every GPU measures both grouped partitions and
their corresponding dense controls; partition order and backend order are
counterbalanced.  After ten direct warm-up calls, calibration chooses an
adaptive CUDA Graph work unit targeting approximately 100~ms per replay.  Each
backend then receives a three-second pilot and a final batch containing at
least eight seconds of uninterrupted same-path conditioning immediately
followed by at least 1.5 seconds of measurement.  The host submits both phases
behind one device gate before releasing the GPU, with no synchronization or
bookkeeping at the phase boundary.

\begin{table}[!htb]
  \centering
  \small
  \setlength{\tabcolsep}{5pt}
  \begin{tabular}{rccccc}
    \toprule
    $T$ & \textbf{Dense} & \textbf{4 experts} & \textbf{4 / dense} &
      \textbf{8 experts} & \textbf{8 / dense} \\
    \midrule
    1      & 7    & 4    & 0.588 & 4    & 0.620 \\
    16     & 111  & 67   & 0.601 & 68   & 0.608 \\
    128    & 854  & 520  & 0.608 & 548  & 0.643 \\
    256    & 1126 & 1086 & 0.964 & 1085 & 0.963 \\
    512    & 1293 & 1236 & 0.956 & 1230 & 0.951 \\
    1024   & 1381 & 1306 & 0.945 & 1298 & 0.941 \\
    2048   & 1422 & 1356 & 0.953 & 1350 & 0.949 \\
    4096   & 1401 & 1335 & 0.953 & 1331 & 0.950 \\
    8192   & 1416 & 1252 & 0.884 & 1250 & 0.882 \\
    16384  & 1412 & 1186 & 0.840 & 1184 & 0.839 \\
    32768  & 1413 & 1168 & 0.827 & 1169 & 0.827 \\
    65536  & 1411 & 1167 & 0.827 & 1167 & 0.827 \\
    \bottomrule
  \end{tabular}
  \caption{\textbf{Absolute useful throughput for the matched projections, in
  TFLOP/s.}  Each grouped column is the mean over eight paired trials on eight
  B300 GPUs.  Every GPU measures the identical dense control with both grouped
  partitions; the dense column first averages that pair per GPU, then averages
  the eight GPU-level values.  Ratios use the paired dense measurement from the
  corresponding grouped trial. Rows show a subset of the 17 swept token
  counts.}
  \label{tab-grouped-dense-matched-projection}
\end{table}

Both grouped layouts approach the dense projection at $T=256$ and sustain
roughly 94--96\% of dense throughput through $T=4096$.  Their close agreement
shows that dividing the fixed projection into eight half-width GEMMs adds
little cost beyond the four-way partition in this range.  At 32K and 64K, all
paths operate under the software power limit and grouped throughput settles at
approximately 82.7\% of dense, or 1.17 versus 1.41~PFLOP/s.  These oversized
cases diagnose large-matrix behavior;
they are not representative training shapes.

\FloatBarrier
\subsection{Very Large Grouped GEMMs Enter a Power-Limited Clock Regime}
\label{app-grouped-gemm-large-m-power}

The main sweep stops at $M=128\mathrm{K}$ total rows, or $M_e=8\mathrm{K}$
rows per local expert.  We separately stress-tested $M_e=4\mathrm{K}$ through
$32\mathrm{K}$, reaching 512K total rows across 16 experts.  The 16K and 32K
per-expert cases, 256K and 512K rows per rank across 16 local experts, are
deliberately outside realistic Olmo training shapes; they
expose the power-and-clock response behind the large-shape throughput decline.

The diagnostic uses eight counterbalanced trials across four B300 GPUs.  Each
backend and shape receives an uninterrupted three-second pilot followed by an
eight-second conditioning region and a 1.5-second measured region.  Shape and
backend order vary across trials.  The host submitted work at least 25 times
faster than the GPU executed it, so CPU launch gaps were not on the measured
path.  All trials used BF16 operands, with expert weights drawn from a normal
distribution with standard deviation 0.02 and standard-normal activations from
fixed seeds, and grouped GEMM and BMM received identical valid activations and
expert weights.

\begin{figure}[!htb]
  \centering
  \includegraphics[width=\linewidth]{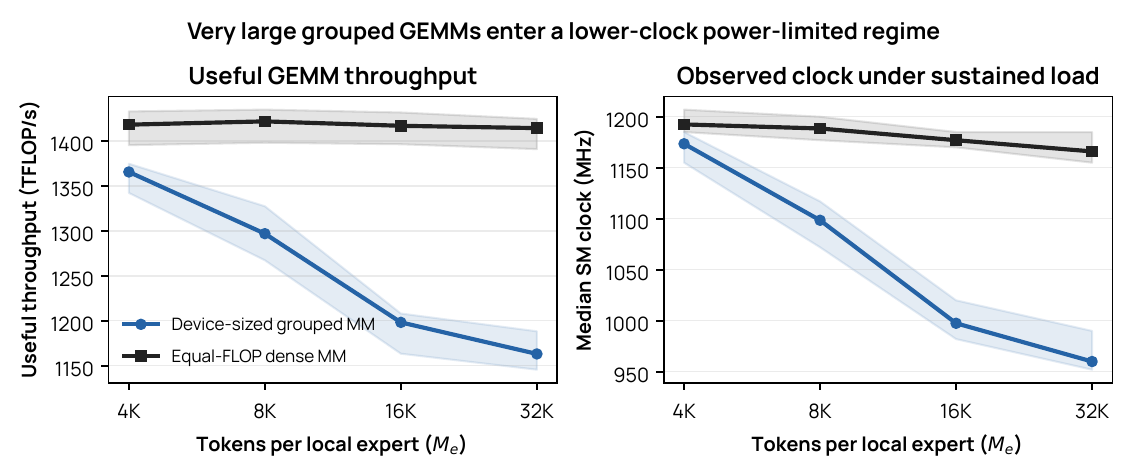}
  \caption{\textbf{The very-large-$M_e$ grouped GEMM slowdown follows SM clock,
  not per-clock efficiency.}  Lines show medians and bands show the full range
  across eight trials.  Both paths execute the same useful FLOPs, although the
  dense control reuses one weight matrix while grouped GEMM reads 16 expert
  matrices.  The software power-cap limiter was active in 99.3\% of telemetry
  samples; no thermal, hardware-brake, or reliability limiter was observed.}
  \label{fig-grouped-gemm-large-m-power}
\end{figure}

Figure~\ref{fig-grouped-gemm-large-m-power} shows the relevant comparison.
From $M_e=4\mathrm{K}$ to $32\mathrm{K}$, median grouped GEMM throughput falls
from 1366 to 1163~TFLOP/s while its median SM clock falls from 1174 to 960~MHz.
Per-trial useful throughput per MHz (each trial's throughput divided by its
mean SM clock) has per-shape medians between 1.16 and 1.19~TFLOP/s/MHz;
dividing the median throughput by the median clock instead gives
1.16--1.21~TFLOP/s/MHz.  The
equal-FLOP dense control remains approximately flat at 1.41--1.42~PFLOP/s,
and the unplotted capacity-BMM control remains approximately flat at
1.21--1.22~PFLOP/s.  Under this sustained load, BMM takes about
6\% longer than grouped GEMM at $M_e=8\mathrm{K}$, compared with 28.9\% in
the short timed runs of Figure~\ref{fig-grouped-gemm-capacity-sweep}, and
it is faster than grouped GEMM at $16\mathrm{K}$ and $32\mathrm{K}$.

Nsight timelines at $M_e=2\mathrm{K}$ and $32\mathrm{K}$, captured in a
separate profiling run of the same 16-expert projection, show the same
CUTLASS grouped-GEMM kernel template.  Together with the nearly constant
throughput per MHz, this rules out a high-$M_e$ dispatch change and provides no
evidence of an algorithmic scaling collapse.  Instead, sustained very-large
grouped GEMMs settle at a lower clock while operating near the 1.1~kW board
power limit.  The precise microarchitectural reason for this lower operating
point is not established.  These extreme-shape timings need to be interpreted
alongside clocks, power, and operand values.

\FloatBarrier

\needspace{16\baselineskip}
\subsection{Grouped Kernels Accept Managed Buffers}

The BF16 grouped GEMM wrapper accepts explicit buffers:

\needspace{13\baselineskip}
\begin{lstlisting}
grouped_mm(
  mat_a,
  mat_b,
  offs=expert_offsets,
  out=optional_forward_output,
  input_grad_out=optional_dgrad_output,
)
\end{lstlisting}

The explicit \texttt{out} and \texttt{input\_grad\_out} buffers matter because expert compute often writes into buffers whose reuse point is chosen by the MoE runtime rather than by the PyTorch allocator.

The MXFP8 grouped GEMM wrapper has two important forms:

\needspace{13\baselineskip}
\begin{lstlisting}
scaled_grouped_mm_q(
  mat_a,
  mat_b,
  offs=expert_offsets,
  prequantized_lhs=optional_lhs_q_scales,
  prequantized_rhs=optional_rhs_q_scales,
  prequantized_rhs_for_dgrad=optional_transposed_rhs_q_scales,
)
\end{lstlisting}

and:

\needspace{13\baselineskip}
\begin{lstlisting}
scaled_grouped_mm_q_fp8_weight(
  mat_a,
  grad_anchor,
  offs=expert_offsets,
  prequantized_lhs=optional_lhs_q_scales,
  prequantized_rhs=required_rhs_q_scales,
  prequantized_rhs_for_dgrad=required_transposed_rhs_q_scales,
  wgrad_sink=optional_fp8_weight_grad_sink,
)
\end{lstlisting}

Forward and dgrad can use the scaled grouped GEMM backend. Wgrad has a different
current policy: the fused rowwise path dequantizes the required operands,
executes BF16 \texttt{F.grouped\_mm}, and accumulates the result into the
FP8-weight gradient buffer. The integrated path has no native MXFP8 grouped
Wgrad.

\subsection{Fused FP8 Autograd Saves the Values Needed by Backward}

The fused rowwise FP8 autograd node primarily prevents communication and
expert buffers from being reused before backward finishes.
When its configurable SwiGLU-recompute mode is enabled, it saves:

\begin{itemize}
\item route maps and route probabilities
\item optional gathered combine routes for router-probability gradients
\item \texttt{up\_gate\_q/scales} for SwiGLU backward and BF16 SwiGLU recompute
\item dispatched input \texttt{dispatch\_out\_q/scales} for up/gate wgrad
\item grouped GEMM offsets
\end{itemize}

In that mode, backward restores BF16 \texttt{up\_gate}, recomputes
\texttt{h = swiglu(up\_gate)}, and uses the recomputed value for down Wgrad.
The configuration default enables this policy, but the production recipe
explicitly disables it. Because the choice changes both retained activations
and backward work, it is part of the measured workload definition.

\paragraph{Base-transformer recompute alternatives.}
\label{app-base-transformer-recompute}
Outside the DDP stack described in Table~\ref{tab-recompute-granularities},
the base transformer additionally exposes module-pattern, operator-save-list,
and compiler-budget modes. The operator save list always retains
flash-attention outputs %
\citep{dao2022flashattention}, reduce-scatter results,
and the \texttt{abs}/\texttt{max} reductions reused as low-precision amax
statistics, and saves every second matrix multiplication. These alternatives
are separate from the DDP policies measured in Section~\ref{sec-recompute}.

\paragraph{PP and TBO composition constraint.}
The pipeline schedule counts one model forward while two TBO lanes may each
acquire a lease (Section~\ref{subsec-rowwise-ep-buffer-reuse}). This report
makes no measured claim about the slot-count factor under combined PP and
TBO.

\needspace{9\baselineskip}
\subsection{Some Kernel Prototypes Are Not Integrated}

The table separates the live kernel paths from retained prototypes and
identifies the conversion costs that remain in the live path.

\begingroup
\small
\begin{longtable}{
  >{\RaggedRight\arraybackslash}p{0.26\linewidth}
  >{\RaggedRight\arraybackslash}p{0.31\linewidth}
  >{\RaggedRight\arraybackslash}p{0.31\linewidth}}
\toprule
\textbf{Boundary} & \textbf{Current path} & \textbf{Prototype or remaining cost} \\
\midrule
\endfirsthead
\toprule
\textbf{Boundary} & \textbf{Current path} & \textbf{Prototype or remaining cost} \\
\midrule
\endhead
\midrule
\multicolumn{3}{r}{\itshape Continued on next page}\\
\endfoot
\bottomrule
\endlastfoot
Weighted FP8 combine backward & Integrated weighting plus row quantization
avoids a BF16 \texttt{[N, K, D]} intermediate; qdata and scales then use
separate PUTs & A direct weighted-quantize+NVSHMEM-PUT kernel exists only as
an unwired benchmark prototype \\
\arrayrulecolor{black!22}\hline\arrayrulecolor{black}
Qdata/scale transport & Separate regular transfers in the integrated path &
Paired and packed transfers are benchmark prototypes, not live training \\
\arrayrulecolor{black!22}\hline\arrayrulecolor{black}
Natural-layout MXFP8 grouped Wgrad & Operands dequantize and use BF16
\texttt{F.grouped\_mm} & Native MXFP8 grouped Wgrad exists only as a
prototype; this report makes no end-to-end claim about it. \\
\arrayrulecolor{black!22}\hline\arrayrulecolor{black}
Scale swizzling and RHS prequantization & FP8 reduces payload size but adds
separate quantization and scale-layout time, which can dominate small
profiles & An RHS cache amortizes only the weight-side conversion;
activation-side quantization and layout recur every microbatch \\
\end{longtable}
\endgroup
\addtocounter{table}{-1}%

\subsection{Kernel Profiles Separate Transport and Compute Costs}

Kernel profiles are interpreted through the following quantities:

\begin{center}
\small
\begin{tabularx}{\linewidth}{>{\RaggedRight\arraybackslash}p{0.34\linewidth}X}
\toprule
\textbf{Measurement} & \textbf{Why it matters} \\
\midrule
Kernel launch count per MoE block & Shows whether fusion reduces Python/autograd fragmentation. \\
\arrayrulecolor{black!22}\hline\arrayrulecolor{black}
Dispatch PUT time & Measures route-indexed one-sided send cost. \\
\arrayrulecolor{black!22}\hline\arrayrulecolor{black}
Gather GET time & Measures remote row pull cost during combine and dispatch backward. \\
\arrayrulecolor{black!22}\hline\arrayrulecolor{black}
Combine reduce time & Separates transport from local weighted reduction. \\
\arrayrulecolor{black!22}\hline\arrayrulecolor{black}
Grouped GEMM time and achieved utilization & Identifies whether expert compute or communication dominates. \\
\arrayrulecolor{black!22}\hline\arrayrulecolor{black}
Padding/capacity overhead & Captures wasted grouped GEMM work from routing imbalance. \\
\arrayrulecolor{black!22}\hline\arrayrulecolor{black}
FP8 quantization time & Shows whether reduced communication payload offsets quantization cost. \\
\arrayrulecolor{black!22}\hline\arrayrulecolor{black}
Scale swizzle/layout time & Detects hidden FP8 layout overhead. \\
\arrayrulecolor{black!22}\hline\arrayrulecolor{black}
BF16 Wgrad time in the MXFP8 path & Quantifies the cost of the current
non-native MXFP8 Wgrad policy. \\
\arrayrulecolor{black!22}\hline\arrayrulecolor{black}
Peak allocated memory by phase & Connects kernel behavior to lease high-water and PP/recompute pressure. \\
\arrayrulecolor{black!22}\hline\arrayrulecolor{black}
Rowwise stream wait time under TBO & Measures whether ordering waits erase overlap. \\
\bottomrule
\end{tabularx}
\end{center}

\section{Reproducible Experiment Configurations and Metrics}
\label{app-experiment-configurations}

This appendix records the hardware and workloads behind the report's
quantitative results and defines their accounting conventions. Experiment
records vary in which configurations, environment details, raw logs, profiles,
and plotting sources were retained; the individual descriptions state that
coverage. Credentials and secret environment variables are excluded.

\subsection{Audited Hardware and Software}

Table~\ref{tab-hardware-pool} records the allocation inspected on 2026-07-10,
which hosted the profiling and kernel experiments; the production-throughput
runs of Section~\ref{sec-production-throughput} used larger allocations of the
same node type (Appendix~\ref{app-production-throughput-runs}).
These are hardware and link properties, not measured collective bandwidth.

\begin{table}[!htb]
\centering
\small
\begin{tabularx}{\linewidth}{lX}
\toprule
\textbf{Item} & \textbf{Audited value} \\
\midrule
Compute & Four homogeneous nodes; eight NVIDIA B300 SXM6 AC GPUs per node; 32
GPUs total \\
\arrayrulecolor{black!22}\hline\arrayrulecolor{black}
GPU memory & 275,040 MiB reported per GPU \\
\arrayrulecolor{black!22}\hline\arrayrulecolor{black}
Driver/toolkit & NVIDIA driver 590.48.01; driver CUDA compatibility 13.1;
installed CUDA toolkit 13.0.88 \\
\arrayrulecolor{black!22}\hline\arrayrulecolor{black}
Intra-node fabric & Every GPU pair reported as NV18; 18 links per GPU at
53.125 GB/s per link \\
\arrayrulecolor{black!22}\hline\arrayrulecolor{black}
Inter-node fabric & Eight active nominal 800-Gb/s XDR InfiniBand rails per
node, one PXB-local rail aligned with each GPU; additional 100-Gb/s HDR ports
present \\
\arrayrulecolor{black!22}\hline\arrayrulecolor{black}
Shared storage & Olmo-core checkout on WekaFS, visible identically from all
four nodes \\
\arrayrulecolor{black!22}\hline\arrayrulecolor{black}
Olmo-core & The DDP training stack of this report at the revision audited
on 2026-07-10, with a clean working tree \\
\bottomrule
\end{tabularx}
\caption{\textbf{Audited hardware pool.} The inventory describes nominal hardware and
link properties; application bandwidth, clocks, power mode, and GPU
temperature vary with the run and are not inferred from it.}
\label{tab-hardware-pool}
\end{table}

Software-environment fields identify the activated interpreter, PyTorch,
NCCL, NVSHMEM, Transformer Engine/torchao, compiler, and extension versions
where recorded. A
non-login shell can select a different Python or omit the CUDA toolkit from
\texttt{PATH}, so the machine inventory alone does not identify the software
that executed a run. Network-interface and HCA selection likewise belongs to
the realized launch environment rather than a machine-level default.

\subsection{Audited Workload Roles}

Table~\ref{tab-appendix-benchmark-configs} describes a 30-block reference recipe
and its reduced profiling counterpart, distinct from the headline run
configurations in Section~\ref{sec-production-throughput}. The two exercise
the same core BF16 rowwise path but are not interchangeable performance points.

\begin{table}[!htb]
\centering
\small
\setlength{\tabcolsep}{4pt}
\begin{tabular}{
  >{\RaggedRight\arraybackslash\bfseries}p{0.13\linewidth}
  >{\RaggedRight\arraybackslash}p{0.23\linewidth}
  >{\RaggedRight\arraybackslash}p{0.24\linewidth}
  >{\RaggedRight\arraybackslash}p{0.29\linewidth}}
\toprule
\textbf{Role} & \textbf{Model and batch} & \textbf{Topology/precision} &
\textbf{Experiment purpose} \\
\midrule
Profiling & 8 blocks, 32 routed experts, top-8, $d=1536$, sequence 8192,
0.75-Mi-token global batch & EP-MP 8; 1\textendash{}4 nodes; BF16 rowwise by default; no
PP, CP, TP, TBO, or recompute & Host-visible route-shape profile;
synchronized/rowwise/DeepEP topology comparison; initial multi-node scaling \\
\arrayrulecolor{black!22}\hline\arrayrulecolor{black}
30-block reference & 30 blocks, 128 routed experts, top-8, $d=1536$, sequence
8192, 18-Mi-token global batch & EP-MP 8; BF16 rowwise; no PP, CP, TP, TBO,
or recompute in the base recipe & Full-reshard FSDP versus DDP-based training;
production-scale throughput, efficiency, traffic, variance, and reliability \\
\bottomrule
\end{tabular}
\caption{\textbf{Reference and profiling workload roles.} The profiling
workload is a reduced configuration for mechanism studies, not a throughput
measurement for the 30-block reference.}
\label{tab-appendix-benchmark-configs}
\end{table}

\subsection{The Main FSDP Comparison Uses Full Resharding}
\label{app-fsdp-reshard-policy}

The main-text FSDP comparison uses the no-PP FSDP2 path with \textbf{full
resharding}:
each wrapped block all-gathers its full weights for forward, reshards them,
all-gathers them again for backward, and reduce-scatters its gradients. This is
the minimum-persistent-parameter-memory endpoint compared with Olmo's DDP path,
which keeps each rank's assigned weights in device memory. FSDP2 can retain
unsharded weights through backward or across intermediate accumulation
backwards, but the latter policy makes peak parameter memory approach that of
the DDP path with sharded optimizer state; it is therefore a mechanism
ablation rather than a primary competing design in this report
\citep{zhao2023pytorchfsdp,pytorch2026fsdp2docs}.

Those retention policies lie between the two endpoints in parameter memory
and communication; the main text analyzes only the full-reshard endpoint. FSDP and
hybrid sharded data parallelism (HSDP) are also distinct configurations rather
than interchangeable labels.

\subsection{Reduced Profiling Changes the Workload}

The 30-block reference recipe and its reduced profiling sibling exercise the same
general BF16 rowwise path, but they are different sparse workloads. Based on
configuration-derived counts, the reference recipe has
$\alpha_P=10.89\%$, a $9.18\times$ capacity-to-active gap, and $N/K=16$.
Its 18-Mi-token optimizer step gives a balanced-routing mean of
$1.125\,\mathrm{Mi}$ assignments per routed expert. The profiling workload has
$\alpha_P=49.40\%$, a $2.02\times$ gap, $N/K=4$, and only
$0.1875\,\mathrm{Mi}$ balanced assignments per routed expert. A profiling
result therefore cannot be presented as throughput for the 30-block reference.

\begin{table}[!htb]
\centering
\small
\begin{tabularx}{\linewidth}{l>{\RaggedRight\arraybackslash}X>{\RaggedRight\arraybackslash}X}
\toprule
\textbf{Quantity} & \textbf{30-block reference} &
\textbf{Profiling workload} \\
\midrule
Blocks and width & 30 blocks; first 2 dense, remaining 28 routed; $d=1536$ &
8 blocks; first 2 dense, remaining 6 routed; same width and mixer family \\
\arrayrulecolor{black!22}\hline\arrayrulecolor{black}
Routed layer & $N=128$, $K=8$, routed intermediate width 1536, one shared expert &
$N=32$, $K=8$, with the same expert widths and shared branch \\
\arrayrulecolor{black!22}\hline\arrayrulecolor{black}
Routing and capacity & Softmax top-$K$, normalized route weights, capacity factor
1.25 & Same \\
\arrayrulecolor{black!22}\hline\arrayrulecolor{black}
Sequence and global batch & 8192 tokens; 18 Mi tokens per optimizer step &
8192 tokens; 0.75 Mi tokens per optimizer step \\
\arrayrulecolor{black!22}\hline\arrayrulecolor{black}
Configuration-derived size & 26.688B total, 2.906B active & 2.014B total,
0.995B active \\
\arrayrulecolor{black!22}\hline\arrayrulecolor{black}
Parallel/precision path & EP-MP 8; BF16 rowwise; no PP, CP, TP, TBO, or
recompute & Same core path, with profiling callbacks and deliberately reduced
model and batch \\
\bottomrule
\end{tabularx}
\caption{\textbf{30-block reference and reduced profiling workload
at the audited revision.} Their different model and batch sizes prevent
direct throughput comparison, even though they share the same core execution
path.}
\label{tab-running-config}
\end{table}

The Introduction's wall-time examples came from separate early
comparisons in Megatron-LM and legacy Olmo-core. Their complete
software and hardware environments have not been reconstructed. They therefore
illustrate the observed slowdown that motivated this work, but do not provide a
matched cross-framework benchmark.

For the profiling workload, the default rank microbatch is three sequences
(24,576 tokens).
The 0.75-Mi-token global batch yields four accumulation microbatches on one
eight-GPU node and one on the 32-GPU pool. The profiling callbacks, reduced
layers/experts, and smaller batch are profiling-specific; the BF16 rowwise
routed-token path, EP-MP degree, lack of PP/CP/TP/TBO/recompute, and FP32
gradient and optimizer policy are shared with the 30-block reference.

\subsection{Captured Submission-Regime Profiles}
\label{app-sync-profile-pair}

\begin{figure}[!htb]
\centering
\begin{subfigure}{\linewidth}
  \centering
  \includegraphics[width=0.82\linewidth]{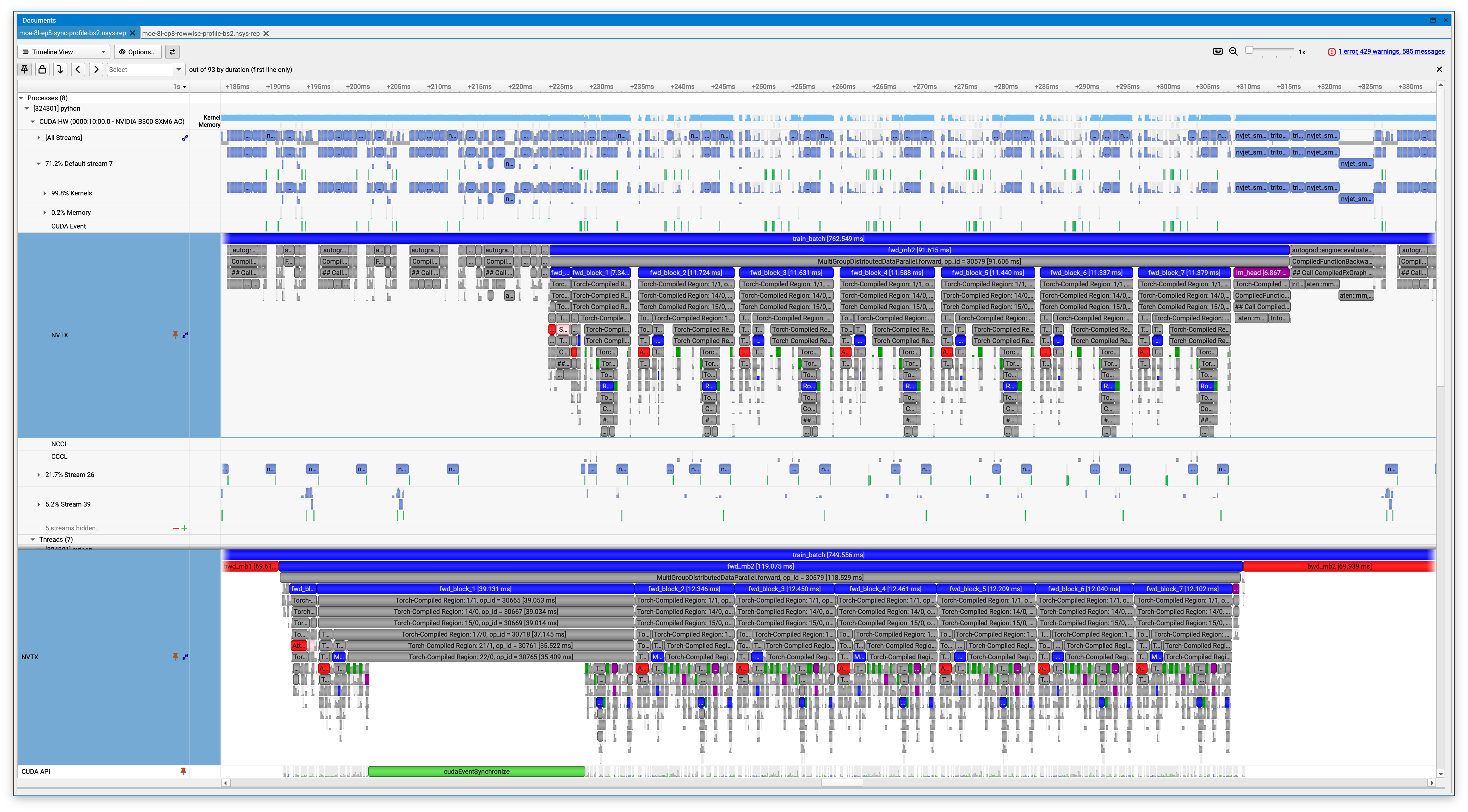}
  \caption{Synchronized EP8 path.}
\end{subfigure}

\vspace{0.4em}

\begin{subfigure}{\linewidth}
  \centering
  \includegraphics[width=0.82\linewidth]{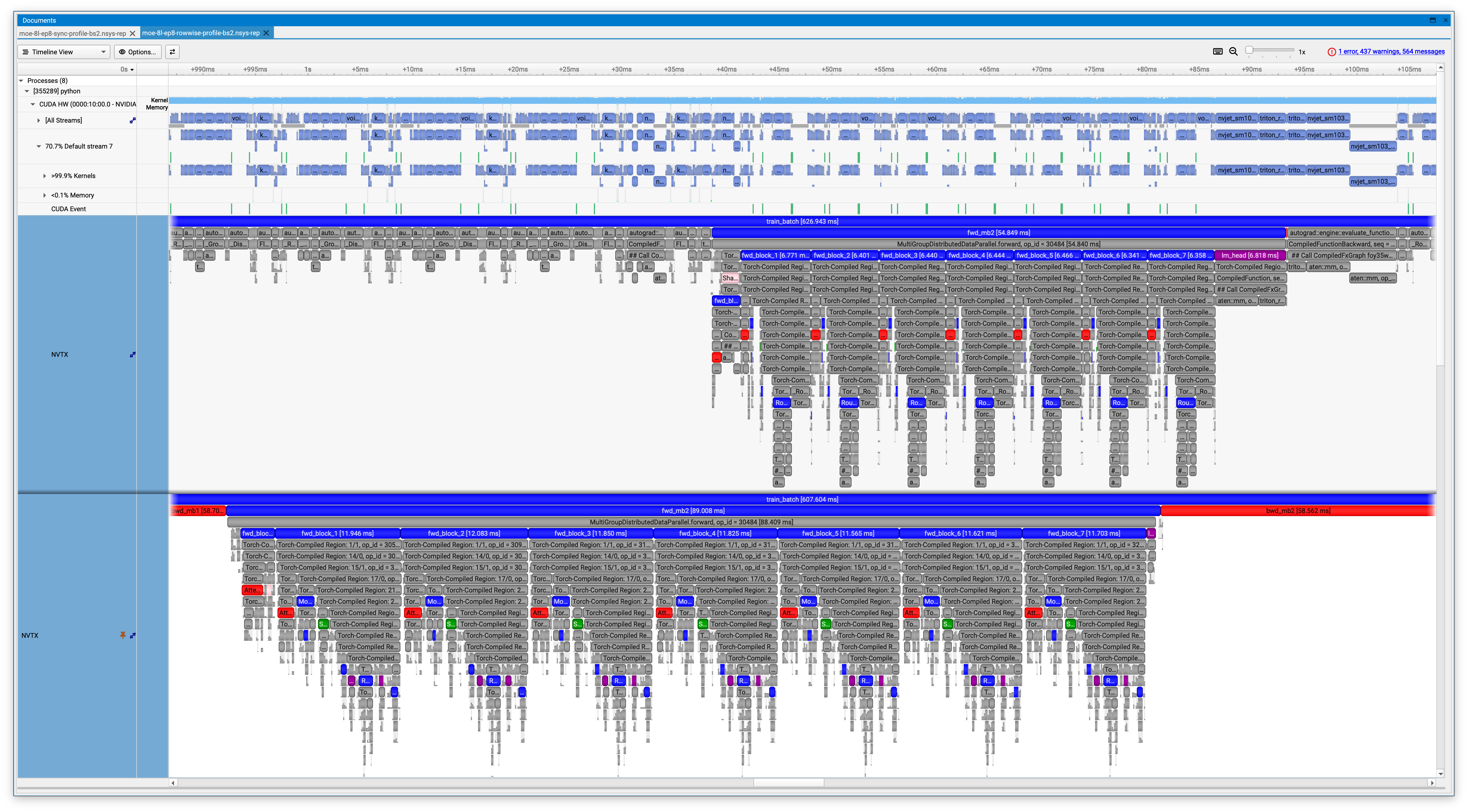}
  \caption{Rowwise EP8 path.}
\end{subfigure}
\caption{\textbf{Captured timelines show the two submission regimes.} Both
Nsight Systems profiles run the same eight-layer development model (128
routed experts, top-8, one shared expert, EP8, BF16, two sequences per rank
microbatch) on eight B300 GPUs; only the expert-parallel path differs. This
capture predates and differs from the reduced profiling workload of
Table~\ref{tab-appendix-benchmark-configs}. In the
synchronized trace (a), the per-block host wait appears as
\texttt{cudaEventSynchronize} in the CUDA API lane, the CPU and GPU ranges of
each forward block stay in lock step, and the GPU kernel rows show recurring
gaps. In the rowwise trace (b), no host wait interrupts submission: the CPU
finishes submitting the microbatch while the GPU is still executing earlier
work, and the GPU-side forward blocks run back to back. Each panel is one
captured instance of one rank; the two traces are separate runs, so this
figure illustrates the submission mechanism rather than a controlled
throughput comparison.}
\label{fig-sync-profile-pair}
\end{figure}

\begin{figure}[!htb]
\centering
\begin{subfigure}{\linewidth}
  \centering
  \includegraphics[width=\linewidth]{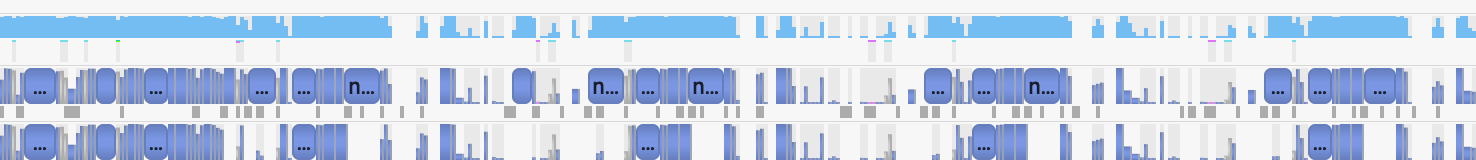}
  \caption{Synchronized EP8 path.}
\end{subfigure}

\vspace{0.5em}

\begin{subfigure}{\linewidth}
  \centering
  \includegraphics[width=\linewidth]{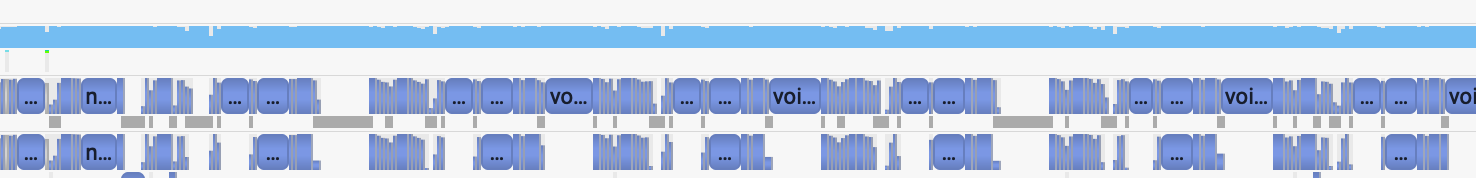}
  \caption{Rowwise EP8 path.}
\end{subfigure}
\caption{\textbf{Magnified GPU kernel rows of the two captures.}  Each strip
crops the CUDA HW utilization band and its two All-Streams kernel rows from
the corresponding panel of Figure~\ref{fig-sync-profile-pair}, over an
equal-width sub-window of each timeline.  In the synchronized capture~(a),
the utilization band is repeatedly interrupted and the kernel rows show
recurring idle gaps where the device waits for host submission.  In the
rowwise capture~(b), the band is continuous and the kernel rows are densely
packed.  The two strips come from separate runs whose timeline zoom levels
differ, so bar widths are not comparable across panels; the crops
illustrate the submission mechanism, not a throughput measurement.}
\label{fig-sync-profile-gpu-zoom}
\end{figure}

Section~\ref{sec-sync-cpu-ahead} describes two submission regimes: a host
wait inside every routed block that holds the CPU and GPU in lock step, and a
synchronization-free path in which the CPU runs ahead of the device.
Figure~\ref{fig-sync-profile-pair} shows one captured Nsight Systems timeline
for each regime, and Figure~\ref{fig-sync-profile-gpu-zoom} magnifies their
GPU kernel rows.

\subsection{Verifying the Sync-Free Path}
\label{app-sync-verification}

To check the sync-free property described in Section~\ref{sec-sync-device-scheduling},
we inspect the executed timeline. A
qualifying steady-state trace contains no blocking D2H call and no
submission-bound idle interval, which is what the captured pair in
Figures~\ref{fig-sync-profile-pair} and~\ref{fig-sync-profile-gpu-zoom}
shows for the rowwise path. Step time alone cannot identify a submission
problem. A useful profile reports CPU API and framework time, GPU kernel and
communication time, the distance from CPU launch to GPU start, GPU idle
intervals, blocking D2H calls, and end-to-end tokens/s, aligned at both the
kernel level and larger NVTX ranges such as a transformer block or
microbatch. PyTorch provides an experimental synchronization debug mode, but
its coverage is incomplete for distributed and sparse operations; it is an
audit aid rather than a substitute for a timeline
\citep{pytorchSyncDebugMode}.

A standalone single-GPU grouped-linear microbenchmark from the MXFP8 study
(Section~\ref{sec-mxfp8-training}) isolates the submission tax that this
checklist is meant to expose. Graph replay appears in this report only as a
measuring instrument; the production training path never captures or
replays a graph (Section~\ref{subsec-static-cudagraph-lesson}).
Table~\ref{tab-sync-graph-seed} compares ordinary eager submission with CUDA
Graph replay for a locally adapted OLMoE-family benchmark
\citep{muennighoff2024olmoe}. Its $1536\rightarrow3072$ up projection and 16
balanced local expert groups are settings of this single-B300 benchmark,
not the published OLMoE configuration. ``Prequantized'' begins with MXFP8 input;
``including Q'' also charges the activation-quantization (Q) step. Every number is a
speedup over the matching BF16 grouped GEMM.

\begin{table}[!htb]
\centering
\small
\renewcommand{\arraystretch}{1.15}
\setlength{\tabcolsep}{5pt}
\begin{tabular}{rcccc}
\toprule
& \multicolumn{2}{c}{\textbf{Prequantized input}} &
  \multicolumn{2}{c}{\textbf{Including Q}} \\
\cmidrule(lr){2-3}\cmidrule(lr){4-5}
\textbf{Rows $M$} & \textbf{Eager} & \textbf{Graph} &
                     \textbf{Eager} & \textbf{Graph} \\
\midrule
128    & $0.66\times$ & $2.17\times$ & $0.34\times$ & $2.03\times$ \\
4,096  & $0.65\times$ & $1.26\times$ & $0.32\times$ & $1.06\times$ \\
32,768 & $1.43\times$ & $1.42\times$ & $1.21\times$ & $1.22\times$ \\
\bottomrule
\end{tabular}
\caption{\textbf{Eager submission can hide the device-level benefit of short
MXFP8 expert operations.} Each value is the speed of the MXFP8 operation
relative to a matching BF16 grouped GEMM at the same $M$ and submission mode:
$1\times$ means equal time, and larger is faster than BF16. Results are from
one GPU and one random seed.  Eager values use 25-ms warm-up and
100-ms timed windows; graph values replay 32 captured copies with three warm-up and five timed
replays.  CUDA Graph replay is a warm-cache
device-service-time ceiling, not training wall time.}
\label{tab-sync-graph-seed}
\end{table}

At $M=128$, the captured device sequence is more than $2\times$ faster than
the BF16 baseline while ordinary eager execution is slower. At $M=32{,}768$,
the eager and graph results converge. The convergence indicates that host submission affects short-operator
timings. The eager--graph gap is still not a clean measure of CPU launch
time: replay changes the cache regime and excludes
graph-launch latency, and the benchmark omits routing, communication,
autograd, and the rest of the training step. The relevant end-to-end metric
therefore remains eager training wall time.

Removing the split-list round trip is also only one side of the trade-off:
direct row movement adds its own route construction,
capacity storage, and completion costs. Several experiments examine
different parts of that trade-off. The profiled example of
Section~\ref{subsec-rowwise-ep-motivation} localizes the shorter
dispatch-and-combine path, the captured timelines above show the two
submission regimes, the microbenchmark isolates the submission tax on short
expert operations,
and the feasibility point of Table~\ref{tab-ep-memory-throughput-seed} shows
one complete operating point without attributing its throughput to a single
mechanism.

\subsection{Checkpoint-Branched LBL-Coefficient Runs}
\label{app-token-gerrymandering-run}

Figures~\ref{fig-token-gerrymandering-training}
and~\ref{fig-token-gerrymandering-objective} derive from archived metric
histories of two trajectories of the production pretraining project.  The
experiment package bundled with this report holds the run manifests with
full effective configurations, the merged metric histories and their merge
report, and the plotting script.

\begin{itemize}[leftmargin=*]
\item \textbf{Model.} 48 transformer blocks at $d_{\mathrm{model}}=2048$;
block 0 is dense, and each of the 47 MoE blocks uses 64 experts, top-4
selection, sigmoid gating, dropless dispatch, and expert intermediate width 2048;
grouped-query attention %
\citep{ainslie2023gqa} with 16 query and 4 key/value
heads; \texttt{dolma2} tokenizer %
\citep{allenai2024dolma2tokenizer}.
\item \textbf{Data and batch.} \texttt{OLMo-mix-0625} at sequence length
8{,}192; global batch 4{,}194{,}304 tokens per step.
\item \textbf{Objective.} Instance-granularity LBL without per-layer
rescaling; LM-head z-loss $10^{-5}$; router z-loss disabled.  The branched
quantity is the LBL coefficient: $0.05$ on the original trajectory and
$0.005$ on the branch.
\item \textbf{Optimizer.} Skip-step AdamW %
\citep{loshchilov2019decoupled} at
learning rate $3\times10^{-5}$
with betas $(0.9,0.95)$, weight decay $0.1$ (zero on embeddings), a
warmup--stable--decay schedule %
\citep{hu2024minicpm,hagele2024scaling} with a
2{,}000-step warmup, and gradient
clipping at norm 1.
\item \textbf{Execution.} HSDP data parallelism with BF16 parameters; FP8 is
disabled.  These trajectories run the production pretraining stack rather
than the expert-parallel paths of this report; the two branches differ only
in the LBL coefficient, and no systems claim rests on them.
\item \textbf{Branch protocol.} The branch reloads the original run's model
and optimizer checkpoint at step 17{,}000 (71.3 billion tokens) and changes
only the coefficient.  The archive covers steps 1--25{,}669 of the original
run and 17{,}001--81{,}084 of the branch.  The
Section~\ref{sec-routing-load-balancing} windows compare medians over steps
17{,}001--17{,}999 and 24{,}000--25{,}669, before the branch's batch-size
transition at step 40{,}000.
\end{itemize}

\subsection{Batch-Size Warmup Run}
\label{app-lr-batch-warmup-run}

This subsection records the training run behind
Figure~\ref{fig-lr-batch-warmup} in Section~\ref{sec-lr-batch-scaling}.

\begin{itemize}[leftmargin=*]
\item \textbf{Model.} 24 transformer blocks; 48 routed experts with top-4
selection and one shared expert; 3.21B active and 23.10B total parameters.
\item \textbf{Batch schedule.} The global batch increases through 1, 2, 3,
6, 9, 15, and 24 Mi tokens over approximately 2.07 trillion training
tokens; the later transitions fall near 409B, 777B, and 1.375T cumulative
tokens.
\item \textbf{Learning rate.} Square-root batch scaling
(Equation~\ref{eq-lr-batch-scaling} with $\alpha=\tfrac{1}{2}$) applied at
each batch increase.
\item \textbf{Optimizer.} Skip-step AdamW with betas $(0.9,0.95)$ and weight
decay $0.1$, zero on embeddings, normalization weights, and the LM head; Muon
is disabled.
\item \textbf{Provenance.} The figure stitches 26 run segments of one
training campaign from its archived metric histories; at every overlapping
step the newest segment's value is retained.  The artifact bundled with this
report contains the raw exported metric histories, a per-segment run
manifest, the stitched and plotted series, and the scripts that regenerate
the figure from the archive alone.
\item \textbf{Scope.} The bundle preserves metric histories and figure
provenance. The optimizer settings above come from the run's training recipe
as committed during the campaign; the serialized run configurations of the
underlying segments are not re-audited here.
\end{itemize}

\subsection{Experiment Artifact Contents}

The following fields describe the configuration and evidence relevant to
replicating a run. They define how this report describes its experiment
records, not a claim that every archived artifact contains every field.
Availability is stated with each experiment.

\begin{itemize}[leftmargin=*]
\item \textbf{Provenance.} Repository and dependency revisions, patch/diff,
command, run name, and configuration after all overrides;
\item \textbf{Model and workload.} Total and active parameters, layers, mixers,
experts, top-$K$, shared experts, capacity/drop policy, sequence, local/global
useful batch, and random seeds;
\item \textbf{Topology.} Nodes/GPUs and DP, CP, EP-DP, EP-MP, PP, VPP,
microbatch, accumulation, and schedule;
\item \textbf{Precision and layout.} Parameter, activation, accumulation,
reduction, optimizer, communication, cache, and checkpoint dtypes and layouts;
\item \textbf{Environment.} GPU clocks/power mode, driver/toolkit, activated
software versions, NCCL/NVSHMEM interfaces and HCAs, compiler/cache controls,
and relevant non-secret environment;
\item \textbf{Measurement protocol.} Data-prewarm, compile, symmetric-pool
prewarm, warmup, measured steps, repetitions, synchronization points,
exclusions, and failure/retry policy;
\item \textbf{Raw outputs.} Raw logs, rank-level metric tables, profiler
captures, allocator traces, checkpoint metadata, and plotting scripts.
\end{itemize}

The effective serialized configuration, rather than the command string alone,
determines the realized topology and workload after command-line overrides.

\subsection{Production-Throughput Run Details}
\label{app-production-throughput-runs}

This subsection is the run-level companion to
Table~\ref{tab-production-throughput-headline}. All rows use sequence length
8192, top-4 routing, and one shared expert. The main table selects one
random-routing systems reference per operating point; the tables below preserve the
learned-routing companions and the alternate batches and topologies. A
tracking-system status of ``crashed'' for these short systems runs can reflect
manual termination after the desired observations and is not, by itself,
evidence that training failed.

\begingroup
\scriptsize
\setlength{\tabcolsep}{3pt}
\renewcommand{\arraystretch}{1.15}
\newcommand{\rowline}{\arrayrulecolor{black!22}\hline\arrayrulecolor{black}}
\begin{longtable}{
  >{\RaggedRight\arraybackslash\bfseries}m{0.165\linewidth}
  >{\RaggedRight\arraybackslash}m{0.215\linewidth}
  >{\RaggedRight\arraybackslash}m{0.215\linewidth}
  >{\centering\arraybackslash}m{0.115\linewidth}
  >{\RaggedRight\arraybackslash}m{0.20\linewidth}}
\toprule
\textbf{Configuration} & \textbf{Model} & \textbf{Batch and scale} &
\makecell{\textbf{Parallel}\\\textbf{topology}} & \makecell{\textbf{Numeric/runtime}\\\textbf{path}} \\
\midrule
\endfirsthead
\toprule
\textbf{Configuration} & \textbf{Model} & \textbf{Batch and scale} &
\makecell{\textbf{Parallel}\\\textbf{topology}} & \makecell{\textbf{Numeric/runtime}\\\textbf{path}} \\
\midrule
\endhead
\midrule
\multicolumn{5}{r}{\itshape Continued on next page}\\
\endfoot
\bottomrule
\noalign{\vskip 5pt}
\caption{\textbf{Resolved configuration matrix for the production-throughput
runs.} GBS and rank microbatch are useful tokens; Mi denotes $2^{20}$. The topology
column reports dense-DP / expert-DP / EP / PP degrees. The serialized run
configuration, not a subsequently edited local recipe, is authoritative.}
\label{tab-production-resolved-configs}\\
\endlastfoot
Tiny-64E, no EP & 16 layers; 64 routed experts; 1.59B active, 12.91B total &
GBS 8 Mi; rank microbatch 65,536; 2 nodes, 16 B300s & 16 / 16 / 1 / 1 &
BF16; no recompute; local expert execution \\
\rowline
Tiny-64E, EP8 & Same model as Tiny-64E, no EP & GBS 8 Mi; rank microbatch
65,536; 2 nodes, 16 B300s & 16 / 2 / 8 / 1 & BF16; rowwise NVSHMEM; no
recompute \\
\rowline
Small-64E, 24 Mi & 32 layers; 64 routed experts; 4.29B active, 40.86B total &
GBS 24 Mi; rank microbatch 24,576; 16 nodes, 128 B300s & 128 / 16 / 8 / 1 &
BF16; rowwise NVSHMEM; no recompute \\
\rowline
Small-64E, 16 Mi & Same Small-64E model & GBS 16 Mi; rank microbatch 16,384;
16 nodes, 128 B300s & 128 / 16 / 8 / 1 & BF16; rowwise NVSHMEM; no recompute \\
\rowline
Medium-64E & 39 layers; 64 routed experts; 7.37B active, 70.22B total & GBS
16 Mi; rank microbatch 16,384; 16 nodes, 128 B300s & 64 / 8 / 8 / 2 & BF16;
rowwise NVSHMEM; no recompute \\
\rowline
Medium-96E & 39 layers; 96 routed experts; 7.38B active, 103.75B total & GBS
24 Mi; rank microbatch 16,384; 16 nodes, 128 B300s & 64 / 8 / 8 / 2 & BF16;
rowwise NVSHMEM; no recompute \\
\rowline
Medium-128E & 39 layers; 128 routed experts; 7.38B active, 137.27B total & GBS
24 Mi; rank microbatch 16,384; 16 nodes, 128 B300s & 32 / 4 / 8 / 4 & BF16;
rowwise NVSHMEM; no recompute \\
\rowline
Large-128E, PP4 & 47 layers; 128 routed experts; 15.14B active, 295.99B total &
GBS 32 Mi; rank microbatch 8192; 64 nodes, 512 B300s & 128 / 16 / 8 / 4 &
BF16; rowwise NVSHMEM; no recompute \\
\rowline
Large-128E, PP8 & Same Large-128E model & GBS 24 Mi; rank microbatch 8192;
64 nodes, 512 B300s & 64 / 8 / 8 / 8 & BF16; rowwise NVSHMEM; no recompute \\
\rowline
Ultra-128E & 63 layers; 128 routed experts; 58.36B active, 1.200T total & GBS
64 Mi; rank microbatch 8192; 64 nodes, 512 B300s & 64 / 8 / 8 / 8 & MXFP8
attention and MLP compute; rowwise NVSHMEM; per-layer recompute \\
\rowline
Ultra-256E & 63 layers; 256 routed experts; 58.41B active, 2.380T total & GBS
64 Mi; rank microbatch 16,384; 64 nodes, 512 B300s & 64 / 2 / 32 / 8 &
MXFP8 attention and MLP compute; DeepEP v2; per-layer recompute \\
\end{longtable}
\endgroup

\begingroup
\scriptsize
\setlength{\tabcolsep}{3pt}
\renewcommand{\arraystretch}{1.15}
\newcommand{\rowline}{\arrayrulecolor{black!22}\hline\arrayrulecolor{black}}
\needspace{30\baselineskip}
\begin{longtable}{
  >{\RaggedRight\arraybackslash\bfseries}m{0.19\linewidth}
  >{\centering\arraybackslash}m{0.09\linewidth}
  >{\centering\arraybackslash}m{0.09\linewidth}
  >{\RaggedRight\arraybackslash}m{0.55\linewidth}}
\toprule
\textbf{Configuration} & \textbf{Routing} &
\makecell{\textbf{TFLOP/s}\\\textbf{/GPU}} & \textbf{Observation and use} \\
\midrule
\endfirsthead
\toprule
\textbf{Configuration} & \textbf{Routing} &
\makecell{\textbf{TFLOP/s}\\\textbf{/GPU}} & \textbf{Observation and use} \\
\midrule
\endhead
\midrule
\multicolumn{4}{r}{\itshape Continued on next page}\\
\endfoot
\bottomrule
\noalign{\vskip 5pt}
\caption{\textbf{Run observations and headline-selection record.} Throughput values are
representative observations from the runs' throughput curves (for the nine-step
Ultra-256E run, its best observed step), not arithmetic
means across repeated launches. ``Headline'' identifies the
random-routing value carried into Table~\ref{tab-production-throughput-headline}.}
\label{tab-production-run-observations}\\
\endlastfoot
Tiny-64E, no EP & Learned & 895 & Began near 870 and stabilized after roughly 100 steps; learned-routing companion \\
\rowline
Tiny-64E, no EP & Random & 903 & Headline; random-routing curve was nearly stable from the first step \\
\rowline
Tiny-64E, EP8 & Learned & 820 & Began near 799, then stabilized; learned-routing EP8 run with no random-routing pair \\
\rowline
Small-64E, 24 Mi & Learned & 768 & Learned-routing companion \\
\rowline
Small-64E, 24 Mi & Random & 841 & Headline \\
\rowline
Small-64E, 16 Mi & Random & 821 & Alternate smaller GBS and rank microbatch; not a repeat of the 24-Mi run \\
\rowline
Medium-64E & Random & 853 & Headline \\
\rowline
Medium-96E & Random & 855 & Headline \\
\rowline
Medium-96E & Random & 781 & Repeat with identical resolved configuration; not selected \\
\rowline
Medium-96E & Random & 740 & Repeat with identical resolved configuration; not selected \\
\rowline
Medium-128E & Random & 843 & Headline \\
\rowline
Medium-128E & Random & 792 & Repeat with identical resolved configuration; not selected \\
\rowline
Large-128E, PP4 & Random & 768 & Headline \\
\rowline
Large-128E, PP8 & Random & 757 & Alternate PP8 and 24-Mi GBS; not a repeat of the headline run \\
\rowline
Ultra-128E & Random & 858 & Headline \\
\rowline
Ultra-256E & Random & 689 & Headline short-run observation; nine steps and no stabilized plateau \\
\end{longtable}
\endgroup

Slower repeat launches of the Medium-96E and Medium-128E configurations appear
in Table~\ref{tab-production-run-observations} but are omitted from the
headline selection. Environmental problems were suspected, but the
specific cause was not established. The headline table reports the highest
rate observed for each configuration, which is neither a repetition mean
nor an estimate of the hardware limit.

\subsection{Metrics Track Useful Work, Time, and Memory}
\label{app-metric-definitions}

Table~\ref{tab-appendix-metric-definitions} defines the quantities used to
compare useful work, time, and memory. Each experiment specifies which
metrics and supplemental diagnostics were retained; a definition here does
not imply that the quantity was measured in every run.

\begin{table}[!tb]
\centering
\footnotesize
\begin{tabularx}{\linewidth}{lX}
\toprule
\textbf{Metric} & \textbf{Definition} \\
\midrule
Steady-state step time & Wall time from the start of an optimizer step's first
forward to completion of optimizer update and required cache refresh, after
declared startup/warmup. Depending on the retained samples, summaries may
include median, mean, standard deviation, and tails \\
\arrayrulecolor{black!22}\hline\arrayrulecolor{black}
Useful tokens/s/GPU & Global non-padding tokens contributing to the normalized
loss per measured step, divided by synchronized step time and GPU count;
capacity-dropped routes are a separate quantity rather than silently removed \\
\arrayrulecolor{black!22}\hline\arrayrulecolor{black}
Useful-model MFU & Useful active-model FLOPs divided by elapsed time, GPU
count, and the stated per-GPU compute peak. Sections~\ref{sec-olmo-ddp-training-stack}
and~\ref{sec-production-throughput} use the B300 dense BF16 peak of
2250 TFLOP/s/GPU for every row, including MXFP8, and label this common
BF16-reference normalization explicitly. It is not utilization of the FP8
peak. Recompute and inactive experts add no useful FLOPs to the numerator \\
\arrayrulecolor{black!22}\hline\arrayrulecolor{black}
D2H route-shape wall time & Count and duration of host waits/copies causally between
router metadata production and dependent dispatch/compute launches, plus the
corresponding CPU launch gap and GPU idle interval from the same timeline \\
\arrayrulecolor{black!22}\hline\arrayrulecolor{black}
Communication exposure & Dispatch/combine or weight/gradient communication
time not hidden by useful compute, with bytes, primitive, process group, and
forward/backward scope identified \\
\arrayrulecolor{black!22}\hline\arrayrulecolor{black}
Route/drop statistics & Offered and accepted assignments, total and per-source
drop rate, per-expert load distribution, capacity utilization, and rank skew \\
\arrayrulecolor{black!22}\hline\arrayrulecolor{black}
Peak device memory & Maximum allocated and reserved bytes per rank after a
declared reset point. Supplemental diagnostics, where recorded, give max/median
ranks and whether compiler/graph memory pools, symmetric pools, derived
caches, and checkpoint staging were allocated at reset and remained allocated
through the measured window \\
\arrayrulecolor{black!22}\hline\arrayrulecolor{black}
Lease high-water & Maximum concurrently checked-out saved-payload slots by
block/payload/pipeline part. A schedule-derived prediction and transient shared
scratch usage are separate diagnostics where available \\
\arrayrulecolor{black!22}\hline\arrayrulecolor{black}
Checkpoint peak/time & Pre-save and peak allocated/reserved bytes, and elapsed
save or load time. Supplemental diagnostics include bytes written/read,
retries, and caches allocated at the measurement boundary, where recorded \\
\arrayrulecolor{black!22}\hline\arrayrulecolor{black}
Scaling efficiency & Measured useful throughput relative to the declared
one-node or smallest valid topology reference under a stated strong/weak-scaling
batch rule; rank variance is a supplemental diagnostic where recorded \\
\bottomrule
\end{tabularx}
\caption{\textbf{Metric definitions for quantitative comparisons in this
report.} Profiler queries and calculation scripts are part of the
corresponding experiment artifact when available.}
\label{tab-appendix-metric-definitions}
\end{table}

For an optimizer step with $F_{\mathrm{step}}$ useful model FLOPs, elapsed
time $T_{\mathrm{step}}$ in seconds, and $G$ GPUs, the report uses
\begin{equation}
\mathrm{MFU}_{\mathrm{BF16\mbox{-}ref}}
= \frac{F_{\mathrm{step}}}
{G\,T_{\mathrm{step}}\,C_{\mathrm{BF16}}},
\qquad C_{\mathrm{BF16}}=2.25\times10^{15}\ \mathrm{FLOP/s}
\quad\text{per B300 GPU}.
\label{eq-intro-mfu}
\end{equation}
This is the dense BF16 Tensor Core peak, not NVIDIA's doubled 2:4-sparse
peak %
\citep{nvidiaHGXB200spec}; selecting a subset of experts does not make
their weight matrices 2:4 sparse. The numerator counts active-model work,
excluding activation recomputation, while elapsed time includes the complete
step. We retain this BF16 reference for MXFP8 runs so the normalization is
consistent; those values are not utilization of the FP8 peak.

\FloatBarrier

\section{Glossary of Parallelism, Routing, and Training Storage}
\label{app-glossary-notation}

The same object can have a model meaning, a runtime placement, and a
checkpoint representation. This glossary fixes the terms used for those
layers.

\newcommand{\glossaryrowline}{\arrayrulecolor{black!22}\hline\arrayrulecolor{black}}
\newenvironment{glossarytable}[1]{%
  \begingroup
  \small
  \renewcommand{\theHtable}{glossary.\thesubsection}%
  \begin{longtable}{
    >{\RaggedRight\arraybackslash}p{0.26\linewidth}
    >{\RaggedRight\arraybackslash}p{0.66\linewidth}}
  \toprule
  \textbf{#1} & \textbf{Meaning} \\
  \midrule
  \endfirsthead
  \toprule
  \textbf{#1} & \textbf{Meaning} \\
  \midrule
  \endhead
  \midrule
  \multicolumn{2}{r}{\itshape Continued on next page}\\
  \endfoot
  \bottomrule
  \endlastfoot
}{%
  \end{longtable}
  \endgroup
  \addtocounter{table}{-1}%
}

\needspace{6\baselineskip}
\subsection{Parallel Dimensions and Mesh Axes}

\begin{glossarytable}{Term}

$P$ / PP & Pipeline degree. PP partitions layers and changes execution order,
not which data-parallel ranks replicate each parameter. \\
\glossaryrowline
$D$ / DP & Dense data-parallel degree. With CP, ordinary dense gradients reduce
over the folded DP$\times$CP replica pool; dense optimizer state is sharded or
replicated over DP and duplicated across CP. \\
\glossaryrowline
$C$ / CP & Context-parallel degree. CP shards sequence before the PP schedule
and participates in the folded expert pool. \\
\glossaryrowline
$D_E$ / EP-DP & Expert replica degree. Ranks on this axis hold the same local
EP-MP expert shard and reduce its ordinary gradients together. \\
\glossaryrowline
$M_E$ / EP-MP & Expert-sharding degree. Ranks on this axis hold different
routed-expert shards; it is not a gradient all-reduce group. \\
\glossaryrowline
EP & Expert parallelism, used generically for distributing routed experts and
moving token rows to ranks that store the selected experts. \\
\glossaryrowline
TP & Tensor parallelism, which partitions individual parameter tensors and
their operations. \\
\glossaryrowline
dense mesh & Device-mesh view \texttt{(pp?, dp, cp?)}. \\
\glossaryrowline
MoE mesh & View of the same ranks as
\texttt{(pp?, ep\_dp, ep\_mp)}. Within a PP stage,
$DC=D_E M_E$. \\
\end{glossarytable}

\subsection{Model Size and Routing Quantities}

\begin{glossarytable}{Symbol or term}
$d$ / \texttt{d\_model} & Transformer model width. \\
\glossaryrowline
$h_e$ & Routed expert intermediate width. \\
\glossaryrowline
$m_s$ & Total shared-expert intermediate width
(Section~\ref{sec-ungated-shared-experts}); distinct from the microbatch count $m$. \\
\glossaryrowline
$N$ / \texttt{num\_experts} & Total routed-expert count in an MoE layer. \\
\glossaryrowline
$K$ / \texttt{top\_k} & Number of routed experts selected per token. \\
\glossaryrowline
$S$ & Sequence length before context-parallel sharding. \\
\glossaryrowline
$T$ & Number of input tokens before top-$K$ expansion and capacity handling. \\
\glossaryrowline
$B$ / $m$ & Useful tokens processed per rank in one optimizer step / number of
microbatches in that step, when used in the cost ledger before model
parallelism. \\
\glossaryrowline

\texttt{num\_local\_experts} & Routed experts stored on one EP-MP rank, normally
$N/M_E$. \\
\glossaryrowline
active parameters $P_{\mathrm{active}}$ & Parameters used by one token; routed
expert work scales with $K$, not $N$. \\
\glossaryrowline
total parameters $P_{\mathrm{total}}$ & Full persistent parameter pool;
optimizer tensors and checkpoint size scale with this quantity. \\
\glossaryrowline
$X\text{B}@Y\text{B}$ & Model-size shorthand for $X$ billion active and $Y$
billion total parameters (active@total), for example
$3.21\text{B}@23.10\text{B}$. \\
\glossaryrowline
$P_{\mathrm{local}}$ & Parameter size held by one rank after PP and EP
placement. This is neither the model-wide total parameter count nor the per-token
active parameter count. \\
\glossaryrowline
$P_E$ / $P_{\mathrm{dense}}$ & Routed-expert and dense (non-expert) parameter
counts of the whole model; one rank holds about $P_E/M_E$ of the expert part. \\
\glossaryrowline
expert activation ratio $\alpha_E$ & Fraction of routed experts selected for
one token, $K/N$. \\
\glossaryrowline
parameter activation ratio $\alpha_P$ & Fraction of total model parameters
active for one token, $P_{\mathrm{active}}/P_{\mathrm{total}}$. It generally
differs from $\alpha_E$ because part of the model is always active. \\
\glossaryrowline
$\Gamma_{\mathrm{capacity}}$ & Model-wide capacity-to-active gap,
$P_{\mathrm{total}}/P_{\mathrm{active}}=1/\alpha_P$. It is not itself an
overhead or a conversion from bytes or launches to elapsed time. \\
\glossaryrowline
routed experts & Sparse branch selected by the router and partitioned over
EP-MP. \\
\glossaryrowline
shared expert & Always-active branch that remains in the dense parameter
family. \\
\glossaryrowline
router & Projection and policy producing selected expert indices, route
weights, and load-balancing metadata. \\
\glossaryrowline
expert offsets/counts & GPU or host metadata delimiting unequal per-expert row
ranges for grouped computation. \\
\end{glossarytable}

\needspace{6\baselineskip}
\subsection{Runtime Buffers and Reuse Terms}

\begin{glossarytable}{Term}
persistent training state & Compute weights, accumulated gradients, FP32 main
weights, and optimizer states that must remain available across microbatches
or update boundaries. \\
\glossaryrowline
execution working memory & Activations, routing and communication buffers, and
workspaces used by execution. A preallocated fixed-address buffer can still
belong to this class. \\
\glossaryrowline
static placement / shape & A stable placement, buffer identity or address,
capacity, shape, or graph structure. Static values need not be immutable. \\
\glossaryrowline
symmetric memory & Collectively established storage that ranks in one team can
address through one-sided operations. \\
\glossaryrowline
Olmo symmetric memory & Current rowwise backend built around Olmo's
CUDA/NVSHMEM wrapper rather than the legacy Torch symmetric-memory path. \\
\glossaryrowline
dispatch / combine & Movement from token order to expert-buffer order / return
and router-weighted accumulation into token order. \\
\glossaryrowline
lease & Reservation of a symmetric payload that must remain valid for a later
backward. Autograd releases it after the consumer finishes. \\
\glossaryrowline
shared transient slot & Bounded scratch reused across blocks or lanes; its existence does not indicate when a saved payload may be overwritten. \\
\glossaryrowline
fixed per-block cache & Stable cached storage used when sharing is disabled or
unsafe; a cache hit alone does not show that backward has finished with the
previous contents. \\
\glossaryrowline
high-water & Maximum simultaneous live payload count predicted or observed for
a stage and storage family. \\
\glossaryrowline
qdata\textendash{}scale payload & One MXFP8 value represented by quantized data
and its block scales. Both components must remain valid and be released together. \\
\glossaryrowline
BF16 anchor & Temporary initialization value and cache seed for an
\texttt{FP8WeightStore}-backed parameter. It may be
released after optimizer setup and is not an independent checkpoint source. \\
\glossaryrowline
MXFP8 compute-weight cache & Derived orientation-specific qdata/scale compute
weight, refreshed from the FP32 main weight. \\
\glossaryrowline
MXFP8 runtime payload & Derived qdata/scale representation for activation or
communication data. \\
\end{glossarytable}

\subsection{Optimizer and Checkpoint Terms}

\begin{glossarytable}{Term}
\texttt{MultiGroup\allowbreak Distributed\allowbreak Data\allowbreak Parallel} & Olmo reducer assigning a
process group per parameter family and keeping persistent bucket views. \\
\glossaryrowline
dense bucket & Gradient bucket reduced over the folded DP$\times$CP replica
group. \\
\glossaryrowline
expert bucket & Routed-expert bucket reduced over EP-DP, the group holding
replicas of the same expert shard. \\
\glossaryrowline
\texttt{OLMoDDPOptimizer} & Olmo optimizer that manages FP32 main weights,
optimizer states, gradient intake, global gradient norm, copy-back, and checkpoint views. \\
\glossaryrowline
compute weight & Weight representation consumed by forward or backward
computation: a BF16 tensor for an ordinary parameter or a derived,
orientation-specific MXFP8 qdata/scale cache for a selected linear. \\
\glossaryrowline
FP32 main weight & FP32 parameter value updated by the optimizer and used to
refresh its compute weights. \\
\glossaryrowline
\texttt{main\_grad} & Optimizer-facing high-precision gradient slice. \\
\glossaryrowline
copy-back & Reconstruction of compute weights after a successful update or load. \\
\glossaryrowline
\texttt{FP8WeightStore} & Trainable FP8-weight interface with
high-precision gradients and derived orientation-specific MXFP8 compute-weight caches; not an
FP8 \texttt{nn.Parameter}. \\
\glossaryrowline
checkpoint-view DTensor & Global tensor view exposed to distributed
checkpointing, normally without a full GPU expert tensor. \\
\glossaryrowline
topology portability & Ability to restore complete model and optimizer
tensors under a changed DP, EP-MP, EP-DP, or PP parameter-placement map. \\
\end{glossarytable}

\section{System Method Trade-Off Matrix}
\label{app-system-method-tradeoffs}
\begin{sidewaysfigure}[p]
\centering
\includegraphics[width=\linewidth]{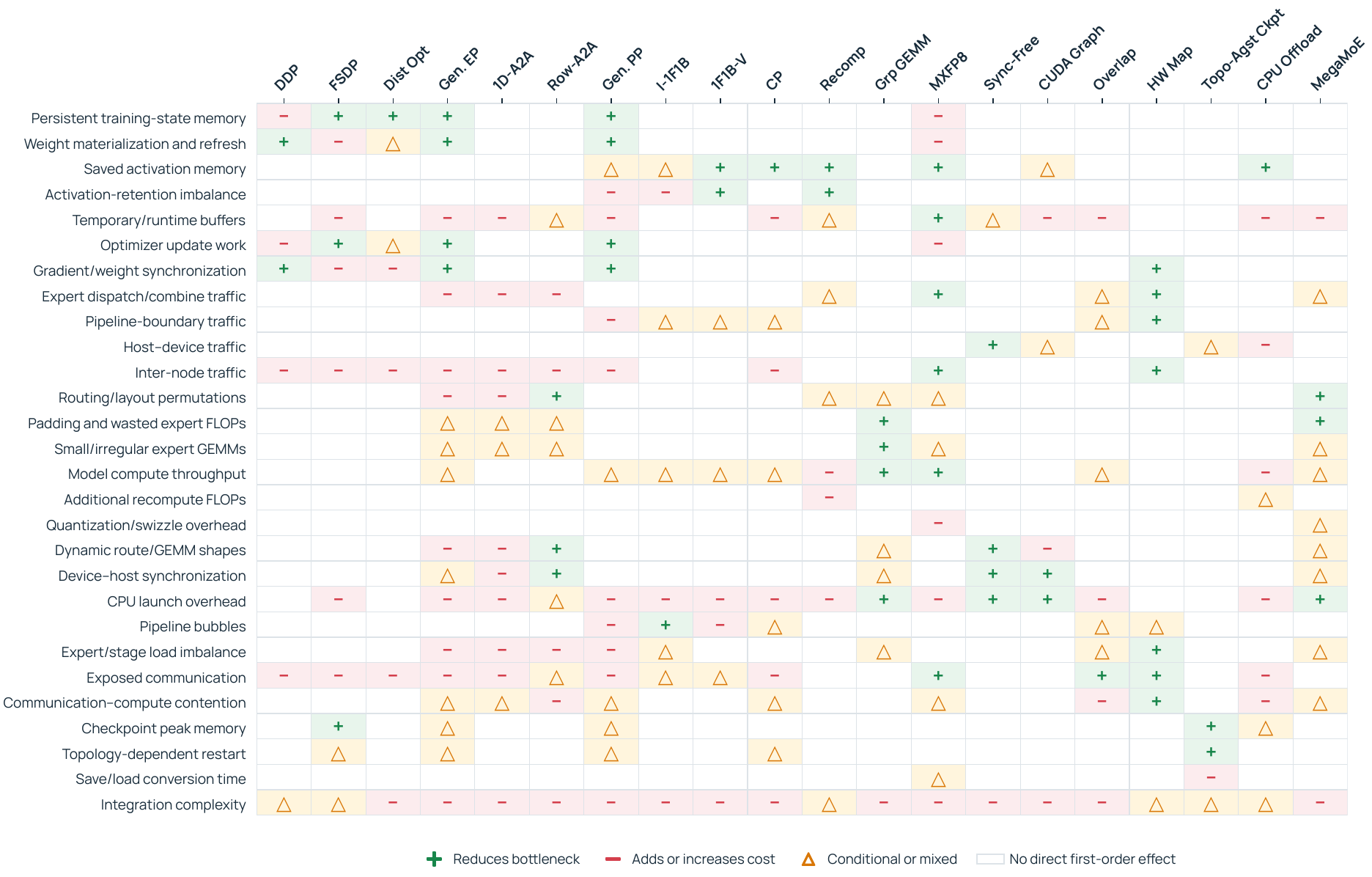}
\caption[System methods and their first-order trade-offs.]{\textbf{First-order
effects of the methods in this report.} General EP and PP columns show the effect of the sharding mechanism;
the adjacent transport and schedule columns show implementation-specific
effects. Signs describe the usual first-order direction for the designs in
this report. Exact outcomes depend on model, batch, topology, hardware, and
implementation; the qualitative signs do not predict end-to-end speedup.}
\label{fig-system-method-tradeoffs}
\end{sidewaysfigure}

The methods in this report solve different parts of the MoE systems problem.
Figure~\ref{fig-system-method-tradeoffs} collects their direct first-order
effects in one place. The matrix separates a general parallelism mechanism
from its implementation or schedule. General Expert Parallelism (Gen. EP), for
example, describes expert sharding and its unavoidable token movement;
one-dimensional all-to-all (1D-A2A) and rowwise all-to-all (Row-A2A) describe
two implementations of that movement. Similarly, general Pipeline Parallelism
(Gen. PP) describes layer sharding, while Interleaved-1F1B (I-1F1B) and
1F1B-V describe two schedules.

The remaining headers abbreviate Distributed Data Parallel (DDP), Fully
Sharded Data Parallel (FSDP), distributed optimizer (Dist Opt), Context
Parallelism (CP), activation recomputation (Recomp), grouped general matrix
multiplication (Grp GEMM), microscaling FP8 (MXFP8), synchronization-free
execution (Sync-Free), CUDA Graph execution (CUDA Graph),
communication--compute overlap (Overlap), hardware-aware topology mapping (HW
Map), topology-agnostic checkpointing (Topo-Agst Ckpt), CPU activation offload
(CPU Offload), and the MegaMoE megakernel prototype (MegaMoE;
Section~\ref{subsec-megamoe-lesson}).

A green plus means that the method directly reduces the named bottleneck. A
red minus means that it introduces or increases that cost. An amber triangle
marks an effect that depends materially on model shape, topology, schedule,
hardware, or implementation. An empty cell means that the method has no direct
first-order effect; it does not mean that the two are independent in a complete
training run. The matrix is a qualitative design guide, not a performance
ranking, and combinations of methods can change the individual effects.

\FloatBarrier

\section{Expert-Merge Normalization in Public MoE Models}
\label{app-expert-merge-practices}

Section~\ref{sec-merge-expert-activations} motivates Olmo's routed/shared
merge from its dense-MLP decomposition and initialization scale. This appendix
places that choice beside public model and framework implementations
(Table~\ref{tab-public-expert-merge-practices}). The
comparison is based on the cited reports, released configurations, and
source code.

For consistency with Equation~\ref{eq-merge-routed-shared}, let $p_i$ denote
selected routed scores when they are renormalized to sum to one, let $\alpha$
be any additional routed scale, so that the routed coefficients are
$r_i=\alpha p_i$, and let $s(x)$ be the shared-expert coefficient. A rule
that instead retains the top-$K$ probabilities $q_i(x)$ of a softmax over all
$N$ experts has token-dependent routed mass
$m_K(x)=\sum_{i\in\mathcal{S}(x)}q_i(x)<1$. These are separate design choices:
normalizing the selected routed scores does not determine either $\alpha$ or
$s(x)$.

\begin{table}[!htb]
  \centering
  \footnotesize
  \renewcommand{\arraystretch}{1.18}
  \setlength{\tabcolsep}{4pt}
  \begin{tabularx}{\linewidth}{>{\raggedright\arraybackslash}p{0.18\linewidth}
      >{\raggedright\arraybackslash}p{0.27\linewidth}
      >{\raggedright\arraybackslash}p{0.20\linewidth} >{\RaggedRight\arraybackslash}X}
    \toprule
    \textbf{Model or implementation} & \textbf{Routed coefficients} &
    \textbf{Shared branch} & \textbf{Interpretation} \\
    \midrule
    Olmo, this report
      & Selected normalization followed by $\alpha=K$; hence
        $\sum_i r_i=K$.
      & Unit gain, $s=1$.
      & Unit-average selected coefficients; matches the additive dense-block
        reference under uniform selected routing. \\
\arrayrulecolor{black!22}\hline\arrayrulecolor{black}

    Megatron-Core default, when shared experts are enabled
      & Post-top-$K$ softmax with $\alpha=1$ by default; $\alpha$ is
        configurable.
      & Unit gain by default; an optional learned sigmoid gate is available.
      & A configurable framework rather than one prescribed model recipe
\citep{nvidia2026megatroncoremoe,nvidia2026megatronlm}. \\
\arrayrulecolor{black!22}\hline\arrayrulecolor{black}

    DeepSeek-V3 released reference
      & Selected sigmoid affinities are renormalized and multiplied by
        $\alpha=2.5$.
      & Unit gain.
      & A model-specific routed scale between one and $K=8$
\citep{deepseekai2024deepseekv3,deepseekai2024deepseekv3code}. \\
\arrayrulecolor{black!22}\hline\arrayrulecolor{black}

    Qwen1.5-MoE and Qwen2-MoE released models
      & Full softmax over all experts followed by top-$K$ without selected
        renormalization; routed mass is $m_K(x)<1$.
      & Learned sigmoid gate.
      & An upcycled recipe in which routed mass and shared gain are both
        token dependent %
\citep{qwenteam2024qwen15moe,qwen2024qwen2,qwenteam2024qwen15moeconfig,qwenteam2024qwen2moeconfig,huggingface2024qwen2moeimplementation}. \\
\arrayrulecolor{black!22}\hline\arrayrulecolor{black}

    Qwen3.5 MoE public implementation
      & Selected top-$K$ probabilities are renormalized to sum to one.
      & Learned sigmoid gate.
      & Unit routed mass with a token-dependent shared contribution
\citep{qwenteam2026qwen35modelcard,huggingface2026qwen35moe}. \\
\arrayrulecolor{black!22}\hline\arrayrulecolor{black}

    Mixtral
      & Softmax over the selected logits; routed mass is one.
      & No shared expert.
      & A routed-only weighted-average convention
\citep{jiang2024mixtral,mistralai2024mistralinference}. \\
\arrayrulecolor{black!22}\hline\arrayrulecolor{black}

    OLMoE-1B-7B
      & Full softmax over 64 experts followed by top-8 without selected
        renormalization; routed mass is below one.
      & No shared expert.
      & The merge rule in this report is not inherited from the earlier OLMoE
        model %
\citep{muennighoff2024olmoe}. \\
\arrayrulecolor{black!22}\hline\arrayrulecolor{black}

    MegaBlocks options
      & Selected route weights may remain unnormalized or receive an
        $L_p$ normalization.
      & Direct addition by default; an option uses routed coefficient
        $K/(K+1)$ and shared coefficient $1/(K+1)$.
      & Exposes normalization as a configurable recipe choice rather than a
        fixed framework identity %
\citep{megablocks2026source}. \\
    \bottomrule
  \end{tabularx}
  \caption{\textbf{Public MoE implementations use different routed and
  shared-expert merge rules.} The table records framework defaults and
  released inference implementations; these sources do not establish which
  merge rule was used in every original pretraining run.}
  \label{tab-public-expert-merge-practices}
\end{table}

Megatron-Core illustrates the distinction between a framework default and a
model recipe particularly clearly. Its current examples use routed scales of
2.5 for DeepSeek-V3, 2.827 for Kimi K2, and either 2.5 or 5.0 for the surveyed
Nemotron 3 variants, rather than deriving one scale from $K$ or expert width.
The examples establish that the mechanism supports model-specific choices;
they are not a controlled ablation showing that one value is preferable
\citep{nvidia2026megatronlm}.

The released DeepSeek-V3 reference requires a similar qualification. Its
inference code applies the routed scale 2.5, while the DeepSeek-V3 report's
routing equations describe normalized selected sigmoid affinities and an additive
shared branch. The original large-scale training implementation is not public,
so the released inference setting does not establish the exact pretraining merge formula.

Across these systems, routed mass may be below one, equal to one, multiplied
by a model-specific constant, or equal to $K$; the shared contribution may be
absent, have unit gain, or use a learned gate. Expert widths and initialization
histories also differ, and Qwen's upcycling changes the relevant starting
point. The initialization-scale rationale in
Section~\ref{sec-merge-expert-activations} applies to Olmo's architecture.
These public implementations do not establish a universal or uniquely
optimal merge normalization.

\section{Context Parallelism and MoE Parallel Folding}
\label{app-context-parallelism}

Long sequences create an activation-memory and attention-computation problem
that Expert Parallelism (EP) does not address. \textbf{Context Parallelism (CP)} divides the
sequence dimension among ranks that process the same training examples. With a
CP degree $C$, each rank enters a Transformer block with approximately $S/C$
of the sequence positions. CP therefore reduces the local size of
sequence-shaped activations, but it does not itself partition parameters or
create additional data-parallel examples. Routed-expert parameter placement remains the
job of EP.

Olmo-core uses the \textbf{Ulysses} form of CP for the MoE path
\citep{jacobs2023deepspeedulysses}. Before attention, each rank holds queries,
keys, and values in a context-sharded layout. An all-to-all operation changes
that layout as follows for batch size $B$, sequence length $S$, head count $H$,
and head width $d_h$:

\begin{equation}
  [B, S/C, H, d_h]
  \quad\longrightarrow\quad
  [B, S, H/C, d_h],
  \label{eq-ulysses-layout-transform}
\end{equation}

\noindent so each rank evaluates the full sequence for a subset of attention
heads, which requires $C$ to divide $H$. A second all-to-all returns the output to the context-sharded layout.
The remaining operations in the Transformer block continue on local token
rows. In particular, the MoE router receives the local
$S/C$ rows directly; the model does not gather the full sequence before
routing.

CP ranks are not independent data-parallel replicas because they contribute
different sequence slices of the same examples. This distinction matters when
CP is combined with EP. Let $D$ and $C$ be the dense Data Parallelism (DP) and
CP degrees within one pipeline stage, and let $D_E$ and $M_E$ be the
expert-replica and expert-sharding degrees. Olmo-core forms two
views of the same per-stage rank set:

\begin{equation}
  \mathcal{G}_{\mathrm{stage}}
    = \mathcal{A}_{\mathrm{DP}} \times \mathcal{A}_{\mathrm{CP}}
    = \mathcal{A}_{\mathrm{EP\text{-}DP}}
      \times \mathcal{A}_{\mathrm{EP\text{-}MP}},
  \qquad
  DC = D_E M_E.
  \label{eq-cp-ep-parallel-folding}
\end{equation}

\noindent The dense view keeps CP visible to attention and input preparation.
Dense gradients reduce over the combined DP$\times$CP group. The MoE view
instead uses EP-MP to partition routed experts and EP-DP to synchronize copies
of the same expert shard. Dispatch and combine operate over the folded expert
groups, so routed tokens from all DP and CP coordinates can reach the ranks
that store their selected experts.

This construction is a focused application of \textbf{MoE parallel folding},
which allows dense attention and sparse expert layers to use different
parallel mappings on the same devices
\citep{liu2025moeparallelfolding}. For example, four ranks can form a dense
mesh with $D=2$ and $C=2$, then be viewed by the MoE layers as $D_E=1$ and
$M_E=4$. Enabling CP${}=2$ and EP-MP${}=4$ therefore still uses four ranks within
the stage, rather than allocating an independent four-rank expert group for
each CP coordinate. More generally, parallel folding prevents CP and EP from
being treated as orthogonal multipliers when their layer-local work can share
the same ranks.

Pipeline Parallelism (PP) adds an outer partition over layers. Within each PP
stage, inputs and labels are sequence-sharded before the pipeline schedule
starts, while the original sequence length is retained for positional
encoding. Later stages receive already-sharded activations and do not apply CP
a second time. Equation~\ref{eq-cp-ep-parallel-folding} then holds
independently inside every pipeline stage. Within this layout, CP assigns
sequence rows, EP assigns routed experts, and PP assigns layers to ranks.

CP lowers the per-rank footprint of many
sequence-shaped activations and makes longer contexts feasible, while Ulysses
adds attention-layout all-to-alls. Parallel folding does not remove that
communication; it prevents CP from unnecessarily multiplying the device count
required by EP and gives attention and routed experts the process groups that
match their respective computation.

\section{Author Contributions}
\label{app:author-contributions}

\textbf{Tianhua Tao} led the project and carried out its core research and
development: conceptualization, system design and implementation, experimental
evaluation, profiling, analysis, visualization, and manuscript preparation.

\textbf{Akshita Bhagia} contributed broadly to the Olmo-core software,
including but not limited to MoE code refactoring and integration, software
maintenance, and testing.

\textbf{Dirk Groeneveld, Pete Walsh, and Tyler Romero} made foundational
contributions to Olmo-core and early MoE prototyping, establishing the software
foundation on which the Olmo-core DDP system described in this report builds.

\textbf{Yashas Samaga and Jacob Morrison} participated in technical
discussions on MoE integration and downstream use. Their experience with the
software informed feedback on design, training workflows, and usability.

\textbf{Iz Beltagy} provided team leadership, project direction, and
strategic coordination.

\textbf{Taira Anderson} provided technical program management, including
compute-resource allocation, scheduling, project administration, and
coordination.

\textbf{Noah A. Smith and Hannaneh Hajishirzi} provided strategic guidance and
research advising. Hannaneh Hajishirzi also served as the project's primary
academic advisor.

\subsection{Use of AI Tools}
\label{app:ai-assistance}

We used AI tools to assist with developing the codebase and GPU kernels,
drafting and revising this report, managing experiments, and creating
visualizations. Building and documenting the training stack required repeated
iteration across distributed systems code, kernel behavior, experiment
orchestration, and technical explanation. We used AI assistance to support
that iteration, explore implementation alternatives, and reduce repetitive
coding and editorial work. This was a practical part of a human-led research
process: the research direction, design decisions, and interpretation of
results remained the responsibility of the authors.

AI-generated suggestions were assessed through code inspection, testing,
profiling, experimental comparisons, and source checks, as appropriate to the
task. We also kept a citation audit that records source evidence and review
notes for individual cited claims. These checks can themselves miss errors;
AI assistance does not transfer responsibility for the work. The authors
remain responsible for the methods, results, and statements in this report.

\clearpage
\renewcommand{\bibfont}{\small\raggedright}
\bibliographystyle{ai2style/plainnat}
\phantomsection
\addcontentsline{toc}{section}{References}
\bibliography{references}

\end{document}